\documentstyle[titlepage,preprint,aps,floats,epsfig]{revtex}
\newcommand{\beq}{\begin{equation}}
\newcommand{\eeq}{\end{equation}}
\newcommand{\eq}[1]{eq.(\ref{#1})}
\newcommand{\crossed}[1]{#1\hspace{-.5em}\slash}
\begin{document}
\draft
\tighten
\preprint{PSU/TH/226; hep-ph/0002158}
\title {Theory of Light Hydrogenlike Atoms}
\medskip
\author {Michael I. Eides \thanks{E-mail address:
eides@phys.psu.edu, eides@thd.pnpi.spb.ru}}
\address{ Department of Physics, Pennsylvania
State University,
University Park, PA 16802, USA\\
and
Petersburg Nuclear Physics Institute,
Gatchina, St.Petersburg 188350, Russia}
\author{Howard Grotch\thanks{E-mail address: asdean@pop.uky.edu}}
\address{College of Arts and Sciences, University of Kentucky,
Lexington, KY 40506, USA}
\author{Valery A. Shelyuto \thanks{E-mail address:
shelyuto@fn.csa.ru}}
\address{D. I.  Mendeleev Institute of Metrology,
St.Petersburg 198005, Russia}

\maketitle

\begin{abstract}
The present status and recent  developments in the theory of light
hydrogenic atoms, electronic  and muonic, are extensively reviewed. The
discussion is based on the quantum field theoretical approach to
loosely bound composite systems. The basics of the quantum field
theoretical approach, which provide the framework needed for a
systematic derivation of all higher order corrections to the energy
levels, are briefly discussed. The main physical ideas behind the
derivation of all binding, recoil, radiative, radiative-recoil, and
nonelectromagnetic spin-dependent and spin-independent corrections to
energy levels of hydrogenic atoms are discussed and, wherever possible,
the fundamental elements of the derivations of these corrections are
provided.  The emphasis is on new theoretical results which were not
available in earlier reviews. An up-to-date set of all theoretical
contributions to the energy levels is contained in the paper.  The
status of modern theory is tested by comparing the theoretical results
for the energy levels with the most precise experimental results for
the Lamb shifts and gross structure intervals in hydrogen, deuterium,
and helium ion $He^+$, and with the experimental data on the hyperfine
splitting in muonium, hydrogen and deuterium.  \end{abstract}


\newpage
\tableofcontents

\newpage

\part{Introduction}

Light one-electron atoms are a classical subject of quantum physics.
The very discovery and further progress of quantum mechanics is
intimately connected to the explanation of the main features of
hydrogen energy levels. Each step in development of quantum
physics led to a better understanding of the bound state physics. Bohr
quantization rules of the old quantum theory were created in order to
explain the existence of the stable discrete energy levels. The
nonrelativistic quantum mechanics of Heisenberg and Schr\"odinger
provided a self-consistent scheme for description of bound states. The
relativistic spin one half Dirac equation quantitatively described the
main experimental features of the hydrogen spectrum.  Discovery of the
Lamb shift \cite{lr}, a subtle discrepancy between the predictions of
the Dirac equation and the experimental data, triggered development of
modern relativistic quantum electrodynamics, and subsequently the
Standard Model of modern physics.

Despite its long and rich history the theory of atomic bound states
is still very much alive today. New importance to the bound state
physics was given by the development of quantum chromodynamics, the
modern theory of strong interactions. It was realized that all hadrons,
once thought to be the elementary building blocks of matter, are
themselves atom-like bound states of elementary quarks bound by the
color forces. Hence, from a modern point of view, the theory of atomic
bound states could be considered as a theoretical laboratory and
testing ground for exploration of the subtle properties of the bound
state physics, free from further complications connected with the
nonperturbative effects of quantum chromodynamics which play an
especially important role in the case of light hadrons. The quantum
electrodynamics and quantum chromodynamics bound state theories are so
intimately intertwined today that one often finds theoretical research
where new results are obtained simultaneously, say for positronium and
also heavy quarkonium.

The other powerful stimulus for further development of the bound state
theory is provided by the spectacular experimental progress in precise
measurements of atomic energy levels. It suffices to mention
that the relative uncertainty of measurement of the frequency of the
$1S-2S$ transition in hydrogen was reduced during the last decade by
three orders of magnitude from $3\cdot10^{-10}$ \cite{boshier} to
$3.4\cdot10^{-13}$ \cite{udem}. The relative uncertainty in measurement
of the muonium hyperfine splitting was reduced recently by the factor
$3$ from $3.6\cdot10^{-8}$ \cite{mbb} to $1.2\cdot10^{-8}$ \cite{lbdd}.

This experimental development was matched in recent years by rapid
theoretical progress, and we feel that now is a good time to review
bound state theory. The theory of hydrogenic bound states is widely
described in the literature. The basics of nonrelativistic theory is
contained in any textbook on quantum mechanics, and the relativistic
Dirac equation and the Lamb shift are discussed in any textbook on
quantum electrodynamics and quantum field theory. An excellent source
for the early results is the classic book by Bethe and Salpeter
\cite{bs}. The last comprehensive review of the theory \cite{sy} was
published more than ten years ago. A number of reviews were published
recently which contain new theoretical results
\cite{grotch94,dvoegl,eid,kin96,sapir96,mohr96,plwhkh,kin98}.  However,
a coherent discussion of the modern status of the theory, to the best
of our knowledge, is missing in the literature, and we will try to
provide this in the current paper.

\begin{figure}[ht]
\centerline{\epsfig{file=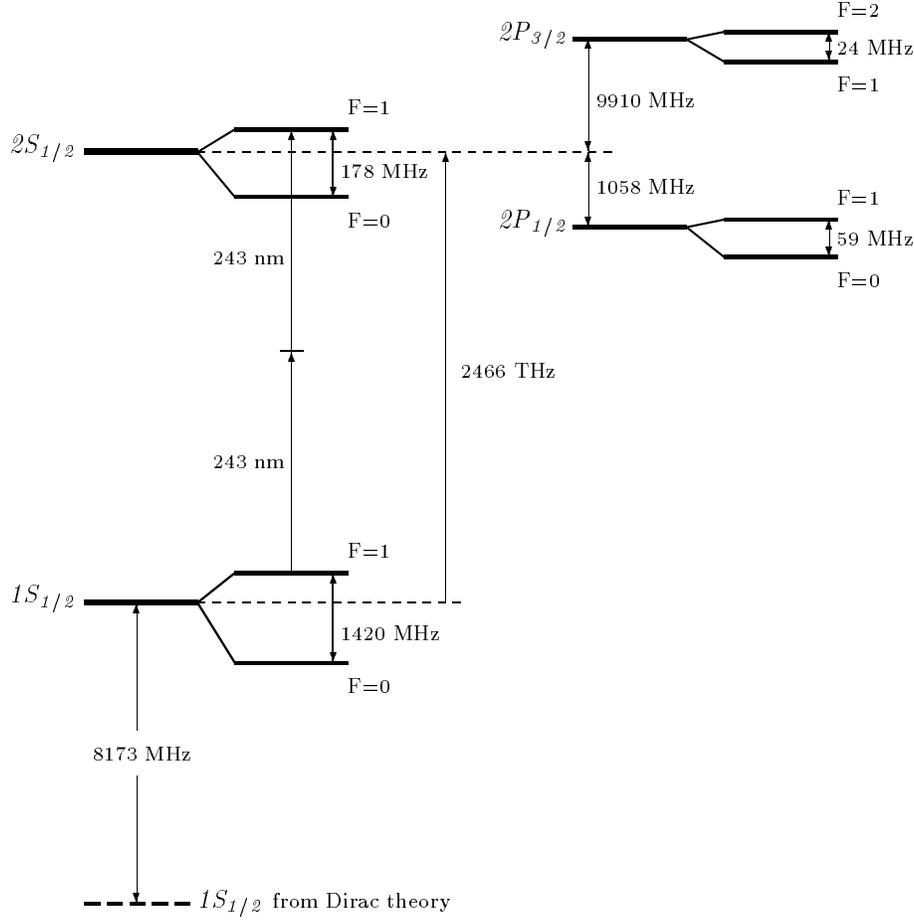}}
\vspace{0.5cm}
\caption{Hydrogen energy levels}
\label{hydr}
\end{figure}

\begin{figure}[ht]
\centerline{\epsfig{file=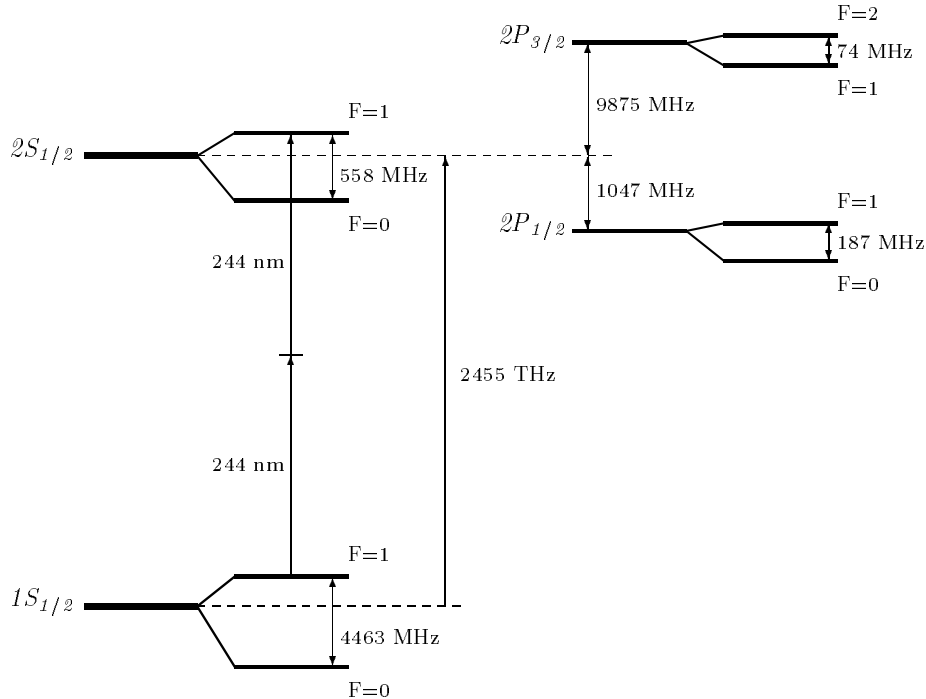}}
\vspace{0.5cm}
\caption{Muonium energy levels}
\label{muonle}
\end{figure}

Our goal here is to present a state of the art discussion of the
theory of the Lamb shift and hyperfine  splitting in light hydrogenlike
atoms. In the body of the paper the spin independent corrections are
discussed mainly as corrections to the hydrogen energy levels (see
Fig.\ \ref{hydr}), and the theory of hyperfine splitting is discussed
in the context of the hyperfine splitting in the ground state of
muonium (see Fig.\ \ref{muonle}). These two simple atomic systems are
singled out for practical reasons, because highly precise experimental
data exists in both cases, and the most accurate theoretical results
are also obtained for these cases. However, almost all formulae in
this review are valid also for other light hydrogenlike systems, and
some of these other applications, including muonic atoms, will be
discussed in the text as well. We will present all theoretical results
in the field, with emphasis on more recent results which either were
not discussed in sufficient detail in the previous theoretical reviews
\cite{bs,sy}, or simply did not exist when the reviews were written.
Our emphasis on the theory means that, besides presenting an exhaustive
compendium of theoretical results, we will also try to present a
qualitative discussion of the origin and magnitude of different
corrections to the energy levels, to give, when possible,
semiquantitative estimates of expected magnitudes, and to describe the
main steps of the theoretical calculations and the new effective
methods which were developed in recent years. We will not attempt to
present a detailed comparison of theory with the latest experimental
results, leaving this task to the experimentalists. We will use the
experimental results only for illustrative purposes.

The paper consists of three main parts. In the introductory part
we briefly remind the reader of the main characteristic features of the
bound state physics. Then follows a detailed discussion of the
corrections to the energy levels which do not depend on the nuclear
spin. The last third of the paper is devoted to a systematic discussion
of the physics of hyperfine splitting. Different corrections to the
energy levels are ordered with respect to the natural small parameters
$\alpha$, $Z\alpha$, $m/M$ and nonelectrodynamic parameters like the
ratio of the nucleon size to the radius of the first Bohr orbit. These
parameters have a transparent physical nature in the light hydrogenlike
atoms. Powers of $\alpha$ describe the order of quantum electrodynamic
corrections to the energy levels, parameter $Z\alpha$ describes the
order of relativistic corrections to the energy levels, and the small
mass ratio of the light and heavy particles is responsible for the
recoil effects beyond the reduced mass parameter present in a
relativistic bound state\footnote{We will return to a more detailed
discussion of the role of different small parameters below.}.
Corrections which depend both on the quantum electrodynamic parameter
$\alpha$ and the relativistic parameter $Z\alpha$ are ordered in a
series over $\alpha$ at fixed power of $Z\alpha$, contrary to the
common practice accepted in the physics of highly charged ions with large
$Z$. This ordering is more natural from the point of view of the
nonrelativistic bound state physics, since all radiative corrections to a
contribution of a definite order in the nonrelativistic expansion originate
from the same distances and describe the same physics, while the radiative
corrections to the different terms in nonrelativistic expansion over
$Z\alpha$ of the same order in $\alpha$ are generated at vastly
different distances and could have drastically different magnitudes.

A few remarks about our notation. All formulae below are written for
the energy shifts. However, not energies but frequencies are measured
in the spectroscopic experiments. The formulae for the energy shifts
are converted to the respective expressions for the frequencies with
the help of the De Broglie relationship $E=h\nu$. We will ignore the
difference between the energy and frequency units in our theoretical
discussion. Comparison of the theoretical expressions with the
experimental data will always be done in the frequency units, since
transition to the energy units leads to loss of accuracy.  All numerous
contributions to the energy levels in different sections of this paper
are generically called $\Delta E$ and as a rule do not carry any
specific labels, but it is understood that they are all different.

Let us mention briefly some of the closely related subjects which are
not considered in this review. The physics of the high $Z$ ions is
nowadays a vast and well developed field of research, with its own
problems, approaches and tools, which in many respects are quite
different from the physics of low $Z$ systems.  We discuss below the
numerical results obtained in the high $Z$ calculations only when they
have a direct relevance for the low $Z$ atoms.  The reader can find a
detailed discussion of the high $Z$ physics in a number of recent
reviews (see, e.g., \cite{mps}).  In trying to preserve a reasonable
size of this review we decided to omit discussion of positronium, even
though many theoretical expressions below are written in such form
that for the case of equal masses they turn into respective corrections
for the positronium energy levels. Positronium is qualitatively
different from hydrogen and muonium not only due to the equality of
the masses of its constituents, but because unlike the other light
atoms there exists a whole new class of corrections to the positronium
energy levels generated by the annihilation channel which is absent in
other cases. Our discussion of the new theoretical methods will be
incomplete due to omission of the recently developed and
now popular nonrelativistic QED (NRQED) \cite{caswelllep} which
was especially useful in the positronium calculations, but was
rarely used in the hydrogen and muonium physics. Very lucid
presentations of NRQED exist in the recent literature (see, e.g.,
\cite{kn10}).

\part{Theoretical Approaches to the Energy Levels of Loosely Bound Systems}

\section{Nonrelativistic Electron in the Coulomb Field}

In the first approximation, energy levels of one-electron atoms are
described by the solutions of the Schr\"odinger equation for an electron
in the field of an infinitely heavy Coulomb center with charge $Z$ in terms
of the proton charge\footnote{We are using the system of units  where
$\hbar=c=1$.}

\beq        \label{schreqcoulinf}
(-\frac{\Delta}{2m}-\frac{Z\alpha}{r})\psi({\bf r})=E_n \psi({\bf r})
\eeq
\[
\psi_{nlm}({\bf r})=R_{nl}(r)Y_{lm}(\frac{\bf r}{r})
\]
\[
E_n=-\frac{m(Z\alpha)^2}{2n^2}, \qquad n=1,2,3\ldots,
\]

\noindent
where $n$ is called the principal quantum number. Besides the principal
quantum number $n$ each state is described by the value of angular momentum
$l=0,1,\ldots,n-1$, and projection of the orbital angular momentum
$m=0,\pm1,\ldots,\pm l$.  In the nonrelativistic Coulomb problem all
states with different orbital angular momentum but the same principal
quantum number $n$ have the same energy, and the energy levels of the
Schr\"odinger equation in the Coulomb field are $n$-fold degenerate
with respect to the total angular momentum quantum number. As in any
spherically symmetric problem, the energy levels in the Coulomb field
do not depend on the projection of the orbital angular momentum on an
arbitrary axis, and each energy level with given $l$ is additionally
$2l+1$-fold degenerate.

Straightforward calculation of the characteristic values of the
velocity, Coulomb potential and kinetic energy in the stationary states
gives

\beq
<n|{\bbox v}^2|n>=<n|\frac{{\bbox p}^2}{m^2}|n>=\frac{(Z\alpha)^2}{n^2},
\eeq
\[
<n|\frac{Z\alpha}{r}|n>
=\frac{m(Z\alpha)^2}{n^2},
\]
\[
<n|\frac{{\bbox p}^2}{2m}|n>=\frac{m(Z\alpha)^2}{2n^2}.
\]

We see that due to the smallness of the fine structure constant
$\alpha$  a one-electron atom is a loosely bound nonrelativistic
system\footnote{We are interested in low-$Z$ atoms in this paper.  High-$Z$
atoms cannot be treated as nonrelativistic systems, since an expansion in
$Z\alpha$ is problematic.} and all relativistic effects may be treated as
perturbations. There are three characteristic scales in the
atom. The smallest is determined by the binding energy $\sim
m(Z\alpha)^2$, the next is determined by the characteristic electron
momenta $\sim mZ\alpha$, and the last one is of order of the electron
mass $m$.

Even in the framework of nonrelativistic quantum mechanics one can
achieve a much better description of the hydrogen spectrum by taking into
account the finite mass of the Coulomb center. Due to the nonrelativistic
nature of the bound system under consideration, finiteness of the nucleus
mass leads to substitution of the reduced mass instead of the electron mass
in the formulae above. The finiteness of the nucleus mass introduces the
largest energy scale in the bound system problem - the heavy particle mass.

\section{Dirac Electron in the Coulomb Field }\label{diraccoul}

The relativistic dependence of the energy of a free classical particle on
its momentum is described by the relativistic square root

\beq      \label{einsqroot}
\sqrt{{\bf p}^2+m^2}\approx m+\frac{{\bf p}^2}{2m}-\frac{{\bf
p}^4}{8m^3}+\ldots.
\eeq

The kinetic energy operator in the Schr\"odinger equation corresponds to the
quadratic term in  this nonrelativistic expansion, and thus the
Schr\"odinger equation  describes only the leading nonrelativistic
approximation to the hydrogen energy levels.

The classical nonrelativistic expansion goes over ${\bf p}^2/m^2$. In the
case of the loosely bound electron, the expansion in ${\bf p}^2/m^2$
corresponds to expansion in $(Z\alpha)^2$; hence, relativistic corrections
are given by the expansion over {\it even} powers of $Z\alpha$. As we have
seen above, from the explicit expressions for the energy levels in the
Coulomb field the same parameter $Z\alpha$ also characterizes the
binding energy.  For this reason, parameter $Z\alpha$ is also often
called the binding parameter, and the relativistic corrections carry
the second name of binding corrections.

Note that the series expansion for the relativistic corrections in the bound
state problem goes literally over the  binding parameter $Z\alpha$, unlike
the case of the scattering problem in QED, where the expansion parameter
always contains an additional factor $\pi$ in the denominator and the
expansion typically goes over $\alpha/\pi$. This absence of the extra factor
$\pi$ in the denominator of the expansion parameter is a typical feature of
the Coulomb problem. As we will see below, in the combined expansions over
$\alpha$ and $Z\alpha$, expansion over $\alpha$  at fixed power of the
binding parameter $Z\alpha$ always goes over $\alpha/\pi$, as in the case of
scattering. Loosely speaking one could call successive terms in the series
over $Z\alpha$ the relativistic corrections, and successive terms in the
expansion over $\alpha/\pi$ the loop or radiative corrections.

For the bound electron, calculation of the relativistic corrections
should also take into account the contributions due to its spin
one half. Account for the spin one half does not change the fundamental
fact that all relativistic (binding) corrections are described by the
expansion in even powers of $Z\alpha$, as in the naive expansion of the
classical relativistic square root in \eq{einsqroot}. Only the
coefficients in this expansion change due to presence of spin. A proper
description of all relativistic corrections to the energy levels is
given by the Dirac equation with a Coulomb source. All relativistic
corrections may easily be obtained from the exact solution of the Dirac
equation in the external Coulomb field (see, e.g., \cite{bd,blp})

\beq \label{naivdirac}
E_{nj}=m f(n,j),
\eeq

\noindent
where

\beq  \label{fnjdef}
f(n,j)=\left[1+\frac{(Z\alpha)^2}{\left(\sqrt{(j+\frac{1}{2})^2
-(Z\alpha)^2}+n-j-\frac{1}{2}\right)^2}\right]^{-\frac{1}{2}}
\eeq
\[
\approx 1-\frac{(Z\alpha)^2}{2n^2}-\frac{(Z\alpha)^4}{2n^3}
(\frac{1}{j+\frac{1}{2}}-\frac{3}{4n})
-\frac{(Z\alpha)^6}{8n^3}\left[\frac{1}{(j+\frac{1}{2})^3}
+\frac{3}{n(j+\frac{1}{2})^2}+\frac{5}{2n^3}-\frac{6}{n^2(j+\frac{1}{2})}
\right]
+\ldots,
\]

\noindent
and $j=1/2,3/2,\ldots,n-1/2$ is the total angular momentum of the state.

In the Dirac spectrum, energy levels with the same principal quantum
number $n$ but different total angular momentum $j$ are split into $n$
components of the fine structure, unlike the nonrelativistic Schr\"odinger
spectrum where all levels with the same $n$ are degenerate. However,
not all degeneracy is lifted in the spectrum of the Dirac equation:
the energy levels corresponding to the same $n$ and $j$ but different
$l=j\pm1/2$ remain doubly degenerate. This degeneracy is lifted by the
corrections connected with the finite size of the Coulomb source, recoil
contributions, and by the dominating QED loop contributions. The respective
energy shifts are called the Lamb shifts (see exact definition in Section
\ref{leadreclambdef}) and will be one of the main subjects of discussion
below. We would like to emphasize that the quantum mechanical (recoil and
finite nuclear size) effects alone do not predict anything of the scale of
the experimentally observed Lamb shift which is thus essentially a quantum
electrodynamic (field-theoretical) effect.

One trivial improvement of the Dirac formula for the energy levels may
easily be achieved if we take into account that, as was already discussed
above, the electron motion in the Coulomb field is essentially
nonrelativistic, and, hence, all contributions to the binding energy
should contain as a factor the reduced mass of the electron-nucleus
nonrelativistic system rather than the electron mass. Below we will
consider the expression with the reduced mass factor

\beq \label{dirac}
E_{nj}=m+m_r [f(n,j)-1],
\eeq

\noindent
rather than the naive  expression in \eq{naivdirac}, as a starting
point for calculation of corrections to the electron energy levels.
In order to provide a solid starting point for further
calculations the Dirac spectrum with the reduced mass dependence in
\eq{dirac} should be itself derived from QED (see Section
\ref{leadreclambdef} below), and not simply postulated on physical
grounds as is done here.

\section{Bethe-Salpeter Equation and the Effective Dirac
Equation}\label{effdir}

\begin{figure}[h]
\centerline{\epsfig{file=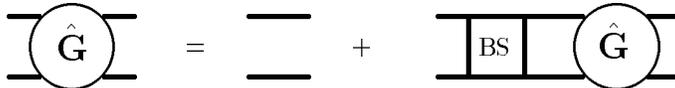}}
\vspace{0.5cm}
\caption{Bethe-Salpeter equation}
\label{bs}
\end{figure}

\begin{figure}
\centerline{\epsfig{file=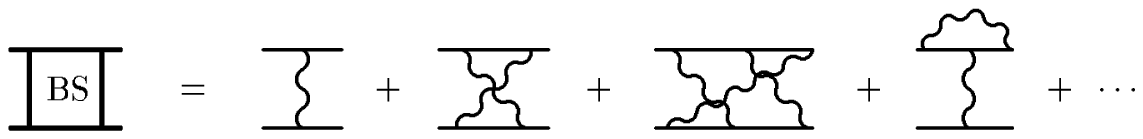}}
\vspace{0.5cm}
\caption{Kernel of the Bethe-Salpeter equation}
\label{bskern}
\end{figure}

Quantum field theory provides an unambiguous way to find energy levels
of any composite system. They are determined by the positions of the poles
of the respective Green functions. This idea was first realized in the form
of the Bethe-Salpeter (BS) equation for the two-particle Green function (see
Fig.\ \ref{bs})  \cite{bethesalp}

\beq \label{bethesalp}
\hat G=S_0+S_0K_{BS}\hat G,
\eeq

\noindent
where $S_0$ is a free two-particle Green function, the kernel $K_{BS}$ is a
sum of all two-particle irreducible diagrams in Fig.\ \ref{bskern},  and
$\hat G$ is the total two-particle Green function.

At first glance the field-theoretical BS equation has nothing in common
with the quantum mechanical Schr\"odinger and Dirac equations discussed
above.  However, it is not too difficult to demonstrate that with selection
of a certain subset of interaction kernels (ladder and crossed ladder),
followed by some natural approximations, the BS eigenvalue equation reduces
in the leading approximation, in the case of one light and one heavy
constituent, to the Schr\"odinger or Dirac eigenvalue equations for a
light particle in a field of a heavy Coulomb center. The basics of the
BS equation are described in many textbooks (see, e.g.,
\cite{blp,itzzub,fgross}), and many important results were obtained in
the BS framework.

However, calculations beyond the leading order in the original BS framework
tend to be rather complicated and nontransparent. The reasons for these
complications can be traced to the dependence of the BS wave function on the
unphysical relative energy (or relative time), absence of the exact solution
in the zero-order approximation, non-reducibility of the ladder
approximation to the Dirac equation, when the mass of the heavy particle
goes to infinity, etc. These difficulties are generated not only by the
nonpotential nature of the bound state problem in quantum field
theory, but also by the unphysical classification of diagrams with the help
of the notion of two-body reducibility. As it was known from the very
beginning \cite{bethesalp} there is a tendency to cancellation between
the contributions of the ladder graphs and the graphs with crossed photons.
However, in the original BS framework, these graphs are treated in
profoundly different ways. It is quite natural, therefore, to seek such
a modification of the BS equation, that the crossed and ladder graphs
play a more symmetrical role. One also would like to get rid of other
drawbacks of the original BS formulation, preserving nevertheless its
rigorous field-theoretical contents.

The BS equation allows a wide range of modifications since one can freely
modify both the zero-order propagation function and the leading order
kernel, as long as these modifications are consistently taken into account
in the rules for construction of the higher order approximations, the latter
being consistent with \eq{bethesalp} for the two-particle Green function. A
number of variants of the original BS equation were developed since its
discovery (see, e.g.,\cite{gy1,gy2,fgr,dulf,lep}). The guiding principle in
almost all these approaches was to restructure  the BS equation in such a
way, that it would acquire a three-dimensional form, a soluble and
physically natural leading order approximation in the form of the
Schr\"odinger or Dirac equations, and more or less transparent and regular
way for selection of the kernels relevant for calculation of the corrections
of any required order.

We will describe, in some detail, one such modification, an
effective Dirac equation (EDE) which was derived in a number of papers
\cite{gy2,fgr,dulf,lep}. This new equation is more convenient in many
applications  than the original BS equation, and we will derive  some
general formulae connected with this equation. The physical idea behind this
approach is that in the case of a loosely bound system of two particles of
different masses, the heavy particle spends almost all its life not far from
its own mass shell. In such case some kind of Dirac equation for the
light particle in an external Coulomb field should be an excellent starting
point for the perturbation theory expansion. Then it is convenient to choose
the free two-particle propagator in the form of the product of the heavy
particle mass shell projector $\Lambda$ and the free electron propagator

\beq                \label{freeprop}
\Lambda S(p,l,E)=2\pi i\delta^{(+)}(p^2-M^2)\frac{\crossed p+M}{E-\crossed
p-m}(2\pi)^4\delta^{(4)}(p-l)
\eeq

\noindent
where $p_\mu$ and $l_\mu$ are the momenta of the incoming and outgoing
heavy particle, $E_\mu-p_\mu$ is the momentum of the incoming electron
($E=(E,{\bf 0})$ - this is the choice of the reference frame), and
$\gamma$-matrices associated with the light and heavy particles act only
on the indices of the respective particle.

The free propagator in \eq{freeprop} determines other building blocks and
the form of a two-body equation equivalent to the BS equation, and the
regular perturbation theory formulae in this case were obtained in
\cite{dulf,lep}.

In order to derive these formulae let us first write the BS equation in
\eq{bethesalp} in an explicit form

\beq
\hat G(p,l,E)=S_0(p,l,E)+\int\frac{d^4k}{(2\pi)^4}
\int\frac{d^4q}{(2\pi)^4}S_0(p,k,E)K_{BS}(k,q,E)\hat G(q,l,E),
\eeq

\noindent
where

\beq
S_0(p,k,E)=\frac{i}{\crossed p-M}\frac{i}{E-\crossed
l-m}(2\pi)^4\delta^{(4)}(p-l).
\eeq

The amputated two-particle Green function $G_T$ satisfies the equation

\beq          \label{amputgreen}
G_T=K_{BS}+K_{BS}S_0G_T,
\eeq

A new kernel corresponding to the free two-particle propagator in
\eq{freeprop} may be defined via this amputated two-particle Green function

\beq      \label{newkern}
G_T=K+K\Lambda SG_T.
\eeq

\begin{figure}[ht]
\centerline{\epsfig{file=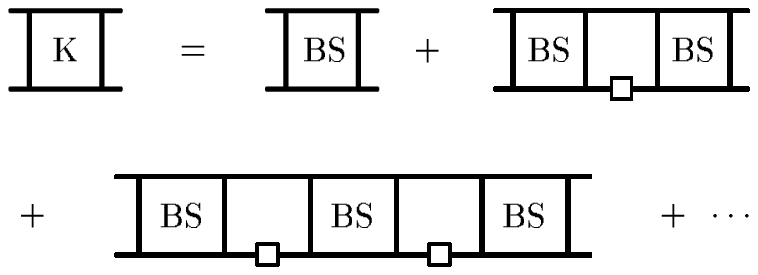}}
\vspace{0.5cm}
\caption{Series for the kernel of the Effective Dirac equation}
\label{kernedefig}
\end{figure}

Comparing \eq{amputgreen} and \eq{newkern} one easily obtains the
diagrammatic series for the new kernel $K$ (see Fig.\ \ref{kernedefig})

\beq
K(q,l,E)=[I-K_{BS}(S_0-\Lambda S)]^{-1}K_{BS}=K_{BS}(q,l,E)+
\eeq
\[
\int\frac{d^4r}{(2\pi)^4}
K_{BS}(q,r,E)\left\{\frac{i}{\crossed r-M}\frac{i}{E-\crossed r-m}
-2\pi i\delta^{(+)}(r^2-M^2)\frac{\crossed r+M}{E-\crossed r-m}\right\}
K_{BS}(r,l,E)+\ldots.
\]

The new bound state equation is constructed for the two-particle Green
function defined by the relationship

\beq
G=\Lambda S+\Lambda SG_T\Lambda S.
\eeq

The two-particle Green function $G$ has the same poles as the initial Green
function $\hat G$ and satisfies the BS-like equation

\beq
G=\Lambda S+\Lambda SKG,
\eeq

\noindent
or, explicitly,

\beq
G(p,l,E)=2\pi i\delta^{(+)}(p^2-M^2)\frac{\crossed p+M}{E-\crossed
p-m}(2\pi)^4\delta^{(4)}(p-l)
\eeq
\[
+2\pi i\delta^{(+)}(p^2-M^2)\frac{\crossed p+M}{E-\crossed
p-m}\int\frac{d^4q}{(2\pi)^4}
K(p,q,E)G(q,l,E).
\]

This last equation is completely equivalent to the original BS equation,
and may be easily written in a three-dimensional form

\beq  \label{nonhom3dim}
\widetilde G({\bf p},{\bf l},E)=\frac{\crossed p+M}{E-\crossed p-m}
\left\{(2\pi)^3\delta^{(3)}({\bf p-l}) +\int\frac{d^3q}{(2\pi)^32E_q}
iK({\bf p},{\bf q},E)\widetilde G({\bf q},{\bf l},E)\right\},
\eeq

\noindent
where all four-momenta are on the mass shell $p^2=l^2=q^2=M^2$,
$E_q=\sqrt{{\bf p}^2+M^2}$, and the three-dimensional two-particle Green
function $\widetilde G$ is defined as follows

\beq
G(p,l,E)=2\pi i\delta^{(+)}(p^2-M^2)\widetilde G({\bf p},{\bf l},E)
2\pi i\delta^{(+)}(l^2-M^2).
\eeq

Taking the residue at the bound state pole with energy $E_n$ we obtain a
homogeneous equation

\beq  \label{sixtcompdir}
(\crossed E_n-\crossed p-m)\phi({\bf p},E_n)=(\crossed p+M)
\int\frac{d^3q}{(2\pi)^32E_q}
iK({\bf p},{\bf q},E_n)\phi({\bf q},E_n)
\eeq

Due to the presence of the heavy particle mass shell projector on the
right hand side the wave function in \eq{sixtcompdir} satisfies a free Dirac
equation with respect to the heavy particle indices

\beq
(\crossed p-M)\phi({\bf p},E_n)=0.
\eeq

Then one can extract a free heavy particle spinor from the wave function in
\eq{sixtcompdir}

\beq
\phi({\bf p},E_n)=\sqrt{2E_n}U({\bf p})\psi({\bf p},E_n)
\eeq

\noindent
where

\beq
U({\bf p})=\left(
\begin{array}{rc}
\sqrt{E_p+M}&I\\
\sqrt{E_p-M}&\frac{\boldmath\mbox{$p\cdot\sigma$}}{|{\boldmath\mbox{$ p$}}|}
\end{array}
\right).
\eeq

\begin{figure}[h]
\centerline{\epsfig{file=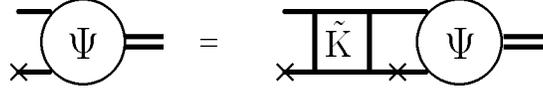}}
\vspace{0.5cm}
\caption{Effective Dirac equation}
\label{edefig}
\end{figure}

Finally, the eight-component wave function $\psi({\bf p},E_n)$ (four
ordinary electron spinor indices, and two extra indices corresponding
to the two-component spinor of the heavy particle) satisfies the
effective Dirac equation (see Fig.\ \ref{edefig})

\beq  \label{ede}
(\crossed E_n-\crossed p-m)\psi({\bf p},E_n)=\int\frac{d^3q}{(2\pi)^32E_q}
i\widetilde K({\bf p},{\bf q},E_n)\psi({\bf q},E_n),
\eeq

\noindent
where

\beq
\widetilde K({\bf p},{\bf q},E_n)=\frac{\bar U({\bf p})K({\bf p},{\bf
q},E_n)U({\bf q})}{\sqrt{4E_pE_q}},
\eeq

\noindent
$k=(E_n-p_0,-{\bf p})$ is the electron momentum, and the crosses on the
heavy line in Fig.\ \ref{edefig} mean that the heavy particle is on
its mass shell.

The inhomogeneous equation \eq{nonhom3dim} also fixes the normalization
of the wave function.

Even though the total kernel in \eq{ede} is unambiguously defined, we
still have freedom to choose the zero-order kernel $K_0$ at our
convenience, in order to obtain a solvable lowest order approximation.
It is not difficult to obtain a regular perturbation theory series for
the corrections to the zero-order approximation corresponding to the
difference between the zero-order kernel $K_0$ and the exact kernel
$K_0+\delta K$

\beq         \label{perturbth}
E_n=E_n^0+(n|i\delta K(E_n^0)|n)\left(1+(n|i\delta K'(E_n^0)|n)\right)
\eeq
\[
+(n|i\delta K(E_n^0)G_{n0}(E_n^0)i\delta K(E_n^0)|n)
\left(1+(n|i\delta K'(E_n^0)|n)\right)+\dots,
\]

\noindent
where the summation of intermediate states goes with the weight
$d^3p/[(2\pi)^32E_p]$ and is realized with the help of the subtracted
free Green function of the EDE with the kernel $K_0$

\beq
G_{n0}(E)=G_{0}(E)-\frac{|n)(n|}{E-E_n^0},
\eeq

\noindent
conjugation is understood in the Dirac sense, and $\delta K'(E_n^0)\equiv
(dK/dE)_{|E=E_n^0}$.

The only apparent difference of the EDE \eq{ede} from the regular Dirac
equation is connected with the dependence of the interaction kernels on
energy. Respectively the perturbation theory series in \eq{perturbth}
contain, unlike the regular nonrelativistic perturbation series, derivatives
of the interaction kernels over energy. The presence of these derivatives is
crucial for cancellation of the ultraviolet divergences in the expressions
for the energy eigenvalues.

\begin{figure}[h]
\centerline{\epsfig{file=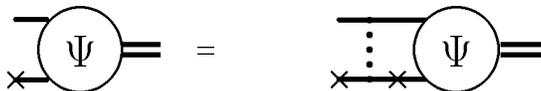}}
\vspace{0.5cm}
\caption{Effective Dirac equation in the external Coulomb field }
\label{edeextfield}
\end{figure}

A judicious choice of the zero-order kernel (sum of the Coulomb and Breit
potentials, for more detail see, e.g, \cite{gy1,gy2,lep}) generates a
solvable unperturbed EDE in the external Coulomb field in Fig.\
\ref{edeextfield}\footnote{Strictly speaking the external field in this
equation is not exactly Coulomb but also includes a transverse
contribution.}. The eigenfunctions of this equation may be found
exactly in the form of the Dirac-Coulomb wave functions (see, e.g,
\cite{lep}).  For practical purposes it is often sufficient to
approximate  these exact wave functions by the product of the
Schr\"odinger-Coulomb wave functions with the reduced mass and the free
electron spinors which depend on the electron mass and not on the
reduced mass. These functions are very convenient for calculation of
the high order corrections, and while below we will often skip some
steps in the derivation of one or another high order contribution from
the EDE, we advise the reader to keep in mind that almost all
calculations below are done with these unperturbed wave functions.

\part{General Features of the Hydrogen Energy Levels}

\section{Classification of Corrections}

The zero-order effective Dirac equation with a Coulomb source provides only
an approximate description of loosely bound states in QED, but the spectrum
of this Dirac equation may serve as a good starting point for obtaining more
precise results.

The magnetic moment of the heavy nucleus is completely ignored in the Dirac
equation with a Coulomb source, and, hence, the hyperfine splitting of the
energy levels is missing in its spectrum. Notice that the magnetic
interaction between the nucleus and the electron may be easily described
even in the framework of the nonrelativistic quantum mechanics, and the
respective calculation of the leading contribution to the hyperfine
splitting was done a long time ago by Fermi \cite{ef}.

Other corrections to the Dirac energy levels do not arise in the
quantum mechanical treatment with a potential, and for their calculation, as
well as for calculation of the corrections to the hyperfine splitting,
field-theoretical methods are necessary. All electrodynamic corrections to
the energy levels may be written in the form of the power series expansion
over three small parameters $\alpha$, $Z\alpha$ and ${m}/{M}$ which
determine the properties of the  bound state. Account for the additional
corrections of nonelectromagnetic origin induced by the strong and weak
interactions introduces additional small parameters, namely, the ratio of
the nuclear radius and the Bohr radius, the Fermi constant, etc.   It should
be noted that the coefficients in the power series for the energy levels
might themselves be slowly varying functions (logarithms) of these
parameters.

Each of the small parameters above plays an important and unique role.
In order to organize further discussion of different contributions to the
energy levels it is convenient to classify corrections in accordance with
the small parameters on which they depend.

Corrections which depend only on the parameter $Z\alpha$ will be called
{\em relativistic} or {\em binding corrections}.  Higher powers of
$Z\alpha$ arise due to deviation of the theory from a nonrelativistic
limit, and thus represent a relativistic expansion. All such
contributions are contained in the spectrum of the effective Dirac
equation in the external Colomb field.

Contributions to the energy which depend only on the small parameters
$\alpha$ and $Z\alpha$  are called {\em radiative corrections}. Powers
of $\alpha$ arise only from the quantum electrodynamics loops, and all
associated corrections have a quantum field theory nature. Radiative
corrections do not depend on the recoil factor $m/M$ and thus may be
calculated in the framework of QED for a bound electron in an external
field. In respective calculations one deals only with the complications
connected with the presence of quantized fields, but the two-particle
nature of the bound state and all problems connected with the
description of the bound states in relativistic quantum field theory
still may be ignored.

Corrections which depend on the mass ratio $m/M$ of the light and heavy
particles reflect a deviation from the theory with an infinitely heavy
nucleus. Corrections to the energy levels which depend on ${m}/{M}$
and $Z\alpha$ are called {\em recoil corrections}. They describe
contributions to the energy levels which cannot be taken into account
with the help of the reduced mass factor. The presence of these
corrections signals that we are dealing with a truly two-body problem,
rather than with a one-body problem.

Leading recoil corrections in $Z\alpha$ (of order $(Z\alpha)^4(m/M)^n$)
still may be taken into account with the help of the effective Dirac
equation in the external field since these corrections are induced by
the one-photon exchange. This is impossible for the higher order recoil
terms which reflect the truly relativistic two-body nature of the bound
state problem. Technically, respective contributions are induced by the
Bethe-Salpeter kernels with at least two-photon exchanges and the whole
machinery of relativistic QFT is necessary for their calculation.
Calculation of the recoil corrections is simplified by the absence of
ultraviolet divergences, connected with the purely radiative  loops.

{\em Radiative-Recoil corrections} are the expansion terms in the
expressions for the energy levels which depend simultaneously on the
parameters $\alpha$, ${m}/{M}$ and $Z\alpha$.  Their calculation
requires application of all the heavy artillery of QED, since we have
to account both for the purely radiative loops and for the
relativistic two-body nature of the bound states.

The last class of corrections contains {\em nonelectromagnetic corrections},
effects of weak and strong interactions.  The largest correction induced by
the strong interaction is connected with the finiteness of the nuclear size.

Let us emphasize once more that  hyperfine structure, radiative, recoil,
radiative-recoil, and nonelectromagnetic corrections are all missing in
the Dirac energy spectrum. Discussion of their calculations is the main
topic of this review.

\section{Physical Origin of the Lamb Shift} \label{originlamb}

\begin{figure}[h]
\centerline{\epsfig{file=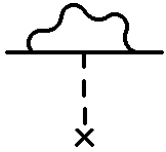,height=2cm}}
\vspace{0.5cm}
\caption{Leading order contribution to the electron radius}
\label{elradfig}
\end{figure}

According to QED an electron continuously emits and absorbs virtual
photons (see the leading order diagram in Fig.\ \ref{elradfig}) and as
a result its electric charge is spread over a finite volume instead of
being pointlike\footnote{The numerical factor in \eq{electrradfirt}
arises due to the common relation between the expansion of the form
factor and the mean square root radius

\beq \label{electronff}
F(-{\bf k}^2)=1-\frac{1}{6}<r^2>{\bf k}^2.
\eeq}

\beq     \label{electrradfirt}
<r^2>=
-6\;\frac{dF_1({-\bf k}^2)}{d{\bf k}^2}_{|{\bf k}^2=0}\approx
\frac{1}{m^2}\frac{2\alpha}{\pi}\ln\frac{m}{\rho}
\approx\frac{1}{m^2}\frac{2\alpha}{\pi}\ln(Z\alpha)^{-2}.
\eeq

In order to obtain this estimate of the electron radius we have taken
into account that the electron is slightly off mass shell in the bound
state.  Hence, the would be infrared divergence in the electron charge
radius is cut off by its virtuality $\rho=(m^2-p^2)/m$ which is of
order of the nonrelativistic binding energy $\rho\approx  m(Z\alpha)^2$.

The finite radius of the electron generates a correction to the Coulomb
potential (see, e.g.,  \cite{bd})

\beq \label{ffinsertt}
\delta V=\frac{1}{6}<r^2>\Delta V=\frac{2\pi}{3}Z\alpha<r^2>\delta({\bf r}),
\eeq

\noindent
where $V=-Z\alpha/r$ is the Coulomb potential.

The respective correction to the energy levels is simply given by the matrix
element of this perturbation. Thus we immediately discover that the finite
size of the electron produced by the QED radiative corrections leads to a
shift of the hydrogen energy levels. Moreover, since this perturbation is
nonvanishing only at the source of the Coulomb potential, it influences
quite differently the energy levels with different orbital angular momenta,
and, hence, leads to splitting of the levels with the same total angular
momenta but different orbital momenta. This splitting lifts the degeneracy
in the spectrum of the Dirac equation in the Coulomb field, where the energy
levels depend only on the principal quantum number $n$ and the total angular
momentum $j$.

It is very easy to estimate this splitting (shift of the $S$ level energy)

\beq                   \label{qff}
\Delta
E=<nS|\delta V|nS>
\approx|\Psi_n(0)|^2\frac{2\pi(Z\alpha)}{3}<r^2>
\eeq
\[
\approx
\frac{4m(Z\alpha)^4}{n^3}\frac{\alpha}{3\pi}\ln[(Z\alpha)^{-2}]
\delta_{l0}\;_{_{|n=2}}\approx1330~\mbox{MHz}.
\]

This result should be compared with the experimental number of about
$1040$ MHz and the agreement is satisfactory for such a crude estimate.
There are two qualitative features of this result to which we would
like to attract the reader's attention. First, the sign of the energy
shift may be obtained without calculation. Due to the finite radius of
the electron its charge in the $S$ state is on the average more spread
out around the Coulomb source than in the case of the pointlike
electron. Hence, the binding is weaker than in the case of the
pointlike electron and the energy of the level is higher. Second,
despite the presence of nonlogarithmic contributions missing in our
crude calculation, their magnitude is comparatively small, and the
logarithmic term above is responsible for the main contribution to the
Lamb shift. This property is due to the would be infrared divergence of
the considered contribution, which is cutoff by the small (in
comparison with the electron mass) binding energy. As we will see
below, whenever a correction is logarithmically enhanced, the
respective logarithm gives a significant part of the correction, as is
the case above.

\begin{figure}[h]
\centerline{\epsfig{file=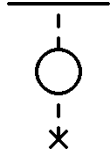,height=2cm}}
\vspace{0.5cm}
\caption{Leading order polarization insertion}
\label{polradfig}
\end{figure}

Another obvious contribution to the Lamb shift of the same leading
order is connected with the polarization insertion in the photon propagator
(see Fig.\ \ref{polradfig}).  This correction  also induces a correction to
the Coulomb potential

\beq      \label{leadcoulpollambnaiv}
\delta V=-\frac{\Pi(-{\bf k}^2)}{{\bf k}^4}_{|{\bf k}^2=0}\Delta V=
\frac{\alpha}{15\pi m^2}\Delta V
=-\frac{4}{15}\frac{\alpha(Z\alpha)}{m^2}\delta({\bf r}),
\eeq

\noindent
and the respective correction to the $S$-level energy is equal to

\beq                 \label{qpol}
\Delta
E=<nS|\delta V|nS>=
-|\Psi_n(0)|^2\frac{4\alpha(Z\alpha)}{15m^2}
\eeq
\[
=-\frac{4m(Z\alpha)^4}{n^3}\frac{\alpha}{15\pi}
\delta_{l0}\;_{_{|n=2}}\approx-30~\mbox{MHz}.
\]

Once again the sign of this correction is evident in advance. The
polarization correction may be thought of as a correction to the
electric charge of the nucleon induced by the fact that the electron
sees the proton from a finite distance\footnote{We remind the reader
that according to the common renormalization procedure the electric
charge is defined as a charge observed at a very large distance.}.
This means that the electron, which has penetrated in the polarization
cloud, sees effectively a larger charge and experiences a stronger
binding force, which lowers the energy level. Experimental observation
of this contribution to the Lamb shift played an important role in the
development of modern quantum electrodynamics since it explicitly
confirmed the very existence of the closed electron loops. Numerically
vacuum polarization contribution is much less important than the
contribution connected with the electron spreading due to quantum
corrections, and the total shift of the level is positive.

\section{Natural Magnitudes of Corrections to the
Lamb Shift} \label{lambqualmagn}

Let us emphasize that the main contribution to the Lamb
shift is a radiative correction itself (compare \eq{qff},\eq{qpol})  and
contains a logarithmic enhancement factor. This is extremely
important when one wants to get a qualitative understanding of the magnitude
of the higher order corrections to the Lamb shift discussed below.  Due to
the presence of this accidental logarithmic enhancement it is
impossible to draw conclusions about the expected magnitude of higher
order corrections to the Lamb shift simply by comparing them to the
magnitude of the leading order contribution.  It is more reasonable to
extract from this leading order contribution the term which can be
called the skeleton factor and to use it further as a normalization
factor. Let us write the leading order contributions in
\eq{qff},\eq{qpol} obtained above in the form

\beq
\frac{4m(Z\alpha)^4}{n^3}\times\mbox{\sl radiative correction},
\eeq

\noindent
where the radiative correction is either the slope of the Dirac form factor,
roughly speaking equal to $m^2{dF_1(-{\bf k}^2)}/{d{\bf k}^2}_{|{\bf
k}^2=0}=\alpha/(3\pi)\ln(Z\alpha)^{-2}$, or the polarization correction
$m^2{\Pi(-{\bf k}^2)}/{{\bf k}^4}_{|{\bf
k}^2=0}={\alpha}/({15\pi})$.

It is clear now that the scale setting factor which should be used for
qualitative estimates of the high order corrections to the Lamb shift is
equal to  $4m(Z\alpha)^4/n^3$. Note the characteristic dependence on the
principal quantum number $1/n^3$ which originates from the square of the
wave function at the origin $|\psi(0)|^2\sim1/n^3$. All corrections induced
at small distances (or at high virtual momenta) have this
characteristic dependence and are called {\it state-independent}. Even the
coefficients before the leading powers of the low energy logarithms
$\ln(Z\alpha)^2$ are state-independent since these leading logarithms
originate from integration over the wide intermediate momenta region from
$m(Z\alpha)^2$ to $m$, and the respective factor before the logarithm is
determined by the high momenta part of the integration region. Estimating
higher order corrections to the Lamb shift it is necessary to
remember, as mentioned above, that unlike the case of radiative
corrections to the scattering amplitudes, in the bound state problem factors
$Z\alpha$ are not accompanied by an extra factor $\pi$ in the denominator.
This well known feature of the Coulomb problem provides one more reason to
preserve $Z$ in analytic expressions (even when $Z=1$), since in this way
one may easily separate powers of $Z\alpha$ not accompanied by $\pi$ from
powers of $\alpha$ which always enter formulae in the combination
$\alpha/\pi$.

\part{External Field Approximation}

We will first discuss corrections to the basic Dirac energy levels which
arise in the external field approximation. These are leading
relativistic corrections with exact mass dependence and radiative
corrections.

\section{Leading Relativistic Corrections with Exact Mass
Dependence}\label{leadreclambdef}

We are considering a loosely bound two-particle system. Due to the
nonrelativistic nature of this bound state it is clear that the main
$(Z\alpha)^2$ contribution to the binding energy depends only on one
mass parameter, namely, on the nonrelativistic reduced mass, and does not
depend separately on the masses of the constituents. Relativistic
corrections to the energy levels of order $(Z\alpha)^4$, describing the
fine structure of the hydrogen spectrum, are missing in the
nonrelativistic Schr\"odinger equation approach. The correct
description of the fine structure for an infinitely heavy Coulomb
center is provided by the relativistic Dirac equation, but it tells us
nothing about the proper mass dependence of these corrections for the
nucleus of finite mass. There are no reasons to expect that in the case
of a system of two particles with finite masses relativistic
corrections of order $(Z\alpha)^4$ will depend only on the reduced
mass. The dependence of these corrections on the masses of the
constituents should be more complicated.

The solution of the problem of the proper mass dependence of the
relativistic corrections of order $(Z\alpha)^4$ may be found in the
effective Hamiltonian framework. In the center of mass system the
nonrelativistic Hamiltonian for a system of two particles with Coulomb
interaction has the form

\beq     \label{twopartcoul}
H_0=\frac{\bf p^2}{2m}+\frac{\bf p^2}{2M}-\frac{Z\alpha}{r}.
\eeq

\noindent
In a nonrelativistic loosely bound system expansion over $(Z\alpha)^2$
corresponds to the nonrelativistic expansion over $v^2/c^2$. Hence, we
need an effective Hamiltonian including the terms of the first order in
$v^2/c^2$ for proper description of the corrections of relative order
$(Z\alpha)^2$ to the nonrelativistic energy levels. Such a Hamiltonian
was first considered by Breit \cite{breit29}, who realized that all
corrections to the nonrelativistic two-particle Hamiltonian of the
first order in $v^2/c^2$ may be obtained from the sum of the free
relativistic Hamiltonians of each of the particles and the relativistic
one-photon exchange. This conjecture is intuitively obvious since extra
exchange photons lead to at least one extra factor of $Z\alpha$, thus
generating contributions to the binding energy of order $(Z\alpha)^5$
and higher.

An explicit expression for the Breit potential was derived in \cite{bg}
from the one-photon exchange amplitude with the help of the
Foldy-Wouthuysen transformation\footnote{We do not consider hyperfine
structure now and thus omit in \eq{breitpot} all terms in the Breit
potential which depend on the spin of the heavy particle.}

\beq      \label{breitpot}
V_{Br}=\frac{\pi Z\alpha}{2}\left(\frac{1}{m^2}
+\frac{1}{M^2}\right)\delta^3({\bf r})-\frac{Z\alpha }{2mMr}
\left({\bf p^2}+\frac{{\bf r}({\bf r\cdot p})\cdot{\bf p}}{r^2}\right)
\eeq
\[
+\frac{Z\alpha}{r^3}\left(\frac{1}{4m^2}+\frac{1}{2mM}\right)[{\bf
r\times p}] \cdot\bbox\sigma.
\]

\noindent
A simplified derivation of the Breit interaction potential may be found
in many textbooks (see, e.g., \cite{blp}).

\begin{figure}[ht]
\centerline{\epsfig{file=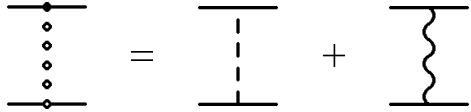,height=1.5cm}}
\vspace{0.5cm}
\caption{Sum of the Coulomb and Breit kernels}
\label{coulbreitfig}
\end{figure}

All contributions to the energy levels up order $(Z\alpha)^4$ may be
calculated from the total Breit Hamiltonian

\beq   \label{totbreitnaiv}
H_{Br}=H_0+V_{Br},
\eeq

\noindent
where the interaction potential is the sum of the instantaneous Coulomb
and Breit potentials in  Fig.\ \ref{coulbreitfig}.

The corrections of order $(Z\alpha)^4$ are just the first order matrix
elements of the Breit interaction between the Coulomb-Schr\"odinger
eigenfunctions of the Coulomb Hamiltonian $H_0$ in \eq{twopartcoul}.
The mass dependence of the Breit interaction is known exactly, and the
same is true for its matrix elements. These matrix elements and, hence,
the exact mass dependence of the contributions to the energy levels of
order $(Z\alpha)^4$, beyond the reduced mass, were first obtained a
long time ago \cite{bg}

\beq          \label{barkglov}
E_{nj}^{tot}=(m+M)-\frac{m_r(Z\alpha)^2}{2n^2}
-\frac{m_r(Z\alpha)^4}{2n^3}\left(\frac{1}{j+\frac{1}{2}}-\frac{3}{4n}
+\frac{m_r}{4n(m+M)}\right)
\eeq
\[
+\frac{(Z\alpha)^4m_r^3}{2n^3M^2}
\left(\frac{1}{j+\frac{1}{2}}-\frac{1}{l+\frac{1}{2}}\right)(1-\delta_{l0}).
\]

\noindent
Note the emergence of the last term in \eq{barkglov}  which
lifts the characteristic degeneracy in the Dirac spectrum between
levels with the same $j$ and $l=j\pm1/2$.  This means that the
expression for the energy levels in \eq{barkglov} already predicts a
nonvanishing contribution to the classical Lamb shift
$E(2S_\frac{1}{2})-E(2P_\frac{1}{2})$. Due to the smallness of the
electron-proton mass ratio this extra term is extremely small in
hydrogen.  The leading contribution to the Lamb shift, induced by the
QED radiative correction, is much larger.

In the Breit Hamiltonian in \eq{breitpot} we have omitted all terms
which depend on spin variables of the heavy particle. As a result the
corrections to the energy levels in \eq{barkglov} do not depend on the
relative orientation of the spins of the heavy and light particles
(in other words they do not describe hyperfine splitting). Moreover,
almost all contributions in \eq{barkglov} are independent not only of
the mutual orientation of spins of the heavy and light particles but
also of the magnitude of the spin of the heavy particle. The only
exception is the small contribution proportional to the term
$\delta_{l0}$, called the Darwin-Foldy contribution. This term arises
in the matrix element of the Breit Hamiltonian only for the spin
one-half nucleus and should be omitted for spinless or spin one nuclei.
This contribution combines naturally with the nuclear size correction,
and we postpone its discussion to Section \ref{darwinfoldyspin} dealing
with the nuclear size contribution.

In the framework of the effective Dirac equation in the external Coulomb
field \footnote{We remind the reader that the external field in this
equation also contains a transverse contribution.} (see Fig.\
\ref{edeextfield}) the result in \eq{barkglov} was first obtained in
\cite{gy1} (see also \cite{gy2,lep}) and rederived once again in \cite{sy},
where it was presented in the form

\beq          \label{leadrecoil}
E_{nj}^{tot}=(m+M)+m_r [f(n,j)-1]
-\frac{m_r^2}{2(m+M)}[f(n,j)-1]^2
\eeq
\[
+\frac{(Z\alpha)^4m_r^3}{2n^3M^2}
\left(\frac{1}{j+\frac{1}{2}}-\frac{1}{l+\frac{1}{2}}\right)(1-\delta_{l0}).
\]

This equation has the same contributions of order $(Z\alpha)^4$ as in
\eq{barkglov}, but formally this expression also contains nonrecoil and
recoil corrections of order $(Z\alpha)^6$ and higher. The nonrecoil part of
these contributions is definitely correct since the Dirac energy spectrum is
the proper limit of the spectrum of a two-particle system in the nonrecoil
limit $m/M=0$. As we will discuss later the first-order mass ratio
contributions in \eq{leadrecoil} correctly reproduce recoil corrections of
higher orders in $Z\alpha$ generated by the Coulomb and Breit exchange
photons. Additional first order mass ratio recoil contributions of order
$(Z\alpha)^6$ will be calculated below. Recoil corrections of order
$(Z\alpha)^8$ were never calculated and at the present stage the mass
dependence of these terms should be considered as completely unknown.

Recoil corrections depending on {\it odd} powers of $Z\alpha$ are also
missing in \eq{leadrecoil}, since as was explained above all corrections
generated by the one-photon exchange necessarily depend on the even powers
of $Z\alpha$. Hence, to calculate recoil corrections of order $(Z\alpha)^5$
one has to consider the nontrivial contribution of the box diagram. We
postpone discussion of these corrections until Section \ref{lambreczalpha5}.

It is appropriate to give an exact definition of what is called
the Lamb shift. In the early days of the Lamb shift studies,
experimentalists measured not a shift but the classical Lamb splitting
$E(2S_\frac{1}{2})-E(2P_\frac{1}{2})$ between the energy levels which
are degenerate according to the naive Dirac equation in the Coulomb
field. This splitting is an experimental observable defined
independently of any theory. Modern experiments now produce high
precision experimental data for the nondegenerate $1S$ energy level,
and the very notion of the Lamb shift in this case, as well as in the
case of an arbitrary energy level, does not admit an unambiguous
definition. It is most natural to call the Lamb shift the sum of all
contributions to the energy levels which lift the double degeneracy of
the Dirac-Coulomb spectrum with respect to $l=j\pm1/2$ (see Section
\ref{diraccoul}).  There emerged an almost universally adopted
convention to call the Lamb shift the sum of all contributions to the
energy levels beyond the first three terms in \eq{leadrecoil} and
excluding, of course, all hyperfine splitting contributions. This means
that we define the Lamb shift by the relationship

\beq          \label{lambdef}
E_{njl}^{tot}=(m+M)+m_r [f(n,j)-1]
-\frac{m_r^2}{2(m+M)}[f(n,j)-1]^2+L_{njl}
\equiv E_{nj}^{DR}+L_{njl}.
\eeq

\noindent
We will adopt this definition below.

\section{Radiative Corrections of Order
$\alpha^{\lowercase{n}}(Z\alpha)^4\lowercase{m}$}

Let us turn now to the discussion of radiative corrections which may be
calculated in the external field approximation.

\subsection{Leading Contribution to the Lamb Shift}

\subsubsection{Radiative Insertions in the Electron line and the Dirac
Form Factor Contribution} \label{leadinglamb}

\begin{figure}[ht]
\centerline{\epsfig{file=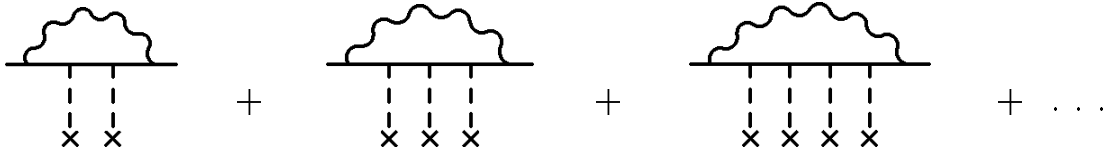,height=2cm}}
\vspace{0.5cm}
\caption{Kernels with many spanned Coulomb photons}
\label{spanfig}
\end{figure}

The main contribution to the Lamb shift was first estimated in the
nonrelativistic approximation by Bethe \cite{bethe}, and calculated by Kroll
and Lamb \cite{kl}, and by French and Weisskopf \cite{fw}. We have  already
discussed above qualitatively the nature of this contribution. In the
effective Dirac equation framework the apparent perturbation kernels to be
taken into account are the diagrams with the radiative photon spanning any
number of the exchanged Coulomb photons in Fig.\ \ref{elradfig} and Fig.\
\ref{spanfig}. The dominant logarithmic contribution to the
Lamb shift is produced by the slope of the Dirac form factor $F_1$, but
superficially all these kernels can lead to corrections of order
$\alpha(Z\alpha)^4$ and one cannot discard any of them.  An additional
problem is connected with the infrared divergence of the kernels on the mass
shell. There cannot be any true infrared divergence in the bound state
problem since all would be infrared divergences are cut off either at
the inverse Bohr radius or by the electron binding energy. Nevertheless
such spurious on-shell infrared divergences can complicate the
calculations.

An important step which greatly facilitates treatment of all these problems
consists in separation of the radiative photon integration region
with the help of auxiliary parameter $\sigma$ in such way that
$m(Z\alpha)^2\ll\sigma\ll m(Z\alpha)$. It is easy to see that in the high
momentum region each additional Coulomb photon produces an extra factor
$Z\alpha$, so it is sufficient to consider only the kernel with one Coulomb
photon in this region. Moreover, the auxiliary parameter $\sigma$ provides
an infrared cutoff for the vertex graph and thus solves the problem of the
would be infrared divergence. Due to the choice of the parameter
$\sigma\gg m(Z\alpha)^2$ one may ignore the binding energy in the high
momentum region. The main contribution to the Lamb
shift is induced by the Dirac form factor $F_1(k^2)-1$ which is
proportional to the transferred momentum squared at small momentum
transfer.  This transferred momentum squared factor exactly cancels the
Coulomb photon propagator attached to the Dirac form factor, and
momentum space integrations over wave function momenta factorize,
thereby producing the wave function squared at the origin in the
coordinate space. It is clear that if one would take into account the
small virtuality of either external electron line, expanding the
integrand in this virtuality, it would lead to an extra factor of
momentum squared in the integrand, and after integration with the wave
function would lead to an extra factor $(Z\alpha)^2$ in the
contribution to the energy shift. Hence, since we are interested in the
contribution of order $\alpha(Z\alpha)^4$, we may freely ignore the
virtuality of the electron line in the kernel in the high momentum region.
It is also clear even at this stage, that the high momentum region does not
produce any contribution for the non-$S$ states because the wave function
vanishes at the origin for such states, and, hence, the logarithmic
contribution is missing for the non-$S$ states.

Of course, all approximations made above are invalid in the
low momentum integration region, where one cannot consider only the kernel
with the radiative photon spanning only one Coulomb exchange. For soft
radiative photons with characteristic momenta of order $m(Z\alpha)^2$ graphs
with any number of spanned exchanged photons in Fig.\ \ref{spanfig} have the
same order of magnitude and one has to take these graphs into account
simultaneously. This means that one has to calculate the matrix element of
the exact self-energy operator in the external Coulomb field between
Dirac-Coulomb wave functions. This problem may seem formidable at first
sight, but it can be readily solved with the help of old-fashioned
perturbation theory by inserting a complete set of intermediate states and
performing calculations in the dipole approximation, which is adequate to
accuracy $\alpha(Z\alpha)^4m$.  It should be mentioned that the magnitude of
the upper boundary of the interval for the auxiliary parameter $\sigma$ was
chosen in order to provide validity of the dipole approximation.

The most important fact about the auxiliary parameter $\sigma$ is that one
can use different approximations for calculations of the high- and
low-momentum contributions. In the high-momentum region the factor
$m(Z\alpha)^2/k\leq(m(Z\alpha)^2/\sigma\ll 1$ plays the role of a small
parameter and one can consider binding effects as small corrections. In
the low-momentum region $k/(mZ\alpha)\leq\sigma/(mZ\alpha)\ll 1$ one
may use the nonrelativistic multipole expansion, and the main
contribution in this region is given by the dipole contribution. The
crucial point is that for $k\sim\sigma$ both expansions are valid
simultaneously and one can match them without loss in accuracy.
Matching the high- and low-momentum contributions one obtains the
classical result for the shift of the energy level generated by the
slope of the Dirac form factor

\beq  \label{diracslope}
\Delta E=
\left\{\left[\frac{1}{3}\ln\frac{m(Z\alpha)^{-2}}{m_r}+\frac{11}{72}
\right]\delta_{l0}
-\frac{1}{3}\ln k_0(n,l)\right\}\frac{4\alpha(Z\alpha)^4}{\pi
n^3}\left(\frac{m_r}{m}\right)^3m,
\eeq

\noindent
where $m_r=mM/(m+M)$ is the reduced mass and $\ln k_0(n,l)$ is the so called
Bethe logarithm.  The factor $m/m_r$ arises in the argument of the would
be infrared divergent logarithm $\ln(m/\lambda)$ since in the
nonrelativistic approximation the energy levels of an atom depend only on
the reduced mass and, hence, the infrared divergence is cut off by the
binding energy $m_r(Z\alpha)^2$ \cite{sy}.

The mass dependence of the correction of order $\alpha(Z\alpha)^4$
beyond the reduced mass factor is properly described by the expression
in \eq{diracslope} as was proved in \cite{bg85,bg87}.  In the same way
as for the case of the leading relativistic correction in
\eq{barkglov}, the result in \eq{diracslope} is exact in the small mass
ratio $m/M$, since in the framework of the effective Dirac equation all
corrections of order $(Z\alpha)^4$  are generated by the kernels with
one-photon exchange. In some earlier papers the reduced mass factors in
\eq{diracslope} were expanded up to first order in the small mass ratio
$m/M$. Nowadays it is important to preserve an exact mass dependence in
\eq{diracslope} because current experiments may be able to detect
quadratic mass corrections (about $2$ kHz for the $1S$ level in
hydrogen) to the leading nonrecoil Lamb shift contribution.

The Bethe logarithm is formally defined as a certain normalized
infinite sum of matrix elements of the coordinate operator over the
Schr\"odinger-Coulomb wave functions. It is a pure number which can in
principle be calculated with arbitrary accuracy, and high accuracy
results for the Bethe logarithm can be found in the literature (see,
e.g. \cite{hm,drakeswain} and references therein). For convenience we
have collected some values for the Bethe logarithms \cite{drakeswain}
in Table I.

\begin{center}
\underline{Table I. Bethe Logarithms for some Lower Levels
\cite{drakeswain}}
\nopagebreak

\begin{tabular}{|l|r|r|}
\hline
&$\ln k_0(n,l)$
&$\Delta E=-\frac{4}{3} \ln k_0(n,l)\frac{\alpha(Z\alpha)^4}{\pi
n^3}(\frac{m_r}{m})^3m$~\mbox{kHz}
\\ \hline\hline
$1S$&$2.984~128~555~765~498$& $-23~591.92$\\
\hline
$2S$&$2.811~769~893~120~563$&$-2~778.66$   \\
\hline
$2P$&$-0.030~016~708~630~213$&$29.66$        \\
\hline
$3S$&$2.767~663~612~491~822$ &$-810.39$\\
\hline
$3P$&$-0.038~190~229~385~312$&$11.18$       \\
\hline
$3D$&$-0.005~232~148~140~883$&$1.53$       \\
\hline
$4S$&$2.749~811~840~454~057$ &$-339.68$      \\
\hline
$4P$&$-0.041~954~894~598~086$&$5.18$       \\
\hline
$4D$&$-0.006~740~938~876~975$&$0.83$       \\
\hline
$4F$&$-0.001~733~661~482~126$&$0.21$       \\
\hline
$5S$&$2.740~823~727~854~572$ &$-173.35$      \\
\hline
$5P$&$-0.044~034~695~591~878$&$2.79$       \\
\hline
$5D$&$-0.007~600~751~257~947$&$0.48$       \\
\hline
$5F$&$-0.002~202~168~381~486$&$0.14$       \\
\hline
$5G$&$-0.000~772~098~901~537$&$0.05$       \\
\hline
$6S$&$2.735~664~206~935~105$ &$-100.13$      \\
\hline
$6P$&$-0.045~312~197~688~974$&$1.66$       \\
\hline
$6D$&$-0.008~147~203~962~354$&$0.30$       \\
\hline
$6F$&$-0.002~502~179~760~279$&$0.09$       \\
\hline
$6G$&$-0.000~962~797~424~841$&$0.04$       \\
\hline
$6H$&$-0.000~407~926~168~297$&$0.01$       \\
\hline
$7S$&$2.732~429~129~187~092$ &$-62.98$      \\
\hline
$7P$&$-0.046~155~177~262~915$&$1.06$       \\
\hline
$7D$&$-0.008~519~223~293~658$&$0.20$       \\
\hline
$7F$&$-0.002~709~095~727~000$&$0.06$       \\
\hline
$7G$&$-0.001~094~472~739~370$&$0.03$       \\
\hline
$7H$&$-0.000~499~701~854~766$&$0.01$       \\
\hline
$7I$&$-0.000~240~908~258~717$&$0.01$       \\
\hline
$8S$&$2.730~267~260~690~589$ &$-42.16$      \\
\hline
$8P$&$-0.046~741~352~003~557$&$0.72$       \\
\hline
$8D$&$-0.008~785~042~984~125$&$0.14$       \\
\hline
$8F$&$-0.002~859~114~559~296$&$0.04$       \\
\hline
$8G$&$-0.001~190~432~043~318$&$0.02$       \\
\hline
$8H$&$-0.000~566~532~724~12$&$0.01$       \\
\hline
$8I$&$-0.000~290~426~172~391$&$$       \\
\hline
$8K$&$-0.000~153~864~500~961$&$$       \\
\hline
\end{tabular}
\end{center}

\subsubsection{Pauli Form Factor Contribution}

The Pauli form factor $F_2$ also generates a small contribution to the Lamb
shift. This form factor does not produce any contribution if one
neglects the lower components of the unperturbed wave functions, since the
respective matrix element is identically zero between the upper
components in the standard representation for the Dirac matrices which we
use everywhere. Taking into account lower components in the nonrelativistic
approximation we easily obtain an explicit expression for the respective
perturbation

\beq
\delta V=-\frac{1}{4m^2}\left[\Delta V
+2~\mbox{\boldmath$\sigma$}~\frac{m}{m_r} ~[\mbox{\boldmath$\nabla$}
V\times{\bf p}]\right]F_2(0),
\eeq

\noindent
where $V=-Z\alpha/r$ is the Coulomb potential. Note the appearance of an
extra factor $m/m_r$ in the coefficient before the second term. This is
readily obtained from an explicit consideration of the radiatively corrected
one photon exchange. In momentum space the term with the Laplacian of
the Coulomb potential depends only on the exchanged momentum, while the
second term contains explicit dependence on the electron momentum. Since the
Pauli form factor depends explicitly on the electron momentum and not on the
relative momentum of the electron-proton system, the transition to
relative momentum, which is the argument of the unperturbed wave functions,
leads to emergence of an extra factor $m/m_r$.

The interaction potential above generated by the Pauli form factor may  be
written in terms of the spin-orbit interaction

\beq
\delta V=
\left[\frac{Z\alpha\pi}{m^2}\delta^3({\bf r})+\frac{Z\alpha\pi}{r^3mm_r}
({\bf s}{\bf l})\right]~F_2(0),
\eeq

\noindent
where

\beq
{\bf s}=\frac{\mbox{\boldmath$\sigma$}}{2},\;\;\;\;{\bf l}
={\bf r}\times{\bf p}.
\eeq

The respective contributions to the Lamb shift are given by

\beq          \label{pauli}
\Delta E_{|l=0}=\frac{(Z\alpha)^4m}{n^3}F_2(0)
\left(\frac{m_r}{m}\right)^3
=\frac{\alpha(Z\alpha)^4m}{2\pi n^3}\left(\frac{m_r}{m}\right)^3,
\eeq
\[
\Delta E_{|l\neq
0}=\frac{(Z\alpha)^4m}{n^3}F_2(0)
\frac{j(j+1)-l(l+1)-3/4}{l(l+1)(2l+1)}
\left(\frac{m_r}{m}\right)^2
\]
\[
=\frac{\alpha(Z\alpha)^4m}{2\pi n^3}
\frac{j(j+1)-l(l+1)-3/4}{l(l+1)(2l+1)}\left(\frac{m_r}{m}\right)^2,
\]

\noindent
where we have used the Schwinger value \cite{sch} of the anomalous magnetic
moment $F_2(0)=\alpha/(2\pi)$. Correct reduced mass factors have been
retained in this expression instead of expanding in $m/M$.

\subsubsection{Polarization Operator Contribution}

The leading polarization operator contribution to the Lamb shift in Fig.\
\ref{polradfig} was already calculated above in \eq{qpol}. Restoring the
reduced mass factors which were omitted in that qualitative discussion, we
easily obtain

\beq                 \label{polariz}
\Delta E=4\pi(Z\alpha)|\Psi_n(0)|^2
\frac{\Pi(-{\bf k}^2)}{{\bf k}^4}_{|{\bf k}^2=0}
=-\frac{4\alpha(Z\alpha)^4m}{15\pi
n^3}\left(\frac{m_r}{m}\right)^3\delta_{l0}.
\eeq

This result was originally obtained in \cite{uehling} long before the advent
of modern QED, and was the only known source for the $2S-2P$
splitting. There is a certain historic irony that for many years it was
the common wisdom "that this effect is .. much too small and has, in
addition, the wrong sign" \cite{bethe} to explain the $2S-2P$ splitting.

\subsection{Radiative Corrections of Order
$\alpha^2(Z\alpha)^4\lowercase{m}$}

From the theoretical point of view, calculation of the corrections of
order $\alpha^2(Z\alpha)^4$ contains nothing fundamentally new in
comparison with the corrections of order $\alpha(Z\alpha)^4$. The scale
for these corrections  is provided by the factor
$4\alpha^2(Z\alpha)^4/(\pi^2n^3)m$, as one may easily see from the
respective discussion above in Section \ref{lambqualmagn}. Corrections
depend only on the values of the form factors and their derivatives at
zero transferred momentum and the only challenge is to calculate
respective radiative corrections with sufficient accuracy.

\subsubsection{Dirac Form Factor Contribution}

\begin{figure}
\centerline{\epsfig{file=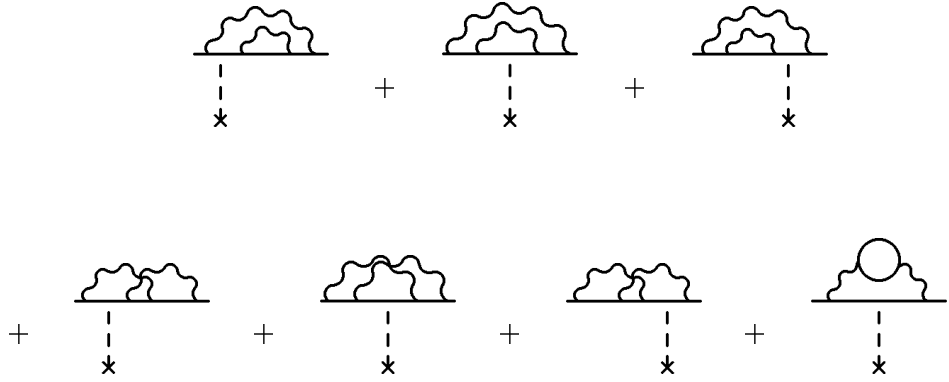,height=5cm}}
\vspace{0.5cm}
\caption{Two-loop electron formfactor}
\label{secondirslfig}
\end{figure}

Calculation of the contribution of order
$\alpha^2(Z\alpha)^4$ induced by the radiative photon insertions in the
electron line is even simpler than the respective calculation of the leading
order contribution. The point is that the second and higher order
contributions to the slope of the Dirac form factor are infrared finite, and
hence, the total contribution of order $(Z\alpha)^4$ to the Lamb shift is
given by the slope of the Dirac form factor. Hence, there is no need to
sum an infinite number of diagrams. One readily obtains for the
respective contribution

\beq \label{ff}
\Delta E_{F_1}=-4\pi
(Z\alpha)|\Psi_n(0)|^2\frac{dF^{(2)}_1(-{\bf k}^2)}{d{\bf k}^2}_{|{\bf
k}^2=0} =0.469~941~4\ldots\frac{4\alpha^2(Z\alpha)^4}{\pi^2n^3}
\left(\frac{m_r}{m}\right)^3m
\:\delta_{l0},
\eeq
\[
\]

\noindent
where we have used the second order slope of the Dirac form factor
generated by the diagrams in Fig.\ \ref{secondirslfig}

\beq
\frac{dF^{(2)}_1(-{\bf k}^2)}{d{\bf k}^2}_{|{\bf k}^2=0}=
\left[\frac{3}{4}\zeta(3)-\frac{\pi^2}{2}\ln2+\frac{49}{432}\pi^2
+\frac{4~819}{5~184}\right]\frac{1}{m^2}\left(\frac{\alpha}{\pi}\right)^2
\approx -\frac{0.469~941~4\ldots}{m^2}\left(\frac{\alpha}{\pi}\right)^2.
\eeq

\noindent
The two-loop slope was considered in the early pioneer works
\cite{weneser,soto}, and for the first time the correct result was
obtained numerically in \cite{ab}. This last work triggered a flurry of
theoretical activity \cite{lautrup,bmrlett,petermann,fox}, followed by
the first completely analyticall calculation in \cite{bmr}. The same
anlytical result for the slope of the Dirac form factor was derived in
\cite{kurlipmer} from the total $e^+e^-$ cross section and the
unitarity condition.

\subsubsection{Pauli Form Factor Contribution}

Calculation of the Pauli form factor contribution follows closely the
one which was performed in order $\alpha(Z\alpha)^4$, the only difference
being that we have to employ the second order contribution to the Pauli
form factor (see Fig.\ \ref{secondirslfig}) calculated a long time ago
in \cite{karpkroll,pet,somm} (the result of the first calculation
\cite{karpkroll} turned out to be in error)

\beq  \label{pauli2}
F^{(2)}_2(0)=\left[\frac{3}{4}\zeta(3)-\frac{\pi^2}{2}\ln2
+\frac{\pi^2}{12}+\frac{197}{144}\right]\left(\frac{\alpha}{\pi}\right)^2
\approx-0.328~478~9\ldots\left(\frac{\alpha}{\pi}\right)^2
\eeq

Then one readily obtains for the Lamb shift contribution

\beq       \label{pauli2lamb}
\Delta E_{|l=0}
=-0.328~478~9\ldots\frac{\alpha^2(Z\alpha)^4m}{\pi^2n^3}
\left(\frac{m_r}{m}\right)^3,
\eeq
\[
\Delta E_{|l\neq
0}=-0.328~478~9\ldots\frac{\alpha^2(Z\alpha)^4m}{\pi^2n^3}
\frac{j(j+1)-l(l+1)-3/4}{l(l+1)(2l+1)}\left(\frac{m_r}{m}\right)^2.
\]

\subsubsection{Polarization Operator Contribution} \label{lambalpha2zal4}

\begin{figure}[ht]
\centerline{\epsfig{file=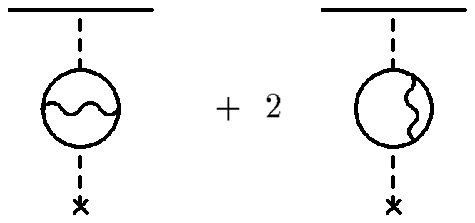,height=2cm}}
\vspace{0.5cm}
\caption{Insertions of two-loop polarization operator}
\label{secondpolfig}
\end{figure}

Here we use well known low momentum asymptotics of the second order
polarization operator \cite{bds,kalsab,schwinger} in Fig.\
\ref{secondpolfig}

\beq
\frac{\Pi(-{\bf k}^2)}{{\bf k}^4}_{|{\bf k}^2=0}=-
\frac{41}{162m^2}\left(\frac{\alpha}{\pi}\right)^2,
\eeq

\noindent
and obtain \cite{bds}

\beq     \label{pol2}
\Delta E=
-\frac{82}{81}\frac{\alpha^2(Z\alpha)^4m}{\pi^2n^3}
\left(\frac{m_r}{m}\right)^3
\delta_{l0}.
\eeq

\subsection{Corrections of Order $\alpha^3(Z\alpha)^4\lowercase{m}$}

\subsubsection{Dirac Form Factor Contribution} \label{3loopdirac}

\begin{figure}[h]
\centerline{\epsfig{file=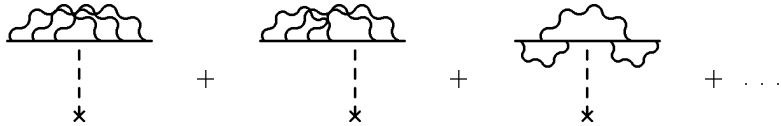}}
\vspace{0.5cm}
\caption{Examples of the three-loop contributions for the electron form
factor}
\label{thirdformfig}
\end{figure}

Calculation of the corrections of order $\alpha^3(Z\alpha)^4$ is similar to
calculation of  the contributions of order $\alpha^2(Z\alpha)^4$.
Respective corrections depend only on the values of the three-loop form
factors or their derivatives at vanishing transferred momentum.
The three-loop contribution to the slope of the Dirac form
factor (Fig.\ \ref{thirdformfig}) was recently calculated analytically
\cite{melrit}

\beq
\frac{dF^{(2)}_1(-{\bf k}^2)}{d{\bf k}^2}_{|{\bf k}^2=0}=-
\left[\frac{25}{8}\zeta(5)-\frac{17}{24}\pi^2\zeta(3)-
\frac{2929}{288}\zeta(3)-\frac{217}{9}a_4-\frac{217}{216}\ln^4{2}
\right.
\eeq
\[
\left.
-\frac{103}{1080}\pi^2\ln^2{2}
+ \frac{41671}{2160}\pi^2\ln2+\frac{3899}{25920}\pi^4-
\frac{454979}{38880}\pi^2-\frac{77513}{186624}\right]
\frac{1}{m^2}\left(\frac{\alpha}{\pi}\right)^3
\]
\[
\approx -\frac{0.171~72\ldots}{m^2}\left(\frac{\alpha}{\pi}\right)^3,
\]

\noindent
where

\beq
a_4=\sum_{n=1}^\infty\frac{1}{2^nn^4}.
\eeq

\noindent
The respective contribution to the Lamb shift is equal to

\beq \label{ff3}
\Delta E_{F_1}
=0.171~72\ldots\frac{4\alpha^3(Z\alpha)^4}{\pi^3n^3}
\left(\frac{m_r}{m}\right)^3m\:
\delta_{l0},
\eeq

\subsubsection{Pauli Form Factor Contribution}

For calculation of the Pauli form factor contribution to the Lamb shift
the third order contribution to the Pauli form factor (Fig.\
\ref{thirdformfig}), calculated numerically in \cite{knan}, and
analytically in \cite{laprem} is used

\beq    \label{pauli3}
F^{(3)}_2(0)=
\left\{\frac{83}{72}\pi^2\zeta(3)-\frac{215}{24}\zeta(5)
+\frac{100}{3}\left[(a_4+\frac{1}{24}\ln^4 2)
-\frac{1}{24}\pi^2\ln^22\right]-\frac{239}{2~160}\pi^4
\right.
\eeq
\[
\left.
+\frac{139}{18}\zeta(3)-\frac{298}{9}\pi^2\ln2+\frac{17~101}{810}\pi^2
+\frac{28~259}{5~184}\right\}\left(\frac{\alpha}{\pi}\right)^3
\approx 1.181~241~456\ldots~\left(\frac{\alpha}{\pi}\right)^3.
\]

Then one obtains for the Lamb shift

\beq      \label{pauli3lamb}
\Delta E_{|l=0}
=1.181~241~456\ldots~\frac{\alpha^3(Z\alpha)^4m}{\pi^3n^3}
\left(\frac{m_r}{m}\right)^3,
\eeq
\[
\Delta E_{|l\neq
0}=1.181~241~456\ldots~\frac{\alpha^3(Z\alpha)^4m}{\pi^3n^3}
\frac{j(j+1)-l(l+1)-3/4}{l(l+1)(2l+1)}\left(\frac{m_r}{m}\right)^2.
\]

\subsubsection{Polarization Operator Contribution}

\begin{figure}[ht]
\centerline{\epsfig{file=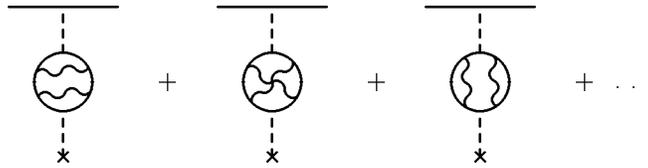}}
\vspace{0.5cm}
\caption{Examples of the three-loop contributions to the polarization
operator}
\label{thirdpolfig}
\end{figure}

In this case the analytic result for the low frequency asymptotics
of the third order polarization operator (see Fig.\
\ref{thirdpolfig}) \cite{baibroad} is used

\beq   \label{pol3}
\frac{\Pi(-{\bf k}^2)}{{\bf k}^4}_{|{\bf
k}^2=0}=-\left(\frac{8~135}{9~216}\zeta(3) -\frac{\pi^2\ln
2}{15}+\frac{23\pi^2}{360}-\frac{325~805}{373~248}\right)\frac{1}{m^2}
\left(\frac{\alpha}{\pi}\right)^3
\eeq
\[
\approx
-\frac{0.362~654~440\ldots}{m^2}\left(\frac{\alpha}{\pi}\right)^3,
\]

\noindent
and one obtains \cite{egpol}

\beq         \label{pol3lamb}
\Delta E=
-1.450~617~763\ldots\frac{\alpha^3(Z\alpha)^4m}{\pi^3n^3}
\left(\frac{m_r}{m}\right)^3
\delta_{l0}.
\eeq

\subsection{Total Correction of Order
$\alpha^{\lowercase{n}}(Z\alpha)^4\lowercase{m}$}

The total contribution of order $\alpha^n(Z\alpha)^4m$ is given by the sum
of corrections in \eq{diracslope}, \eq{pauli}, \eq{polariz}, \eq{ff},
\eq{pauli2lamb}, \eq{pol2}, \eq{ff3}, \eq{pauli3lamb}, \eq{pol3lamb}.
It is equal to

\beq
\Delta E_{|l=0}
=\left\{\left(\frac{4}{3}\ln\frac{m(Z\alpha)^{-2}}{m_r}-\frac{4}{3}\ln
k_0(n,0)+\frac{38}{45}\right)
\right.
\eeq
\[
+\left[-\frac{9}{4}\zeta(3)+\frac{3}{2}\pi^2\ln2
-\frac{10}{27}\pi^2-\frac{2~179}{648}\right]\left(\frac{\alpha}{\pi}\right)
\]
\[
+\left[\frac{85}{24}\zeta(5)
-\frac{121}{72}\pi^2\zeta(3)-\frac{84~071}{2~304}\zeta(3)
-\frac{71}{27}\ln^42
-\frac{239}{135}\pi^2\ln^22+\frac{4~787}{108}\pi^2\ln2
\right.
\]
\[
\left. \left.
-\frac{568}{9}a_4+\frac{1591}{3~240}\pi^4
-\frac{252~251}{9~720}\pi^2+\frac{679~441}{93~312}\right]
\left(\frac{\alpha}{\pi}\right)^2\right\}
\frac{\alpha(Z\alpha)^4m}{\pi n^3}\left(\frac{m_r}{m}\right)^3
\]
\[
=\left\{\left(\frac{4}{3}\ln\frac{m(Z\alpha)^{-2}}{m_r}-\frac{4}{3}\ln
k_0(n,0)+\frac{38}{45}\right)+0.538~95\ldots\left(\frac{\alpha}{\pi}\right)
\right.
\]
\[
\left.
+0.417~504\ldots\left(\frac{\alpha}{\pi}\right)^2\right\}
\frac{\alpha(Z\alpha)^4m}{\pi n^3}\left(\frac{m_r}{m}\right)^3,
\]

\noindent
for the $S$-states, and

\beq
\Delta E_{|l\neq0}
=\left\{-\frac{4}{3}\ln k_0(n,l)\left(\frac{m_r}{m}\right)^3
+\left[\frac{1}{2}+\left(\frac{3}{4}\zeta(3)-\frac{\pi^2}{2}\ln2
+\frac{\pi^2}{12}+\frac{197}{144}\right)\left(\frac{\alpha}{\pi}\right)
\right.\right.
\eeq
\[
+\left(\frac{83}{72}\pi^2\zeta(3)-\frac{215}{24}\zeta(5)
+\frac{100}{3}a_4+\frac{25}{18}\ln^4 2
-\frac{25}{18}\pi^2\ln^22-\frac{239}{2~160}\pi^4
\right.
\]
\[
\left.\left.
+\frac{139}{18}\zeta(3)-\frac{298}{9}\pi^2\ln2+\frac{17~101}{810}\pi^2
+\frac{28~259}{5~184}\right)\left(\frac{\alpha}{\pi}\right)^2\right]
\]
\[
\left.
\frac{j(j+1)-l(l+1)-3/4}{l(l+1)(2l+1)}
\left(\frac{m_r}{m}\right)^2 \right\}
\frac{\alpha(Z\alpha)^4m}{\pi n^3}
\]
\[
=\left\{-\frac{4}{3}\ln k_0(n,l)\left(\frac{m_r}{m}\right)^3
+\left[\frac{1}{2}-0.328~478~9\ldots\left(\frac{\alpha}{\pi}\right)
+1.181~241~456\ldots~\left(\frac{\alpha}{\pi}\right)^2\right]
\right.
\]
\[
\left.
\frac{j(j+1)-l(l+1)-3/4}{l(l+1)(2l+1)}
\left(\frac{m_r}{m}\right)^2 \right\}
\frac{\alpha(Z\alpha)^4m}{\pi n^3}
\]

\noindent
for the non-$S$-states.

Numerically corrections of order $\alpha^n(Z\alpha)^4m$ for the
lowest energy levels give

\beq
\Delta E(1S)=8~115~785.64~ \mbox{kHz},
\eeq
\[
\Delta E(2S)=1~037~814.43~ \mbox{kHz},
\]
\[
\Delta E(2P)=-12~846.46~ \mbox{kHz}.
\]

Contributions of order $\alpha^4(Z\alpha)^4m$ are suppressed by an
extra factor $\alpha/\pi$ in comparison with the corrections of order
$\alpha^3(Z\alpha)^4m$. Their expected magnitude is at the level of
hudredths of kHz even for the $1S$ state in hydrogen, and they are too
small to be of any phenomenological significance.

\subsection{Heavy Particle Polarization Contributions of Order
$\alpha(Z\alpha)^4\lowercase{m}$}\label{heavypartlambpol}

\begin{figure}[h]
\centerline{\epsfig{file=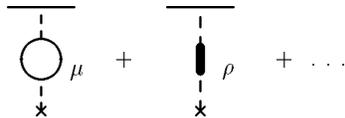,height=1.5cm}}
\vspace{0.5cm}
\caption{Muon-loop and hadron contributions to the polarization
operator}
\label{muonhardpolfig}
\end{figure}

We have considered above only radiative corrections containing virtual
photons and electrons. However, at the current level of accuracy one has to
consider also effects induced by the virtual muons and lightest strongly
interacting particles. The respective corrections to the electron anomalous
magnetic moment are well known \cite{knan} and are still too small to be of
any practical interest for the Lamb shift calculations. Heavy particle
contributions to the polarization operator numerically have the same
magnitude as polarization corrections of order $\alpha^3(Z\alpha)^4$.
Corrections to the low-frequency asymptotics of the polarization
operator are generated by the diagrams in Fig.\ \ref{muonhardpolfig}.
The muon loop contribution to the polarization operator

\beq   \label{muonpol}
\frac{\Pi(-{\bf k}^2)}{{\bf k}^4}_{|{\bf k}^2=0}=
-\frac{\alpha}{15\pi m_\mu^2}
\eeq

\noindent
immediately leads (compare \eq{polariz}) to an additional contribution to
the Lamb shift \cite{karmupol,es}

\beq        \label{muonpollmab}
\Delta E=-\frac{4}{15}\left(\frac{m_r}{m_\mu}\right)^2
\frac{\alpha(Z\alpha)^4}{\pi n^3}\:m_r\:\delta_{l0}.
\eeq

The hadronic polarization contribution to the Lamb shift was estimated
in a number of papers \cite{karmupol,es,fms99}. The light hadron
contribution to the polarization operator may easily be estimated with
the help of vector dominance

\beq   \label{lighthadronpol}
\frac{\Pi(-{\bf k}^2)}{{\bf k}^4}_{|{\bf k}^2=0}=-\sum_i
\frac{4\pi \alpha}{f_{v_i}^2 m_{v_i}^2}
\eeq

\noindent
where $m_{v_i}$ are the masses of the three lowest vector mesons and the
vector meson-photon vertex has the form $em_{v_i}^2/f_{v_i}$.

Estimating contributions of the heavy quark flavors with
the help of the free quark loops one obtains the total hadronic vacuum
polarization  contribution to the Lamb shift in the form \cite{es}

\beq        \label{nonelectr}
\Delta E=-4\left(\Sigma_{v_i}\frac{4\pi^2}{f^2_{v_i}
m^2_{v_i}}+\frac{2}{3}\frac{1}{\mbox{1 GeV}^2}\right)
\frac{\alpha(Z\alpha)^4}{\pi n^3}\:m\:\delta_{l0},
\eeq

Numerically this correction is $-3.18$ kHz for the $1S$-state and
$-0.04$ kHz for the $2S$-state in hydrogen. A compatible but a
more accurate estimate for the heavy particle contribution to the $1S$
Lamb shift $-3.40(7)$ kHz was obtained in \cite{fms99} from the
analysis of the experimental data on the low energy $e^+e^-$
annihilation\footnote{It is not obvious that this contribution should
be included in the phenomenological analysis of the Lamb shift
measurements, since experimentally it is indistinguishable from an
additional contribution to the proton charge radius. We will return to
this problem below in Section \ref{empnuklrad}.}.

\begin{center}
\underline{Table II. Contributions of Order $\alpha^n(Z\alpha)^4m$}
\nopagebreak

\begin{tabular}{|l|rl|c|c|}
\hline
& $4\frac{\alpha(Z\alpha)^4}{\pi
n^3}(\frac{m_r}{m})^3m\approx\frac{3~250~137.65(4)}{n^3}$ kHz &   	&
$\Delta E(1S)$ kHz &$\Delta E(2S)$  kHz
\\
\hline\hline
Bethe(1947)\cite{bethe} &   $$ & &&\\
French,Weisskopf(1949)\cite{fw}&
$\left[\frac{1}{3}\ln\frac{m(Z\alpha)^{-2}}{m_r}+\frac{11}{72}
\right]\delta_{l0}
-\frac{1}{3}\ln k_0(n,l)$    & &$7~925~175.26(9)$&$1~013~988.13(1)$\\
Kroll,Lamb(1949)\cite{kl}&   $$       & &&
\\ \hline
&   $$       & &&\\
Pauli FF $l=0$&$\frac{1}{8}$& & $$&\\
&   $$       & &$406~267.21$&$50~783.40$\\
\hline
Pauli FF $l\neq0$&$\frac{j(j+1)-l(l+1)-3/4}{8l(l+1)(2l+1)}\frac{m}{m_r}$&
&$$&\\
&   $$       & &&\\
\hline
Vacuum Polarization&   $$       & &&\\
Uehling (1935)\cite{uehling}&$-\frac{1}{15}\delta_{l0}$ & &
$-216~675.84$&$-27~084.48$\\
&   $$       & &&
\\
\hline Dirac FF&   $$       & &&\\
Appelquist,&   $[-\frac{3}{4}\zeta(3)
+\frac{\pi^2}{2}\ln2-\frac{49}{432}\pi^2
-\frac{4~819}{5~184}]\frac{\alpha}{\pi}\delta_{l0}$       & &&\\
Brodsky(1970)\cite{ab}&   $$       & &&\\
Barbieri,Mignaco,&
$\approx {0.469~941~4\dots}\frac{\alpha}{\pi}\delta_{l0}$       &
&$3~547.82$&$443.48$
\\
Remiddi(1971)\cite{bmr}&   $$       & &&
\\ \hline
Pauli FF $l=0$ &$(\frac{3}{16}\zeta(3)-\frac{\pi^2}{8}\ln2
+\frac{\pi^2}{48}+\frac{197}{576})\frac{\alpha}{\pi}$ &&&\\
Sommerfield (1957)\cite{somm}&   $$       & &&\\
Peterman (1957)\cite{pet} &$\approx-0.082~119~7\dots\frac{\alpha}{\pi}$&
&$-619.96$&$-77.50$ \\
&   $$       && &\\
\hline
Pauli FF $l\neq0$ &$(\frac{3}{16}\zeta(3)-\frac{\pi^2}{8}\ln2
+\frac{\pi^2}{48}+\frac{197}{576})
$ &&&\\
Sommerfield (1957)\cite{somm}&   $\frac{j(j+1)-l(l+1)-3/4}{l(l+1)(2l+1)}
\frac{\alpha}{\pi}\frac{m}{m_r}$       & &&\\
Peterman (1957)\cite{pet} &
$\approx-0.082~119~7\ldots\frac{j(j+1)-l(l+1)-3/4}{l(l+1)(2l+1)}
\frac{m}{m_r}\frac{\alpha}{\pi}$&
&$$ &\\
&   $$       & &&\\
\hline
Vacuum Polarization&   $$       & &&\\
Baranger,Dyson,&$-\frac{41}{162}(\frac{\alpha}{\pi})
\delta_{l0}$ & &$-1~910.67$ &$-238.83$\\
Salpeter(1952)\cite{bds}&   $$       & &&
\\ \hline
Dirac FF&   $$       & &&\\
Melnikov,&   $[\frac{25}{8}\zeta(5)-\frac{17}{24}\pi^2\zeta(3)-
\frac{2929}{288}\zeta(3)-\frac{217}{9}a_4$  & &&\\
van Ritbergen(1999)\cite{melrit}
&   $-\frac{217}{216}\ln^4{2}-\frac{103}{1080}\pi^2\ln^2{2}
+ \frac{41671}{2160}\pi^2\ln2$       & &&\\
&   $+\frac{3899}{25920}\pi^4-
\frac{454979}{38880}\pi^2-\frac{77513}{186624}]
(\frac{\alpha}{\pi})^2\delta_{l0}$       & &&\\
&
$\approx {0.171~72~\dots}(\frac{\alpha}{\pi})^2\delta_{l0}$       &
&$3.01$&$0.38$
\\
&   $$       & &&
\\ \hline
&   $$       & &&\\
Pauli FF $l=0$ &$\{\frac{83}{288}\pi^2\zeta(3)
-\frac{215}{96}\zeta(5)+\frac{25}{3}[(a_4+\frac{1}{24}\ln^4 2)$ &&&\\
Kinoshita(1990)\cite{knan}&   $$       & &&\\
Laporta,Remiddi(1996)\cite{laprem}&
$-\frac{1}{24}\pi^2\ln^22]-\frac{239}{8~640}\pi^4
+\frac{139}{72}\zeta(3)$
&&&\\
&   $$       & &&\\
&$-\frac{149}{18}\pi^2\ln2+\frac{17~101}{3~240}\pi^2
+\frac{28~259}{20~736}\}(\frac{\alpha}{\pi})^2$&
&&\\
&   $$       & &&\\
&$
\approx 0.295~310~3\ldots~(\frac{\alpha}{\pi})^2$&&$5.18$&$0.65$\\
&   $$       & &&
\\ \hline
\end{tabular}
\end{center}

\newpage
\begin{center}
\underline{Table II. Contributions of Order $\alpha^n(Z\alpha)^4m$
(continuation)}
\nopagebreak
\begin{tabular}{|l|rl|c|c|}
\hline
&   $$       & &&\\
$l\neq0$,Kinoshita(1990)\cite{knan}&$\{\frac{83}{288}\pi^2\zeta(3)
-\frac{215}{96}\zeta(5)+\frac{25}{3}[(a_4+\frac{1}{24}\ln^4 2)$ &&&\\
&   $$       & &&\\
Laporta,Remiddi(1996)\cite{laprem}&
$-\frac{1}{24}\pi^2\ln^22]-\frac{239}{8~640}\pi^4
+\frac{139}{72}\zeta(3)-\frac{149}{18}\pi^2\ln2$
&&&\\
&   $$       & &&\\
&$+\frac{17~101}{3~240}\pi^2
+\frac{28~259}{20~736}\}
\frac{j(j+1)-l(l+1)-3/4}{l(l+1)(2l+1)}
\frac{m}{m_r}(\frac{\alpha}{\pi})^2
$&&&\\
&   $$       & &&\\
&$\approx 0.295~310~3\ldots~\frac{j(j+1)-l(l+1)-3/4}{l(l+1)(2l+1)}
\frac{m}{m_r}(\frac{\alpha}{\pi})^2$&&&\\
&   $$    &   & &
\\ \hline
Vacuum Polarization&   $$       & &&\\
Baikov,Broadhurst(1995)\cite{baibroad}&$-(\frac{8~135}{9~216}\zeta(3)
-\frac{\pi^2\ln 2}{15}+\frac{23\pi^2}{360}-\frac{325~805}{373~248})
\left(\frac{\alpha}{\pi}\right)^2\delta_{l0}$&&&\\
Eides,Grotch(1995)\cite{egpol}&$\approx
-0.362~654~4\ldots
\left(\frac{\alpha}{\pi}\right)^2\delta_{l0}$ & & $-6.36$&$-0.79$\\
&   $$       & &&
\\ \hline
Muonic Polarization &   $$       & &&\\
Karshenboim (1995) \cite{karmupol}&&&&\\
Eides, Shelyuto (1995) \cite{es}&$
-\frac{1}{15}(\frac{m}{m_\mu})^2$&&$-5.07$&$-0.63$\\
 \hline
Hadronic Polarization&   $$       & &&\\
&   $$       & &&\\
Karshenboim (1995) \cite{karmupol}&&&&\\
Eides, Shelyuto (1995) \cite{es}&$
-\Sigma_{v_i}
\frac{4\pi^2}{f^2_{v_i}m^2_{v_i}}-\frac{2}{3}\frac{1}{\mbox{1 GeV}^2}
$&&$-3.18$&$-0.40$\\
Friar,Martorell,Sprung (1999)\cite{fms99}&          & &&\\
\hline
\end{tabular}
\end{center}

\section{Radiative Corrections of Order $\alpha^{\lowercase{n}}
(Z\alpha)^5\lowercase{m}$}\label{radcoralnzan5}

\subsection{Skeleton Integral Approach to Calculations of Radiative
Corrections} \label{skeletonbas}

We have seen  above that calculation of the corrections of order
$\alpha^n(Z\alpha)^4m$  ($n>1$) reduces to calculation of higher order
corrections to the properties of a free electron and to the photon
propagator, namely to calculation of the slope of the electron Dirac
form factor and anomalous magnetic moment, and to calculation of the
leading term in the low-frequency expansion of the polarization
operator. Hence, these contributions to the Lamb shift are independent
of any features of the bound state. A nontrivial interplay between
radiative corrections and binding effects arises first in calculation
of contributions of order $\alpha(Z\alpha)^5m$, and in calculations of
higher order terms in the combined expansion over $\alpha$ and
$Z\alpha$.

\begin{figure}[h]
\centerline{\epsfig{file=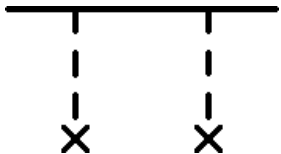,height=2cm}}
\vspace{0.5cm}
\caption{Skeleton diagram with two exchanged Coulomb photons}
\label{coulombskelfig}
\end{figure}

Calculation of the contribution of order $\alpha^n(Z\alpha)^5m$ to the
energy shift is even simpler than calculation of the leading order
contribution to the Lamb shift because the scattering approximation is
sufficient in this case \cite{kks1,kks2,bbf}. Formally this correction is
induced by kernels with at least two-photon exchanges, and in analogy
with the leading order contribution one could also anticipate the appearance
of irreducible kernels with higher number of exchanges. This does not
happen, however, as can be proved formally, but in fact no formal proof is
needed. First one has to realize that for high exchanged momenta expansion
in $Z\alpha$ is valid, and addition of any extra exchanged photon
always produces an extra power of $Z\alpha$.  Hence, in the high-momentum
region only diagrams with two exchanged photons are relevant. Treatment of
the low-momentum region is greatly facilitated by a very general feature
of the Feynman diagrams, namely that the infrared behavior of
any radiatively corrected Feynman diagram (or more accurately any gauge
invariant sum of Feynman diagrams) is milder than the behavior of the
skeleton diagram. Consider the matrix element in momentum space of the
diagrams in Fig.\ \ref{coulombskelfig} with two exchanged Coulomb photons
between the Schr\"odinger-Coulomb wave functions. We will take the external
electron momenta to be on-shell and to have vanishing space components. It
is then easy to see that the contribution of such a diagram to the Lamb
shift is given by the infrared divergent integral

\beq        \label{nonrecskel}
-\frac{16(Z\alpha)^5}{\pi
n^3}\left(\frac{m_r}{m}\right)^3\:m\int_0^\infty\frac{dk}{k^4}
\:\delta_{l0},
\eeq

\noindent
where $k$ is the dimensionless momentum of the exchanged photon
measured in the units of the electron mass. This divergence has a
simple physical interpretation. If we do not ignore small virtualities
of the external electron lines and the external wave functions this
two-Coulomb exchange adds one extra rung to the Coulomb wave function
and should simply reproduce it. The naive infrared divergence above
would be regularized at the characteristic atomic scale $mZ\alpha$.
Hence, it is evident that the kernel with two-photon exchange is
already taken into account in the effective Dirac equation above and
there is no need to try to consider it as a perturbation. Let us
consider now radiative photon insertions in the electron line (see
Fig.\ \ref{elineradinscoulfig}). Account of these corrections
effectively leads to insertion of an additional factor $L(k)$ in the
divergent integral above, and while this factor has at most a
logarithmic asymptotic behavior at large momenta and does not spoil the
ultraviolet convergence of the integral, in the low momentum region it
behaves as $L(k)\sim k^2$ (again up to logarithmic factors), and
improves the low frequency behavior of the integrand.  However, the
integrand is still divergent even after inclusion of the radiative
corrections because the two-photon-exchange box diagram, even with
radiative corrections, contains a contribution of the previous order in
$Z\alpha$, namely the main contribution to the Lamb shift induced by
the electron form factor. This spurious contribution may be easily
removed by subtracting the leading low momentum term from $L(k)/k^4$.
The result of the subtraction is a convergent integral which is
responsible for the correction of order $\alpha(Z\alpha)^5$. As an
additional bonus of this approach one does not need to worry about the
ultraviolet divergence of the one-loop radiative corrections. The
subtraction automatically eliminates any ultraviolet divergent terms
and the result is both ultraviolet and infrared finite.

Due to radiative insertions low integration momenta (of atomic order
$mZ\alpha$) are suppressed in the exchange loops and the effective
integration momenta are of order $m$. Hence, one may neglect the small
virtuality of external fermion lines and calculate the above diagrams with
on-mass-shell external momenta. Contributions to the Lamb shift are given
by the product of the square of the Schr{\"o}dinger-Coulomb wave function at
the origin $|\psi(0)|^2$ and the diagram. Under these conditions
the diagrams in Fig.\ \ref{elineradinscoulfig} comprise a gauge
invariant set and may easily be calculated.

Contributions of the diagrams with more than two exchanged Coulomb photons
are of higher order in $Z\alpha$. This is obvious for the high
exchanged momenta integration region. It is not difficult to
demonstrate that in the Yennie gauge \cite{abr,friedyennie,eksyennie}
contributions from the low exchanged momentum region to the matrix
element with the on-shell external electron lines remain infrared
finite, and hence, cannot produce any correction of order
$\alpha(Z\alpha)^5$. Since the sum of diagrams with the on-shell
external electron lines is gauge invariant this is true in any gauge.
It is also clear that small virtuality of the external electron lines
would lead to an additional suppression of the matrix element under
consideration, and, hence, it is sufficient to consider only two-photon
exchanges for calculation of all corrections of order
$\alpha(Z\alpha)^5$.

The magnitude of the correction of order $\alpha(Z\alpha)^5$ may be easily
estimated before the calculation is carried out. We need to take into
account the skeleton factor $4m(Z\alpha)^4/n^3$ discussed above in
Section \ref{lambqualmagn}, and multiply it by an extra factor
$\alpha(Z\alpha)$. Naively, one could expect a somewhat smaller factor
$\alpha(Z\alpha)/\pi$. However, it is well known that a convergent
diagram with two external photons always produces an extra factor $\pi$
in the numerator, thus compensating the factor $\pi$ in the denominator
generated by the radiative correction.  Hence, calculation of the
correction of order $\alpha(Z\alpha)^5$ should lead to a numerical
factor of order unity multiplied by $4m\alpha(Z\alpha)^5/n^3$.

\subsection{Radiative Corrections of Order
$\alpha(Z\alpha)^5\lowercase{m}$}\label{alphazalpha5nonrec}

\subsubsection{Correction Induced by the Radiative Insertions in the
Electron Line}

\begin{figure}[h]
\centerline{\epsfig{file=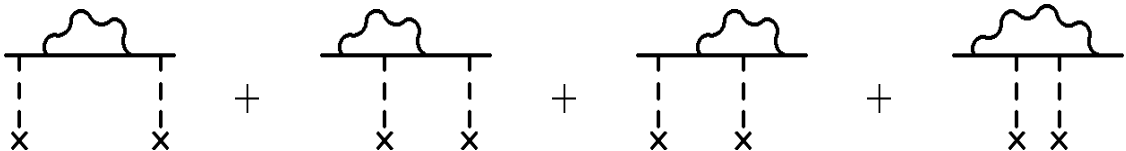,height=2cm,width=15cm}}
\vspace{0.5cm}
\caption{Radiative insertions in the electron line}
\label{elineradinscoulfig}
\end{figure}

This correction is generated by the sum of all possible radiative insertions
in the electron line in Fig.\ \ref{elineradinscoulfig}. In the approach
described above one has to calculate the electron factor corresponding to
the sum of all radiative corrections in the electron line, make the
necessary subtraction of the leading infrared asymptote, insert the
subtracted expression in the intgrand in \eq{nonrecskel}, and then
integrate over the exchanged momentum.  This leads to the result

\beq   \label{aza5}
\Delta E=4\left(1+\frac{11}{128}-\frac{1}{2}\ln
2\right)\frac{\alpha(Z\alpha)^5}{n^3}
\left(\frac{m_r}{m}\right)^3m\:\delta_{l0},
\eeq

\noindent
which was first obtained in \cite{kks1,kks2,bbf} in other approaches.
Note that numerically $1+11/128-1/2\ln 2\approx 0.739$ in excellent
agreement with the qualitative considerations above.

\subsubsection{Correction Induced by the Polarization Insertions in the
External Photons}

\begin{figure}[ht]
\centerline{\epsfig{file=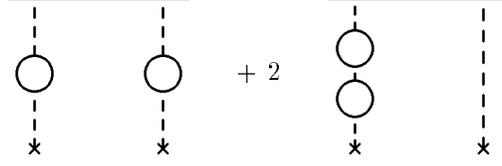}}
\vspace{0.5cm}
\caption{Polarization insertions in the Coulomb lines}
\label{polarizcoulfig}
\end{figure}

The correction of order $\alpha(Z\alpha)^5$ induced by the polarization
operator insertions in the external photon lines in Fig.\
\ref{polarizcoulfig} was obtained in \cite{kks1,kks2,bbf} and may again
be calculated in the skeleton integral approach. We will use the
simplicity of the one-loop polarization operator, and perform this
calculation in more detail in order to illustrate the general
considerations above. For calculation of the respective contribution
one has to insert the polarization operator in the skeleton integrand in
\eq{nonrecskel}

\beq                    \label{sub}
\frac{1}{k^2}\rightarrow \frac{\alpha}{\pi}I_{1}(k),
\eeq

\noindent
where

\beq
{I_1(k)}= \int_0^1 dv \frac{v^2(1-v^2/3)}{4+(1-v^2)k^2}\:.
\eeq

\noindent
Of course, the skeleton integral still diverges in the infrared  after
this substitution since

\beq
I_{1}(0)=\frac{1}{15}.
\eeq

\noindent
This linear infrared divergence $dk/k^2$ is effectively cut off at the
characteristic atomic scale $mZ\alpha$, it lowers the power of the
factor $Z\alpha$, respective would be divergent contribution turns out
to be of order $\alpha(Z\alpha)^4$, and corresponds to the polarization
part of the leading order contribution to the Lamb shift. We carry out
the subtraction of the leading low frequency asymptote of the
polarization operator insertion, which corresponds to the subtraction
of the leading low frequency asymtote in the integrand for the
contribution to the energy shift

\beq
{\tilde I}_{1}(k)\equiv I_{1}(k)-I_{1}(0)
=-\frac{k^2}{4}\int_0^1 dv \frac{v^2(1-v^2)(1-v^2/3)}{4+(1-v^2)k^2}
\eeq

\noindent
and substitute the subtracted expression in the formula for the Lamb
shift in \eq{nonrecskel}. We also insert an additional factor
$2$ in order to take into account possible insertions of the
polarization operator in both photon lines. Then

\beq        \label{aza5pol}
\Delta E=-\:m\:\left(\frac{m_r}{m}\right)^3
\frac{\alpha(Z\alpha)^5}{\pi^2n^3}
\frac{32}{(1-\frac{m^2}{M^2})}\int_0^\infty dk\frac{{\tilde I}_{1}(k)}{k^2}
\delta_{l0}
\eeq
\[
=\:m\:\left(\frac{m_r}{m}\right)^3
\frac{\alpha(Z\alpha)^5}{\pi^2
n^3}\frac{8}{(1-\frac{m^2}{M^2})}\int_0^1dv \int_0^\infty dk
\frac{v^2(1-v^2)(1-v^2/3)}{4+(1-v^2)k^2}\delta_{l0}
\]
\[
=\frac{5}{48}\frac{\alpha(Z\alpha)^5}{n^3}\:\left(\frac{m_r}{m}\right)^3
\:m\:
\delta_{l0}.
\]

We have restored in \eq{aza5pol} the characteristic factor
${1}/(1-{m^2}/{M^2})$ which was omitted in \eq{nonrecskel}, but
which naturally arises in the skeleton integral. However, it is easy to see
that an error generated by the omission of this factor is only about $0.02$
kHz even for the electron-line contribution to the $1S$ level shift, and,
hence, this correction may be safely omitted at the present level of
experimental accuracy.

\subsubsection{Total Correction of Order $\alpha(Z\alpha)^5\lowercase{m}$}

The total correction of order $\alpha(Z\alpha)^5m$ is given by the sum of
contributions in \eq{aza5},\eq{aza5pol}

\beq   \label{aza5tot}
\Delta E=4\left(1+\frac{11}{128}+\frac{5}{192}
-\frac{1}{2}\ln 2\right)\frac{\alpha(Z\alpha)^5}{n^3}
\left(\frac{m_r}{m}\right)^3m\:\delta_{l0}
\eeq
\[
=3.061~622\ldots\frac{\alpha(Z\alpha)^5}{n^3}
\left(\frac{m_r}{m}\right)^3m\:\delta_{l0}
\]
\[
=57~030.70~{\rm kHz}_{|n=1},
\]
\[
=7~128.84~{\rm kHz}_{|n=2}.
\]

\subsection{Corrections of Order $\alpha^2(Z\alpha)^5\lowercase{m}$}
\label{a2za5chap}

\begin{figure}
\centerline{\epsfig{file=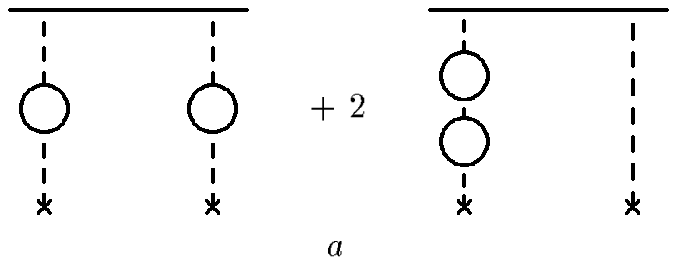,height=2.5cm,width=7cm}
\hskip1cm\epsfig{file=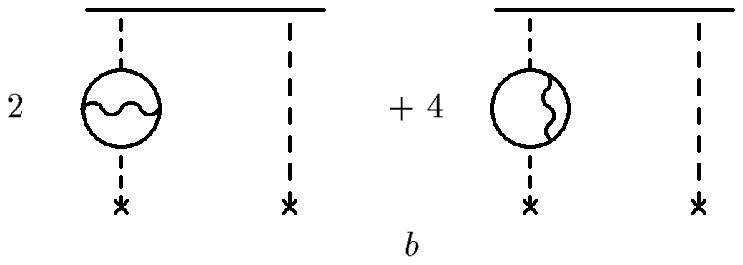,height=2.5cm,width=7cm}}
\vspace{0.8cm}
\centerline{\epsfig{file=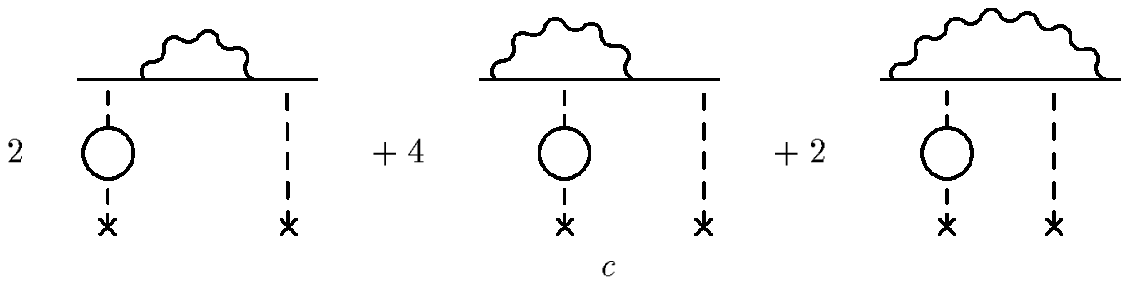,height=3cm}}
\vspace{0.8cm}
\centerline{\epsfig{file=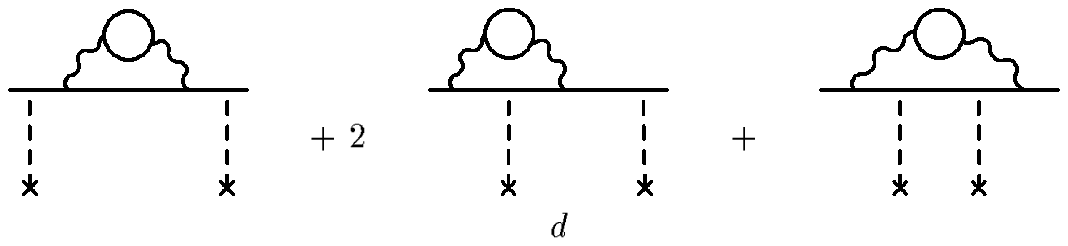,height=2.8cm,width=14cm}}
\vspace{0.8cm}
\centerline{\epsfig{file=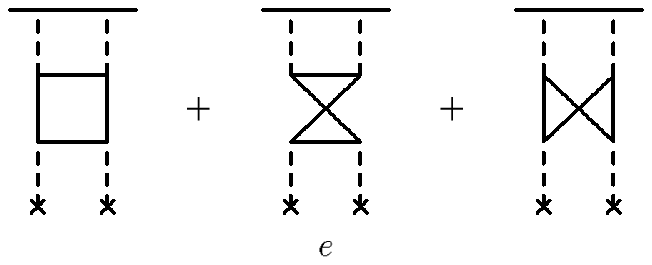,height=3cm}\hskip3cm
\epsfig{file=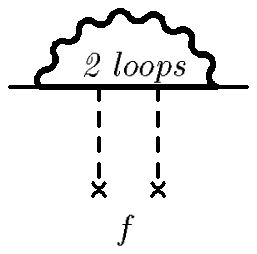,height=3cm}}
\vspace{0.5cm}
\caption{Six gauge invariant sets of diagrams for corrections of order
$\alpha^2(Z\alpha)^5m$}
\label{6setscoulfig}
\end{figure}

Corrections of order $\alpha^2(Z\alpha)^5$ have the same physical
origin as corrections of  order $\alpha(Z\alpha)^5$, and the scattering
approximation is sufficient for their calculation \cite{ego}. We
consider now corrections of higher order in $\alpha$ than in the
previous section and there is a larger variety of relevant graphs. All
six gauge invariant sets of diagrams \cite{ego} which produce
corrections of order $\alpha^2(Z\alpha)^5$  are presented in Fig.\
\ref{6setscoulfig}. The blob called "$2~~loops$" in Fig.\
\ref{6setscoulfig}  $(f)$ means the gauge invariant sum of diagrams
with all possible insertions of two radiative photons in the lectron
line. All diagrams in Fig.\ \ref{6setscoulfig} may be obtained  from
the skeleton diagram in Fig.\ \ref{coulombskelfig} with the help of
different two-loop radiative insertions. As in the case of the
corrections of order $\alpha(Z\alpha)^5$, corrections to the energy
shifts are given by the matrix elements of the diagrams in Fig.\
\ref{6setscoulfig} calculated between free electron spinors with all
external electron lines on the mass shell, projected on the respective
spin states, and multiplied by the square of the
Schr\"{o}dinger-Coulomb wave function  at the origin \cite{ego}.

It should be mentioned that some of the diagrams under consideration
contain contributions of the previous order in $Z\alpha$. These
contributions  are produced by the terms proportional to the exchanged
momentum squared in the low-frequency asymptotic expansion of the radiative
corrections, and are connected with integration over external photon
momenta of characteristic atomic scale $mZ\alpha$. The
scattering approximation is inadequate for their calculation. In the
skeleton integral approach these previous order contributions arise as
powerlike infrared divergences in the final integration over the exchanged
momentum. We subtract leading low-frequency terms in the low-frequency
asymptotic expansions of the integrands, when necessary, and thus remove
the spurious previous order contributions.

\subsubsection{One-Loop Polarization Insertions in the Coulomb Lines}

The simplest correction is induced by the diagrams in Fig.\
\ref{6setscoulfig} $(a)$ with two insertions of the one-loop vacuum
polarization in the external photon lines. The contribution to the Lamb
shift is given by the insertion of the one-loop polarization operator
squared $I_1^2(k)$ in the skeleton integral in \eq{nonrecskel}, and
taking into account the multiplicity factor 3 one easily obtains
\cite{ego,pach1,lap}

\beq     \label{onelooppol}
\Delta E=-\frac{48\alpha^2(Z\alpha)^5}{\pi^3
n^3}\left(\frac{m_r}{m}\right)^3m\int_0^\infty dk
I^2_1(k)\:\delta_{l0}=-\frac{23}{378}\frac{\alpha^2(Z\alpha)^5}{\pi n^3}
\left(\frac{m_r}{m}\right)^3m\:\delta_{l0}.
\eeq

\subsubsection{Insertions of the Irreducible Two-Loop Polarization in the
Coulomb Lines}

The naive insertion $1/k^2\rightarrow I_2(k)$ of the irreducible two-loop
vacuum polarization operator $I_2(k)$ \cite{kalsab,schwinger} in the
skeleton integral in \eq{nonrecskel} would lead to an infrared divergent
integral for the diagrams in Fig.\ \ref{6setscoulfig} $(b)$. This divergence
reflects the existence of the correction of the previous order in $Z\alpha$
connected with the two-loop irreducible polarization. This contribution of
order $\alpha^2(Z\alpha)^4m$ was discussed in Section \ref{lambalpha2zal4},
and as we have seen the respective contribution to the Lamb shift is given
simply by the product of the Schr\"odinger-Coulomb wave function
squared at the origin and the leading low-frequency term of the
function $I_2(0)$. In terms of the loop momentum integration this means
that the relevant loop momenta are of the atomic scale $mZ\alpha$.
Subtraction of the value $I_2(0)$ from the function $I_2(k)$
effectively removes the previous order contribution (the low momentum
region) from the loop integral and one obtains the radiative correction
of order $\alpha^2(Z\alpha)^5m$ generated by the irreducible two-loop
polarization operator \cite{ego,pach1,lap}

\beq       \label{twolooppol}
\Delta E=-\frac{32\alpha^2(Z\alpha)^5}{\pi^3
n^3}\left(\frac{m_r}{m}\right)^3m\int_0^\infty \frac{dk}{k^2}
[I_2(k)-I_2(0)]\:\delta_{l0}
\eeq
\[
=\left(\frac{52}{63}\ln2-\frac{25}{63}\pi+\frac{15~647}{13~230}\right)
\frac{\alpha^2(Z\alpha)^5}{\pi n^3}
\left(\frac{m_r}{m}\right)^3m\:\delta_{l0}.
\]

\subsubsection{Insertion of One-Loop Electron Factor in the Electron Line
and of the One-Loop Polarization in the Coulomb Lines}

The next correction of order $\alpha^2(Z\alpha)^5$ is generated by the
gauge invariant set of diagrams in Fig.\ \ref{6setscoulfig} $(c)$. The
respective analytic expression is obtained from the skeleton integral by
simultaneous insertion in the integrand of the one-loop polarization
function $I_1(k)$ and of the expressions corresponding to all possible
insertions of the radiative photon in the electron line. It is simpler first
to obtain an explicit analytic expression for the sum of all these radiative
insertions in the electron line, which we call the one-loop electron factor
$L(k)$ (explicit expression for the electron factor in different forms may
be found in \cite{bg87,eg,eg4,egs}), and then to insert this electron factor
in the skeleton integral. It is easy to check explicitly that the resulting
integral for the radiative correction is both ultraviolet and infrared
finite.  The infrared finiteness nicely correlates with the physical
understanding that for these diagrams there is no correction of order
$\alpha^2(Z\alpha)^4$ generated at the atomic scale. The respective
integral for the radiative correction was calculated both numerically and
analytically \cite{eg,pach1,egs}, and the result has the following elegant
form

\beq         \label{radphotpol}
\Delta E=-\frac{32\alpha^2(Z\alpha)^5}{\pi^3
n^3}\left(\frac{m_r}{m}\right)^3m\int_0^\infty dk
L(k)I_1(k)\:\delta_{l0}
\eeq
\[
=
\left(\frac{8}{3}\ln^2\frac{1+\sqrt5}{2}-\frac{872}{63}
\sqrt5\ln\frac{1+\sqrt5}{2}  +\frac{628}{63}\ln2-
\frac{2\pi^2}{9}
+\frac{67~282}{6~615}\right)
\frac{\alpha^2\left(Z\alpha\right)^5}{\pi
n^3}\left(\frac{m_r}{m}\right)^3\:{m}\:\delta_{l0}.
\]

\subsubsection{One-Loop Polarization Insertions in the Radiative Electron
Factor}

This correction is induced by the gauge invariant set of diagrams in
Fig.\ \ref{6setscoulfig} $(d)$ with the polarization operator insertions in
the radiative photon. The respective radiatively corrected electron factor
is given by the expression \cite{eg4}

\beq  \label{polloopinsrt}
{\cal L}(k)=\int_0^1dv\frac{v^2(1-\frac{v^2}{3})}{1-v^2}L(k,\lambda),
\eeq

\noindent
where $L(k,\lambda)$ is just the one-loop electron factor used in
\eq{radphotpol} but with a finite photon mass $\lambda=4/(1-v^2)$.

Direct substitution of the radiatively corrected electron factor ${\cal
L}(k)$ in the skeleton integral in \eq{nonrecskel} would lead to an infrared
divergence. This divergence reflects existence in this case of the
correction of the previous order in $Z\alpha$ generated by the two-loop
insertions in the electron line. The magnitude of this previous order
correction is determined by the nonvanishing value of the electron factor
${\cal L}(k)$ at zero

\beq
{\cal L}(0)=-2F'_1(0)-\frac{1}{2}F_2(0),
\eeq

\noindent
which is simply a linear combination of the slope of the two-loop
Dirac form factor and the two-loop contribution to the electron anomalous
magnetic moment.

Subtraction of the radiatively corrected electron factor removes this
previous order contribution which was already considered above, and leads to
a finite integral for the correction of order $\alpha^2(Z\alpha)^5$
\cite{eg4,pach1}

\beq            \label{radelefact}
\Delta E=-\frac{16\alpha^2(Z\alpha)^5}{\pi^3
n^3}\left(\frac{m_r}{m}\right)^3m\int_0^\infty dk\frac{{\cal
L}(k)-{\cal L}(0)}{k^2}\:\delta_{l0}
\eeq
\[
=-0.072~90\ldots\frac{\alpha^2\left(Z\alpha\right)^5}{\pi
n^3}\left(\frac{m_r}{m}\right)^3\:{m}\:\delta_{l0}.
\]

\subsubsection{Light by Light Scattering Insertions in the External
Photons}

The diagrams in Fig.\ \ref{6setscoulfig} $(e)$ with the light by light
scattering insertions in the external photons do not generate corrections of
the previous order in $Z\alpha$. They are both ultraviolet and infrared
finite and respective calculations are in principle quite straightforward
though technically involved. Only numerical results were obtained for the
contributions to the Lamb shift \cite{pach1,eg5}

\beq         \label{lbl}
\Delta E=
-0.122~9\ldots\frac{\alpha^2\left(Z\alpha\right)^5}{\pi
n^3}\left(\frac{m_r}{m}\right)^3\:{m}\:\delta_{l0}.
\eeq

\subsubsection{Diagrams with Insertions of Two Radiative Photons in the
Electron Line}

As we have already seen, contributions of the diagrams with radiative
insertions in the electron line always dominate over the contributions of
the diagrams with radiative insertions in the external photon lines. This
property of the diagrams is due to the gauge invariance of QED. The
diagrams (radiative insertions) with the external photon lines should be
gauge invariant, and as a result transverse projectors correspond to each
external photon. These projectors are rational functions of external
momenta, and they additionally suppress low momentum integration regions in
the integrals for energy shifts. Respective projectors are of course missing
in the diagrams with insertions in the electron line. The low momentum
integration region is less suppressed in such diagrams, and hence they
generate larger contributions to the energy shifts.

\begin{figure}
\centerline{\epsfig{file=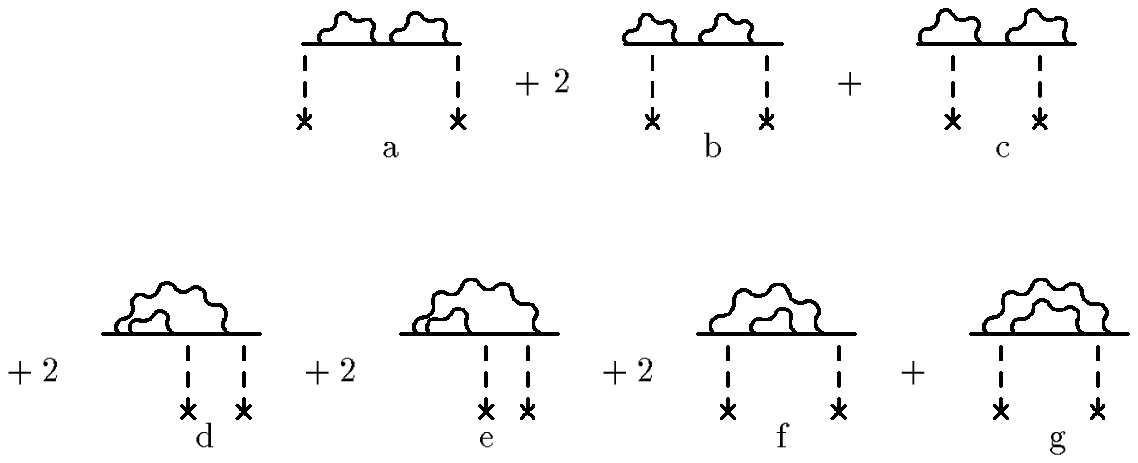,height=6cm}}
\vspace{1cm}
\centerline{\epsfig{file=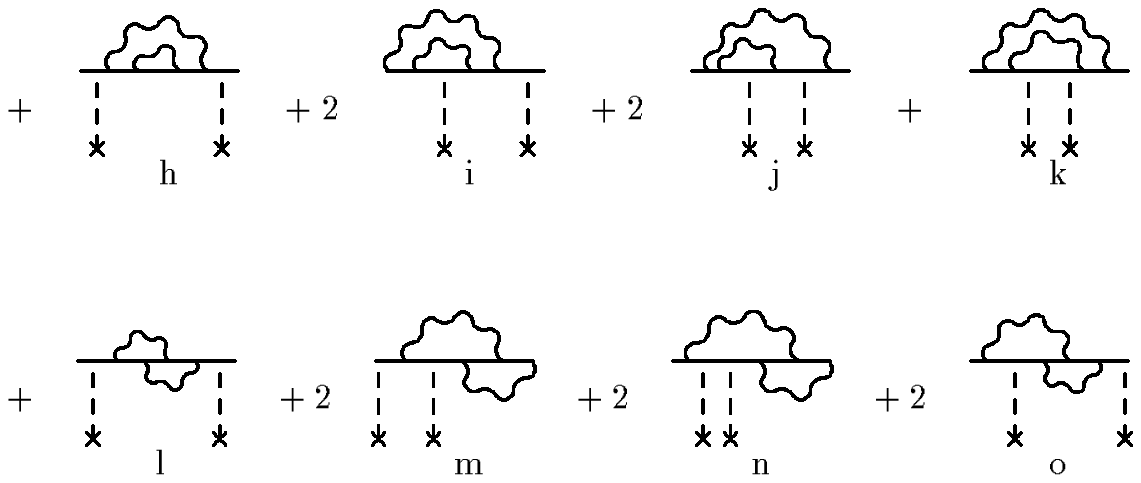,height=6cm}}
\vspace{1cm}
\centerline{\epsfig{file=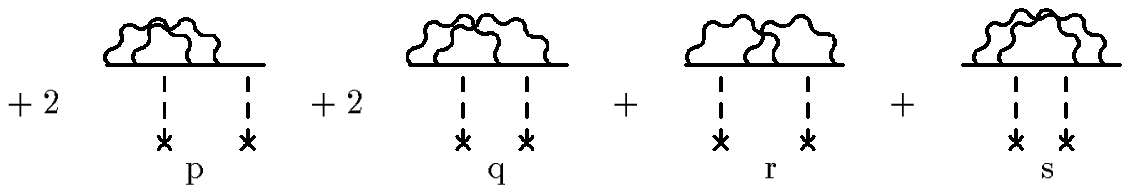,height=2.5cm}}
\vspace{0.5cm}
\caption{Nineteen topologically different diagrams with two
radiative photons insertions in the electron line}
\label{twoloopradcoulfig}
\end{figure}

This general property of radiative corrections clearly  manifests itself in
the case of six gauge invariant sets of diagrams in Fig.\
\ref{6setscoulfig}. By far the largest contribution of order
$\alpha^2(Z\alpha)^5$ to the Lamb shift is generated by the last gauge
invariant set of diagrams in Fig.\ \ref{6setscoulfig} $(f)$, which consists
of nineteen topologically different diagrams \cite{eksl1} presented in Fig.\
\ref{twoloopradcoulfig}. These nineteen graphs may be obtained from the
three graphs for the two-loop electron self-energy by insertion of two
external photons in all possible ways. Graphs in Fig.\
\ref{twoloopradcoulfig} $(a-c)$ are obtained from the two-loop reducible
electron self-energy diagram, graphs in Fig.\ \ref{twoloopradcoulfig}
$(d-k)$ are the result of all possible insertions of two external photons in
the rainbow self-energy diagram, and diagrams in Fig.\
\ref{twoloopradcoulfig} $(l-s)$ are connected with the overlapping two-loop
self-energy graph.  Calculation of the respective energy shift was initiated
in \cite{eksl1,eksl2}, where contributions induced by the diagrams  in Fig.\
\ref{twoloopradcoulfig} $(a-h)$ and in Fig.\ \ref{twoloopradcoulfig}
$(l)$ were obtained. Contribution of all nineteen diagrams to the Lamb
shift was first calculated in \cite{pach2}. In the framework of the skeleton
integral approach the calculation was completed in \cite{esjetp,es}
with the result

\beq         \label{tworad}
\Delta E=
-7.725(1)\ldots\frac{\alpha^2\left(Z\alpha\right)^5}{\pi
n^3}\left(\frac{m_r}{m}\right)^3\:{m}\:\delta_{l0}
\eeq

\noindent
which confirmed the one in \cite{pach2} but is about two orders of magnitude
more precise than the result in \cite{pach2,plwhkh}.

A few comments are due on the magnitude of this important result. It is
sometimes claimed in the literature that it has an unexpectedly large
magnitude. A brief glance at Table III is sufficient to convince oneself
that this is not the case. For the reader who followed closely the
discussion of the scales of different contributions above, it should be
clear that the natural scale for the correction under discussion is set
by the factor $4\alpha^2(Z\alpha)^5/(\pi n^3)m$. The coefficient before
this factor obtained in \eq{tworad} is about $-1.9$ and there is
nothing unusual in its magnitude for a numerical factor corresponding
to a radiative correction. It should be compared  with the respective
coefficient $0.739$ before the factor $4\alpha(Z\alpha)^5/n^3m$ in the
case of the electron-line contribution of the previous order in
$\alpha$.

The misunderstanding about the magnitude of the correction of order
$\alpha^2(Z\alpha)^5m$ has its roots in the idea that the expansion of
energy in a series over the parameter $Z\alpha$  at fixed power of
$\alpha$ should have coefficients of order one. As is clear from the
numerous discussions above, however natural such expansion might seem
from the point of view of calculations performed without expansion over
$Z\alpha$, there are no real reasons to expect that the coefficients
would be of the same order of magnitude in an expansion of this kind.
We have already seen that quite different physics is connected with
the different terms in expansion over $Z\alpha$. The terms of order
$\alpha^n(Z\alpha)^4$ (and $\alpha^n(Z\alpha)^6$, as we will see below) are
generated at large distances (exchanged momenta of order of the atomic
scale $mZ\alpha$) while terms of order $\alpha^n(Z\alpha)^5$ originate
from the small distances (exchanged momenta of order of the electron
mass $m$).  Hence, it should not be concluded that there would be a
simple way to figure out the relative magnitude of the successive
coefficients in an expansion over $Z\alpha$.  The situation is
different for expansion over $\alpha$ at fixed power of $Z\alpha$ since
the physics is the same independent of the power of $\alpha$, and
respective coefficients are all of order one, as in the series for
the radiative corrections in scattering problems.

\subsubsection{Total Correction of Order $\alpha^2(Z\alpha)^5\lowercase{m}$}

The total contribution of order $\alpha^2(Z\alpha)^5$ is given by the sum of
contributions in \eq{onelooppol}, \eq{twolooppol}, \eq{radphotpol},
\eq{radelefact}, \eq{lbl}, \eq{tworad} \cite{egs}

\begin{equation}  \label{total}
\Delta E=\left(\frac{8}{3}\ln^2\frac{1+\sqrt5}{2}-\frac{872}{63}
\sqrt5\ln\frac{1+\sqrt5}{2}  +\frac{680}{63}\ln2-
\frac{2\pi^2}{9}-\frac{25\pi}{63}\right.
\end{equation}
\[
\left.
+\frac{24~901}{2~205}
-7.921(1)\right)
\frac{\alpha^2\left(Z\alpha\right)^5}{\pi
n^3}\left(\frac{m_r}{m}\right)^3\:{m}\:\delta_{l0}
=-6.862(1)\;\frac{\alpha^2\left(Z\alpha\right)^5}{\pi
n^3}\left(\frac{m_r}{m}\right)^3\:{m}\:\delta_{l0}.
\]

\beq
\Delta
E=-6.862(1)\frac{\alpha^2(Z\alpha)^5}{\pi
n^3}\left(\frac{m_r}{m}\right)^3m\:\delta_{l0}
\eeq
\[
=-296.92~(4)~{\rm kHz}_{|n=1},
\]
\[
=-37.115~(5)~{\rm kHz}_{|n=2}.
\]

\subsection{Corrections of Order $\alpha^3(Z\alpha)^5\lowercase{m}$}

Corrections of order $\alpha^3(Z\alpha)^5$ have not been considered in the
literature. From the preceding discussion it is clear that their natural
scale is determined by the factor $4\alpha^3(Z\alpha)^5/(\pi^2 n^3)m$, which
is equal about $0.4$ kHz for the $1S$-state and about $0.05$ kHz for the
$2S$-state. Taking into account the rapid experimental progress in the field
these theoretical calculations may become necessary in the future, if
experimental accuracy in the measurement of the $1S$ Lamb shift at the
level of $1$ kHz, is achieved.

\begin{center}
\underline{Table III. Radiative Corrections of Order $\alpha^n(Z\alpha)^5m$}
\nopagebreak

\begin{tabular}{|l|rl|c|c|}    \hline
     &$4\frac{\alpha(Z\alpha)^5}{n^3}(\frac{m_r}{m})^3m$ &   &$\Delta E(1S)$
kHz&$\Delta E(2S)$  kHz
\\ \hline \hline
Electron-Line Insertions&$$&&&\\
&&&&\\
Karplus,Klein,Schwinger(1951)\cite{kks1,kks2}&$
(1+\frac{11}{128}-\frac{1}{2}\ln 2)\delta_{l0}$ & &$55~090.31$&$6~886.29$\\
Baranger,Bethe,Feynman(1951)\cite{bbf}&$$&&&\\
&&& &
\\ \hline
Polarization Contribution&&&&\\
&&&&\\
Karplus,Klein,Schwinger(1951)\cite{kks1,kks2}&$$ & & $$&\\
&&&&\\
Baranger,Bethe,Feynman(1951)\cite{bbf}&$\frac{5}{192}\delta_{l0}$&
&$1~940.38$&$242.55$\\
&&&  &
\\ \hline
One-Loop Polarization&$$&&&\\
Eides, Grotch,Owen (1992)\cite{ego}
&& & $$ &\\
Pachucki;Laporta(1993)\cite{pach1,lap}&
$-\frac{23}{1~512}\frac{\alpha}{\pi}\delta_{l0}$&&$-2.63$&$-0.33$\\
\hline
Two-Loop Polarization&$$&&&\\
Eides, Grotch,Owen (1992)\cite{ego} &$(\frac{13}{63}\ln2-\frac{25}{252}\pi
+\frac{15~647}{52~920})
\frac{\alpha}{\pi}\delta_{l0}$& & $21.99$ &$2.75$
\\
Pachucki;Laporta(1993)\cite{pach1,lap}&$$&&&\\
\hline
One-Loop Polarization&$$&&&\\
and Electron Factor&$$&&&\\
 &$$&&&\\
Eides, Grotch,(1993)\cite{eg}&$(\frac{2}{3}\ln^2\frac{1+\sqrt5}{2}
-\frac{218}{63}\sqrt5\ln\frac{1+\sqrt5}{2}  $& & $$&
\\
Pachucki(1993)\cite{pach1}&$+\frac{157}{63}\ln2-
\frac{\pi^2}{18}
+\frac{33~641}{13~230})\frac{\alpha}{\pi}\delta_{l0}$&&$26.45$&$3.31$\\
Eides,Grotch,Shelyuto(1997)\cite{egs} &$$&&&\\
\hline
Polarization insertion  &$$&&&\\
in the Electron Factor&$$&&&\\
Eides, Grotch,(1993)\cite{eg4}&$$&&&\\
 &$-0.018~2\frac{\alpha}{\pi}\delta_{l0}$& & $-3.15$&$-0.39$
\\
Pachucki(1993)\cite{pach1}&&&&\\
\hline
Light by Light Scattering&$$&&&\\
Pachucki(1993)\cite{pach1},&$$&&&\\
Eides, Grotch,Pebler(1994)\cite{eg5}&$$&&&\\
 &$-0.030~7\frac{\alpha}{\pi}\delta_{l0}$& & $-5.31$&$-0.66$
\\ \hline
Insertions of Two Radiative &$$&&&\\
Photons in the Electron Line&$$&&&\\
&$$&&&\\
Pachucki(1994)\cite{pach2},&$$&&&\\
Eides, Shelyuto,(1995)\cite{esjetp,es}&$-1.931~2(3)\frac{\alpha}{\pi}
\delta_{l0}$ &
& $-334.24(5)$&
$-41.78$\\
\hline
&$$&&&\\
&$(\pm1?)(\frac{\alpha}{\pi})^2\delta_{l0}$& & $\pm0.4$ &$\pm0.05$\\
\hline
\end{tabular}
\end{center}

\section{Radiative Corrections of Order $\alpha^{\lowercase{n}}
(Z\alpha)^6\lowercase{m}$}

\subsection{Radiative Corrections of Order $\alpha(Z\alpha)^6\lowercase{m}$}

\subsubsection{Logarithmic Contribution Induced by the Radiative Insertions
in the Electron Line}

\begin{figure}[h]
\centerline{\epsfig{file=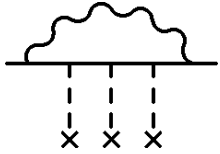,height=2cm}}
\vspace{0.5cm}
\caption{Diagram with three spanned Coulomb photons}
\label{3spancoulfig}
\end{figure}

Unlike the corrections of order $\alpha^n(Z\alpha)^5$, corrections of order
$\alpha^n(Z\alpha)^6$ depend on the large distance behavior of the wave
functions. Roughly speaking this happens because in order to produce a
correction containing six factors of $Z\alpha$ one needs at least three
exchange photons like in Fig.\ \ref{3spancoulfig}. The radiative photon
responsible for the additional factor of $\alpha$ does not suppress
completely the low-momentum region of the exchange integrals. As usual, long
distance contributions turn out to be state-dependent.

The leading correction of order $\alpha(Z\alpha)^6$ contains a logarithm
squared, which can be compared to the first power of logarithm in the
leading order contribution to the Lamb shift. One can understand
the appearance of the logarithm squared factor qualitatively. In the
leading order contribution to the Lamb shift the logarithm was completely
connected with the logarithmic infrared singularity of the electron form
factor. Now we have two exchanged loops and one should anticipate the
emergence of an exchanged logarithm generated by these loops. Note that the
diagram with one exchange loop (e.g, relevant for the correction of order
$\alpha(Z\alpha)^5$) cannot produce a logarithm, since in the external field
approximation the loop integration measure $d^3k$ is odd in the exchanged
momentum, while all other factors in the exchanged integral are even in the
exchanged momentum. Hence, in order to produce a logarithm which can only
arise from the dimensionless integrand it is necessary to consider an even
number of exchanged loops.  These simple remarks may also be understood
in another way if one recollects that in the relativistic corrections
to the Schr\"odinger-Coulomb wave function each power of logarithm is
multiplied by the factor $(Z\alpha)^2$ (this is evident if one expands
the exact Dirac wave function near the origin).

The logarithm squared term is, of course, state-independent since the
coefficient before this term is determined by the high momentum integration
region, where the dependence on the principal quantum number may enter only
via the value of the wave function at the origin squared. Terms linear in
the large logarithm are already state dependent. Logarithmic terms
were first calculated in \cite{layzer,frdyenn,eryen1,eryen2}. For the
$S$-states the logarithmic contribution is equal to

\beq   \label{loga6}
\Delta E_{log}|_{l=0}=
\left\{-\frac{1}{4}\ln^2[\frac{m(Z\alpha)^{-2}}{m_r}]
+\left[\frac{4}{3}\ln2+\ln\frac{2}{n}+\psi(n+1)-\psi(1)
\right.\right.
\eeq
\[
\left. \left.
-\frac{601}{720}-\frac{77}{180n^2}\right]\ln[\frac{m(Z\alpha)^{-2}}{m_r}]
\right\}
\frac{4\alpha(Z\alpha)^6}{\pi n^3}\left(\frac{m_r}{m}\right)^3\:{m},
\]

\noindent
where

\beq     \label{eulerpsi}
\psi(n)=\sum_1^{n-1}\frac{1}{k}+\psi(1),
\eeq

\noindent
is the logarithmic derivative of the Euler $\Gamma$-function
$\psi(x)=\Gamma'(x)/\Gamma(x)$, $\psi(1)=-\gamma$.

For non-$S$-states the state-independent logarithm squared term disappears
and the single-logarithmic contribution has the form

\beq
\Delta E_{log}|_{l\neq0}=
\left[(1-\frac{1}{n^2})(\frac{1}{30}+\frac{1}{12}\delta_{j,\frac{1}{2}})
\delta_{l1}+\frac{6-2l(l+1)/n^2}{3(2l+3)l(l+1)(4l^2-1)}\right]
\ln[\frac{m(Z\alpha)^{-2}}{m_r}]
\eeq
\[
\frac{4\alpha(Z\alpha)^6}{\pi n^3}\left(\frac{m_r}{m}\right)^3\:{m}.
\]

Calculation of the state-dependent nonlogarithmic contribution of order
$\alpha(Z\alpha)^6$ is a difficult task, and has not been done for an
arbitrary principal quantum number $n$. The first estimate of this
contribution was made in \cite{eryen2}. Next the problem was attacked from a
different angle \cite{erickson1,mohr}. Instead of calculating corrections
of order $\alpha(Z\alpha)^6$ an exact numerical calculation of all
contributions with one radiative photon, without expansion over
$Z\alpha$, was performed for comparatively large values of $Z$
($n=2$), and then the result was extrapolated to $Z=1$. In this way
an estimate of the sum of the contribution of order $\alpha(Z\alpha)^6$
and higher order contributions $\alpha(Z\alpha)^7$ was obtained (for
$n=2$ and $Z=1$). We will postpone discussion of the results obtained
in this way up to Section \ref{lambalphazalpha7}, dealing with
corrections of order $\alpha(Z\alpha)^7$, and will consider here only
the direct calculations of the contribution of order
$\alpha(Z\alpha)^6$.

\begin{figure}[ht]
\centerline{\epsfig{file=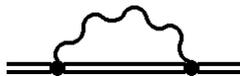,height=1cm}}
\vspace{0.5cm}
\caption{Exact second order self-energy operator}
\label{selfenfig}
\end{figure}

An exact formula in $Z\alpha$ for all nonrecoil corrections of order
$\alpha$ had the form

\beq  \label{seconselfen}
\Delta E=<n|\Sigma^{(2)}|n>,
\eeq

\noindent
where $\Sigma^{(2)}$ is an "exact" second-order self-energy operator for the
electron in the Coulomb field (see Fig.\ \ref{selfenfig}), and hence
contains the unmanageable exact Dirac-Coulomb Green function.  The real
problem with this formula is to extract useful information from it despite
the absence of a convenient expression for the Dirac-Coulomb Green function.
Numerical calculation without expansion over $Z\alpha$, mentioned in the
previous paragraph, was performed directly with the help of this formula.

A more precise than in \cite{eryen2} value of the nonlogarithmic
correction of order $\alpha(Z\alpha)^6$ for the $1S$-state was obtained
in \cite{sapir,palch}, with the help of a specially developed
"perturbation theory" for the Dirac-Coulomb Green function which
expressed this function in terms of the nonrelativistic
Schr\"odinger-Coulomb Green function \cite{hostler,schprop}. But the
real breakthrough was achieved in \cite{pacha6,pachalpha6}, where a new
very effective method of calculation was suggested and very precise
values of the nonlogarithmic corrections of order $\alpha(Z\alpha)^6$
for the $1S$- and $2S$-states were obtained. We will briefly discuss
the approach of papers \cite{pacha6,pachalpha6} in the next subsection.

\subsubsection{New Approach to Separation of the High- and Low-Momentum
Contributions. Nonlogarithmic Corrections}\label{pachappr}

Starting with the very first nonrelativistic consideration of the main
contribution to the Lamb shift \cite{bethe} separation of the contributions
of high- and low-frequency radiative photons became a characteristic feature
of the Lamb shift calculations. The main idea of this approach was already
explained in Section \ref{leadinglamb}, but we skipped over two obstacles
impeding effective implementation of this idea. Both problems are connected
with the effective realization of the matching procedure. In real
calculations it is not always obvious how to separate the two integration
regions in a consistent way, since in the high-momenta region one uses
explicitly relativistic expressions, while the starting point of the
calculation in the low-momenta region is the nonrelativistic dipole
approximation. The problem is aggravated by the inclination to use different
gauges in different regions, since the explicitly covariant Feynman gauge is
the simplest one for explicitly relativistic expressions in the high-momenta
region, while the Coulomb gauge is the gauge of choice in the
nonrelativistic region. In order to emphasize the seriousness of these
problems it suffices to mention that incorrect matching of high- and
low-frequency contributions in the initial calculations of Feynman and
Schwinger led to a significant delay in the publication of the first fully
relativistic Lamb shift calculation of French and Weisskopf
\cite{fw}\footnote{See fascinating description of this episode in
\cite{schweber}.}! It was a strange irony of history that due to these
difficulties it became common wisdom in the sixties that it is better
to try to avoid the separation of the contributions coming from
different momenta regions (or different distances) than to try to
invent an accurate matching procedure. A few citations are appropriate
here. Bjorken and Drell \cite{bd} wrote, having in mind the separation
procedure: "The reader may understandably be unhappy with this
procedure \ldots we recommend the recent treatment of Erickson and
Yennie \cite{eryen1,eryen2}, which avoids the division into soft and
hard photons". Schwinger \cite{schwinger} wrote:  "\ldots  there is a
moral here for us. The artificial separation of high and low
frequencies, which are handled in different ways, must be avoided." All
this was written even though it was understood that the separation of
the large and small distances was physically quite natural and the
contributions coming from large and small distances have a different
physical nature.  However, the distrust to the methods used for
separation of the small and large distances was well justified by the
lack of a regular method of separation.  Apparently different methods
were used for calculation of the high and low frequency contributions,
high frequency contributions being commonly treated in a covariant
four-dimensional approach, while old-fashioned nonrelativistic
perturbation theory was used for calculation of the low-frequency
contributions. Matching these contributions obtained in different
frameworks was an ambiguous and far from obvious procedure, more art
than science. As a result, despite the fact that the methods based on
separation of long and short distance contributions had led to some
spectacular results (see, e.g., \cite{sal,fm}), their self-consistency
remained suspect, especially when it was necessary to calculate the
contributions of higher order than in the classic works. It seemed more
or less obvious that in order to facilitate such calculations one
needed to develop uniform methods for treatment of both small and large
distances.

The actual development took, however, a different direction. Instead of
rejecting the separation of high and low frequencies, more elaborate methods
of matching respective contributions were developed in the last decade, and
the general attitude to separation of small and large distances radically
changed. Perhaps the first step to carefully separate the long and
short distances was done in \cite{gy2}, where the authors had rearranged the
old-fashioned perturbation theory in such a way that one contribution
emphasized the small momentum contributions and led to a Bethe
logarithm, while in the other the small momentum integration region was
naturally suppressed. Matching of both contributions in this approach was
more natural and automatic. However, the price for this was perhaps too
high, since the high momentum contribution was to be calculated in a
three-dimensional way, thus losing all advantages of the covariant
four-dimensional methods.

Almost all new approaches, the skeleton integral approach described
above in Section \ref{skeletonbas} (\cite{es} and references there),
$\epsilon$-method described in this section \cite{pacha6,pachalpha6},
nonrelativistic approach by Khriplovich and coworkers \cite{khriplovich},
nonrelativistic QED of Caswell and Lepage \cite{caswelllep}) not only make
separation of the small and large distances, but try to exploit it most
effectively. In some cases, when the whole contribution comes only from the
small distances, a rather simple approach to this problem is appropriate
(like in the calculation of corrections of order $\alpha^2(Z\alpha)^4$,
$\alpha^3(Z\alpha)^4$, $\alpha(Z\alpha)^5$ and $\alpha^2(Z\alpha)^5$ above,
more examples below) and the scattering approximation is often
sufficient. In such cases, would-be infrared divergences are powerlike.
They simply indicate the presence of the contributions of the previous
order in $Z\alpha$ and may safely be thrown away. In other cases, when
one encounters logarithms which get contributions both from the small
and large distances, a more accurate approach is necessary such as the
one described below. In any case "the separation of low and high
frequencies, which are handled in different ways" not only should not
be avoided but turns out to be a very convenient calculational tool and
clarifies the physical nature of the corrections under consideration.

An effective method to separate contributions of low- and high-momenta
avoiding at the same time the problems discussed above was suggested in
\cite{pacha6,pachalpha6}. Consider in more detail the exact expression
\eq{seconselfen} for the sum of all corrections of orders
$\alpha(Z\alpha)^nm$ ($n\geq1$) generated by the insertion of one
radiative photon in the electron line

\beq
\Delta E=e^2\int\frac{d^4k}{(2\pi)^4i}D_{\mu\nu}(k) <n|\gamma_\mu
G(p'-k;p-k)\gamma_\nu|n>,
\eeq

\noindent
where $G(p'-k;p-k)$ is the exact electron Green function in the
external Coulomb field. As was noted in \cite{pacha6,pachalpha6} one
can rotate the integration contour over the frequency of the radiative
photon in such a way that it encloses singularities along the positive
real axes in the $\omega~(k^0)$ plane. Then one considers separately
the region $\mbox{Re}\,\omega\leq\sigma$ (region I) and
$\mbox{Re}\,\omega\geq\sigma$ (region II), where
$m(Z\alpha)^2\ll\sigma\ll m(Z\alpha)$.  It is easy to see that due to the
structure of the singularities of the integrand, integration over $\bf
k$ in the region I also goes only over the momenta smaller than
$\sigma$ ($|\bf k|\leq\sigma$), while in the region II the final
integration over $\omega$ cuts off all would be infrared divergences of
the integral. Hence, effective separation of high- and low-momenta
integration regions is achieved in this way and, as was explained
above,  due to the choice of the magnitude of the parameter $\sigma$
all would be divergences should exactly cancel in the sum of
contributions of these regions. This cancellation provides an
additional effective method of control of the accuracy of all
calculations. It was also shown in \cite{pachalpha6}  that a change of gauge
in the low-frequency region changes the result of the calculations by a term
linear in $\sigma$. But anyway one should discard such contributions
matching high- and low-frequency contributions.  The matrix element of the
self-energy operator between the exact Coulomb-Dirac wave functions is gauge
invariant with respect to changes of gauge of the radiative photon
\cite{yenniefox}.  Hence, it is possible to use the simple Feynman gauge for
calculation of the high-momenta contribution, and the physical Coulomb gauge
in the low-momenta part. It should be clear now that this method resolves
all problems connected with the separation of the high- and low-momenta
contributions and thus provides an effective tool for calculation of all
corrections with insertion of one radiative photon in the electron line.
The calculation performed in \cite{pacha6,pachalpha6,pach99} successfully
reproduced all results of order $\alpha(Z\alpha)^4$ and $\alpha(Z\alpha)^5$
and produced a high precision result for the constant of order
$\alpha(Z\alpha)^6$

\beq
\Delta E_{non-log}(1S)= -30.924~15~(1)~
\frac{\alpha(Z\alpha)^6}{\pi }\left(\frac{m_r}{m}\right)^3\:{m},
\eeq
\[
\Delta E_{non-log}(2S)=-31.840~47~(1)~
\frac{\alpha(Z\alpha)^6}{8\pi}\left(\frac{m_r}{m}\right)^3\:{m}.
\]

Besides the high accuracy of this result two other features should be
mentioned. First, the state dependence of the constant is very weak, and
second, the scale of the constant is just of the magnitude one should
expect. In order to make this last point more transparent let us write the
total electron-line contribution of order $\alpha(Z\alpha)^6$ to the $1S$
energy shift in the form

\beq
\Delta E(1S)=
\left\{-\ln^2[\frac{m(Z\alpha)^{-2}}{m_r}]
+\left[\frac{28}{3}\ln2-\frac{21}{20}\right]\ln[\frac{m(Z\alpha)^{-2}}{m_r}]
-30.928~90
\right\}
\frac{\alpha(Z\alpha)^6}{\pi }\left(\frac{m_r}{m}\right)^3\:{m}
\eeq
\[
\approx
\left\{-\ln^2[\frac{m(Z\alpha)^{-2}}{m_r}]
+5.42\ln[\frac{m(Z\alpha)^{-2}}{m_r}]
-30.93
\right\}
\frac{\alpha(Z\alpha)^6}{\pi }\left(\frac{m_r}{m}\right)^3\:{m}.
\]

Now we see that the ratio  of the nonlogarithmic term and the coefficient
before the single-logarithmic term is about
$31/5.4\approx5.7\approx0.6\pi^2$.  It is well known that the logarithm
squared terms in QED are always accompanied by the single-logarithmic and
nonlogarithmic terms, and the nonlogarithmic terms are of order $\pi^2$ (in
relation with the current problem see, e.g., \cite{eryen1,eryen2}). This is
just what happens in the present case, as we have demonstrated.

Nonlogarithmic contributions of order $\alpha(Z\alpha)^6$ to the
energies of the $2P$, $3P$ and $4P$-states induced by the radiative photon
insertions in the electron line were obtained in the same framework in
\cite{jenpach,jsm}. We have collected the respective results in Table IV
in terms of the traditionally used coefficient $A_{60}$ \cite{eryen1} which
is defined by the relationship

\beq
\Delta E= A_{60}\:
\frac{\alpha(Z\alpha)^6}{\pi n^3}\left(\frac{m_r}{m}\right)^3\:{m}.
\eeq

\begin{center}
\underline{Table IV. Nonlogarithmic Coefficient $A_{60}$ }
\nopagebreak

\begin{tabular}{|l|rl|c|}
\hline
&$\frac{\alpha(Z\alpha)^6}{\pi n^3}(\frac{m_r}{m})^3m$ &   & kHz
\\ \hline  \hline
$1S$&$$&&$$ \\
Pachucki(1993)\cite{pacha6,pachalpha6,pach99}
&$-30.924~15(1)$&&$-1338.04$         \\
\hline
$2S$&&$$      &   \\
Pachucki(1993)\cite{pacha6,pachalpha6}
&$-31.840~47(1)
$&&$-172.21$         \\
\hline
$2P_\frac{1}{2}$&&$$        & \\
Jentschura,Pachucki(1996)\cite{jenpach}
&$-0.998~91(1)
$&&$-5.40$         \\
\hline
$2P_\frac{3}{2}$&&$$        & \\
Jentschura,Pachucki(1996)\cite{jenpach}
&$-0.503~37(1)
$&&$-2.72$         \\
\hline
$3P_\frac{1}{2}$&&$$  &       \\
Jentschura,Soff,Mohr(1997)\cite{jsm}
&$-1.147~68(1)
$&&$-1.84$         \\
\hline
$3P_\frac{3}{2}$&&$$  &       \\
Jentschura,Soff,Mohr(1997)\cite{jsm}
&$-0.597~56(1)
$&&$-0.96$         \\
\hline
$4P_\frac{1}{2}$&&$$  &       \\
Jentschura,Soff,Mohr(1997)\cite{jsm}
&$-1.195~68(1)
$&&$-0.81$         \\
\hline
$4P_\frac{3}{2}$&&$$  &       \\
Jentschura,Soff,Mohr(1997)\cite{jsm}
&$-0.630~94(1)
$&&$-0.43$         \\
\hline
\end{tabular}
\end{center}

\subsubsection{Correction Induced by the Radiative Insertions in the
External Photons} \label{lambalzal6pol}

\begin{figure}[h]
\centerline{\epsfig{file=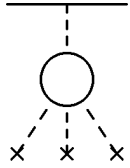}}
\vspace{0.5cm}
\caption{Wichmann-Kroll potential}
\label{wichkrolfig}
\end{figure}

\begin{figure}[h]
\centerline{\epsfig{file=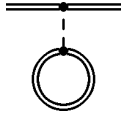,height=1.5cm}}
\vspace{0.5cm}
\caption{Total one-loop polarization potential in the external field}
\label{polpotfig}
\end{figure}

There are two kernels with radiative insertions in the external photon lines
which produce corrections of order $\alpha(Z\alpha)^6$ to the Lamb shift.
First is our old acquaintance -- one-loop polarization insertion in the
Coulomb line in Fig.\ \ref{polradfig}. Its Fourier transform is called the
Uehling potential \cite{uehling,serber}. The second kernel contains the
light-by-light scattering diagrams in Fig.\ \ref{wichkrolfig} with three
external photons originating from the Coulomb source.  The sum of all closed
electron loops in Fig.\ \ref{polpotfig} with one photon connected with the
electron line and an arbitrary number of Coulomb photons originating from
the Coulomb source may be considered as a radiatively corrected Coulomb
potential $V$. It generates a shift of the atomic energy levels

\beq
\Delta E=<n|V|n>.
\eeq

This potential and its effect on the energy levels were first considered
in \cite{wichkr}. Since each external Coulomb line brings an extra factor
$Z\alpha$ the energy shift generated by the Wichmann-Kroll potential
increases for large $Z$. For practical reasons the effects of the Uehling
and Wichmann-Kroll potentials were investigated mainly numerically and
without expansion in $Z\alpha$, since only such results could be compared
with the experiments. Now there exist many numerical results for vacuum
polarization contributions. In accordance with our emphasis on the analytic
results we will discuss here only analytic contributions of order
$\alpha(Z\alpha)^6$, and will return to numerical results in Section
\ref{lambalphzapl7coul}.

\paragraph{Uehling Potential Contribution.}
It is not difficult to present an exact formula containing all corrections
produced by the Uehling potential in Fig.\ \ref{polradfig} (compare with the
respective expression for the self-energy operator above)

\beq             \label{polarint}
\Delta E=4\pi(Z\alpha)<n|\frac{\Pi({\bf k}^2)}{{\bf k}^4}|n>.
\eeq

We have already seen that the matrix element of the first term of the
low-momentum expansion of the one-loop polarization operator between
the nonrelativistic Schr\"odinger-Coulomb wave functions produces
a correction of order $\alpha(Z\alpha)^4$. The next term in the low-momentum
expansion of the polarization operator pushes characteristic momenta in the
integrand to relativistic values, where the very nonrelativistic expansion
is no longer valid, and even makes the integral divergent if one tries to
calculate it between the nonrelativistic wave functions. Due to this effect
we preferred to calculate the correction of order $\alpha(Z\alpha)^5$
induced by the one-loop polarization insertion (as well as the correction of
order $\alpha^2(Z\alpha)^5$ induced by the two-loop polarization) in the
skeleton integral approach in Section \ref{radcoralnzan5}. Note that
all these corrections contribute only to the $S$-states. It is useful
to realize that both these calculations are, from another point of
view, simply results of approximate calculation of the integral in
\eq{polarint} with accuracy $(Z\alpha)^4$ and $(Z\alpha)^5$.  Our next
task is to calculate this integral with accuracy $(Z\alpha)^6$. In this
order both small atomic ($\sim mZ\alpha$) and large relativistic ($\sim
m$) momenta produce nonvanishing contributions to the integral, and as
a result we get nonvanishing contributions to the energy shifts also
for states with nonvanishing angular momenta.

Consider first corrections to the energy levels with nonvanishing angular
momenta. The respective wave functions vanish at the origin in
coordinate space, hence only small photon momenta contribute to the
integral, and one can use the first two terms in the nonrelativistic
expansion of the polarization operator

\beq  \label{2ndorderpol}
\frac{\Pi({\bf k}^2)}{{\bf k}^4}\approx
-\frac{\alpha}{15\pi m^2}+\frac{{\alpha\bf k}^2}{35m^4}
\eeq

\noindent
for calculation of these contributions \cite{ik97,jsm} (compare \eq{polariz}
above). Corrections of order $\alpha(Z\alpha)^6$ turn out to be nonvanishing
only for $l\leq1$. For $2P$-states these corrections were first calculated
in \cite{mohr}, and the result for arbitrary $P$-states \cite{mnf} has the
form

\beq
\Delta
E(nP_j)=-\frac{4}{15}\left(1-\frac{1}{n^2}\right)\left(\frac{1}{14}
+\frac{1}{4}\delta_{j\frac{1}{2}}\right)
\frac{\alpha(Z\alpha)^6}{\pi n^3}\left(\frac{m_r}{m}\right)^3\:{m}.
\eeq

The respective correction to the energy levels of $S$-states originates both
from the large and small distances since the Schr\"odinger-Coulomb wave
function in the $S$-states does not vanish at small distances. Hence, one
cannot immediately apply low-momenta expansion of the polarization operator
for calculation of the matrix element in \eq{polarint}. The leading
logarithmic state-independent contribution to the energy shift is still
completely determined by the first term in the low-momentum expansion of the
polarization operator in \eq{2ndorderpol}, but one has to consider the exact
expression for the polarization operator in order to obtain the
nonlogarithmic contribution.

The logarithmic term originates from the logarithmic correction to the
Schr\"odinger-Coulomb wave function which arises when one takes into
account the Darwin ($\delta$-function) potential which arises in the
nonrelativistic expansion of the Dirac Hamiltonian in the Coulomb field
(see, e.g., \cite{bd,blp} and \eq{darwin} below). Of course, the same
logarithm arises if, instead of calculating corrections to the
Schr\"odinger-Coulomb wave function, one expands the singular factor in
the Dirac-Coulomb wave function over $Z\alpha$. The correction to the
Schr\"odinger-Coulomb wave function $\Psi$ at the distances of order
$1/m$ has the form \cite{friedyennie}

\beq             \label{darwincor}
\delta\Psi=-\frac{1}{2}(Z\alpha)^2\ln(Z\alpha)\: \Psi,
\eeq

\noindent
and substituting this correction in \eq{polarint} one easily obtains the
leading logarithmic contribution to the energy shift
\cite{layzer,eryen1,eryen2}

\beq
\Delta E_{log}(nS)=-\frac{2}{15}\frac{\alpha(Z\alpha)^6}{\pi
n^3}\ln[\frac{m(Z\alpha)^{-2}}{m_r}]\left(\frac{m_r}{m}\right)^3m.
\eeq

Note that the numerical factor before the leading logarithm here is
simply the product of the respective numerical factors in the
correction to the wave-function in \eq{darwincor}, the low-frequency
asymptote of the one-loop polarization $-1/(15\pi)$ in
\eq{leadcoulpollambnaiv}, the factor $4\pi(Z\alpha)$ in \eq{polarint},
and factor $2$ which reflects that both wave functions in the matrix
element in \eq{polarint} have to be corrected.

Calculation of the nonlogarithmic contributions requires more effort.
Complete analytic results for the lowest states were first obtained by P.
Mohr \cite{mohr}

\beq  \label{vac1s}
\Delta
E(1S)=\frac{1}{15}\left[-\frac{1}{2}\ln[\frac{m(Z\alpha)^{-2}}{m_r}]
+\ln2-\frac{1289}{420}\right]
\frac{4\alpha(Z\alpha)^6}{\pi }\left(\frac{m_r}{m}\right)^3m,
\eeq
\[
\Delta E(2S)=\frac{1}{15}\left[-\frac{1}{2}\ln[\frac{m(Z\alpha)^{-2}}{m_r}]
-\frac{743}{240}\right]
\frac{4\alpha(Z\alpha)^6}{8\pi }\left(\frac{m_r}{m}\right)^3m.
\]

As was mentioned above, short distance contributions are state
independent and always cancel in the differences of the form $\Delta
E(1S)-n^3\Delta E(nS)$. This means that such state dependent
differences of energies contain only contributions of large distances
and are much easier to calculate, since one may employ a nonrelativistic
approximation\footnote{This well known feature was often used in the past.
For example, differences of the hyperfine splittings $\Delta E(1S)-n^3\Delta
E(nS)$ were calculated much earlier \cite{mittleman,zwanz1} than the
hyperfine splittings themselves.}. The Uehling potential contributions to
the difference of level shifts $\Delta E(1S)-n^3\Delta E(nS)$ were
calculated in \cite{ik97}  with the help of the nonrelativistic expansion of
the polarization operator in \eq{2ndorderpol}. The result of this
calculation, in conjunction with the Mohr result in \eq{vac1s}, leads to an
analytic expression for the Uehling potential contribution to the Lamb
shift

\beq
\Delta
E(nS)=\frac{1}{15}\left[-\frac{1}{2}\ln[\frac{m(Z\alpha)^{-2}}{m_r}]
-\frac{431}{105} +  (\psi(n+1)-\psi(1)) - \frac{2(n-1)}{n^2}
\right.
\eeq
\[
\left.
+\frac{1}{28n^2} - \ln\frac{n}{2}\right]
\frac{4\alpha(Z\alpha)^6}{\pi n^3}\left(\frac{m_r}{m}\right)^3m,
\]

\noindent
where $\psi(x)$ is the logarithmic derivative of the Euler
$\Gamma$-function, see \eq{eulerpsi}.

\paragraph{Wichmann-Kroll Potential Contribution.}

The only other contribution of order $\alpha(Z\alpha)^6$ connected with
the radiative insertions in the external photons is produced by the term
trilinear in $Z\alpha$ in the Wichmann-Kroll potential in Fig.\
\ref{wichkrolfig}. One may easily check that the first term in the small
momentum expansion of the Wichmann-Kroll potential has the form
\cite{wichkr,mohr83}

\beq       \label{wichmankrsmallmom}
V_{WK}(k)=\left(\frac{19}{45}-\frac{\pi^2}{27}\right)
\frac{\alpha(Z\alpha)^3}{m^2}.
\eeq

This potential generates the energy shift \cite{mohr76,mohr83}

\beq        \label{wichkrcalc}
\Delta
E=\left(\frac{19}{45}-\frac{\pi^2}{27}\right)
\frac{\alpha(Z\alpha)^6}{\pi n^3}\left(\frac{m_r}{m}\right)^3m\:\delta_{l0},
\eeq

\noindent
which is nonvanishing only for the $S$-states.

\subsection{Corrections of Order $\alpha^2(Z\alpha)^6\lowercase{m}$}
\label{alpha2zalpha6}

Corrections of order $\alpha^2(Z\alpha)^6$ originate from large
distances, and their calculations should follow the same path as
calculation of corrections of order $\alpha(Z\alpha)^6$. It is not
difficult to show that the total contribution of order
$\alpha^2(Z\alpha)^6$ is a polynomial in $\ln(Z\alpha)^{-2}$, starting with
the cube of the logarithm. Only the factor before the leading
logarithm cubed and the contribution of the logarithm squared terms to the
difference $\Delta E_L(1S)-8\Delta E_L(2S)$ are known now.
Calculation of these contributions is relatively simple because large
logarithms always originate from the wide region of large virtual
momenta ($mZ\alpha<<k<<m$) and the respective matrix elements of the
perturbation potentials depend only on the value of the
Schr\"odinger-Coulomb wave function or its derivative at the origin, as
we have already seen above in discussion of the main contribution to
the Lamb shift and corrections of order $\alpha(Z\alpha)^6$.

\subsubsection{Electron-Line Contributions}\label{lambcube}

\begin{figure}[ht]
\centerline{\epsfig{file=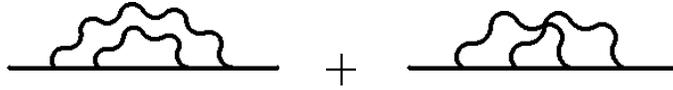,height=1.2cm}}
\vspace{0.5cm}
\caption{Two-loop self-energy operator}
\label{twoloopselfenfig}
\end{figure}

\begin{figure}[ht]
\centerline{\epsfig{file=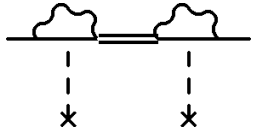,height=1.7cm}}
\vspace{0.5cm}
\caption{Second order perturbation theory contribution with two one-loop
self-energy operators}
\label{twooneloopselfenfig}
\end{figure}

Let us turn now to the general expression for the energy shift in
\eq{perturbth}. Corrections of order $\alpha^2(Z\alpha)^6$ are generated
not only by the term with the irreducible electron two-loop self-energy
operator (see Fig.\ \ref{twoloopselfenfig}) like in \eq{seconselfen} but
also by the second-order perturbation theory term with two one-loop electron
self-energy operators in the external Coulomb field in Fig.\
\ref{twooneloopselfenfig}. It is not hard to check that the contribution
containing the highest power of the logarithm is generated exactly by this
term. Note first that a logarithmic matrix element of the first-order
electron self-energy operator (like the one producing the leading
contribution to the Lamb shift) may be considered in the framework of
perturbation theory as a matrix element of an almost local (it depends on
the momentum transfer only logarithmically) operator since it is induced by
the diagram with relativistic virtual momenta.  Then one may use the same
local operator in order to calculate the higher order perturbation theory
contributions \cite{khriplovich}. We will need only the ordinary
second-order perturbation theory expression

\beq                       \label{secoredr}
\Delta E=2\sum_{m,m\neq n}\frac{<n|V_1|m><m|V_2|n>}{E_n-E_m},
\eeq

\noindent
where $V_1$ and $V_2$ are the perturbation operators, and the factor $2$
is due to two possible orders of the perturbation operators; it is not
present when $V_1=V_2$. Summation over the intermediate states above
includes integration over the continuous spectrum with the weight
$\int{d^3k}/{(2\pi)^3}$.

In order to obtain the maximum power of the large logarithm we take the
quasilocal effective perturbation potential ($V_1=V_2$) in momentum
space in the form

\beq            \label{leadlamb}
V=-\frac{8\alpha(Z\alpha)}{3m^2}\ln\frac{k}{m}.
\eeq

This is just the perturbation potential which generates the leading
contribution to the Lamb shift. It is evident that this potential
leads to a logarithm squared contribution of order
$\alpha^2(Z\alpha)^6$ after substitution in \eq{secoredr}. One may gain
one more logarithm from the continuous spectrum contribution in
\eq{secoredr}. Due to locality of the potential, matrix elements reduce
to the products of the values of the respective wave functions at the
origin and the potentials in \eq{leadlamb}. The value of the continuous
spectrum Coulomb wave function at the origin is well known (see,
e.g.,\cite{landlif}), and

\beq       \label{coulsquarecont}
|\psi_k(0)|^2=\frac{2\pi\gamma_r}{k(1-e^{-\frac{2\pi\gamma_r}{k}})}
\approx1+\frac{\pi\gamma_r}{k}+O\left(\left(\frac{\gamma_r}{k}\right)^2
\right),
\eeq

\noindent
where $\gamma_r=m_rZ\alpha$. The leading term in the large momentum
expansion in \eq{coulsquarecont} generates an apparently linearly
ultraviolet divergent contribution to the energy shift, but this
ultraviolet divergence is due to our nonrelativistic approximation, and
it would be cut off at the electron mass in a truly relativistic
calculation.  What is more important this correction is of order
$(Z\alpha)^5$, and may be safely omitted in the discussion of the
corrections of order $(Z\alpha)^6$. Logarithmic corrections of order
$(Z\alpha)^6$ are generated by the second term in the high momentum
expansion in \eq{coulsquarecont}

\beq
\Delta E=-2\frac{(m_rZ\alpha)^4m_r}{\pi^2n^2}\int
dk\frac{<n|V_1|m><m|V_2|n>}{k}.
\eeq

With the help of this formula one immediately obtains \cite{kar93}

\beq               \label{logcube}
\Delta E=-\frac{8}{27}\frac{\alpha^2(Z\alpha)^6}{\pi^2 n^3}
\ln^3[\frac{m(Z\alpha)^{-2}}{m_r}]\left(\frac{m_r}{m}\right)^5m.
\eeq

Note that the scale of this contribution is once again exactly of the
expected magnitude, namely, this contribution is suppressed by the
factor $(\alpha/\pi)\ln(Z\alpha)^{-2}$ in comparison with the leading
logarithm squared contribution of order $\alpha(Z\alpha)^6$. Of course,
the additional numerical suppression factor $8/27$ could not be
obtained without real calculation.  Numerically, the correction in
\eq{logcube} is about $-28$ kHz for the $1S$-state and calculation of
other corrections of order $\alpha^2(Z\alpha)^6$ is clearly warranted.

\begin{figure}[ht]
\centerline{\epsfig{file=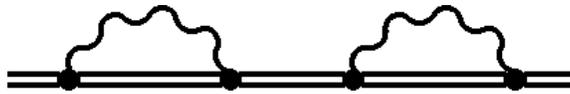,height=1.3cm}}
\vspace{0.5cm}
\caption{Reducible two-loop radiative insertions in the electron line}
\label{reduciblefig}
\end{figure}

Corrections induced by the one-particle reducible two-loop radiative
insertions in the electron line were calculated numerically without
expansion in $Z\alpha$ in recent works \cite{malsap,glnps,yerokhin}.
Effectively, a subset of the diagrams in Fig.\ \ref{reduciblefig}
generating corrections of order $\alpha^2(Z\alpha)^nm$ to the Lamb
shift was summed in \cite{malsap,glnps,yerokhin}. This subset contains
all diagrams which generate the leading logarithm cubed contribution to
the Lamb shift. In the case of $Z=1$ an additional contribution to the
Lamb shift obtained in \cite{malsap} is equal $-71$ kHz for the
$1S$-state in hydrogen, and is much larger than the leading logarithm
contribution in \eq{logcube}, while the result in \cite{glnps} is in
agreement with \eq{logcube}.

The numerical results in \cite{malsap,glnps,yerokhin} were parametrized
as a polynomial in the low energy logarithm $\ln(Z\alpha)^{-2}$, namely
the corrections of order $\alpha^2(Z\alpha)^6m$ were written in the
form

\beq
\Delta E=\left[c_1\ln^3(Z\alpha)^{-2}
+c_2\ln^2(Z\alpha)^{-2}+c_3\ln(Z\alpha)^{-2}\right]
\frac{\alpha^2(Z\alpha)^6}{\pi^2n^3}\:m.
\eeq

Fit of the numerical results leads to the value $c_1=-0.9$
\cite{malsap} for the coefficient before the logarithm cubed to be
compared with the analytic result $c_1=-8/27\approx-0.3$ in
\eq{logcube}. The leading logarithmic result in \eq{logcube} is
generated just by the diagrams considered in \cite{malsap}, there are
no other sources for the logarithm cubed, and, hence, the result in
\cite{malsap}  contradicts the leading perturbation theory contribution
\eq{logcube}. The result of \cite{malsap} was confirmed in a later
independent numerical calculation \cite{yerokhin}.

Meanwhile, the leading perturbation theory contribution in \eq{logcube}
was recently reproduced with the help of the renormalization group
equations in the approach based on nonrelativistic QED \cite{mahohar}.
Results of one more numerical calculation \cite{glnps} are
also consistent with the leading perturbation theory contribution in
\eq{logcube}, and predict a value for the coefficient $c_2=-1.0(1)$
which seems to be reasonable from the perturbation theory point of
view. Clearly, resolution of the contradiction between the perturbation
theory result in \eq{logcube} and numerical result in \cite{glnps} on
one hand, and the numerical result in \cite{malsap,yerokhin} on the
other hand is an urgent problem.  We will use the perturbation theory
result in \eq{logcube} for comparison between theory and experiment  in
Section \ref{compthexp}.

The value of the factor before the logarithm squared term
$c_2=-1.0\pm0.1$ obtained in \cite{glnps} generates a contribution of
about $-10$ kHz to the $1S$ Lamb shift. However, at the present stage
we cannot accept this value of the factor $c_2$ as the true value of
the coefficient before the logarithm squared term because only a
subset of all diagrams with two radiative photon insertions in the
electron line was calculated in \cite{glnps}. While the omitted
diagrams do not contribute to the leading logarithm cubed term, they
generate unknown logarithm squared contributions, and we have to wait
for the completion of the numerical calculation of the remaining
diagrams. It is a remarkable achievement that there is now a real
perspective that the numerical calculations without expansion in
$Z\alpha$ would produce in the near future the value of the logarithm
squared contribution of order $\alpha^2(Z\alpha)^6m$.

Another indication on the magnitude of the logarithm squared
contributions to the energy shift is provided by the logarithm squared
contribution to the difference $\Delta E(1S)-8\Delta E(2S)$ (see
discussion below).  Taking into account this result, as well as the
partial numerical estimate of the logarithm squared term in
\cite{glnps}, we come to the conclusion that a fair estimate of the
logarithm squared corrections to the individual energy levels is given
by one half of the leading logarithm cubed contribution in
\eq{logcube}, and constitutes $14$ kHz for the $1S$-state and $2$ kHz
for the $2S$-state.

The analysis of the contributions of order $\alpha^2(Z\alpha)^n$  confirms
once again, as also emphasized in \cite{malsap}, that there is
no regular rule for the magnitude of the coefficients before the
successive terms in the series over $Z\alpha$ at fixed $\alpha$. This
happens because the terms, say of relative order $Z\alpha$ and
$(Z\alpha)^2$, correspond to completely different physics at small and
large distances and, hence, there is no reason to expect a regular law
for the coefficients in these series.  This should be compared with the
series over $\alpha$ at fixed $Z\alpha$. As we have shown above,
different terms in these series correspond to the same physics and
hence the coefficients in these series change smoothly and may easily
be estimated. This is why we have organized the discussion in this
review in terms of such series. Note that the best way to estimate an
unknown correction of order, say $\alpha^2(Z\alpha)^6m$, which
corresponds to the long distance physics, is to compare it with the
long distance correction of order $\alpha(Z\alpha)^6m$, and not with
the correction of order $\alpha^2(Z\alpha)^5m$ which corresponds to the
short distance physics.  Of course, such logic contradicts the spirit
of the numerical calculations made without expansion over $Z\alpha$ but
it reflects properly the physical nature of different contributions at
small $Z$.

Perturbation theory calculation of logarithm squared contributions to the
energy shift of $S$-levels is impeded by the fact that such contributions
arise both from the discrete and continuous spectrum intermediate
states in \eq{secoredr}, and a complicated interplay of contributions
from the different regions occurs. Hence, in such a calculation it is
necessary to consider the contributions of the one-loop electron
self-energy operators more accurately and the local approximation used
above becomes inappropriate.

\begin{figure}[h]
\centerline{\epsfig{file=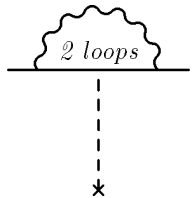}}
\vspace{0.5cm}
\caption{Effective potential corresponding to two-loop vertex}
\label{twoloopvertfig}
\end{figure}

The case of the logarithm squared contributions to the
energy levels with nonvanishing angular momenta is much simpler
\cite{kar96zetp,karjp96}. The second order perturbation theory term with two
one-loop self-energy operators does not generate any logarithm squared
contribution for the state with nonzero angular momentum  since
the respective nonrelativistic wave function vanishes at the origin. Only
the two-loop vertex in Fig.\ \ref{twoloopvertfig} produces a logarithm
squared term in this case. The respective perturbation potential determined
by the second term in the low-momentum expansion of the two-loop Dirac
form factor \cite{yeniefrautchisuur} has the form

\beq
V_2=-\frac{2\alpha^2(Z\alpha)k^2}{9\pi m^4}\ln^2(Z\alpha)^{-2}.
\eeq

\noindent

Calculation of the matrix element of this effective perturbation with the
nonrelativistic wave functions for the $P$-states yields
\cite{kar96zetp,karjp96}

\beq     \label{twolooplod2}
\Delta E(nP)=\frac{4(n^2-1)}{27n^2}\frac{\alpha^2(Z\alpha)^6m}{\pi^2
n^3}\ln^2(Z\alpha)^{-2},
\eeq

\noindent
while for $l>1$ there are no logarithm squared contributions.

\subsubsection{Polarization Operator Contributions}

\begin{figure}[ht]
\centerline{\epsfig{file=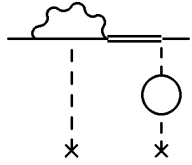}}
\vspace{0.5cm}
\caption{One of the logarithm squared contributions of order
$\alpha^2(Z\alpha)^6$}
\label{vertpolfig}
\end{figure}

Logarithmic contributions corresponding to the diagrams with at least one
polarization insertion may be calculated by the methods described above.
The leading logarithm squared term in Fig.\ \ref{vertpolfig} is
generated when we combine the perturbation potential in \eq{leadlamb}
which corresponds to the one-loop electron vertex and the perturbation
potential in \eq{leadcoulpollambnaiv} which corresponds to the
polarization operator contribution to the Lamb shift \cite{kar96zetp}

\beq
\Delta E(nS)=\frac{8}{45}\frac{\alpha^2(Z\alpha)^6m}{\pi^2
n^3}\ln^2(Z\alpha)^{-2}.
\eeq

Let us remind the reader immediately that the logarithm squared terms
induced by two radiative insertions in the electron line remain
uncalculated.

In the case of $P$-states the leading contribution corresponding to the
diagrams with at least one polarization insertion is linear in the large
logarithm and it is equal to \cite{kar96zetp}

\beq       \label{singllogaza6pol}
\Delta E(nP)=-\frac{8(n^2-1)}{135n^2}\frac{\alpha^2(Z\alpha)^6m}{\pi^2
n^3}\ln(Z\alpha)^{-2},
\eeq

But again there are uncalculated contributions linear in the large
logarithm induced by the diagrams without polarization insertions.

\begin{figure}[h]
\centerline{\epsfig{file=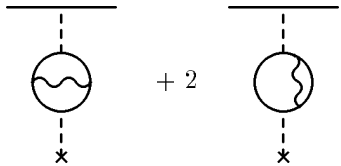}}
\vspace{0.5cm}
\caption{Two-loop polarization potential}
\label{twoloooppolfig}
\end{figure}

The linear in the large logarithm contribution to the energy of the
$S$-level induced by the two-loop vacuum polarization in Fig.\
\ref{twoloooppolfig} is also known \cite{egpol}

\beq
\Delta E=-\frac{41}{81}\frac{\alpha^2(Z\alpha)^6}{\pi^2
n^3}\ln(Z\alpha)^{-2}(\frac{m_r}{m})^3m,
\eeq

\noindent
but again there are many uncalculated contributions of the same order,
and this correction may serve only as an estimate of unknown
corrections.

\subsubsection{Corrections of Order $\alpha^2(Z\alpha)^6\lowercase{m}$ to
$\Delta E_L(1S)-\lowercase{n}^3\Delta
E_L(\lowercase{n}S)$}\label{cancellation}

All state-independent contributions  cancel in the energy shifts
combination $\Delta E(1S)-n^3\Delta E(nS)$, which makes calculation of
this energy difference more feasible than calculation of the individual
energy levels themselves. In the situation when many state-independent
corrections of order $\alpha^2(Z\alpha)^6m$ to the individual energy
levels are still unknown, this leads to more accurate theoretical
prediction for this combination of the energy levels than for  each of
the energy levels themselves. This may be extremely useful in
comparison of the theory and experiment (see, e.g.
\cite{plwhkh,kar96}).

\begin{figure}[ht]
\centerline{\epsfig{file=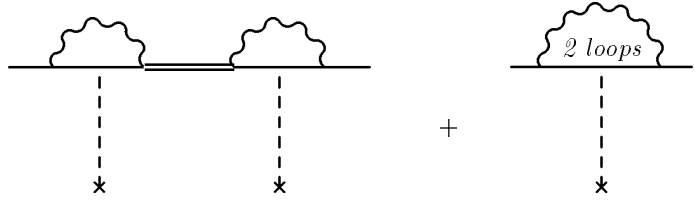}}
\vspace{0.5cm}
\caption{Logarithm squared contributions
to $\Delta E_L(1S)-\lowercase{n}^3\Delta E_L(\lowercase{n}S)$}
\label{secordsqfig}
\end{figure}

The logarithm squared contributions to the difference of energies are
generated in the second order of perturbation theory by two one-loop
vertex operators, and in the first order of perturbation theory by the
one two-loop vertex (see diagrams in Fig.\ \ref{secordsqfig}). Due to
cancellation of the state-independent terms in the difference of energy
levels only intermediate continuous spectrum states with momenta of the
atomic scale $mZ\alpha$ give contributions in the second order of
perturbation theory \cite{karyf95,karjetp94}. Then the local
approximation for the one-loop vertices and the nonrelativistic
approximation for the wave functions is sufficient for calculation of
the logarithm squared contribution to the energy difference generated
by the first diagram in Fig.\ \ref{secordsqfig}. Calculation of the
contribution induced by the second order vertex operator (second
diagram in Fig.\ \ref{secordsqfig}) is quite straightforward. Both
contributions were calculated in a series of papers
\cite{kar96zetp,karjp96,karyf95,karjetp94} with the result

\beq         \label{1s82scanc}
\Delta E(1S)-n^3\Delta
E(nS)=\frac{16}{9}\left[\ln n-\psi(n)+\psi(1)-
\frac{n-1}{n}+\frac{n^2-1}{4n^2}\right]
\frac{\alpha^2(Z\alpha)^6m}{\pi^2 }\ln^2(Z\alpha)^{-2}.
\eeq

Numerically this contribution is equal to $-10.71$ kHz for $\Delta
E(1S)-8\Delta E(2S)$.

Contributions to the difference of energies linear in the large
logarithm are still unknown. Only the linear terms  connected with
polarization insertions were calculated \cite{kar96zetp}

\beq
\Delta E(1S)-n^3\Delta
E(nS)=\frac{32}{45}\left[-\ln n
+\psi(n)-\psi(1)+\frac{n-1}{n}-\frac{n^2-1}{4n^2}\right]
\frac{\alpha^2(Z\alpha)^6m}{\pi^2 }\ln(Z\alpha)^{-2}
\eeq

\noindent
but the contributions of the same order of magnitude induced by the
diagrams without polarization insertions were never calculated. As
usual we expect that the contributions connected exclusively with the
electron line are larger than the polarization contribution above.

In such a situation it seems prudent to assume that the uncertainty in
the difference $\Delta E(1S)-8\Delta E(2S)$, which is due to
uncalculated linear terms, is perhaps about $5$ kHz
\cite{karjp96,kar96}.

The logarithm squared contributions to the individual energy levels are
also unknown. We have assumed  one half of the leading logarithm cubed
contribution in \eq{logcube} as an estimate of all yet uncalculated
corrections of this order (see discussion in Section \ref{lambcube}).
This estimate of the theoretical uncertainties is confirmed by the
magnitude of the logarithm squared contribution to the interval
$E_L(1S)-8E_L(2S)$, which can be considered as an estimate of the scale
of all yet uncalculated corrections of this order. Due to the fact that
we know the logarithm squared contribution \eq{1s82scanc} to the
interval $E_L(1S)-8E_L(2S)$ (see \eq{1s82scanc}) the theoretical
accuracy of this difference is higher than the accuracy of the
expression for $\Delta E(1S)$.

\subsubsection{Corrections of Order $\alpha^3(Z\alpha)^6\lowercase{m}$}

Corrections of order $\alpha^3(Z\alpha)^6$ were never considered in the
literature. They are suppressed in comparison to contributions of
order $\alpha^2(Z\alpha)^6$ by at least an additional factor $\alpha/\pi$
and are too small to be of any phenomenological interest now.


\begin{center}
\underline{Table V. Radiative Corrections of Order $\alpha^n(Z\alpha)^6m$}
\nopagebreak

\begin{tabular}{|l|rl|c|c|}    \hline
     &$4\frac{\alpha(Z\alpha)^6}{\pi
n^3}(\frac{m_r}{m})^3m\approx\frac{173.074}{n^3}$ kHz& &$\Delta E(1S)$
kHz&$\Delta E(2S)$ kHz
\\
\hline \hline
Logarithmic Electron-Line &$$&&$$& \\
Contribution ($l=0$)&$$&&$$& \\
Layzer(1960)\cite{layzer}&$-\frac{1}{4}\ln^2[\frac{m}{m_r}(Z\alpha)^{-2}]
+[\frac{4}{3}\ln2+\ln\frac{2}{n}$& & &\\
Fried,Yennie(1960)\cite{frdyenn}&$+\psi(n+1)-\psi(1)
-\frac{601}{720}-\frac{77}{180n^2}]$&&$$ &\\
Erickson,Yennie(1965)\cite{eryen1,eryen2}&$\ln
[\frac{m}{m_r}(Z\alpha)^{-2}]$ && $-1~882.77$&$-208.16$\\
&$$&&$$ &
\\ \hline
Logarithmic Electron-Line &$$&&$$& \\
Contribution ($l\neq0$)&$$&&$$& \\
&$$&&$$ &\\
Erickson,Yennie(1965)\cite{eryen1,eryen2}&$[(1-\frac{1}{n^2})
(\frac{1}{30}+\frac{1}{12}\delta_{j,\frac{1}{2}})$& & &\\
&$+\frac{6-2l(l+1)/n^2}
{3(2l+3)l(l+1)(4l^2-1)}]\ln[\frac{m}{m_r}(Z\alpha)^{-2}]$ && $$&
\\ \hline
Nonlogarithmic Electron-Line &$$&&$$& \\
Contribution &$$&&$$& \\
Pachucki(1993)($1S,2S$)\cite{pacha6,pachalpha6,pach99}
&$\frac{A_{60}}{4}
$&&$-1~338.04$&$-172.21$         \\
Jentschura,Pachucki(1996)
&$$&&$$         &\\
($2P_\frac{1}{2},2P_\frac{3}{2}$)\cite{jenpach}&&&&\\
Jentschura,Soff,Mohr(1997)
&&&$$ &\\
($3P_\frac{1}{2},3P_\frac{3}{2},4P_\frac{1}{2},4P_\frac{3}{2}$)\cite{jsm}
&&&&
\\ \hline
Logarithmic Polarization&&&$$ &\\
Operator Contribution&&&$$ &\\
&$$ && $$&\\
Layzer(1960)\cite{layzer}&$-\frac{1}{30}\ln[\frac{m}{m_r}(Z\alpha)^{-2}]
\delta_{l0}$&&$-56.77$&$-7.10$\\
Erickson,Yennie(1965)\cite{eryen1,eryen2}&&&$$ &\\
 \hline
Nonlogarithmic Polarization&&&$$ &\\
Operator Contribution ($l=0$)&&&$$ &\\
&$$ && $$&\\
Mohr(1975)\cite{mohr}&$\frac{1}{15}[-\frac{431}{105}+\psi(n+1)-\psi(1)
$& & &\\
Ivanov,Karshenboim(1997)\cite{ik97}&$
-\frac{2(n-1)}{n^2}+\frac{1}{28n^2}-\ln\frac{n}{2}]\delta_{l0}$&&
$-27.41$&$-4.47$ \\
&$$ && $$&
\\ \hline
Nonlogarithmic Polarization&&&$$ &\\
Operator Contribution ($l=1$)&&&$$ &\\
&$$ && $$&\\
Mohr(1975)\cite{mohr}&$-\frac{1}{15}(1-\frac{1}{n^2})[\frac{1}{14}
+\frac{1}{4}\delta_{j,\frac{1}{2}}]$& & &\\
Manakov,Nekipelov,&$$&&$$& \\
Fainstein(1989)\cite{mnf}&$$ && $$&
\\ \hline
Wichmann-Kroll Contribution&$$ && $$&\\
Wichmann,Kroll(1956)\cite{wichkr}&$
(\frac{19}{180}-\frac{\pi^2}{108})\delta_{l0}$& &$2.45$&$0.31$
\\
Mohr(1976)\cite{mohr76,mohr83}&$$&&&
\\
\hline
\end{tabular}
\end{center}


\begin{center}
\underline{Table V. Continuation }
\nopagebreak

\begin{tabular}{|l|rl|c|c|}
\hline
Leading Logarithmic Electron&$$&&$$& \\
-Line Contribution &$$&&$$& \\
&$$&&$$ &\\
Karshenboim(1993)\cite{kar93}&$-\frac{2}{27}(\frac{\alpha}{\pi})
\ln^3(Z\alpha)^{-2}\delta_{l0}$& &  $-28.38$&$-3.55$
\\
\hline
Electron-Line &$$&&$$& \\
Log Squared Term&$$&&$$& \\
$\Delta E(nS)$&&&$$         &\\
&$(?)(\frac{\alpha}{\pi})\ln^2(Z\alpha)^{-2}$& & $\pm10$ &$\pm2$
\\
\hline
Log squared contribution $\Delta E(nP)$&$$ && $$&\\
Karshenboim(1996)\cite{kar96zetp,karjp96}&
$\frac{n^2-1}{27n^2}(\frac{\alpha}{\pi})
\ln^2(Z\alpha)^{-2}$& &  $$&
\\
\hline
Electron-Line Linear &$$&&$$& \\
in Log Term&$$&&$$& \\
$\Delta E(nP)$&&&$$         &\\
&$(?)(\frac{\alpha}{\pi})\ln(Z\alpha)^{-2}$& & $\pm$ &$\pm$
\\
\hline
Log Squared Term&$$&&$$& \\
Connected with Polarization&$$&&$$& \\
$\Delta E(nS)$&&&$$         &\\
Karshenboim(1996)\cite{kar96zetp}&
$\frac{2}{45}(\frac{\alpha}{\pi})\ln^2(Z\alpha)^{-2}$& & $1.73$ &$0.22$
\\
\hline
Linear Log Connected with &&&$$         &\\
Polarization $\Delta E(nP)$&&&$$        & \\
Karshenboim(1996)\cite{kar96zetp}&$-\frac{2(n^2-1)}{135n^2}
(\frac{\alpha}{\pi})\ln(Z\alpha)^{-2}$& & $$ &
\\ \hline
Linear Log Connected with &&&$$         &\\
Two-Loop Polarization $\Delta E(nS)$&&&$$        & \\
Eides,Grotch(1995)\cite{egpol}&$-\frac{41}{324}
(\frac{\alpha}{\pi})\ln(Z\alpha)^{-2}$& & $-0.50$ &$-0.06$\\
\hline
\end{tabular}
\end{center}

\section{Radiative Corrections of Order $\alpha(Z\alpha)^7\lowercase{m}$
and of Higher Orders}

Only partial results are known for corrections of order
$\alpha(Z\alpha)^7m$. However, recent achievements \cite{jms99} in the
numerical calculations without expansion in $Z\alpha$ completely solve
the problem of the corrections of order $\alpha(Z\alpha)^7m$ and of
higher orders in $Z\alpha$.

\subsection{Corrections Induced by the Radiative Insertions
in the Electron Line} \label{lambalphazalpha7}

\begin{figure}[h]
\centerline{\epsfig{file=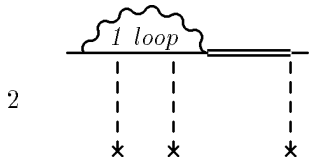}}
\vspace{0.5cm}
\caption{Leading logarithmic contribution of order $\alpha(Z\alpha)^7$
induced by the radiative photon}
\label{leadinglog8fig}
\end{figure}

Consider first corrections of order $\alpha(Z\alpha)^7$
induced by the radiative photon insertions in the electron line. Due to the
Layzer theorem \cite{layzer} the diagram with the radiative photon spanning
four Coulomb photons does not lead to a logarithmic contribution. Hence, all
leading logarithmic contributions of this order may be calculated with the
help of second order perturbation theory in \eq{secoredr}. It is easy
to check that the leading contribution is linear in the large logarithm and
arises when one takes as the first perturbation the local potential
corresponding to the order $\alpha(Z\alpha)^5m$ contribution to the Lamb
shift \eq{aza5}

\beq  \label{aza5lambpot}
V_1=4\left(1+\frac{11}{128}-\frac{1}{2}\ln
2\right)\frac{\pi\alpha(Z\alpha)^2}{m^2},
\eeq

\noindent
and the second perturbation corresponds to the Darwin potential

\beq     \label{darwin}
V_2=-\frac{\pi Z\alpha}{2m_r^2},
\eeq

\noindent
where both potentials are written in momentum space (see Fig.\
\ref{leadinglog8fig}).  Substituting these potentials in \eq{secoredr} one
easily obtains \cite{karjetp94}

\beq                       \label{logalpha7}
\Delta E=\left(2+\frac{11}{64}-\ln 2\right)
\ln[\frac{m(Z\alpha)^{-2}}{m_r}]
\frac{\alpha(Z\alpha)^7}{n^3}\left(\frac{m_r}{m}\right)^3{m}\:\delta_{l0}.
\eeq

The Darwin potential generates the logarithmic correction to the
nonrelativistic Schr\"odinger-Coulomb wave function in \eq{darwincor},
and the result in \eq{logalpha7} could be obtained by taking into
account this correction to the wave function in calculation of the
contribution to the Lamb shift of order $\alpha(Z\alpha)^5m$. This
logarithmic correction is numerically equal $14.43$ kHz for the
$1S$-level in hydrogen, and $1.80$ kHz for the $2S$ level.

The nonlogarithmic contributions of this order were never calculated
directly, but one can obtain a reliable estimate of these contributions
as well as of the contributions of higher orders in $(Z\alpha)$ using
results of the numerical calculations of the contributions of order
$\alpha(Z\alpha)^n$ made without expansion in $Z\alpha$ for $Z$ not too
small. It is convenient to parametrize respective results with the
help of an auxiliary function $G_{SE,7}(Z\alpha)$ defined by the
relation\footnote{The factor $1/\pi$ before the second term in the
square brackets is written here in order to conform with the
traditional notation.}

\beq \label{gsedef}
\Delta E(l,n)=
\left[\left(\frac{139}{64}-\ln 2\right)
\ln[\frac{m(Z\alpha)^{-2}}{m_r}]\delta_{l0}
+\frac{1}{\pi}G_{SE,7}^{l,n}(Z\alpha)\right]
\frac{\alpha(Z\alpha)^7}{n^3}\left(\frac{m_r}{m}\right)^3m.
\eeq

Results for the function $G_{SE}(Z\alpha)$ for small $Z$ may be
obtained with the help of extrapolation from the numerical results for
larger $Z$ \cite{mohr,mohrkim,mohr92}. The results of such extrapolation for
the $1S,2S,2P,4S$- and $4P$-states ($Z=1,2$) are presented in
\cite{mohr91,mohr96} (see also results of an earlier extrapolation in
\cite{wkd}). The extrapolation in \cite{mohr91,mohr96} was done
independently of the exact results in
\cite{pacha6,pachalpha6,jenpach,jsm,karjetp94} for the nonlogarithmic
contribution of order $\alpha(Z\alpha)^6$ and the logarithmic one of order
$\alpha(Z\alpha)^7$. Extrapolations which take into account these exact
results were performed for the hydrogen $1S,2S,2P,3P,4P$-states in
\cite{karyaf95,jenpach,jsm}. We have collected the results of these
extrapolations in Table VI. Taking into account current experimental
accuracy we may conclude from the data in Table VI that the higher order
corrections in $Z\alpha$ of the form $\alpha(Z\alpha)^n$ are small and
the currently known coefficients sufficient for the phenomenological
needs.

A spectacular success was achieved recently in numerical calculation of the
function $G_{SE}(Z\alpha)$\footnote{The function $G_{SE}(Z\alpha)$ is
defined similarly to the function $G_{SE,7}(Z\alpha)$ in \eq{gsedef},
but includes also the nonlogarithmic contribution of order
$\alpha(Z\alpha)^6$.\label{footng7}} for low $Z$ \cite{jms99}. This
self-energy contribution for the $1S$-level in hydrogen was obtained in
this work with numerical uncertainty $0.8$ Hz. This result completely
solves all problems with calculation of the higher order corrections in
$Z\alpha$ of the form $\alpha(Z\alpha)^n$ in the foreseeable future.

\begin{center}
\underline{Table VI. Term $G_{SE,7}$}
\nopagebreak

\begin{tabular}{|l|rl|c|}
\hline
 &$G_{SE,7}$&&$\Delta E$ kHz\\
\hline \hline
$1S$&$$& & \\
Karshenboim(1995)\cite{karyaf95}&$-2.42(15)$&& $-0.76(5)$
\\ \hline
$2P_\frac{1}{2}$&$$& & \\
Jentschura,Pachucki(1996)\cite{jenpach}&$3.1(5)$ & &$0.12(2)$\\
\hline
$2P_\frac{3}{2}$&$$& & \\
Jentschura,Pachucki(1996)\cite{jenpach}&$2.3(5)$ & &$0.09(2)$\\
\hline
$3P_\frac{1}{2}$&$$& & \\
Jentschura,Soff,Mohr(1997)\cite{jsm}&$3.6(5)$ & &$0.04$\\
\hline
$3P_\frac{3}{2}$&$$& & \\
Jentschura,Soff,Mohr(1997)\cite{jsm}&$2.6(5)$ & &$0.03$\\
\hline
$4P_\frac{1}{2}$&$$& & \\
Jentschura,Soff,Mohr(1997)\cite{jsm}&$3.9(5)$ & &$0.02$\\
\hline
$4P_\frac{3}{2}$&$$& & \\
Jentschura,Soff,Mohr(1997)\cite{jsm}&$2.8(5)$ & &$0.01$\\
\hline
\end{tabular}
\end{center}

\subsection{Corrections Induced by the Radiative Insertions
in the Coulomb Lines} \label{lambalphzapl7coul}

There are two contributions of order $\alpha(Z\alpha)^7m$ to the energy
shift induced by the Uehling and the Wichmann-Kroll potentials (see Fig.\
\ref{polarizcoulfig} and Fig.\ \ref{wichkrolfig}, respectively). Respective
calculations go along the same lines as in the case of the Coulomb-line
corrections of order $\alpha(Z\alpha)^6$ considered above.

\paragraph{Uehling Potential Contribution.}

The logarithmic contribution is induced only by the Uehling potential in
Fig.\ \ref{polarizcoulfig}, and may easily be calculated exactly in the
same way as the logarithmic contribution induced by the radiative
photon in \eq{logalpha7}. The only difference is that now the role of
the perturbation potential is played by the kernel which corresponds to
the polarization contribution to the Lamb shift of order
$\alpha(Z\alpha)^5m$

\beq  \label{uelpoteff}
V_1=\frac{5}{48}\frac{\pi\alpha(Z\alpha)^2}{m^2}.
\eeq

\noindent
Then we immediately obtain \cite{mohr}

\beq
\Delta E=\frac{5}{96}
\ln[\frac{m(Z\alpha)^{-2}}{m_r}]
\frac{\alpha(Z\alpha)^7}{n^3}\left(\frac{m_r}{m}\right)^3{m}\:\delta_{l0}.
\eeq

\noindent
It is not difficult to calculate analytically nonlogarithmic corrections of
order $\alpha(Z\alpha)^7$ generated by the Uehling potential. Using the
formulae from \cite{mohr} one obtains for a few lower levels (see also
\cite{karpol99} for the case of $1S$-state)

\beq  \label{polalpha7}
\Delta
E(1S)=\frac{5}{96}\left[\ln[\frac{m(Z\alpha)^{-2}}{m_r}]
+2\ln2+\frac{23}{15}\right]
{\alpha(Z\alpha)^7}\left(\frac{m_r}{m}\right)^3m,
\eeq
\[
\Delta
E(2S)=\frac{5}{96}\left[\ln[\frac{m(Z\alpha)^{-2}}{m_r}]
+4\ln2+\frac{841}{480}\right]
\frac{\alpha(Z\alpha)^7}{ 2^3}\left(\frac{m_r}{m}\right)^3m,
\]
\[
\Delta
E(2P_\frac{1}{2})=\frac{41}{3~072}
\frac{\alpha(Z\alpha)^7}{2^3}\left(\frac{m_r}{m}\right)^3m,
\]
\[
\Delta
E(2P_\frac{3}{2})=\frac{7}{1~024}
\frac{\alpha(Z\alpha)^7}{2^3}\left(\frac{m_r}{m}\right)^3m.
\]

There are no obstacles to exact numerical calculation of the
Uehling potential contribution to the energy shift without expansion over
$Z\alpha$ and such calculations have been performed with high accuracy
(see \cite{mohr,mohr96} and references therein). The results of these
calculations may be conveniently presented with the help of an auxiliary
function $G_{U,7}(Z\alpha)$ defined by the relationship

\beq      \label{gudef}
\Delta
E(l,n)=\left[\frac{5}{96}
\ln[\frac{m(Z\alpha)^{-2}}{m_r}]\delta_{l0}
+\frac{1}{\pi}G_{U,7}^{l,n}(Z\alpha)\right]
\frac{\alpha(Z\alpha)^7}{n^3}\left(\frac{m_r}{m}\right)^3m.
\eeq

\noindent
For the case of atoms with low $Z$ (hydrogen and helium), values of the
function $G_{U,7}(Z\alpha)$ for the states with $n=1,2,4$ are tabulated in
\cite{mohr96} and respective contributions may easily be calculated for
other states when needed. These numerical results may be used for
comparison of the theory and experiment instead of the results of
order $\alpha(Z\alpha)^7$ given above. We may also use the results of
numerical calculations in order to make an estimate of uncalculated
contributions of the Uehling potential of order $\alpha(Z\alpha)^8$ and
higher. According to \cite{mohr96}

\beq
G_U^{l=0,n=1}(\alpha)=0.428~052.
\eeq

\noindent
Comparing this value with the order $\alpha(Z\alpha)^7m$ result in
\eq{polalpha7} we see that the difference between the exact
numerical result and analytic calculation up to order $\alpha(Z\alpha)^7$ is
about $0.015 $ kHz for the $1S$-level in hydrogen, and, taking into
account the accuracy of experimental results, one may use analytic
results for comparison of the theory and experiment without loss of
accuracy. A similar conclusion is valid for other hydrogen levels.

\paragraph{Wichmann-Kroll Potential Contribution.}

Contribution of the Wichmann-Kroll potential in Fig.\ \ref{wichkrolfig} may
be calculated in the same way as the respective contribution of order
$\alpha(Z\alpha)^6m$ in \eq{wichkrcalc}  by taking the next term in
$Z\alpha$ in the small momentum expansion of the Wichmann-Kroll
potential in \eq{wichmankrsmallmom}. One easily finds
\cite{mohr76,mohr83}

\beq
\Delta
E=\left(\frac{1}{16}-\frac{31\pi^2}{2~880}\right)
\frac{\alpha(Z\alpha)^7}{n^3}
\left(\frac{m_r}{m}\right)^3m\:\delta_{l0}.
\eeq

This contribution is very small and it is clear that at the present level of
experimental accuracy calculation of higher order contributions of the
Wichmann-Kroll potential is not necessary.

\subsection{Corrections of Order $\alpha^2(Z\alpha)^7\lowercase{m}$}

Corrections of order $\alpha^2(Z\alpha)^7$ were never considered in the
literature. They should be suppressed in comparison with the corrections
of order $\alpha(Z\alpha)^7$ by at least the factor $\alpha/\pi$. Even
taking into account possible logarithmic enhancements, these corrections
are not likely to be larger than about $1$ kHz for the $1S$-state and about
$0.1$ kHz for $2S$-state in hydrogen. This means that they are not important
today from the phenomenological point of view.

Concluding our discussion of the purely radiative corrections to the Lamb
shift let us mention once more that the main source of the
theoretical uncertainty in these contributions is connected with the
uncalculated contributions of order $\alpha^2(Z\alpha)^6$, which may
be as large as $14$ kHz for $1S$-state and $2$ kHz for the $2S$-state
in hydrogen. All other unknown purely radiative contributions to the
Lamb shift are much smaller. Note also that due to an extra theoretical
information on the logarithm squared contribution of order
$\alpha^2(Z\alpha)^6$, the purely radiative contributions to the
difference $\Delta E(1S)-8\Delta E(2S)$ are known better than the
purely radiative contributions to the individual energy levels. The
uncertainty in the difference $\Delta E(1S)-8\Delta E(2S)$ due to yet
unknown purely radiative terms is about $5$ kHz.

\begin{center}
\underline{Table VII. Radiative Corrections of Order $\alpha^n(Z\alpha)^7m$}
\nopagebreak

\begin{tabular}{|l|rl|c|c|}    \hline
     &$4\frac{\alpha(Z\alpha)^7}{n^3}(\frac{m_r}{m})^3m$&&$\Delta E(1S)$ kHz
&$\Delta E(2S)$ kHz
\\ \hline \hline
Logarithmic Electron-Line&$$& &  $$&\\
Contribution&$$& &  $$&\\
Karshenboim(1994)\cite{karjetp94}&$
(\frac{139}{256}-\frac{1}{4}\ln2)\ln(Z\alpha)^{-2}\delta_{l0}$& &
$14.43$&$1.80$
\\ \hline
Nonlogarithmic $$&&&$$        & \\
Electron-Line Contribution$$&&&$$        & \\
Mohr(1992)\cite{mohr92}$$&&&$$        & \\
Karshenboim(1995)\cite{karyaf95}& &&$-0.76(5)$ &$-0.09(1)$
 \\
\hline
Logarithmic Polarization&$$& &  $$&\\
Operator Contribution&$$& &  $$&\\
Mohr(1975)\cite{mohr}
&$\frac{5}{384}\ln(Z\alpha)^{-2}\delta_{l0}$& & $0.51$ &$0.06$
\\ \hline
Nonlogarithmic Polarization&$$& &  $$&\\
Operator Contribution&$$& &  $$&\\
$\Delta E(nS)$, Mohr(1975)\cite{mohr}&&&$0.15$         &$0.03$\\
\hline
Wichmann,Kroll(1956)\cite{wichkr}&$
(\frac{1}{64}-\frac{31\pi^2}{11~520})\delta_{l0}$& &$-0.04$&$-0.01$
\\
Mohr(1976)\cite{mohr76,mohr83}&$$&&&
\\
\hline
Corrections of order &$$& &  $$&\\
$\alpha^2(Z\alpha)^7$&$(\pm)\frac{\alpha}{\pi}$& &  $\pm 1$&$\pm0.1$
\\ \hline
\end{tabular}
\end{center}

\part{Essentially Two-Particle Recoil Corrections}

\section{Recoil Corrections of Order $(Z\alpha)^5(\lowercase{m}/M)
\lowercase{m}$} \label{lambreczalpha5}

Leading relativistic corrections of order $(Z\alpha)^4$ and their  mass
dependence were discussed above in Section \ref{leadreclambdef} in the
framework of the Breit equation and the effective Dirac equation in the
external field in Fig.\ \ref{edeextfield}. The exact mass dependence of
these corrections could be easily calculated because all these
corrections are induced by the one-photon exchange. The effective Dirac
equation in the external field produces leading relativistic
corrections with correct mass dependence because the one-photon
exchange kernel is properly taken into account in this equation. Some
other recoil corrections of higher orders in $Z\alpha$ are also
partially generated by the effective Dirac equation with the external
source. All such corrections are necessarily of even order in $Z\alpha$
since all expansions for the energy levels of the Dirac equation are
effectively nonrelativistic expansions over $v^2$; they go over
$(Z\alpha)^2$, and, hence, the next recoil correction produced by the
effective Dirac equation in the external field is of order
$(Z\alpha)^6$. The result for the recoil correction of order
$(Z\alpha)^6(m/M)$, obtained in this way, is incomplete and we will
improve it below. First we will consider the even larger recoil
correction of order $(Z\alpha)^5(m/M)$, which is completely missed in
the spectrum of the Breit equation or of the effective Dirac equation
with the Coulomb potential, and which can be calculated only by taking
into account the two-particle nature of the QED bound-state problem.

\begin{figure}[h]
\centerline{\epsfig{file=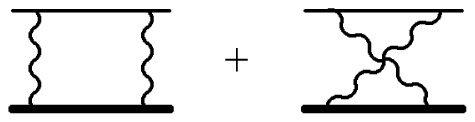,height=1.5cm}}
\vspace{0.5cm}
\caption{Diagrams with two-photon exchanges}
\label{twophotexchfig}
\end{figure}

\begin{figure}[ht]
\centerline{\epsfig{file=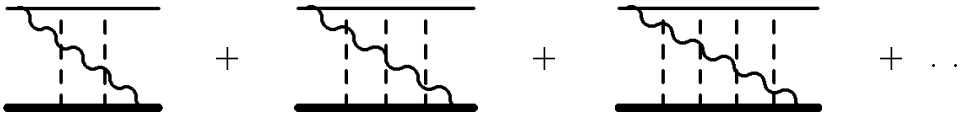,height=1.5cm}}
\vspace{0.5cm}
\caption{Irreducible kernels with arbitrary number of the
exchanged Coulomb photons}
\label{irredkernfig}
\end{figure}

The external field approximation is clearly inadequate for calculation of
the recoil corrections and, in principle, one needs the machinery of the
relativistic two-particle equations to deal with such contributions to the
energy levels. The first nontrivial recoil corrections are generated by
kernels with two-photon exchanges. Naively one might expect that all
corrections of order $(Z\alpha)^5(m/M)m$ are generated only by the
two-photon exchanges in Fig.\ \ref{twophotexchfig}. However, the situation
is more complicated. More detailed consideration shows that the two-photon
kernels are not sufficient and irreducible kernels in Fig.\
\ref{irredkernfig} with arbitrary number of the exchanged Coulomb
photons spanned by a transverse photon also generate contributions of
order $(Z\alpha)^5(m/M)m$. This effect is similar to the case of the
leading order radiative correction of order $\alpha(Z\alpha)^4$
considered in Section \ref{leadinglamb} when, due to a would-be
infrared divergence, diagrams in Fig.\ \ref{spanfig} with any number of
the external Coulomb photons spanned by a radiative photon give
contributions of one and the same order since the apparent factor
$Z\alpha$ accompanying each extra external photon is compensated by a
small denominator connected with the small virtuality of the bound
electron. Exactly the same effect arises in the case of the leading
recoil corrections. All kernels with any number of exchanged Coulomb
photons spanned by an exchanged transverse photon generate
contributions to the leading recoil correction.

Let us describe this similarity between the leading  contribution to the
Lamb shift and the leading recoil correction in more detail following a
nice physical interpretation which was given in \cite{khriplovich}. The
leading contribution to the Lamb shift in \eq{qff} is proportional to the
mean square of the electron radius which may be understood as a
result of smearing of the fluctuating electron coordinate due to its
interaction with the fluctuating electromagnetic field \cite{welton}.
In the considerations leading to \eq{qff} we considered the proton as
an infinitely heavy source of the Coulomb field. If we take into
account the finiteness of the proton mass, then the factor $<r^2>$ in
\eq{qff} will turn into $<(\Delta r_1-\Delta r_2)^2>=<(\Delta
r_1)^2>+<(\Delta r_2)^2>-2<~(\Delta r_1)(\Delta r_2)>$, where $\Delta
r_1$ and $\Delta r_2$ are fluctuations of the coordinates of the
electron and the proton, respectively. Averaging the squares of the
fluctuations of the coordinates of both particles proceeds exactly as
in the case of the electron in the Coulomb field in \eq{electrradfirt}
and generates the leading contribution to the Lamb shift and recoil
correction of relative order $(m/M)^2$. This recoil factor arises
because the average fluctuation of the coordinate squared equal to the
average radius squared of the particle is inversely proportional to
mass squared of this particle.  Hence, it is clear that the average
$<\Delta r_1\Delta r_2>$ generates a recoil correction of the first
order in the recoil factor $m/M$. Note that the correlator $<\Delta
r_1\Delta r_2>$ is different from zero only when averaging goes over
distances larger than the scale of the atom $1/(mZ\alpha)$ or in
momentum space over fluctuating momenta of order $mZ\alpha$ and
smaller. For smaller distances (or larger momenta) fluctuations of the
coordinates of two particles are completely uncorrelated and the
correlator of two coordinates is equal to zero. Hence, the logarithmic
contribution  to the recoil correction originates from the momentum
integration region $m(Z\alpha)^2\ll k\ll m(Z\alpha)$, unlike the
leading logarithmic contribution to the Lamb shift which originates
from a wider region $m(Z\alpha)^2\ll k\ll m$. A new feature of the
leading recoil correction is that the upper cutoff to the logarithmic
integration is determined by the inverse size of the atom. We will see
below how all these qualitative features are reproduced in the exact
calculations.

Complete formal analysis of the recoil corrections in the framework of
the relativistic two-particle equations, with derivation of all relevant
kernels, perturbation theory contributions, and necessary
subtraction terms may be performed along the same lines as was done
for hyperfine splitting in \cite{eksann1}. However, these results
may also be understood without a cumbersome formalism by starting with
the simple scattering approximation. We will discuss recoil
corrections below using this less rigorous but more physically transparent
approach.

As we have already realized from the qualitative discussion above, the
leading recoil correction is generated at large distances, and
small exchanged momenta  are relevant for its calculation. The choice of
gauge of the exchanged photons is, in such a case, determined by the
choice of gauge in the effective Dirac equation with the one photon
potential. This equation was written in the Coulomb gauge and, hence,
we have to use the Coulomb gauge also in the kernels with more than one
exchanged photon.  Since the Coulomb and transverse propagators have
different form in the Coulomb gauge it is natural to consider
separately diagrams with Coulomb-Coulomb, transverse-transverse and
Coulomb-transverse exchanges.

\subsection{Coulomb-Coulomb Term}

\begin{figure}[h]
\centerline{\epsfig{file=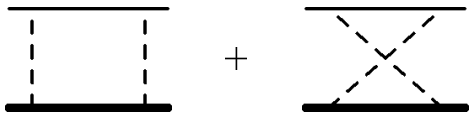,height=1.5cm}}
\vspace{0.5cm}
\caption{Coulomb-Coulomb two-photon exchanges}
\label{coulcoulfig}
\end{figure}

Coulomb exchange is already taken into account in the construction of the
zero-order effective Dirac equation, where the Coulomb source plays the
role of the external potential. Hence, additional contributions of
order $(Z\alpha)^5$ could be connected only with the high-momentum
Coulomb exchanges. Let us start by calculating  the contribution of the
skeleton Coulomb-Coulomb diagrams with on-shell external electron lines
in Fig.\ \ref{coulcoulfig}, with the usual hope that the integrals
would tell us themselves about any possible inadequacy of such an
approximation.

Direct calculation of the two Coulomb exchange photon contribution leads to
the integral

\beq          \label{skelcoul}
\Delta
E=-\frac{4}{(1-\mu^2)}\frac{(Z\alpha)^5m}{\pi
n^3}\left(\frac{m_r}{m}\right)^3\int_0^\infty\frac{dk}{k^4}[f(\mu
k)-\mu f(k)],
\eeq

\noindent
where

\beq
f(k)=3\sqrt{1+k^2}+\frac{1}{\sqrt{1+k^2}},
\eeq

\noindent
and $\mu=m/M$. The apparent asymmetry of the expression in \eq{skelcoul}
with respect to masses of the heavy and light particle emerged because the
dimensionless momentum $k$ in this formula is measured in terms of the
electron mass.

At small momenta the function $f(k)$ behaves as

\beq                     \label{infrarediv}
f(\mu k)-\mu f(k)\approx 4(1-\mu)-\mu(1-\mu)k^2+O(k^6),
\eeq

\noindent
and the skeleton integral in \eq{skelcoul}  diverges as $\int dk/k^4$ in the
infrared region. The physical meaning of this low-momenta infrared
divergence is clear; it corresponds to the Coulomb exchange
contribution to the Schr\"odinger-Coulomb wave function. The
Coulomb wave function graphically includes a sum of Coulomb ladders and
the addition of an extra rung does not change the wave function.
However, if one omits the binding energy, as we have effectively done
above, one would end up with an infrared divergent integral instead of
the self-reproducing Schr\"odinger-Coulomb wave function. A slightly
different way to understand the infrared divergence in \eq{skelcoul} is
to realize that the terms in \eq{infrarediv} which generate the
divergent contribution correspond to the residue of the heavy proton
pole in the box diagram. Once again these heavy particle pole contributions
build the Coulomb wave function and we have to subtract them not only to
avoid an apparent divergence in the approximation when we neglect the
binding energy, but in order to avoid double counting.

We would like to emphasize here that, even if one would forget about
the threat of double counting, an emerging powerlike infrared
divergence would remind us of its necessity. Any powerlike infrared
divergence is cutoff by the binding energy, and has a well defined
order in the parameter $Z\alpha$. It is most important that the integral in
\eq{skelcoul} does not contain any logarithmic infrared divergence at small
momenta. In such a case one can unambiguously subtract in the integrand the
powerlike infrared divergent terms and the remaining integral will be
completely convergent. Then only high intermediate momenta of the order of
the electron mass contribute to the subtracted integral, the respective
diagram is effectively local in coordinate space, and the contribution
to the energy shift of order $(Z\alpha)^5$ is simply given by the
product of this integral and the nonrelativistic Schr\"odinger-Coulomb
wave function squared at the origin. Any attempt to take into account
small virtuality of the external electron lines (equivalent to taking
into account nonlocality of the diagram in the coordinate space) would
lead to additional factors of $Z\alpha$, which we do not consider yet.
Direct calculation, after subtraction, of the first two leading
low-frequency terms in  the integrand in \eq{skelcoul} immediately
gives

\beq                   \label{totcul}
\Delta
E_{sub}=-\frac{4}{(1-\mu^2)}\frac{(Z\alpha)^5m}{\pi
n^3}\left(\frac{m_r}{m}\right)^3\int_0^\infty\frac{dk}{k^4}\left\{f(\mu
k)-\mu f(k)-[4(1-\mu)-\mu(1-\mu)k^2]\right\}
\eeq
\[
=-\frac{4\mu}{3}\frac{(Z\alpha)^5m}{\pi
n^3}\left(\frac{m_r}{m}\right)^3,
\]

\noindent
reproducing the well known result \cite{sal,fm,gy2}.

Let us emphasize once again that an exact calculation (in contrast to the
calculation with the logarithmic accuracy) of the Coulomb-Coulomb
contribution with the help of the skeleton integral turned out to be
feasible due to the absence of the low-frequency logarithmic divergence.
For the logarithmically divergent integrals the low-frequency cutoff
is supplied by the wave function, and in such a case it is impossible to
calculate the constant on the background of the logarithm in the skeleton
approximation. In such cases more accurate treatment of the low-frequency
contributions is warranted.

\subsection{Transverse-Transverse Term}

\begin{figure}[h]
\centerline{\epsfig{file=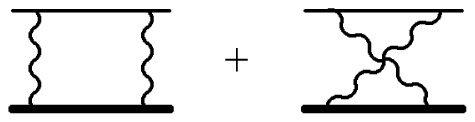,height=1.5cm}}
\vspace{0.5cm}
\caption{Transverse-transverse two-photon exchanges}
\label{trtrfig}
\end{figure}

The kernels with two transverse exchanges in Fig.\ \ref{trtrfig} give
the following contribution to the energy shift in the scattering
approximation

\beq  \label{transtrans}
\Delta E
=-\frac{2\mu}{1-\mu^2}\frac{(Z\alpha)^5m}{\pi
n^3}\left(\frac{m_r}{m}\right)^3 \int_0^\infty{dk}(f(k)-\mu^3f(\mu k)),
\eeq

\noindent
where

\beq
f(k)=\frac{1}{k}-\frac{1}{\sqrt{1+k^2}}.
\eeq

This integral diverges only logarithmically at small momenta. Hence,
this contribution does not contain either corrections of the previous
order or the non-recoil corrections. The main low-frequency logarithmic
divergence produces $\ln Z\alpha$ and the factor before this logarithm may
easily be calculated in the scattering appproximation. This
approximation is unsufficient for calculation of the nonlogarithmic
contribution, and respective calculation requires a more accurate
consideration \cite{sal,fm}. A new feature of the integral in
\eq{transtrans}, as compared with the other integrals discussed so far,
is that the exchanged momenta higher than the electron mass produce a
nonvanishing contribution. This new integration region from the
electron to the proton mass, which was discovered in \cite{fm}, arises
here for the first time in the bound state problem. As we will see
below, especially in discussion of the hyperfine splitting, these high
momenta are responsible for a number of important contributions to the
energy shifts.

The high-momentum contribution to the Lamb shift is suppressed by the
second power of the recoil factor $(m/M)^2$, and is rather small. Let
us note that the result we will obtain below in this section is
literally valid only for an elementary proton, since for the integration
momenta comparable with the proton mass one cannot ignore the composite
nature of the proton and has to take into account its internal structure  as
it is described by the phenomenological form factors.  It is also necessary
to take into account inelastic contributions in the diagrams with the two
exchanged photons. We will consider these additional contributions later in
Section \ref{lambsizestr} dealing with the nonelectromagnetic contributions
to the Lamb shift.

The state-independent high-frequency contribution as well as the
low-frequency logarithmic term are different from zero only for the
$S$-states and may easily be calculated with the help of \eq{transtrans}

\beq    \label{transcut}
\Delta E=\left[-\frac{2\mu^2\ln\mu}{1-\mu^2}-2\ln(1+\mu)
+(2\ln Z\alpha+2\ln2)\right]\mu
\frac{(Z\alpha)^5m}{\pi n^3}\left(\frac{m_r}{m}\right)^3\delta_{l0},
\eeq

\noindent
in complete accord with the well known result \cite{sal,fm,gy2}. Note
that despite its appearance this result is symmetric under permutation of
the heavy and light particles, as expected beforehand, since the
diagrams with two transverse exchanges are  symmetric. In order to preserve
this symmetry we cut the integral from below at momenta of order
$m_rZ\alpha$, calculating the contribution in \eq{transcut}.

In order to obtain the state-dependent low-frequency contribution of the
double transverse exchange it is necessary  to restore the  dependence of
the graphs with two exchanged photons on the external momenta and calculate
the matrix elements of these diagrams between the momentum dependent wave
functions. Respective momentum integrals should be cut off from above at
$m_rZ\alpha$. The wave function momenta provide an effective lower cutoff
for the loop integrals and one may get rid of the upper cutoff by matching
the low- and high-frequency contributions. The calculation for an
arbitrary principal quantum number is rather straightforward but
tedious \cite{sal,fm,eryen1,gy2,erickson2,erickgrotch} and leads to the
result

\beq
\Delta      \label{tottrtr}
E=\left\{\left[2\ln\frac{2Z\alpha}{n}+2[\psi(n+1)-\psi(1)]+\frac{n-1}{n}
+\frac{8(1-\ln2)}{3}-2\frac{M^2\ln\frac{m}{m_r}-m^2\ln\frac{M}{m_r}}{M^2-m^2}
\right] \delta_{l0}
\right.
\eeq
\[
\left.
-\frac{1-\delta_{l0}}{l(l+1)(2l+1)}\right\}
\frac{m}{M}
\frac{(Z\alpha)^5m}{\pi n^3}\left(\frac{m_r}{m}\right)^3.
\]

Let us emphasize that the total contribution of the double transverse
exchange is given by the matrix element of the two-photon exchanges
between the Schr\"odinger-Coulomb wave functions, and no kernels with
higher number of exchanges arise in this case, unlike the case of the
main contribution to the Lamb shift discussed in Section
\ref{leadinglamb} and the case of the transverse-Coulomb recoil
contribution which we will discuss next.

\subsection{Transverse-Coulomb Term}

\begin{figure}[ht]
\centerline{\epsfig{file=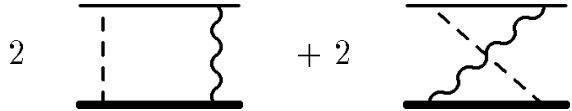,height=1.5cm}}
\vspace{0.5cm}
\caption{Transverse-Coulomb two-photon exchanges}
\label{trcoulfig}
\end{figure}

One should expect that the contribution of the transverse-Coulomb
diagrams in Fig.\ \ref{trcoulfig} would vanish in the scattering
approximation because, in this approximation, there are no external
vectors which are needed in order to contract the transverse photon
propagator. The only available vector in the scattering approximation
is the exchanged momentum itself, which turns into zero after
contraction with the transverse photon propagator. It is easy to check
that this is just what happens, the electron and proton traces are
proportional to the exchanged momentum $k_i$  in the scattering
approximation and vanish, being dotted with the transverse photon
propagator.

This does not mean, however, that the diagrams with one transverse
exchange do not contribute to the energy shift. We still have to
explore if any contributions could be generated by the exchange of the
transverse photon, with a small momentum between $m_r(Z\alpha)$ and
$m_r(Z\alpha)^2$, when one clearly cannot neglect the
momenta of the external wave functions which are of the same order of
magnitude.  Hence, we have to consider all kernels in Fig.\
\ref{irredkernfig} with a transverse exchanged photon spanning an
arbitrary number of Coulomb exchanges. As we have already discussed in
the beginning of this section one might expect that when the momentum
of the transverse photon is smaller than the characteristic atomic
momentum $m_rZ\alpha$ (in other words when the wavelength of the
transverse quantum is larger than the size of the atom) the
contribution to the Lamb shift generated by such a photon would only
differ by an additional factor $m/M$ from the leading contribution to
the Lamb shift of order $\alpha(Z\alpha)^4m$, simply because such a
photon cannot tell the electron from the proton. The extra factor $m/M$
is due simply to the smaller velocity of the heavy particle in the atom
(we remind the reader that the transverse photon interaction vertex
with a charged particle in the nonrelativistic approximation is
proportional to the velocity of the particle). Old-fashioned
perturbation theory is more suitable for exploration of such small
intermediate momenta contributions.  Correction due to the exchange of
the transverse photon is described in this framework simply as a second
order perturbation theory contribution where the role of the
perturbation potential plays the transverse photon emission
(absorption) vertex. In this framework the Coulomb potential plays the role
of the unperturbed potential, so the simple second order contribution
which we just described takes into account all kernels of the
relativistic two-body equation in Fig.\ \ref{irredkernfig} with any
number of the Coulomb exchanges spanned by the transverse photon.
Summation over intermediate states in the nonrelativistic perturbation
theory in our case means integration over all intermediate momenta. It
is clear that for momenta larger than the characteristic atomic
momentum $m_rZ\alpha$ integration over external wave function momenta
decouples (and we obtain instead of the wave functions their value at
the origin) and one may forget about the binding energies in the
intermediate states. Then the contribution of this high (larger than
$m_rZ\alpha$) region of momenta reduces to the matrix element of the
Breit interaction (transverse quanta exchange). As we have explained
above, this matrix element does not give any contribution to the Lamb shift
(but it gives the main contribution to hyperfine splitting, see below). All
this means that the total recoil correction of order $(Z\alpha)^5(m/M)m$ may
be calculated in the nonrelativistic approximation.  Calculations go exactly
in the same way as calculation of the leading low energy contribution to the
Lamb shift in Section \ref{leadinglamb}. Due to validity of the
nonrelativistic approximation the matrix elements of the Dirac matrices
corresponding to the emission of transverse quanta by the constituent
particles reduce to the velocities $\alpha_i\rightarrow p_i/m_i$, where
$p_i$ and $m_i$ are the momenta of the constituents (of the same magnitude
and with opposite directions in the center of mass system) and their masses.
Note that the recoil factor arises simply as a result of kinematics. Again,
as in the case of the main contribution to the Lamb shift in Section
\ref{leadinglamb} one introduces an auxiliary parameter $\sigma$
($m_r(Z\alpha)^2\ll\sigma\ll m_r(Z\alpha)/n$) in order to facilitate further
calculations. Then the low-frequency contribution coincides up to an extra
factor $2Zm/M$ (extra factor $2$ arises due to two ways for emission of the
transverse quanta, by the first and the second constituents), with the
respective low-frequency contribution of order $\alpha(Z\alpha)^4m$, and we
may simply borrow the result from that calculation (compare
\eq{diracslope})

\beq
\Delta E^<=\frac{8}{3}\left[\ln\frac{2\sigma}{m_r(Z\alpha)^2}\delta_{l0}
-\ln k_0(n,l)\right]\frac{m}{M}\frac{(Z\alpha)^5m}{\pi
n^3}\left(\frac{m_r}{m}\right)^3
\eeq

In the region where $k\geq \sigma$ we may safely neglect the binding
energies in the denominators of the second-order perturbation theory and
thus simplify the integrand. After integration one obtains

\beq
\Delta E^>=\frac{8}{3}\left[\ln\frac{m_r(Z\alpha)}{n\sigma}
+[\psi(n+1)-\psi(1)]-\frac{1}{2n}+\frac{5}{6}+\ln2
\right.
\eeq
\[
\left.
-\frac{1-\delta_{l0}}{2l(l+1)(2l+1)}\right]
\frac{m}{M}\frac{(Z\alpha)^5m}{\pi
n^3}\left(\frac{m_r}{m}\right)^3\delta_{l0}.
\]

Let us emphasize once more that, as was discussed above, the "high
frequencies" in this formula are effectively cut at the characteristic
bound state momenta $m_r(Z\alpha)/n$. This leads to two specific
features of the formula above.  First, this expression contains the
characteristic logarithm of the principal quantum number $n$, and,
second, the logarithm of the recoil factor $(1+m/M)$ is missing, unlike
the case of the nonrecoil correction of order $\alpha(Z\alpha)^4m$ in
\eq{diracslope}. The source of this difference is easy to realize. In
the nonrecoil case the effective upper cutoff was supplied by the
electron mass $m$, while at low frequency only the reduced mass enters
all expressions. This mismatch between  masses leads to appearance of
the logarithm of the recoil factor. In the present case, the effective
upper cutoff $m_r(Z\alpha)/n$ also depends only on the reduced mass,
and, hence, an extra factor under the logarithm does not arise.

Matching both contributions we obtain \cite{sal,fm,eryen1,gy2,erickson2}

\beq           \label{ctcontr}
\Delta E=\frac{8}{3}\left\{\left[\ln\frac{2}{nZ\alpha}
+[\psi(n+1)-\psi(1)]-\frac{1}{2n}+\frac{5}{6}+\ln2\right]\delta_{l0}
-\ln k_0(n,l)
\right.
\eeq
\[
\left.
-\frac{1-\delta_{l0}}{2l(l+1)(2l+1)}\right\}
\frac{m}{M}\frac{(Z\alpha)^5m}{\pi n^3}\left(\frac{m_r}{m}\right)^3.
\]

The total recoil correction of order $(Z\alpha)^5(m/M)m$ is given by the sum
of the expressions in \eq{totcul}, \eq{tottrtr}, and \eq{ctcontr}

\beq      \label{totrecza5lamb}
\Delta E=\left\{\left[\frac{2}{3}\ln(Z\alpha)^{-1}
+\frac{14}{3}\left(\ln\frac{2}{n}+\psi(n+1)-\psi(1)+\frac{2n-1}{2n}\right)
-\frac{1}{9}
\right.\right.
\eeq
\[
\left.\left.
-2\frac{M^2\ln\frac{m}{m_r}-m^2\ln\frac{M}{m_r}}{M^2-m^2}
\right]\delta_{l0}
-\frac{8}{3}\ln k_0(n,l)-\frac{7(1-\delta_{l0})}{3l(l+1)(2l+1)}\right\}
\frac{m}{M}\frac{(Z\alpha)^5m}{\pi n^3}\left(\frac{m_r}{m}\right)^3.
\]

\section{Recoil Corrections of Order $(Z\alpha)^6(\lowercase{m}/M)
\lowercase{m}$}

\subsection{The Braun Formula}

Calculation of the recoil corrections of order $(Z\alpha)^6(m/M)m$ requires
consideration of the kernels with three exchanged photons. As in the case of
recoil corrections of order $(Z\alpha)^5$ low exchange momenta produce
nonvanishing contributions, external wave functions do not decouple,
and exact calculations in the direct diagrammatic framework are rather
tedious and cumbersome. There is a long and complicated history of
theoretical investigation on this correction. The program of diagrammatic
calculation was started in \cite{erickgrotch,dgo,dge}. Corrections
obtained in these works contained logarithm $Z\alpha$ as well as a
constant term.  However, completely independent calculations
\cite{kmy,fkmy} of both recoil and nonrecoil logarithmic contributions
of order  $(Z\alpha)^6$ showed that somewhat miraculously all
logarithmic terms cancel in the final result. This observation required
complete reconsideration of the whole problem. The breakthrough was
achieved in \cite{pg}, where one and the same result was obtained
in two apparently different frameworks.  The first, more traditional
approach, used earlier in \cite{eg,erickgrotch,dgo,dge}, starts
with an effective Dirac equation in the external field.  Corrections to
the Dirac energy levels are calculated with the help of a systematic
diagrammatic procedure.  The other logically independent calculational
framework, also used in \cite{pg}, starts with an exact expression for
all recoil corrections of the first order in the mass ratio of the light and
heavy particles $m/M$.  This remarkable expression, which is exact in
$Z\alpha$, was first discovered by M. A. Braun \cite{braun}, and rederived
and refined later in a number of papers \cite{shab,yelkh,pg}.  A
particularly transparent representation of the Braun formula was obtained in
\cite{yelkh}

\begin{equation}                      \label{braun}
\Delta E_{rec}=-\frac{1}{M}Re\int\frac{d\omega d^3k}{(2\pi)^4 i}
<{n}|\left({\bf p}-{\bf\hat D}(\omega,{\bf
k})\right)G(E+\omega)\left({\bf p}-{\bf\hat D}(\omega,{\bf
k})\right)|n>,
\end{equation}

\noindent
where $G(E+\omega)$ is the total Green function of the Dirac electron
in the external Coulomb field, and ${\bf\hat D}(\omega,{\bf k})$ is the
transverse  photon propagator in the Coulomb gauge

\begin{equation}
{\bf\hat D}(\omega,{\bf k})=
-4\pi Z\alpha\frac{\mbox{\boldmath$\alpha_k$}}{\omega^2-{\bf k}^2+i0}\:,
\end{equation}

\noindent
and

\begin{equation}
\mbox{\boldmath$\alpha_k$}=\mbox{\boldmath$\alpha$}-\frac{{\bf
k}(\mbox{\boldmath$\alpha k$})}{{\bf k}^2},\qquad
\mbox{\boldmath$\alpha$}=\gamma_0\mbox{\boldmath$\gamma$}\:.
\end{equation}

Before returning to the recoil corrections of order $(Z\alpha)^6$ we
will digress to the Braun formula. We will not give a detailed
derivation of this formula, referring the reader instead to the original
derivations \cite{braun,shab,yelkh,pg}. We will however present below
some physically transparent semiquantative arguments which make the
existence and even the exact appearance of the Braun formula very
natural.

\begin{figure}[h]
\centerline{\epsfig{file=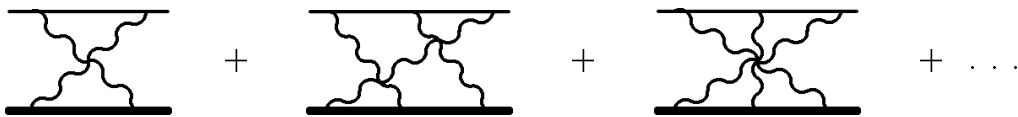,height=1.5cm}}
\vspace{0.5cm}
\caption{Irreducible kernels with crossed exchange photon lines}
\label{crossedextkernfig}
\end{figure}

Let us return to the original Bethe-Salpeter equation
(see \eq{bethesalp}). As we have already discussed there are many
ways to organize the Feynman diagrams which comprise the kernel of this
equation. However, in all common perturbation theory considerations of
this kernel the main emphasis is on presenting the kernel in an
approximate form sufficient for calculation of corrections to the
energy levels of a definite order in the coupling constant $\alpha$.
The revolutionary idea first suggested in \cite{labz72} and
elaborated in \cite{braun}, was to reject such an approach completely,
and instead to organize the perturbation theory with respect to another
small parameter, namely, the mass ratio of the light and heavy
particles. To this end an expansion of the heavy particle propagator
over $1/M$ was considered in \cite{braun}. It is well known (see, e.g.,
\cite{braun}) that in the leading order of this expansion the
Bethe-Salpeter equation reduces to the Dirac equation for the light
particle in the external Coulomb field created by the heavy particle.
This is by itself nontrivial, but well understood by now, since to
restore the Dirac equation in the external field one has to take into
account irreducible kernels of the Bethe-Salpeter equation with
arbitrary number of crossed exchange photon lines (see Fig.\
\ref{crossedextkernfig}). Unlike the solutions of the effective Dirac
equation, considered above in section \ref{effdir}, the solutions of
the Dirac equation obtained in this way contain as a mass parameter the
mass of the light particle, and not the reduced mass of the system.
This is the price one has to pay in the Braun approach for summation of
all corrections in the expansion  over $Z\alpha$.  The zero-order Green
function in this approach is simply the Coulomb-Dirac Green function.
The next step in the derivation of the Braun formula is to consider all
kernels of the Bethe-Salpeter equation which produce corrections of
order $m/M$. The crucial observation which immediately leads to the
closed expression for the recoil corrections of order $m/M$, is that
all corrections linear in the mass ratio are generated by the kernels
where all but one of the heavy particle propagators are replaced by the
leading terms in their large mass expansion, and this remaining
propagator is replaced by the next term in the large mass expansion of
the heavy particle propagator.  Respective kernels with the minimum
number of exchanged photons are the box and the crossed box diagrams in
Fig.\ \ref{twophotexchfig} where the heavy particle propagator is
replaced by the second term in its large mass expansion.  All diagrams
obtained from these two by insertions of any number of external Coulomb
photons between the two exchanges in Fig.\ \ref{twophotexchfig} and/or
of any number of the radiaitive photons in the electron line also
generate linear in the mass ratio corrections.  It is not difficult to
figure out that these are the only kernels which produce corrections
linear in the small mass ratio, all other kernels generate corrections
of higher order in $m/M$, and, hence, are not interesting in this
context.

Then the linear in the small mass ratio contribution to the energy
shift is equal to the matrix element of the two graphs in Fig.\
\ref{twophotexchfig} with the total electron Green function in the
external Coulomb field instead of the upper electron line. This matrix
element which should be calculated between the unperturbed
Dirac-Coulomb wave functions reduces after simple algebraic
transformations to the Braun formula in \eq{braun}. All terms in the
Braun formula have a transparent physical sense. The term containing
product $\bf pp$ (first obtained in \cite{labz72}) originates from the
exchange of two Coulomb photons, the terms with $\bf p\hat D$ and
$\bf\hat Dp$ correspond to the exchange of Coulomb and magnetic
(transverse) quanta, and the term $\bf\hat D \hat D$ is connected with
the double transverse exchange.

Another useful perspective on the Braun formula is provided by the idea,
first suggested in the original work \cite{braun}, and later used as a tool
to rederive \eq{braun} in \cite{yelkh,pg}, that the recoil corrections
linear in the small mass ratio $m/M$ are associated with the matrix element
of the nonrelativistic proton Hamiltonian

\beq                        \label{nonrelham}
H=\frac{({\bf p}-e{\bf A})^2}{2M}.
\eeq

There is a clear one-to one correspondence between the terms in this
nonrelativistic Hamiltonian and the respective terms in \eq{braun}. The
latter could be obtained as matrix elements of the operators which enter the
Hamiltonian in \eq{nonrelham} \cite{yelkh,pg}.

\subsection{Lower Order Recoil Corrections and the Braun formula}

Being exact in the parameter $Z\alpha$ and an expansion in the mass ratio
$m/M$ the Braun formula in \eq{braun} should reproduce with linear accuracy
in the small mass ratio all purely recoil corrections of orders
$(Z\alpha)^2(m/M)m$, $(Z\alpha)^4(m/M)m$, $(Z\alpha)^5(m/M)m$ in
\eq{leadrecoil} which were discussed above.

Corrections of lower orders in $Z\alpha$ are generated by the simplified
Coulomb-Coulomb  and Coulomb-transverse entries in \eq{braun}. The main part
of the Coulomb-Coulomb contribution in Eq.(\ref{braun}) may be written in
the form

\begin{equation}    \label{coulombbr}
\Delta E^{(1)}_{Coul}=
\frac{1}{2M}<{n}|{\bf p}^2|n>,
\end{equation}

\noindent
while the Breit (nonretarded) part of the magnetic contribution has the
form

\begin{equation}      \label{breitbr}
\Delta E_{Br}=
-\frac{1}{2M}
<{n}|{\bf p}{\bf\hat D}(0,k)
+{\bf\hat D}(0,k){\bf p}|n>.
\end{equation}

\noindent
Calculation of the matrix elements in \eq{coulombbr} and \eq{breitbr} is
greatly simplified by the use of the virial relations (see, e.g.,
\cite{ee,shab91}), and one obtains the sum of the contributions in
\eq{coulombbr} and \eq{breitbr} in a very nice form \cite{shab}
(compare \eq{leadrecoil})

\begin{equation}                  \label{gys}
\Delta E^{(1)}_{Coul}+\Delta E_{Br}=\frac{m^2-E_{nj}^2}{2M}
=\left\{-\frac{m}{M}\left[f(n,j)-1\right]-\frac{m}{2M}\left[f(n,j)-1\right]^2
\right\}m,
\end{equation}

\noindent
where $E_{nj}$ and $f(n,j)$ are defined in \eq{naivdirac} and
\eq{fnjdef}, respectively. This representation again emphasizes the
simple physical idea behind the Braun formula that the recoil
corrections of the first order in the small mass ratio $m/M$ are given
by the matrix elements of the heavy particle kinetic energy.

The recoil correction in \eq{gys} is the leading order $(Z\alpha)^4$
relativistic contribution to the energy levels generated by the
Braun formula, all other contributions to the energy levels produced
by the remaining terms in the Braun formula start at least with the
term of order $(Z\alpha)^5$ \cite{shab}.  The expression in \eq{gys}
exactly reproduces all contributions linear in the mass ratio in
\eq{leadrecoil}. This is just what should be expected since it is
exactly Coulomb and Breit potentials which were taken in account in the
construction of the effective Dirac equation which produced
\eq{leadrecoil}.  The exact mass dependence of the terms of order
$(Z\alpha)^2(m/M)m$ and $(Z\alpha)^4(m/M)m$ is contained in
\eq{leadrecoil}, and, hence, terms linear in the mass ratio in \eq{gys} give
nothing new.  It is important to realize at this stage that the
contributions of order $(Z\alpha)^6(m/M)m$ in \eq{leadrecoil} and \eq{gys}
coincide as well, so any corrections of this order obtained with the help of
the entries in the Braun formula not taken care of in \eq{coulombbr} and
\eq{breitbr}, should be added to the order $(Z\alpha)^6(m/M)m$ contribution
in \eq{leadrecoil}.

\subsection{Recoil Correction of Order $(Z\alpha)^6(\lowercase{m}/M)
\lowercase{m}$ to the $S$ Levels}\label{elkfootn}

Calculation of the recoil contribution of order $(Z\alpha)^6(m/M)m$ to
the $1S$ and $2S$ states generated by the Braun formula was first performed
in \cite{pg}. Separation of the high- and low-frequency  contributions was
made with the help of the $\epsilon$-method \cite{pachalpha6} described
above in Section \ref{pachappr}. Hence, not only were contributions of order
$(Z\alpha)^6(m/M)m$ obtained in \cite{pg}, but also parts of recoil
corrections of order $(Z\alpha)^5$ linear in $m/M$, discussed in Section
\ref{lambreczalpha5}, were reproduced for the $1S$-state. The older methods
of Section \ref{lambreczalpha5} lead to a more precise  result for the
recoil corrections of order $(Z\alpha)^5$, and, hence, this calculation in
the framework of the Braun formula is not too interesting on its own.
However, it served as an important check of self-consistency of calculations
in \cite{pg}. Calculations in \cite{pg}, while in principle very
straightforward, turned out to be rather lengthy just because all
corrections of previous orders in $Z\alpha$ were recovered.

The agreement on the magnitude of $(Z\alpha)^6(m/M)m$ contribution for the
$1S$ and $2S$ states obtained in the diagrammatic approach and in the
framework of the Braun formula achieved in \cite{pg} seemed to
put an end to all problems connected with this correction.  However, it was
claimed in a later work \cite{elkh}, that the result of \cite{pg} is in
error. The discrepancy between the results of \cite{pg,elkh} was even
more confusing since the calculation in \cite{elkh} was also performed
with the help of the Braun formula. It was observed in \cite{elkh} that due
to the absence of the logarithmic contributions of order $(Z\alpha)^6(m/M)m$
proved earlier in \cite{fkmy}, the calculations may be organized in a more
compact way than in \cite{pg}. The main idea of \cite{elkh} is that it is
possible to make some approximations which are inadequate for calculation of
the contributions of the previous orders in $Z\alpha$, significantly
simplifying calculation of the correction of order $(Z\alpha)^6(m/M)m$. Due
to absence of the logarithmic contributions of order $(Z\alpha)^6(m/M)m$
proved in \cite{fkmy}, infrared divergences connected with these crude
approximations would be powerlike and can be safely thrown away. Next,
absence of logarithmic corrections of order $(Z\alpha)^6(m/M)m$ means that
it is not necessary to worry too much about matching the low- and
high-frequency (long- and short-distance in terms of \cite{elkh})
contributions, since each region will produce only nonlogarithmic
contributions and correction terms would be suppressed as powers of the
separation parameter. Of course, such an  approach would be doomed if the
logarithmic divergences were present, since in such a case one could not
hope to calculate an additive constant to the logarithm, since the exact
value of the integration cutoff would not be known.

Despite all these nice features of the approach of \cite{elkh} the
result was erroneous and contradicted the result in \cite{pg}. The
discrepancy was resolved in \cite{eg97}, where a new logically
independent calculation of the recoil corrections of order
$(Z\alpha)^6(m/M)m$ was performed.  A subtle error in dealing with
cutoffs in \cite{elkh} was discovered and the result of \cite{pg} for
the $S$-states with $n=1,2$ was confirmed. The recoil correction of
order $(Z\alpha)^6(m/M)m$ for $S$ states with arbitrary principal
quantum number $n$ beyond that which is already contained in
\eq{leadrecoil} has the form \cite{eg97} \footnote{The author of
\cite{elkh} has published later a new paper \cite{elkhnew} where the
error in \cite{elkh} is acknowledged. A new result for the recoil
corrections of order $(Z\alpha)^6(m/M)m$ was obtained in
\cite{elkhnew}, which, being different from the previous result
\cite{elkh} of the same author, contradicts also results of
\cite{pg,eg97}. We think that the manner of separation of the
contributions from the large and small distances in \cite{elkhnew} is
arbitrary and inconsistent, and consider the result of \cite{elkhnew}
to be in error.\label{elkfootnn}}

\begin{equation}                 \label{tot}
\Delta
E_{tot}=\left(4\ln2-\frac{7}{2}\right)\frac{(Z\alpha)^6}{n^3}\frac{m}{M}\:m.
\end{equation}

\noindent
This result was recently confirmed in \cite{sab} where the recoil correction
of the first order in the mass ratio was calculated without expansion over
$Z\alpha$ for $1S$ and $2S$ states in hydrogen. The difference between the
results in \eq{tot} and in \cite{sab} may be considered as an estimate
of the recoil corrections of higher order in $Z\alpha$. This difference
is equal $0.22(1)$ kHz for the $1S$ level, and is about  $0.03$ kHz for
the $2S$ level. It is too small to be important for comparison between
theory and experiment at the current accuracy level.

\subsection{Recoil Correction of Order $(Z\alpha)^6(\lowercase{m}/M)
\lowercase{m}$ to the Non-$S$ Levels}  \label{recza6kh}

Recoil corrections of order $(Z\alpha)^6(m/M)m$ to the energy levels with
nonvanishing orbital angular momentum may also be calculated with the
help of the Braun formula \cite{jenpach}. We would prefer to discuss
briefly another approach, which was used in the first calculation of
the recoil corrections of order $(Z\alpha)^6(m/M)m$ to the $P$ levels
\cite{gemk}.  The idea of this approach (see, e.g., review in
\cite{kmy}) is to extend the standard Breit interaction Hamiltonian
(see, e.g., \cite{blp}) which takes into account relativistic
corrections of order $v^2/c^2$ to the next order in the nonrelativistic
expansion, and also take into account the corrections of order
$v^4/c^4$.  Contrary to the common wisdom, such an approach turns out
to be quite feasible and effective, and it was worked out in a number
of papers \cite{ky,kmy}, and references therein. This nonrelativistic
approach is limited to the calculation of the large distance (small
intermediate momenta) contributions since any short distance correction
leads effectively to an ultraviolet divergence in this framework.
Powerlike ultraviolet divergences demonstrate the presence of the
corrections of the lower order in $Z\alpha$ (in contrast to the
scattering approximation where the presence of such corrections reveals
itself in the form of powerlike infrared divergences), and are not
under control in this approach.  However, the logarithmic ultraviolet
divergences are well under control and produce logarithms of the fine
structure constant. A number of logarithmic contributions to the energy
levels and decay widths were calculated in this approach \cite{ky,kmy}.

In the case of states with nonvanishing angular momenta the small distance
contributions are effectively suppressed by the vanishing of the wave
function at the origin, and the  perturbation theory becomes convergent in
the nonrelativistic region. Then this nonrelativistic approach leads to an
exact result for the recoil correction of order $(Z\alpha)^6(m/M)m$ for the
$P$ states \cite{gemk}

\beq
\Delta E(nP)=\frac{2}{5}\left(1-\frac{2}{3n^2}\right)
\frac{(Z\alpha)^6}{n^3}\frac{m}{M}\:m.
\eeq

Again, this expression contains only corrections not taken into account
in \eq{leadrecoil}. The approach of \cite{gemk} may be generalized for
calculation of the recoil corrections to the energy levels with even
higher orbital angular momenta.

The general expression for the recoil corrections of order
$(Z\alpha)^6(m/M)m$ to the energy level with an arbitrary nonvanishing
angular momentum was obtained in \cite{jenpach}

\beq
\Delta E(nL)=\frac{3}{4(l-\frac{1}{2})(l+\frac{1}{2})
(l+\frac{3}{2})}\left(1-\frac{l(l+1)}{3n^2}\right)
\frac{(Z\alpha)^6}{n^3}\frac{m}{M}\:m.
\eeq

\section{Recoil Correction of Order $(Z\alpha)^7(\lowercase{m}/M)$}
\label{recoilgenza7}

Leading logarithm squared contribution to the recoil correction of
order $(Z\alpha)^7(\lowercase{m}/M)$ was recently independently
calculated in two works \cite{pk99,melelkh99} in the same
framework \cite{khriplovich} as the corrections $(Z\alpha)^6(m/M)m$ to
the $P$ levels above

\beq         \label{recza7}
\Delta E=-\frac{11}{15}
\frac{(Z\alpha)^7m}{\pi n^3}\ln^2(Z\alpha)\frac{m}{M}
\left(\frac{m_r}{m}\right)^3\delta_{l0}.
\eeq

Numerically this contribution is much smaller than the experimental
error bars of the current Lamb shift measurements. However,
due to linear dependence of the recoil correction on the
electron-nucleus mass ratio, the respective contribution to the
hydrogen-deuterium isotope shift (see Section \ref{isshift} below) is
larger than the experimental uncertainty, and should be taken into
account in comparison between theory and experiment. One half of the
leading logarithm squared contribution in \eq{recza7} ($-0.21$ kHz for
the $1S$ level in hydrogen) may be accepted as a fair estimate of the
yet uncalculated single logarithmic and nonlogarithmic contributions of
order $(Z\alpha)^7(\lowercase{m}/M)$.

\begin{center}
\underline{Table VIII. Recoil Corrections to Lamb Shift}
\nopagebreak

\begin{tabular}{|l|rl|c|c|}    \hline
     & $\frac{(Z\alpha)^5}{\pi n^3}\frac{m_r^3}{mM}$ &   & $\Delta E(1S)$ kHz
&$\Delta E(2S)$ kHz
\\ \hline  \hline
Coulomb-Coulomb Term$$&&&$$         &
\\
$$&&&$$         &\\
Salpeter (1952)\cite{sal}&$-\frac{4}{3}\delta_{l0}$&&$-590.03$&$-73.75$\\
Fulton,Martin(1954)\cite{fm}&&&$$ &\\
$$&&&$$         &
\\ \hline
Transverse-Transverse Term$$&&&$$         &
\\
Bulk Contribution$$&&&$$         &\\
Salpeter (1952)\cite{sal}&$\{2\ln\frac{2Z\alpha}{n}+2[\psi(n+1)-\psi(1)]
$&&$$         &\\
Fulton,Martin(1954)\cite{fm}&$+\frac{n-1}{n} +\frac{8(1-\ln2)}{3}\}
\delta_{l0}-\frac{1-\delta_{l0}}{l(l+1)(2l+1)}$&&$-2~494.01$ &$-305.46$\\
Erickson,Yennie (1965)\cite{eryen1}$$&&&$$         &\\
Grotch,Yennie(1969)\cite{gy2}&$$&&$$&\\
Erickson(1977)\cite{erickson2}$$&&&$$        & \\
Erickson,Grotch(1988)\cite{erickgrotch}$$&&&$$&         \\
 \hline
Transverse-Transverse Term$$&&&$$         &\\
Very high momentum $$&&&$$         &\\
Contribution$$&&&$$         &\\
&$-\frac{2}{M^2-m^2}(M^2\ln\frac{m}{m_r}-m^2\ln\frac{M}{m_r})\delta_{l0}$
&&$-0.48$&$-0.06$\\
Fulton,Martin(1954)\cite{fm}&&&$$ &\\
Erickson(1977)\cite{erickson2}$$&&&$$        & \\
$$&&&$$         &
\\ \hline
Coulomb-Transverse Term$$&&&$$         &\\
$$&&&$$         &\\
Salpeter (1952)\cite{sal}&$\frac{8}{3}\{[\ln\frac{2}{nZ\alpha}
+[\psi(n+1)-\psi(1)]$&&$$         &\\
Fulton,Martin(1954)\cite{fm}&$-\frac{1}{2n}+\frac{5}{6}+\ln2]\delta_{l0}
$&&$$ &\\
Erickson,Yennie (1965)\cite{eryen1}
&$-\ln k_0(n,l)-\frac{1-\delta_{l0}}{2l(l+1)(2l+1)}\}$&&$5~494.03$&$720.56$\\
Grotch,Yennie(1969)\cite{gy2}&$$&&$$&\\
Erickson(1977)\cite{erickson2}$$&&&$$        & \\
Erickson,Grotch(1988)\cite{erickgrotch}$$&&&$$&         \\
\hline
$\Delta E(nS)$&&&$$&\\
&&&$$&\\
Pachucki,Grotch(1993)\cite{pg}&$(4\ln2-\frac{7}{2})(\pi Z\alpha)\delta_{l0}$
&&$-7.38$&$-0.92$\\
Eides,Grotch(1997)\cite{eg97}&
&& & \\
\hline
$\Delta E(nL)(l\neq0)$
&&&$$ &\\
Golosov,Elkhovski,&&&$$&\\
Milshtein,Khriplovich(1995)\cite{gemk}&$\frac{3}{4(l-\frac{1}{2})
(l+\frac{1}{2})(l+\frac{3}{2})}
(1-\frac{l(l+1)}{3n^2})(\pi Z\alpha)$&&$$& \\
Jentschura,Pachucki(1996)\cite{jenpach}&&&$$ & \\
\hline
$\Delta E(nS)$&&&$$&\\
&&&$$&\\
Pachucki,Karshenboim(1999)\cite{pk99}&$-\frac{11}{15}
(Z\alpha)^2\ln^2(Z\alpha)\delta_{l0}$&&$-0.42$& $-0.05$\\
Melnikov,Yelkhovsky(1999)\cite{melelkh99}& &&
$$ & \\
\hline
\end{tabular}
\end{center}

\part{Radiative-Recoil Corrections}

In the standard nomenclature the name radiative-recoil is reserved for the
recoil corrections to pure radiative effects, i.e., for corrections of the
form $\alpha^m(Z\alpha)^n(m/M)^k$.

Let us start systematic discussion of such corrections with the
recoil corrections to the leading contribution to the Lamb shift.
The most important observation here is that the mass dependence of all
corrections of order $\alpha^m(Z\alpha)^4$ obtained above is exact, as
was proved in \cite{bg85,bg87}, and there is no additional mass dependence
beyond the one already present in \eq{diracslope}-\eq{pol3}. This conclusion
resembles the similar conclusion about the exact mass dependence of the
contributions to the energy levels of order $(Z\alpha)^4m$ discussed above,
and it is valid essentially for the same reason. The high frequency part of
these corrections is generated only by the one photon exchanges, for which
we know the exact mass dependence, and the only mass scale in the low
frequency part, which depends also on multiphoton exchanges, is the
reduced mass.

\section{Corrections of Order $\alpha(Z\alpha)^5(\lowercase{m}/M)
\lowercase{m}$}

The first nontrivial radiative-recoil correction is of order
$\alpha(Z\alpha)^5$. We have already discussed the nonrecoil
contribution of this order in Section \ref{alphazalpha5nonrec}. Due to
the wave function squared factor this correction naturally contained an
explicit factor $(m_r/m)^3$. Below we will discuss radiative-recoil
corrections of order $\alpha(Z\alpha)^5$ with mass ratio dependence
beyond this factor $(m_r/m)^3$.

\subsection{Corrections Generated by the Radiative Insertions in the
Electron Line} \label{lambradrecel}

\begin{figure}[ht]
\centerline{\epsfig{file=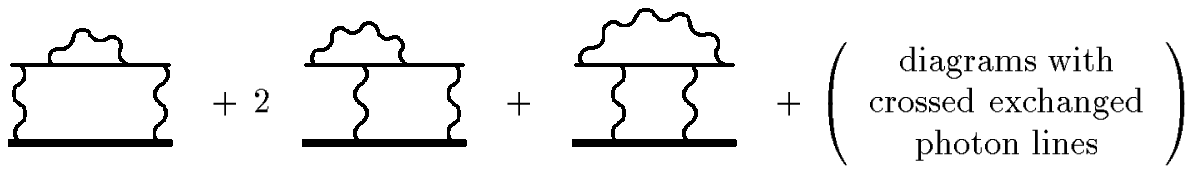,height=2cm}}
\vspace{0.5cm}
\caption{Electron-line radiative-recoil corrections}
\label{ellineradreclamb}
\end{figure}

The diagrammatic calculation of the radiative-recoil correction of order
$\alpha(Z\alpha)^5$ induced by the radiative insertions in the electron
line in Fig.\ \ref{ellineradreclamb} was performed in
\cite{bg85,bg871,bg87}, where it was separated into two steps. First,
the recoil contributions produced by the
nonrelativistic heavy particle pole, which were neglected above in our
discussion of the nonrecoil contributions  in Section
\ref{alphazalpha5nonrec} (see Fig.\ \ref{elineradinscoulfig}) were
calculated. Second, the remaining nonpole contributions of the box and
crossed box diagrams in Fig.\ \ref{ellineradreclamb} were obtained.
Only the high intermediate loop momenta are involved in the
calculations, and the resulting contribution is nonvanishing only for
the $S$ states, and has the form \cite{bg85,bg871,bg87}

\beq                   \label{bgradrec}
\Delta
E=\left(\frac{35}{4}\ln2-8+\frac{1}{5}+\frac{31}{192}-0.415(4)\right)
\frac{\alpha(Z\alpha)^5}{n^3}\frac{m}{M}\:m\:\delta_{l0}
\eeq
\[
=-1.988(4)\:\frac{\alpha(Z\alpha)^5}{n^3}\frac{m}{M}\:m\:\delta_{l0}.
\]

Another viable approach to calculation of this correction is based on the
use of the Braun formula. The Braun formula depends on the total electron
Green function in the external Coulomb field and automatically includes all
radiative corrections to the electron line. Only one-loop insertions of
the radiative photon in the electron line should be preserved in order
to obtain corrections of order $\alpha(Z\alpha)^5$. At first sight
calculation of these corrections with the help of the Braun formula may
seem to be overcomplicated because, as we already mentioned above, the
Braun formula produces a total correction of the first order in the
mass ratio. Hence, exact calculation of the radiative-recoil correction
of order $\alpha(Z\alpha)^5$ with its help would produce not only  the
contributions we called radiative-recoil above, but also the first term
in the expansion over the mass ratio of the purely radiative
contribution in \eq{aza5}. This contribution  should be omitted in
order to avoid double counting. It is not difficult to organize the
calculations based on the Braun formula in such a way that the reduced
mass correction connected with the nonrecoil contribution would not
show up and calculation of the remaining terms would be significantly
simplified \cite{pach95}. The idea is as follows. Purely radiative
corrections of order $\alpha(Z\alpha)^5$, together with the standard
$(m_r/m)^3$ factor were connected with the nonrelativistic heavy
particle pole in the two photon exchange diagrams which corresponds to the
zero order term in the proton propagator expansion over $1/M$. On the other
hand, the Braun formula explicitly picks up the first order term in the
proton propagator heavy mass expansion. This means that the Braun formula
produces the term corresponding to the reduced mass dependence of
the nonrecoil contribution only when the high integration momentum goes
through one of the external wave functions. New radiative-recoil
contributions, which do not reduce to the tail of the mass ratio cubed
factor in \eq{aza5} are generated only by the integration region where
the high momentum goes through the loop described by an explicit
operator in the Braun formula. For calculation of the matrix element in
this regime, it is sufficient to ignore external virtualities and to
approximate the external wave functions by their values at the origin.
The respective calculation reduces therefore to a variant of the
scattering approximation calculation, the only difference being that
the form of the skeleton structure is now determined by the Braun
formula. As usual, in the scattering approximation approach the
integral under consideration contains powerlike infrared divergence,
which corresponds to the recoil contribution of the previous order in
$Z\alpha$ and should simply be subtracted. Explicit calculation of the
Braun formula contribution in this regime was performed in
\cite{pach95}

\beq                   \label{pachradrec}
\Delta
E=\left(2\ln2-\frac{79}{32}-\frac{2.629~46(1)+0.245~23(1)}{\pi^2}\right)
\frac{\alpha(Z\alpha)^5}{n^3}\frac{m}{M}\:m\:\delta_{l0}
\eeq
\[
=-1.374\:\frac{\alpha(Z\alpha)^5}{n^3}\frac{m}{M}\:m\:\delta_{l0}.
\]

The results in \eq{bgradrec} and \eq{pachradrec} contradict each
other. Since both approaches to calculations are quite safe at least
one of them should contain an arithmetic error. Numerically the
discrepancy is about 6 kHz for $1S$ state. This discrepancy is not too
important for the $1S$ Lamb shift measurements, since the error bars of
even the best current experimental results are a few times larger (see
Table XXI below). What is much more important from the phenomenological
point of view, this radiative-recoil correction, linear in the
electron-nucleus mass ratio, directly contributes to the
hydrogen-deuterium isotope shift (see Section \ref{isshift} below), and
the respective discrepancy in the isotope shift is about $18$ times
larger than the experimental uncertainty of the isotope shift. A new
independent calculation of the radiative-recoil contribution of order
$\alpha(Z\alpha)^5(m/M)$ is needed in order to resolve this discrepancy.

The radiative-recoil correction of order $\alpha(Z\alpha)^5$ is clearly
connected with the high integration momenta region in the two-photon
exchange kernels. In this situation there is no need to turn to the Braun
formula, one may directly use the scattering approximation approach which is
ideally suited for such calculation. A new calculation in this framework
is under way now \cite{egs97}.

\subsection{Corrections Generated by the Polarization Insertions in the
Photon Lines}

\begin{figure}[ht]
\centerline{\epsfig{file=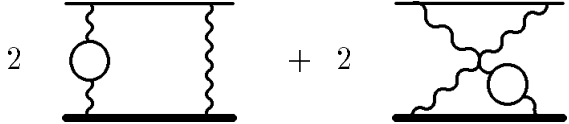,height=1.7cm}}
\vspace{0.5cm}
\caption{Photon-line radiative-recoil corrections}
\label{photlineradreclambfir}
\end{figure}

Calculation of the radiative-recoil correction generated by the one-loop
polarization insertions in the exchanged photon lines in Fig.\
\ref{photlineradreclambfir} follows the same path as calculation of the
correction induced by the insertions in the electron line. The
respective correction was independently calculated analytically both in
the skeleton integral approach \cite{egradrecoil} and with the help of
the Braun formula \cite{pach95}.

Due to the simplicity of the photon polarization operator the calculation
based on the scattering approximation \cite{egradrecoil} is so
straightforward that we can present here all relevant formulae without
making the text too technical.  The skeleton integral for the recoil
corrections corresponding to the diagrams with two exchanged photons in
Fig.\ \ref{twophotexchfig} has the form

\beq                       \label{skelrec}
\Delta E
=\frac{16(Z\alpha)^2|\psi(0)|^2}{m^2(1-\mu^2)}\int_0^\infty
\frac{kdk}{(k^2+\lambda^2)^2}
\left\{\mu\sqrt{1+\frac{k^2}{4}}\left(\frac{1}{k}+\frac{k^3}{8}\right)
\right.
\eeq
\[
\left.
-\sqrt{1+\frac{\mu^2k^2}{4}}\left(\frac{1}{k}+\frac{\mu^4k^3}{8}\right)
-\frac{\mu k^2}{8}\left(1+\frac{k^2}{2}\right)
+\frac{\mu^3k^2}{8}\left(1+\frac{\mu^2k^2}{2}\right)
+\frac{1}{k}\right\},
\]

\noindent
where $\mu=m/M$.

We have already subtracted in \eq{skelrec} the nonrecoil part of the
skeleton integral. This subtraction term is given by the nonrelativistic
heavy particle pole contribution \eq{nonrecskel} in the two photon
exchange. Next, we insert the polarization operator in the integrand in
\eq{skelrec} according to the rule in  \eq{sub}. The skeleton
integrand in \eq{skelrec} behaves as $\mu/k^4$ at small momenta and naive
substitution in \eq{sub} leads to a linear infrared divergence. This
divergence $\int dk/k^2$ would be cut off at the atomic scale $1/(mZ\alpha)$
by the wave function momenta in an exact calculation. The low-momentum
contribution would clearly be of order $\alpha(Z\alpha)^4$ and we may simply
omit it since we already know this correction. Thus, to obtain the recoil
correction of order $\alpha(Z\alpha)^5m$ it is sufficient to subtract the
leading low-frequency asymptote in the radiatively corrected skeleton
integrand. The subtracted integral for the radiative-recoil correction
(the integral in \eq{skelrec} with inserted polarization operator and the
low-frequency asymptote subtracted) has to be multiplied by an additional
factor $2$ needed in order to take into account that the polarization may be
inserted in each of the two photon lines in the skeleton diagrams in
Fig.\ \ref{twophotexchfig}. It can easily be calculated analytically if
one neglects contributions of higher order in the small mass ratio
\cite{egradrecoil}

\beq                   \label{polradrec}
\Delta E=\left(\frac{2\pi^2}{9}-\frac{70}{27}\right)
\frac{\alpha(Z\alpha)^5}{\pi^2n^3}\frac{m}{M}\:m\:\delta_{l0}.
\eeq

Calculation of the same contribution with the help of the Braun formula
was made in \cite{pach95}. In the Braun formula approach one also
makes the substitution in \eq{sub} in the propagators of the exchange
photons, factorizes external wave functions as was explained above (see
Section \ref{lambradrecel}), subtracts the infrared divergent part of the
integral corresponding to the correction of previous order in $Z\alpha$, and
then calculates the integral. The result of this calculation \cite{pach95}
nicely coincides with the one in \eq{polradrec}\footnote{The
radiative-recoil correction to the Lamb shift induced by the
polarization insertions in the exchanged photons was also calculated in
\cite{elkhpol}. The result of that work contradicts the results in
\cite{egradrecoil,pach95}. The calculations in \cite{elkhpol} are made
in the same way as the calculation of the recoil correction of order
$(Z\alpha)^6(m/M)m$ in \cite{elkh}, and lead to a wrong result for the
same reason (see discussion in footnote \ref{elkfootnn} in section
\ref{elkfootn}). }.

\subsection{Corrections Generated by the Radiative Insertions in the Proton
Line} \label{radprot}

We have discussed above insertions of radiative corrections either in the
electron line or in the exchanged photon line in the skeleton diagrams in
Fig.\ \ref{twophotexchfig}. One more option, namely, insertions of a
radiative photon in the heavy particle line also should be considered.
The leading order correction generated by such insertions is a
radiative-recoil contribution of order $(Z^2\alpha)(Z\alpha)^4(m/M)m$.
Note that this correction contains one less power of the parameter
$\alpha$ than the first nontrivial radiative-recoil correction of order
$\alpha(Z\alpha)^5(m/M)m$ generated by the radiative insertions in the
electron line considered above.  There is nothing enigmatic about this
apparent asymmetry, since the mass dependence of the leading order
contribution to the Lamb shift of order $\alpha(Z\alpha)^4m$ in
\eq{diracslope} is known exactly, and thus the would-be radiative
correction of order $\alpha(Z\alpha)^4(m/M)m$ is hidden in the leading
order contribution to the Lamb shift.

No new calculation is needed to obtain the correction of order
$(Z^2\alpha)(Z\alpha)^4(m/M)m$ generated by the radiative insertions in
the proton line. It is almost obvious that to order $(Z\alpha)^4$ the
contributions to the energy level generated by the radiative insertions in
the fermion lines are symmetric with respect to interchange of the light and
heavy lines \cite{fm}. Then in the case of an elementary proton
radiative-recoil correction generated by the radiative photon insertions in
the heavy line may be obtained from the leading order contributions to the
Lamb shift in \eq{diracslope} and \eq{pauli} by the substitutions
$m\rightarrow M$ and $\alpha\rightarrow(Z^2\alpha)$. Both
substitutions are obvious, the first one arises because the leading term in
the infrared expansion of the first order radiative corrections to
the fermion vertex contains the mass of the particle under consideration,
and the second simply  reminds us that the charge of the heavy particle is
$Ze$. Hence, the Dirac form factor contribution is equal to

\beq  \label{diracnucl}
\Delta E=
\left\{\left[\frac{1}{3}\ln\frac{M(Z\alpha)^{-2}}{m_r}+\frac{11}{72}
\right]\delta_{l0}
-\frac{1}{3}\ln k_0(n,l)\right\}\frac{4(Z^2\alpha)(Z\alpha)^4}{\pi
n^3}\frac{m_r^3}{M^2},
\eeq

\noindent
and the Pauli form factor leads to the correction

\beq          \label{pauliprpton}
\delta E_{|l=0}
=\frac{(Z^2\alpha)(Z\alpha)^4}{2\pi n^3}\frac{m_r^3}{M^2},
\eeq
\[
\delta E_{|l\neq
0}
=\frac{j(j+1)-l(l+1)-3/4}{l(l+1)(2l+1)}
\frac{(Z^2\alpha)(Z\alpha)^4m}{2\pi n^3}
\frac{m_r^2}{M}.
\]

These formulae are derived for an elementary heavy particle,
and do not take into account the composite nature of the proton. The
presence of the logarithm of the heavy particle mass $M$ in \eq{diracnucl}
indicates that the logarithmic loop integration in the form factor integral
goes up to the mass of the particle where one could no longer ignore the
composite nature of the proton. For the composite proton the
integration would be cut from above not by the proton mass but by the size
of the proton. The usual way to account for the proton structure is to
substitute the proton form factor in the loop integral.  After calculation
we obtain instead of \eq{diracnucl}

\beq  \label{diracnuclreal}
\Delta E=
\left\{\left[\frac{1}{3}\ln\frac{\Lambda(Z\alpha)^{-2}}{m_r}+\frac{11}{72}
+(-\frac{1}{24}-\frac{7\pi}{32}\frac{\Lambda^2}{4M^2}
+\frac{2}{3}(\frac{\Lambda^2}{4M^2})^2+\ldots)\right]\delta_{l0}
\right.
\eeq
\[
\left.
-\frac{1}{3}\ln k_0(n,l)\right\}\frac{4(Z^2\alpha)(Z\alpha)^4}{\pi
n^3}\frac{m_r^3}{M^2},
\]

\noindent
where $\Lambda^2=0.71$ GeV$^2$ is the parameter in the proton dipole
form factor. As we have expected it replaces the proton mass in the
role of the upper cutoff for the logarithmic loop integration. Note
also that we have obtained an additional constant in
\eq{diracnuclreal}.

The anomalous magnetic moment contribution in \eq{pauliprpton} also
would be modified by inclusion of a nontrivial form factor, but
the contribution to the proton magnetic moment should be
considered together with the nonelectromagnetic contributions to the
proton magnetic moment. The anomalous magnetic moment of the nucleus
determined experimentally includes the electromagnetic contribution
and, hence, even modified by the nontrivial form factor contribution in
\eq{pauliprpton} should be ignored in the phenomenological analysis.
Usually the total contribution of the proton anomalous magnetic moment
is hidden in the main proton charge radius contribution defined via the
Sachs electric form factor \footnote{This topic will be discussed in
more detail below in Section \ref{sachs}.}.

From the practical point of view, the difference between the results in
\eq{diracnucl} and \eq{diracnuclreal} is about $0.18$ kHz for $1S$
level in hydrogen and at the current level of experimental precision
the distinctions between the expressions in \eq{diracnucl} and
\eq{diracnuclreal} may be ignored in the discussion of the Lamb shift
measurements. These distinctions should be however taken into account
in the discussion of the hydrogen-deuterium isotope shift (see below
Section \ref{isshift}).

An alternative treatment of the correction of order
$(Z^2\alpha)(Z\alpha)^4(m/M)m$ was given in \cite{pach95}. The idea of
this work was to modify the standard definition of the proton charge
radius, and include the first  order quantum electrodynamic radiative
correction into the proton radius determined by the strong
interactions. From the practical point of view for the $nS$ levels in
hydrogen the recipe of \cite{pach95} reduces to elimination of the
constant $11/72$ in \eq{diracnucl} and omission of the Pauli correction
in \eq{pauliprpton}. Numerically such a modification reduces the
contribution to the $1S$ energy level in hydrogen by $0.14$ kHz in
comparison with the naive result in \eq{diracnucl}, and increases it by
$0.03$ kHz in comparison with the result in \eq{diracnuclreal}. Hence,
for all practical needs at the current level of experimental precision
there is no contradictions between our result above in
\eq{diracnuclreal}, and the result in \cite{pach95}.

However, the approach of \cite{pach95} from our point of view is
unattractive; we prefer to stick with the standard definition of the
intrinsic characteristics of the proton as determined by the strong
interactions. Of course, in such an approach one has to extract the values
of the proton parameters from the experimental data, properly taking into
account quantum electrodynamic corrections. Another advantage of the
standard approach advocated above is that in the case of the absence
of a nontrivial nuclear form factor (as, for example, in the muonium
atom with an  elementary nucleus) the formula in \eq{diracnuclreal}
reduces to the classical expression in \eq{diracnucl}.

\section{Corrections of Order $\alpha(Z\alpha)^6(\lowercase{m}/M)
\lowercase{m}$}\label{recgenaza6}

Leading logarithm squared contribution of order
$\alpha(Z\alpha)^6(m/M)$ was independenly calculated in
\cite{pk99,melelkh99} in the framework of the approach
developed in \cite{khriplovich} (see discussion in Section
\ref{recza6kh})

\beq       \label{aza7lmabrec}
\Delta E=\frac{2}{3}\frac{\alpha(Z\alpha)^6}{\pi n^3}\ln^2(Z\alpha)^{-2}
\frac{m_r^3}{mM}\:\delta_{l0}
\eeq

This correction is at the present time not too important in the
phenomenological discussion of the $1S$ and $2S$ Lamb shifts. However,
due to the usual linear dependence of the radiative-recoil correction
on the electron-nucleus mass ratio, the double logarithm contribution
in \eq{aza7lmabrec} is already at the present level of experimental
accuracy quite significant as a contribution to the hydrogen-deuterium
isotope shift (see Section \ref{isshift}).

Single logarithmic and nonlogarithmic contributions may be estimated as
one half of the leading logarithm squared contribution, this
constitutes about $0.8$ kHz and $0.1$ kHz for the $1S$ and $2S$ levels
in hydrogen, respectively. In view of the of the rapid experimental
progress in the isotope shift measurements (see Table XXII below)
calculation of these remaining corrections of order
$\alpha(Z\alpha)^6(\lowercase{m}/M) \lowercase{m}$ deserves
further theoretical efforts.


\begin{center}
\underline{Table IX. Radiative-Recoil Corrections}
\nopagebreak

\begin{tabular}{|l|rl|c|c|}    \hline
     & $ $ &   &$\Delta E(1S)$  kHz&$\Delta E(2S)$ kHz
\\ \hline \hline
Radiative insertions&$$&&$$& \\
in the electron line&$$&&$$& \\
&$$&&$$& \\
Bhatt,Grotch(1987)\cite{bg85,bg87,bg871}&$-1.988(4)
\frac{\alpha(Z\alpha)^5}{n^3}\frac{m^2}{M}\delta_{l0}$&&$-20.17$&$-2.52$\\
&$$&&$$& \\
\hline
&$$&&$$& \\
Radiative insertions&$$&&$$& \\
in the electron line&$$&&$$& \\
&$$&&$$& \\
Pachucki(1995)\cite{pach95}&$-1.374
\frac{\alpha(Z\alpha)^5}{n^3}\frac{m^2}{M}\delta_{l0}$&&$-13.94$&$-1.74$\\
&$$&&$$&
\\ \hline
Polarization insertions&$$&&$$& \\
&$$&&$$& \\
Eides,Grotch(1995)\cite{egradrecoil}&
$(\frac{2\pi^2}{9}-\frac{70}{27})
\frac{\alpha(Z\alpha)^5}{\pi^2 n^3}\frac{m^2}{M}\delta_{l0}$&&$-0.41$&$-0.05$\\
Pachucki(1995)\cite{pach95}&$$&&&
\\ \hline
&$$&&$$& \\
Dirac FF insertions&$$&&$$& \\
in the heavy line
&$\left\{\left[\frac{1}{3}\ln\frac{\Lambda(Z\alpha)^{-2}}{m_r}
+(-\frac{1}{24}-\frac{7\pi}{32}\frac{\Lambda^2}{4M^2}\right.\right.$
&&$$& \\
&$$&&$$&\\
&$\left.
+\frac{2}{3}(\frac{\Lambda^2}{4M^2})^2+\ldots)+\frac{11}{72}\right]
\delta_{l0}$
&&$$& \\
&$$&&$$&\\
&$\left.
-\frac{1}{3}\ln k_0(n)\right\}\frac{4(Z^2\alpha)(Z\alpha)^4}{\pi n^3}
\frac{m_r^3}{M^2}$&&$4.58$&$0.58$ \\
&$$&&$$& \\
\hline
&&&$$&\\
Pachucki,Karshenboim(1999)\cite{pk99}&$\frac{2}{3}\frac{\alpha(Z\alpha)^6}
{\pi n^3}\ln^2(Z\alpha)^{-2}
\frac{m_r^3}{mM}\delta_{l0}$&&$1.52$& $0.19$\\
Melnikov,Yelkhovsky(1999)\cite{melelkh99}& &&
$$ & \\
&&&$$&\\
\hline
\end{tabular}
\end{center}

\part{ Nuclear Size and Structure Corrections}
\label{nonelectromagnetic}

The one-electron atom is a composite nonrelativistic system loosely bound by
electromagnetic forces. The characteristic size of the atom is of the order
of the Bohr radius $1/(mZ\alpha)$, and this scale is too large for effects
of other interactions (weak and strong, to say nothing about gravitational)
to play a significant role. Nevertheless, in high precision experiments
effects connected with the composite nature of the nucleus can become
observable. By far the most important nonelectomagnetic contributions
are connected with the finite size of the nucleus and its structure.
Both the finite radius of the proton and its structure constants do not
at present admit precise calculation from first principles  in the
framework of QCD - the modern theory of strong interactions.
Fortunately, the main nuclear parameters affecting the atomic energy
levels may be either measured directly, or admit almost model
independent calculation in terms of other experimentally measured
parameters.

Besides the strong interaction effects connected with the nucleus,
strong interactions affect the energy levels of atoms via nonleptonic
contributions to the photon polarization operator. Once again, these
contributions admit calculation in terms of experimental data, as we
have already discussed above in Setion \ref{heavypartlambpol}. Minor
contribution to the energy shift is also generated by the weak gauge
boson exchange to be discussed below.

\section{Main Proton Size Contribution}\label{lambnuclsize}

The nucleus is a bound system of strongly interacting particles.
Unfortunately, modern QCD does not provide us with the tools to calculate
the bound state properties of the proton (or other nuclei) from first
principles, since the QCD perturbation expansion does not work at
large (from the QCD point of view) distances which are characteristic for
the proton structure, and the nonperturbative methods are not mature enough
to produce good results.

Fortunately, the characteristic scales of the strong and electromagnetic
interactions are vastly different, and at the large distances which are
relevant for the atomic problem the influence of the proton
(or nuclear) structure may be taken into account with the help of a few
experimentally measurable proton properties. The largest and by far the
most important correction to the atomic energy levels connected with the
proton structure is induced by its finite size.

\begin{figure}[h]
\centerline{\epsfig{file=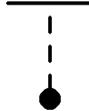,height=1.5cm}}
\vspace{0.5cm}
\caption{Proton radius contribution to the Lamb shift. Bold dot
corresponds to the form factor slope}
\label{ffslope}
\end{figure}

The leading nuclear structure contribution to the energy shift is completely
determined by the slope of the nuclear form factor  in Fig.\
\ref{ffslope} (compare \eq{electronff})

\beq   \label{formfactor}
F(-{\bf k}^2)\approx 1-\frac{\bf k^2}{6}<r^2>.
\eeq

The respective perturbation potential is given by the form factor slope
insertion in the external Coulomb potential (see \eq{ffinsertt})

\beq
-\frac{4\pi(Z\alpha)}{\bf k^2}\rightarrow \frac{2\pi(Z\alpha)}{3}<r^2>,
\eeq

\noindent
and we immediately obtain

\beq          \label{chargerad}
\Delta E=\frac{2\pi(Z\alpha)}{3}<r^2>|\psi(0)|^2
=\frac{2(Z\alpha)^4}{3n^3}m_r^3<r^2>\delta_{l0}.
\eeq

We see that the correction to the energy level induced by the finiteness of
the proton charge radius shifts the energy level upwards, and is
nonvanishing only for the $S$ states\footnote{Note the similarity of this
discussion to the consideration of the level shift induced by the
polarization insertion in the external Coulomb photon in Section
\ref{originlamb}. However, unlike the present case the polarization
insertion leads to a negative contribution to the energy levels since the
polarization cloud screens the source charge.} Physically the finite radius
of the proton means that the proton charge is smeared over a finite volume,
and the electron spends some time inside the proton charge cloud and
experiences a smaller attraction than in the case of the pointlike nucleus
(Compare similar arguments in relation with the finite radius of the
electron below \eq{qff}).

The result in \eq{chargerad} needs some clarification. In the derivation
above it was implicitly assumed that the photon-nucleus vertex is determined
by the expression in \eq{formfactor}. However, for a nucleus with spin this
interaction depends on more than one form factor, and the effective slope of
the photon-nucleus vertex contains in the general case some additional terms
besides the nucleus radius. We will consider the real situation for
nuclei of different spins below.

\subsection{Spin One-Half Nuclei} \label{sachs}

The photon-nucleus interaction vertex is described by the Dirac ($F_1$) and
Pauli ($F_2$) form factors

\beq
\gamma_\mu F_1(k^2)-\frac{1}{2M}\sigma_{\mu\nu}k^\nu F_2(k^2),
\eeq

\noindent
where at small momentum transfers

\beq
F_1(-{\bf k^2})\approx 1-\frac{<r^2>_F}{6}{\bf k^2},
\eeq
\[
F_2(0)=\frac{g-2}{2}.
\]

Hence, at low momenta the photon-nucleus interaction vertex (after the
Foldy-Wouthuysen transformation and transition to the two-component nuclear
spinors) is described by the expression

\beq                        \label{currentexp}
1-{\bf k}^2\frac{1+8F_1'M^2+2F_2(0)}{8M^2}=
1-{\bf k}^2[\frac{1}{8M^2}+\frac{<r^2>_F}{6}+\frac{g-2}{8M^2}].
\eeq

For an elementary proton $<r^2>_F=0$, $g=2$, and only the first term in the
square brackets survives. This term leads to the well known local Darwin
term in the electron-nuclear effective potential (see, e.g., \cite{bg})
and generates the contribution proportional to the factor $\delta_{l0}$
in \eq{barkglov}. As was pointed out in \cite{foldy}, in addition to
this correction, there exists an additional contribution of the same
order produced by the term proportional to the anomalous magnetic
moment in \eq{currentexp}.

However, this is not yet the end of the story, since the proton
charge radius is usually defined via the Sachs electric form factor
$G_E$, rather than the Dirac form factor $F_1$

\beq             \label{electricrad}
G_E(-{\bf k^2})\approx 1-\frac{<r^2>_G}{6}{\bf k^2}.
\eeq

The Sachs electric and magnetic form factors are defined as (see, e.g.
\cite{blp})

\beq          \label{sachsform}
G_E(-{\bf k}^2)=F_1(-{\bf k}^2)-\frac{\bf k^2}{4M^2}F_2(-{\bf k}^2),
\eeq
\[
G_M(-{\bf k}^2)=F_1(-{\bf k}^2)+F_2(-{\bf k}^2).
\]

In terms of this new charge radius the photon-nucleus vertex above has the
form

\beq    \label{sachvertex}
1-{\bf k}^2[\frac{1}{8M^2}+\frac{<r^2>_G}{6}].
\eeq

We now see that for a real proton the charge radius contribution has exactly
the form in \eq{chargerad}, where the charge radius is defined in
\eq{electricrad}. The only other term linear in the momentum  transfer
in the photon-nucleus vertex in \eq{sachvertex} generates the
$\delta_{l0}$ term in \eq{barkglov}.  Hence, if one uses the proton charge
radius defined via the Sachs form factor, the net contribution of order
$(Z\alpha)^4m_r^3/M^2$ has exactly the same form as if the proton were an
elementary particle with $g=2$.

The existing experimental data on the root mean square (rms) proton
charge radius \cite{dhr,ssbw} lead to the proton size correction about
$1100$ kHz for the $1S$ state in hydrogen and about $140$ kHz for the $2S$
state, and is thus much larger than the experimental accuracy of the Lamb
shift measurements. Unfortunately, there is a discrepancy between the
results of \cite{dhr,ssbw}, which influences the theoretical prediction for
the Lamb shifts, and a new experiment on measuring the proton charge radius
is badly needed. A new value of the proton charge radius was
derived recently from the improved theoretical analysis of the low
momentum transfer scattering experiments \cite{rosenfeld99}. The
phenomenological situation connected with the experimental data on the
proton rms charge radius will be discussed in more detail below in
Section \ref{phen1s}.

\subsection{Nuclei with Other Spins}  \label{darwinfoldyspin}

The general result for the nuclear charge radius and the Darwin-Foldy
contribution for a nucleus with arbitrary spin was obtained in \cite{kms}.
It was shown there that one may write a universal formula for the
sum of these contributions irrespective of the spin of the nucleus if the
nuclear charge radius is defined with the help of the same form factor for
any spin. However, for historic reasons, the definitions of the nuclear
charge radius are not universal, and respective formulae have different
appearances for different spins. We will discuss here only the
most interesting cases of the spin zero and spin one nuclei.

The case of the spin zero nucleus is the simplest
one. For an elementary scalar particle the low momentum nonrelativistic
expansion of the photon-scalar vertex starts with the ${\bf k}^4/M^4$  term,
and, hence, the respective contribution to the  energy shift is suppressed
by an additional factor $1/M^2$ in comparison to the spin one-half case.
Hence, in the case of the scalar nucleus there is no Darwin term
$\delta_{l0}$ in \eq{barkglov} \cite{bd,owen}. Interaction of the composite
scalar nucleus with photons is described only by one form factor, and the
slope of this form factor is called the charge radius squared.  Hence, in
the scalar case the charge radius contribution is described by
\eq{chargerad}, and the Darwin term is absent in \eq{barkglov}.

The spin one case is more complicated since for the vector nucleus its
interaction with the photon is described in the general case by three
form factors. The nonrelativistic limit of the photon-nucleus vertex
in this case was considered in \cite{pachkar}, where it was shown that
with the standard definition of the deuteron charge radius (the case of
the deuteron is the only phenomenologically interesting case of the
spin one nucleus) the situation with the Darwin-Foldy and the charge
radius contribution is exactly the same as in the case of the scalar
particle.  Namely, the Darwin-Foldy contribution is missing in
\eq{barkglov} and the charge radius contribution is given by
\eq{chargerad}. It would be appropriate to mention here recent work
\cite{friarmartspr}, where a special choice of definition of the
nuclear charge radius is advocated, namely it is suggested to include
the Darwin-Foldy contribution in the definition of the nuclear charge
radius. While one can use any consistent definition of the nuclear
charge radius, this particular choice seems to us to be unattractive
since in this case even a truly pointlike particle in the sense of
quantum field theory (say an electron) would have a finite charge
radius  even in zero-order approximation. The phenomenological aspects
of the deuteron charge radius contribution to the hydrogen-deuterium
isotope shift will be discussed later in Section \ref{isshift}.

\subsection{Empirical Nuclear Form Factor and the Contributions to the Lamb
Shift}\label{empnuklrad}

In all considerations above we have assumed the most natural
theoretical definition of nuclear form factors, namely, the
form factor was assumed to be an intrinsic property of the nucleus.
Therefore, the form factor is defined via the effective nuclear-photon
vertex in the absence of electromagnetic interaction. Such a form
factor can in principle be calculated with the help of QCD. The
electromagnetic corrections to the form factor defined in this way may
be calculated in the framework of QED perturbation theory. Strictly
speaking all formula above are valid with this definition of the form
factor.

However, in practice, form factors are measured experimentally and
there is no way to switch off the electromagnetic interaction. Hence,
in order to determine the form factor experimentally one has in
principle to calculate electromagnetic corrections to the elastic
electron-nucleus scattering which is usually used to measure the form
factors \cite{dhr,ssbw}. In the usual fit to the experimental data not
all electromagnetic corrections to the scattering amplitude are usually
taken into account (see e.g., discussion in \cite{sy,fms99}). First,
all vacuum polarization insertions, excluding the electron vacuum
polarization, are usually ignored.  This means that respective
contributions to the energy shift in \eq{nonelectr} are swallowed by
the empirical value of the nuclear charge radius squared. They are
effectively taken into account in the contribution to the energy shift
in \eq{chargerad}, and should not be considered separately.  Next there
are the corrections of order $(Z^2\alpha)(Z\alpha)^4$ to the energy
shift\footnote{We have already considered these corrections together
with other radiative-recoil corrections above, in Section
\ref{radprot}.  This discussion will be partially reproduced here in
order to make the present section self-contained.}. The perturbative
electromagnetic contributions to the Pauli form factor should be
ignored, since they definitely enter the empirical value of the nuclear
$g$-factor.  The situation is a bit more involved with respect to the
electromagnetic contribution to the Dirac nuclear form factor. The QED
contribution to the slope of the Dirac form factor is infrared
divergent, and, hence, one cannot simply include it in the empirical
value of the nuclear charge radius. Of course, as is well known, there
is no real infrared divergence in the proper description of the
electron-nucleus scattering if the soft photon radiation is properly taken
into account (see, e.g., \cite{bd,blp}). This means that the proper
determination of the empirical proton form factor, on the basis of the
experimental data, requires account of the electromagnetic radiative
corrections, and the measured value of the nuclear charge radius
squared does not include the electromagnetic contribution.  Hence, the
radiative correction of order $(Z^2\alpha)(Z\alpha)^4$ in
\eq{diracnucl} should be included in the comparison of the theory with
the experimental data on the energy shifts.

\section{~Nuclear Size and Structure Corrections of Order $(Z\alpha)^5
\lowercase{m}$} \label{lambsizestr}

Corrections of relative order $(Z\alpha)^5$ connected with the
nonelementarity of the nucleus are generated by the diagrams with
two-photon exchanges. As usual  all corrections of order $(Z\alpha)^5$,
originate from high (on the atomic scale) intermediate momenta. Due
to the composite nature of the nucleus, besides intermediate elastic nuclear
states, we also have to consider the contribution of the diagrams with
inelastic intermediate states.

\subsection{Nuclear Size Corrections of Order
$(Z\alpha)^5\lowercase{m}$}

Let us consider first the contribution generated only by the elastic
intermediate nuclear states. This means that we will treat the nucleus here
as a particle which interacts with the photons via a nontrivial
experimentally measurable form factor $F_1(k^2)$, i.e. the
electromagnetic interaction of our nucleus is nonlocal, but we will
temporarily ignore its excited states.

\begin{figure}[h]
\centerline{\epsfig{file=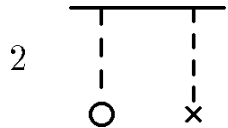,height=1.7cm}}
\vspace{0.5cm}
\caption{Diagrams for elastic nuclear size corrections of order
$(Z\alpha)^5\lowercase{m}$ with one form factor insertion. Empty dot
corresponds to factor $F_1(-k^2)-1$}
\label{ffeltwophotonefig}
\end{figure}

As usual we start with the skeleton integral contribution in
\eq{nonrecskel} corresponding to the two-photon skeleton diagram in
Fig.\ \ref{coulombskelfig}. Insertion of the factor $F_1(-k^2)-1$ in
the proton vertex corresponds to the presence of a nontrivial proton
form factor\footnote{Subtraction is necessary in order to avoid double
counting since the subtracted term in the vertex corresponds to the
pointlike proton contribution, already taken into account in the
effective Dirac equation.}.

\begin{figure}[h]
\centerline{\epsfig{file=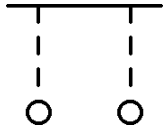,height=1.7cm}}
\vspace{0.5cm}
\caption{Diagrams for elastic nuclear size corrections of order
$(Z\alpha)^5\lowercase{m}$ with two form factor insertions. Empty dot
corresponds to factor $F_1(-k^2)-1$}
\label{ffeltwophottwofig}
\end{figure}

We have to consider diagrams in Fig.\ \ref{ffeltwophotonefig} with insertion
of one factor $F_1(-k^2)-1$ in the proton vertex (there are two such
diagrams, hence an extra factor two below)\footnote{Dimensionless
integration momentum in \eq{nonrecskel} was measured in electron mass. We
return here to dimensionful integration momenta, which results in an
extra factor $m^3$ in the numerators in \eq{oneff}, \eq{twoff} and
\eq{zema} in comparison with the factor in the skeleton integral
\eq{nonrecskel}. Notice also the minus sign before the momentum in the
arguments of form factors; it arises because in the equations below
$k=|{\bf k}|$.}

\beq            \label{oneff}
\Delta E=-32mm_r^3
\frac{(Z\alpha)^5}{\pi n^3}
\int_0^\infty\frac{dk}{k^4}\left(F_1(-k^2)-1\right),
\eeq

\noindent
and the diagrams in Fig.\ \ref{ffeltwophottwofig} with insertion of two
factors $F_1(-k^2)-1$ in two proton vertices

\beq                 \label{twoff}
\Delta E=-16m{m_r}^3\frac{(Z\alpha)^5}{\pi n^3}
\int_0^\infty\frac{dk}{k^4}\left(F_1(-k^2)-1\right)^2.
\eeq

The low momentum integration region in the integral in
\eq{oneff} produces a linearly divergent infrared contribution, which
simply reflects the presence of the correction of order $(Z\alpha)^4$,
calculated in Section \ref{lambnuclsize}. We will subtract this
divergent contribution.  Besides this uninteresting divergent term, the
integral in \eq{oneff} also contains the finite contribution induced by
high intermediate momenta, which should be taken on a par with the
contribution in \eq{twoff}.

The scale of the integration momenta in \eq{twoff}, and \eq{oneff} is
determined by the form factor scale. High momenta in the
present context means momenta of the form factor scale, to be
distinguished from high momenta in other chapters which
often meant momenta of the scale of the electron mass. The
characteristic momenta in the present case are much higher.

The total contribution of order $(Z\alpha)^5$ has the form

\beq   \label{zema}
\Delta E=-16m{m_r}^3
\frac{(Z\alpha)^5}{\pi n^3}
\int_0^\infty\frac{dk}{k^4}\left(F_1^2(-k^2)-1\right),
\eeq

\noindent
and, after subtraction of the divergent contribution and carrying
out the Fourier transformation, one obtains \cite{bortr,friar79}
\footnote{The result in \cite{friar79} has the factor $m_r^4$
instead of $mm_r^3$ before the integral in \eq{bortrfr}. This difference
could become important only after calculation of a recoil correction to
the contribution in \eq{bortrfr}.}

\beq  \label{bortrfr}
\Delta E
=-\frac{m(Z\alpha)^5}{3n^3}m_r^3<r^3>_{(2)},
\eeq

\noindent
where $<r^3>_{(2)}$ is the third Zemach moment \cite{zemach},
defined via weighted convolution  of two nuclear charge densities
$\rho(r)$

\beq  \label{thirdzem}
<r^3>_{(2)}\equiv\int d^3r_1d^3r_2 \rho(r_1)\rho(r_2)|\bf r_1+r_2|^3.
\eeq

Parametrically this result is of order $m(Z\alpha)^5(m/\Lambda)^3$, where
$\Lambda$ is the form factor scale. Hence, this correction is suppressed in
comparison with the leading proton size contribution not only by an extra
factor $Z\alpha$ but also by the extra small factor $m/\Lambda$. This
explains the smallness of this contribution, even in comparison with the
proton size correction of order $(Z\alpha)^6$ (see below
Section \ref{nuclsizezalpha6}), since one factor $m/\Lambda$ in \eq{bortrfr}
is traded for a much larger factor $Z\alpha$ in that logarithmically
enhanced contribution.

The result in \eq{bortrfr} depends on the third Zemach moment, or in other
words, on a nontrivial weighted integral of the product of two charge
densities, and cannot be measured directly, like the rms proton charge
radius. This means that the correction under consideration may only
conditionally be called the proton size contribution. It depends on the
fine details of the form factor momentum dependence, and not only on the
directly measurable low-momentum behavior of the form factor. This feature
of the result is quite natural taking into account high intermediate momenta
characteristic for the integral in \eq{zema}. In the practically important
cases of hydrogen and deuterium, reliable results for this contribution may
be obtained.  Numerically, the nuclear radius correction of order
$(Z\alpha)^5$ is equal to $-35.9$~Hz for the $1S$ state and
$-4.5$~Hz for the $2S$ state in hydrogen. These corrections are rather
small.  A much larger contribution arises in \eq{bortrfr} to
the energy levels of deuterium. The deuteron, unlike the proton, is a
loosely bound system, its radius is much larger than the proton radius, and
the respective correction to the energy levels is also larger.  The
contribution of the correction under consideration \cite{friarmay}

\beq         \label{deuterza5}
\Delta E=0.49~\mbox{kHz}
\eeq

\noindent
to the $2S-1S$ energy splitting in deuterium should be taken into
account in the discussion of the hydrogen-deuterium isotope shift.

We started consideration of the proton size correction of relative
order $(Z\alpha)^5$ by inserting the factors $(F_1(-k^2)-1)$ in the
external field skeleton diagrams in Fig.\ \ref{coulombskelfig}.
Technically the external field diagrams correspond to the heavy
particle pole contribution in the sum of all skeleton diagrams with two
exchanged photons in Fig.\ \ref{coulcoulfig}-Fig.\ \ref{trcoulfig}. In
the absence of form factors the nonpole contributions of the diagrams
in Fig.\ \ref{coulcoulfig}-Fig.\ \ref{trcoulfig} were suppressed by
the recoil factor $m/M$ in comparison with the heavy pole contribution,
and this justified their separate consideration. However, as we have
seen, insertion of the form factors pushes the effective integration
momenta in the  high momenta region $\sim \Lambda$ even in the external
field diagrams. Then even the external field contribution  contains the
recoil factor  $(m/\Lambda)^3$. We might expect that, after insertion of
the proton form factors, the nonpole contribution of the skeleton
diagrams with two exchanged photons in Fig.\ \ref{coulcoulfig}-Fig.\
\ref{trcoulfig} would contain the recoil factor $(m/M)(m/\Lambda)^2$,
and would not be parametrically suppressed in comparison with the pole
contribution in \eq{bortrfr}. The total contribution of the skeleton
diagrams with the proton form factor insertions was calculated in
\cite{faustmart99} for the $2S$ state, and  the difference
between this result and the nonrecoil result in \eq{bortrfr} turned out
to be $-0.25$ Hz. At the current level of theoretical and
experimental accuracy we can safely ignore such tiny differences
between the pole and total proton size contributions of order
$(Z\alpha)^5$.

\subsection{Nuclear Polarizability Contribution of Order $(Z\alpha)^5
\lowercase{m}$ to $S$-Levels}\label{nuclstructza5}

\begin{figure}[h]
\centerline{\epsfig{file=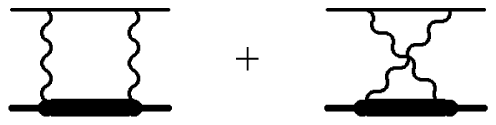,height=1.7cm}}
\vspace{0.5cm}
\caption{Diagrams for nuclear polarizability correction
of order $(Z\alpha)^5\lowercase{m}$}
\label{inelasticfig}
\end{figure}

The description of nuclear structure corrections of order
$(Z\alpha)^5m$ in terms of nuclear size and nuclear polarizability
contributions is somewhat artificial. As we have seen above the nuclear
size correction of this order depends not on the charge radius of the
nucleus but on the third Zemach moment in \eq{thirdzem}. One might
expect the inelastic intermediate nuclear states in Fig.\
\ref{inelasticfig} would generate corrections which are even smaller
than those connected with the third Zemach moment, but this does not
happen. In reality, the contribution of the inelastic intermediate
states turns out to be even larger than the elastic contribution since
the powerlike decrease of the form factor is compensated in this case
by the summation over a large number of nuclear energy levels.  The
inelastic contributions to the energy shift were a subject of intensive
study for a long time, especially for muonic atoms (see, e.g.,
\cite{erichuf,bernjarlsk1,bernjarlsk2,spk,bernjarlsk3,friar77,rosenfeld83}).
Corrections to the energy levels were obtained in these works in the
form of certain integrals of the inelastic nuclear structure functions,
and the dominant contribution is produced by the nuclear electric and
magnetic polarizabilities.

The main feature of the polarizability contribution to the energy shift is
its logarithmic enhancement \cite{spk,berner}. The appearance of the large
logarithm may easily be understood with the help of the skeleton
integral. The heavy particle factor in the two-photon exchange
diagrams is now described by the photon-nucleus inelastic forward Compton
amplitude \cite{khrsen}

\beq         \label{incompton}
M=\bar\alpha(\omega,{\bf k}^2){\bf EE}^*+
\bar\beta(\omega,{\bf k}^2){\bf BB}^*,
\eeq

\noindent
where $\bar\alpha(\omega,{\bf k}^2)$ and $\bar\beta(\omega,{\bf k}^2)$
are proton (nuclear) electric and magnetic polarizabilities.

In terms of this Compton amplitude the two-photon contribution has the
form

\beq
\Delta
E=-\frac{4\alpha(m_rZ\alpha)^3}{n^3}\int\frac{d^4k}{(2\pi)^4i}
\frac{D_{im}D_{jn}}{k^4}\frac{Tr\{\gamma_i((1+\gamma_0)m-\hat
k)\gamma_j\}}{k^2-2mk_0}M_{mn},
\eeq

\noindent
which reduces after elementary transformations to\footnote{Integration
in \eq{khripllogpol} goes over dimensionless momentum $k$ measured in
electron mass.}

\beq  \label{khripllogpol}
\Delta
E=-\frac{\alpha(Z\alpha)^3}{4\pi n^3}{m_r}^3m
\int_0^\infty{dk}
\left\{\left[(k^4+8)\sqrt{k^2+4}-k(k^4+2k^2+6)\right]\bar\alpha(\omega,k^2)
\right.
\eeq
\[
\left.
+\left[k^3(k^2+4)^\frac{3}{2}-k(k^4+6k^2+6)\right]\bar\beta(\omega, k^2)
\right\},
\]

\noindent
and we remind the reader that an extra factor $Z^2\alpha$ is hidden in the
definition of the polarizabilities.

Ignoring the momentum dependence of the polarizabilities, one immediately
comes to a logarithmically ultraviolet divergent integral \cite{khrsen}

\beq     \label{khrsenlog}
\Delta E
=-\frac{\alpha(Z\alpha)^3}{\pi n^3}{m_r}^3m
\left[5\bar\alpha(0,0)-\bar\beta(0,0)\right]
\left(\ln\frac{\Lambda}{m}+O(1)\right),
\eeq

\noindent
where $\Lambda$ is an ultraviolet cutoff. In real calculations the
role of the cutoff is played by the characteristic excitation energy of
the nucleus.

The sign of the energy shift is determined by the electric
polarizability and has a clear physical origin. The electron polarizes
the nucleus, an additional attraction between the induced dipole and
the electron emerges, and shifts the energy level down.

In the case of hydrogen the characteristic excitation energy is about
$300$ MeV, the logarithm is rather large, and the logarithmic
approximation works very well. Using the proton polarizabilities
\cite{macgibbon} one easily obtains the polarizability contribution for
the proton $nS$ state \cite{spk,khrsen,khrsen2,khrsen3}

\beq     \label{khrsenhyd}
\Delta E(nS)=-\frac{70~(11)~(7)}{n^3}~\mbox{Hz},
\eeq

\noindent
where the error in the first parenthesis describes the accuracy of the
logarithmic approximation, and the error in the second parenthesis is
due to the experimental  data on the polarizabilities.

A slightly different numerically polarizability contribution

\beq    \label{rosenfeld99}
\Delta E(1S)=-\frac{95~(7)}{n^3}~\mbox{Hz},
\eeq

\noindent
was obtained in \cite{rosenfelder99}, and also, with a somewhat
larger error bars, in \cite{faustmart99}. Discrepancy between the
results in \eq{khrsenhyd} and \eq{rosenfeld99} is preserved even when
both groups of authors use one and the same data on proton
polarizabilities from \cite{babusci}. Technically the disagreement
between the results in \cite{spk,khrsen,khrsen2,khrsen3} and
\cite{rosenfelder99} is due to the expression for the polarizability
energy shift in the form of an integral of the total photoabsorption
cross section, which was used in \cite{rosenfelder99}. This expression
was derived in \cite{berner} under the assumption that the invariant
amplitudes for the forward Compton scattering satisfy the dispersion
relations without subtractions. The subtraction term in the dispersion
relation for the forward Compton scattering amplitude is also missing
in \cite{faustmart99}. Without this subtraction term the dominant
logarithmic contribution to the energy shift becomes proportional to
the sum of the electric and magnetic polarizabilities
$\bar\alpha+\bar\beta$, while in \cite{spk,khrsen,khrsen2,khrsen3} this
contribution is proportional to another linear combination of
polarizabilities (see \eq{khrsenlog}). Restoring this necessary
subtraction term the authors of \cite{faustmart99} obtained $-82$ Hz
for the polarizability contribution instead of their result in
\cite{faustmart99}\footnote{Private communication from I. B.
Khriplovich and A. P. Martynenko.\label{prvkhripl}}

In the other experimentally interesting case of deuterium, nuclear
excitation energies are much lower and a more accurate account of the
internal structure of the deuteron is necessary. As is well known, due to
smallness of the binding energy the model independent zero-range
approximation provides a very accurate description of the deuteron.
The polarizability contributions to the energy shift in deuterium are
again logarithmically enhanced and in the zero-range approximation
one obtains a model independent result \cite{msz,milkhr}

\beq
\Delta
E=-\frac{\alpha(Z\alpha)^3}{\pi n^3}{m_r}^3m
\left\{5\alpha_d\left[\ln\frac{8I}{m}+\frac{1}{20}\right]
-\beta_d\left[\ln\frac{8I}{m}-1.24\right]\right\},
\eeq

\noindent
where $\kappa=45.7$ MeV is the inverse deuteron size, $I=\kappa^2/m_p$ is
the absolute value of the deuteron binding energy, and the deuteron electric
and magnetic polarizabilities are defined as

\beq
\alpha_d(\omega)=\frac{2}{3}\:(Z^2\alpha)\sum_n\frac{(E_n+I)|<0|{\bf
r}|n>|^2}{\omega^2-(E_n+I)^2},
\eeq
\[
\beta_d(0)=\frac{\alpha(\mu_p-\mu_n)^2}{8m_p\kappa^2}
\frac{1+\frac{\kappa_1}{3\kappa}}{1+\frac{\kappa_1}{\kappa}},
\]

\noindent
where $\kappa_1=7.9$ MeV determines the position of the virtual level in the
neutron-proton $^1S_0$ state.

Numerically the polarizability contribution to the deuterium $1S$ energy
shift in the zero range approximation is equal to \cite{milkhr}

\beq
\Delta E(1S)=(-22.3+0.31)~\mbox{kHz},
\eeq

\noindent
where the first number in the parenthesis is the contribution of
the electric polarizability, and the second is the contribution of the
magnetic polarizability. This zero range contribution results in the
correction

\beq
\Delta E(1S-2S)=19.3~\mbox{kHz},
\eeq

\noindent
to the $1S-2S$ interval, and describes the total polarizability
contribution with an accuracy of about one percent.

New experimental data on the deuterium-hydrogen isotope shift (see
Table XXII below) have an accuracy of about $0.1$ kHz and, hence, a more
accurate theoretical result for the polarizability contribution is
required. In order to obtain such a result it is necessary to go beyond
the zero range approximation, and take the deuteron structure into
account  in more detail. Fortunately, there exist a number of
phenomenological potentials which describe the properties of the
deuteron in all details. Some calculations with realistic
proton-neutron potentials were performed recently
\cite{luros,leiros,friarpayne,friarapr}. The most precise results were
obtained in \cite{friarapr}

\beq      \label{friaraprissh}
\Delta E(1S-2S)=18.58~(7)~\mbox{kHz},
\eeq

\noindent
which are consistent with the results of the other works
\cite{msz,leiros}.

The result in \eq{friaraprissh} is obtained neglecting the
contributions connected with the polarizability of the constituent
nucleons in the deuterium atom, and the polarizability contribution of
the proton in the hydrogen atom. Meanwhile, as may be seen from
\eq{khrsenhyd}, proton polarizability contributions are comparable to
the accuracy of the polarizability contribution in \eq{friaraprissh},
and cannot be ignored. The deuteron is a weakly bound system and it is
natural to assume that the deuteron polarizability is a sum of the
polarizability due to relative motion of the nucleons and their
internal polarizabilities.  The nucleon polarizabilities in the
deuteron coincide with the polarizabilities of the free nucleons well
within the accuracy of the logarithmic approximation \cite{khrsen2}.
Therefore the proton polarizability contribution to the
hydrogen-deuterium isotope shift cancels, and the contribution to this
shift, which is due to the internal polarizabilities of the nucleons, is
completely determined by the neutron polarizability. This neutron
polarizability contribution to the hydrogen-deuterium isotope shift was
calculated in the logarithmic approximation in \cite{khrsen2}

\beq           \label{khrsen2}
\Delta E(1S-2S)=53~(9)~(11)~\mbox{Hz}.
\eeq

\section{Nuclear Size and Structure Corrections of Order $(Z\alpha)^6
\lowercase{m}$}

\subsection{Nuclear Polarizability Contribution to $P$-Levels}
\label{nuclpolarizbza6}

The leading polarizability contribution of order $(Z\alpha)^5$  obtained
above is proportional to the nonrelativistic wave function at the origin
squared, and  hence, exists only for the $S$ states. The leading
polarizability contribution to the non $S$-states is of order  $(Z\alpha)^6$
and may easily be calculated. Consider the Compton amplitude in
\eq{incompton} as the contribution to the bound state energy induced by the
external field of the electron at the nuclear site. Then we
calculate the matrix element in \eq{incompton} between the electron states,
considering the field strengths from the Coulomb field generated by
the orbiting electron. We obtain \cite{erichuf} (the overall factor $1/2$
is due to the induced nature of the nuclear dipole)

\beq                 \label{lpolar}
\Delta E=-\frac{\alpha\bar\alpha}{2}<n,l|\frac{1}{r^4}|n,l>
=-\frac{3n^2-l(l+1)}{2n^5(l+\frac{3}{2})
(l+1)(l+\frac{1}{2})l(l-\frac{1}{2})}\frac{\alpha(Z\alpha)^4\bar\alpha
m_r^4}{2}.
\eeq

This energy shift is negative because when the electron
polarizes the nucleus this leads to an additional attraction of the
induced dipole to the electron.

The contribution induced by the magnetic susceptibility may also  be
easily calculated \cite{erichuf}, but it is even of higher order in
$Z\alpha$ (order $(Z\alpha)^8$) since the magnetic field strength behaves as
$1/r^3$.  This additional suppression of the magnetic effect is quite
reasonable, since the magnetic field itself is a relativistic effect.

Consideration of the $P$-state polarizability contribution provides us with
a new perspective on the $S$-state polarizability contribution.
One could try to consider the matrix element in \eq{lpolar} between the $S$
states.  Due to nonvanishing of the $S$-state wave functions at the origin
this matrix element is linearly divergent at small distances, which once
more demonstrates that the $S$-state contribution is of a lower order
in $Z\alpha$ than for the $P$-state, and that for its calculation one
has to treat small distances more accurately than was done in \eq{lpolar}.

\subsection{Nuclear Size Correction of Order
$(Z\alpha)^6\lowercase{m}$}\label{nuclsizezalpha6}

The nuclear size and polarizability corrections of order $(Z\alpha)^5$
obtained in Section \ref{lambsizestr} are very small. As was explained
there, the suppression of this contribution is due to the large magnitude of
the characteristic momenta responsible for this correction. The nature of
the suppression is especially clear in the case of the Zemach radius
contribution in \eq{bortrfr}, which contains the small factor
$m_r^3<r^3>_{(2)}$. The nuclear size correction of order $(Z\alpha)^6m$
contains an extra factor $Z\alpha$ in comparison with the nuclear size and
polarizability corrections of order $(Z\alpha)^5$, but its main part is
proportional to the proton charge radius squared. Hence, we should expect
that despite an extra power of $Z\alpha$ this correction is numerically
larger than the nuclear size and polarizability contributions of the
previous order in $Z\alpha$. As we will see below, calculations confirm this
expectation and, moreover, the contribution of order $(Z\alpha)^6m$ is
additionally logarithmically enhanced.

Nuclear size corrections of order $(Z\alpha)^6$  may be obtained in a quite
straightforward way in the framework of the quantum mechanical third
order perturbation theory. In this approach one considers the difference
between the electric field generated by the nonlocal charge density
described by the nuclear form factor and the field of the pointlike charge
as a perturbation operator \cite{bortr,friar79}.

The main part of the nuclear size $(Z\alpha)^6$ contribution
which is proportional to the nuclear charge radius squared may also be
easily obtained in a simpler way, which clearly demonstrates the source
of the logarithmic enhancement of this contribution. We will first discuss
in some detail this simple-minded approach, which essentially coincides with
the arguments used above to obtain the main contribution to the Lamb shift
in \eq{qff}, and the leading proton radius contribution in \eq{chargerad}.

The potential of an extended nucleus is given by the expression

\beq
V(r)=-Z\alpha\int d^3r'\frac{\rho(r')}{|{\bf r-r'}|},
\eeq

\noindent
where $\rho(r')$ is the nuclear charge density.

Due to the finite size of the nuclear charge distribution, the
relative distance between the nucleus and the electron is not constant but
is subject to additional fluctuations with probability $\rho(r)$. Hence, the
energy levels experience an additional shift

\beq
\Delta E=
<n|\int d^3r''\rho(r'')[V(r+r'')-V(r)]|n>.
\eeq

Taking into account that the size of the nuclear charge distribution is much
smaller than the atomic scale, we immediately obtain

\beq       \label{naivecharge}
\Delta E=\frac{2\pi}{3}\:(Z\alpha)<r^2>\int
d^3r\rho(r)\psi(r)^+\psi(r).
\eeq

We will now discuss contributions contained in \eq{naivecharge} for
different special cases.

\subsubsection{Correction to the $\lowercase{n}S$-Levels}

In the Schr\"odinger-Coulomb approximation the expression in
\eq{naivecharge} reduces to the leading nuclear size correction in
\eq{chargerad}. New results arise if we take into account Dirac corrections
to the Schr\"odinger-Coulomb wave functions of relative order $(Z\alpha)^2$.
For the $nS$ states the product of the wave functions in \eq{naivecharge}
has the form  (see, e.g, \cite{friar79})

\beq
|\psi(r)|^2=|\psi_{Schr}(r)|^2\left\{1-(Z\alpha)^2\left[
\ln\frac{2mrZ\alpha}{n}
+\psi(n)+2\gamma+\frac{9}{4n^2}-\frac{1}{n}-\frac{11}{4}\right]\right\},
\eeq

\noindent
and the additional contribution to the energy shift is equal to

\beq   \label{za6leading}
\Delta E=-\frac{2(Z\alpha)^6}{3n^3}m_r^3<r^2>\left[<\ln\frac{2mrZ\alpha}{n}>
+\psi(n)+2\gamma+\frac{9}{4n^2}-\frac{1}{n}-\frac{11}{4}\right].
\eeq

This expression nicely illustrates the main qualitative features of the
order $(Z\alpha)^6$ nuclear size contribution. First, we observe a
logarithmic enhancement connected with the singularity of the Dirac wave
function at small distances. Due to the smallness of the nuclear size, the
effective logarithm of the ratio  of the atomic size and the nuclear size
is a rather large number; it is  equal to about $-10$ for the $1S$
level in hydrogen and deuterium. The result in \eq{za6leading} contains
all state-dependent contributions of order $(Z\alpha)^6$.

A tedious third order perturbation theory calculation \cite{bortr,friar79}
produces some additional state-independent terms with the net result being
a few percent different from the naive result above. The additional
state-independent contribution beyond the naive result above has the
form \cite{friarmay}

\beq \label{frairaddterms}
\Delta
E=\frac{2(Z\alpha)^6}{3n^3}m_r^3\{\frac{<r^2>}{2}-\frac{<r^3><\frac{1}{r}>}{3}
+\int
d^3rd^3r'\rho(r)\rho(r')\theta(|r|-|r'|)\left[(r^2+r'^2)\ln\frac{|r'|}{|r|}
\right.
\eeq
\[
\left.
-\frac{r'^3}{3|r|}+\frac{r^3}{3|r'|}+\frac{r^2-r'^2}{3}\right]
+6\int
d^3rd^3r'd^3r''\rho(r)\rho(r')\rho(r'')\theta(|r|-|r'|)\theta(|r'|-|r''|)
\left[\frac{r''^2}{3}\ln\frac{|r'|}{|r''|}
\right.
\]
\[
\left.
-\frac{r''^4}{45|r||r'|}+\frac{r''^3}{9}\left(\frac{1}{|r'|}+\frac{1}{|r|}
\right)
+\frac{r'^2r''^2}{36r^2}-\frac{2r'r''^2}{9r}+\frac{r''^2}{9}\right].
\]

Note that, unlike the leading naive terms in \eq{za6leading}, this
additional contribution depends on more detailed features of the
nuclear charge distribution than simply the charge radius squared.

Detailed numerical calculations in the interesting cases of hydrogen
and deuterium were performed in \cite{friarmay}. Nuclear size
contributions of order $(Z\alpha)^6$ to the energy shifts in hydrogen
are given in Table X\footnote{All numbers in Table X are calculated for
the proton radius $r_p=0.862~(12)$ fm, see discussion on the status of
the proton radius results in Section \ref{phen1s}.}
and, as discussed above, they are an order of magnitude
larger than the nuclear size and polarizability contributions of the
previous order in $Z\alpha$.

Respective corrections to the energy levels in deuterium are even much
larger than in hydrogen due to the large radius of the deuteron.  The
nulear size contribution of order $(Z\alpha)^6$ to the $2S-1S$
splitting in deuterium is equal to (we have used in this calculation
the value of the deuteron charge radius obtained in \cite{sitr} from
the analysis of all available experimental data)

\beq     \label{deutradza6}
\Delta E=-3.43~\mbox{kHz},
\eeq

\noindent
and in hydrogen

\beq
\Delta E=-0.61~\mbox{kHz}.
\eeq

We see that the difference of these corrections gives an important
contribution to the hydrogen-deuterium isotope shift.

\subsubsection{Correction to the $\lowercase{n}P$-Levels}

Corrections to the energies of $P$-levels may easily be obtained from
\eq{naivecharge}. Since the $P$-state wave functions vanish at the
origin there are no charge radius squared contributions of lower order,
unlike the case of $S$ states, and we immediately obtain
\cite{friar79}

\beq     \label{nuclsizeza6p}
\Delta E(nP_j)=\frac{(n^2-1)(Z\alpha)^6}{6n^5}
m_r^3<r^2>\delta_{j\frac{1}{2}}.
\eeq

There exist also additional terms of order $(Z\alpha)^6$ proportional to
$<r^4>$ \cite{friar79} but they are suppressed by an additional factor
$m^2<r^2>$ in comparison with the result above and may safely be omitted.

\section{Radiative Correction of Order $\alpha(Z\alpha)^5<\lowercase{r}^2>
\lowercase{m}_{\lowercase{r}}^3$ to the Finite Size
Effect}\label{radcorrfintitesize}

Due to the large magnitude of the leading nuclear size correction in
\eq{chargerad} at the current level of experimental accuracy one also has to
take into account radiative corrections to this effect. These radiative
corrections were first discussed and greatly overestimated in
\cite{borieprl81}.  The problem was almost immediately clarified in
\cite{lye}, where it was shown that the contribution is generated by
large intermediate momenta states and is parametrically a small
correction of order $\alpha(Z\alpha)^5m_r^3<r^2>$.  On the basis of the
estimate in \cite{lye} the authors of \cite{sy} expected the radiative
correction to the leading nuclear charge radius contribution to be of
order 10 Hz for the $1S$-state in hydrogen.

The large magnitude of the characteristic integration momenta \cite{lye} is
quite clear. As we have seen above, in the calculation of the main proton
charge contribution, the exchange momentum squared factor in the
numerator connected with the proton radius cancels with a similar factor in
the denominator supplied by the photon propagator. Any radiative correction
behaves as ${\bf k}^2$ at small momenta, and the presence of such a
correction additionally suppresses small integration momenta and pushes the
characteristic integration momenta into the relativistic region of
order of the electron mass. Hence, the corrections may be calculated
with the help of the skeleton integrals in the scattering
approximation. The characteristic integration momenta in the skeleton
integral are of order of the electron mass, and are still much smaller
than the scale of the proton form factor. As a result respective
contribution to the energy shift depends only on the slope of the form
factor.

The actual calculation essentially coincides with the calculation of the
corrections of order $\alpha^2(Z\alpha)^5$ to the Lamb shift in chapter
\ref{a2za5chap} but is technically simpler due to the triviality of the
proton form factor slope contribution in \eq{formfactor}.

\begin{figure}[ht]
\centerline{\epsfig{file=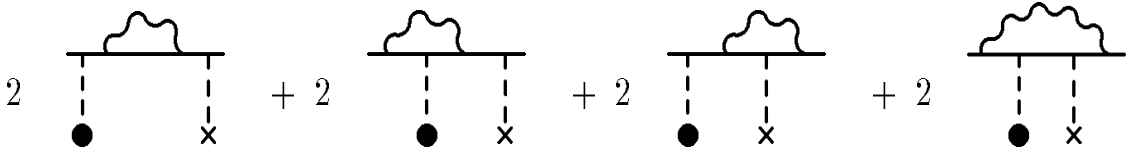,height=2cm,width=15cm}}
\vspace{0.5cm}
\caption{Electron-line radiative correction to the nuclear size effect.
Bold dot corresponds to proton form factor slope}
\label{ellineradcorradfig}
\end{figure}

\begin{figure}[ht]
\centerline{\epsfig{file=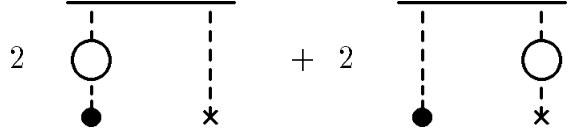,height=1.7cm}}
\vspace{0.5cm}
\caption{Coulomb-line radiative correction to the nuclear size effect.
Bold dot corresponds to proton form factor slope}
\label{coullineradcorradfig}
\end{figure}

There are two sources of radiative corrections to the leading nuclear size
effect, namely, the diagrams with one-loop radiative insertions in the
electron line in Fig.\ \ref{ellineradcorradfig}, and the diagrams with
one-loop polarization insertions in one of the external Coulomb lines in
Fig.\ \ref{coullineradcorradfig}.

\subsection{Electron-Line Correction}

Inserting the electron line factor \cite{eg4,egs} and the proton slope
contribution \eq{formfactor} in the skeleton integral in \eq{nonrecskel},
one immediately obtains \cite{eg97rad}

\beq          \label{eline}
\Delta
E_{e-line}=-1.985~(1)\frac{\alpha(Z\alpha)^5}{n^3}m_r^3<r^2>\delta_{l0}.
\eeq

\noindent
where an additional factor 2 was also inserted in the skeleton integral in
order to take into account all possible ways to insert the slope of the
proton form factor in the Coulomb photons.

In principle, this integral also admits an analytic evaluation in the same
way as it was done for a more complicated integral in \cite{egs}.

\subsection{Polarization Correction}

Calculation of the diagrams with the polarization operator insertion
proceeds exactly as in the case of the electron factor insertion. The
only difference is that one inserts an additional factor 4 in the
skeleton integral to take into account all possible ways to insert the
polarization operator and the slope of the proton form factor in the
Coulomb photons. After an easy analytic calculation one obtains
\cite{friarpol,hylton,eg97rad}

\beq                       \label{pol}
\Delta E_{pol}=\frac{\alpha(Z\alpha)^5}{2n^3}m_r^3<r^2>\delta_{l0}.
\eeq

\subsection{Total Radiative Correction}

The total radiative correction to the proton size effect is given by the sum
of contributions in \eq{pol} and \eq{eline}

\beq          \label{totrad}
\Delta E=-1.485~(1)\frac{\alpha(Z\alpha)^5}{n^3}m_r^3<r^2>\delta_{l0}.
\eeq

This contribution was also considered in \cite{pach93}. Correcting an
apparent misprint in that work, one finds the value $-1.43$ for the
numerical coefficient in \eq{totrad}. The origin of the minor discrepancy
between this value and the one in \eq{totrad} is unclear, since the
calculations in \cite{pach93} were done without separation of the
polarization operator and electron factor contributions.

Numerically the total radiative contribution in \eq{totrad} for
hydrogen is equal to

\begin{equation}
\Delta E(1S)=-0.138\; \mbox{kHz},
\end{equation}
\[
\Delta E(2S)=-0.017\; \mbox{kHz},
\]

and for deuterium

\begin{equation}
\Delta E(1S)=-0.841\; \mbox{kHz},
\end{equation}
\[
\Delta E(2S)=-0.105\; \mbox{kHz}.
\]

These contributions should be taken into account in discussion of the
hydrogen-deuterium isotope shift.


\begin{center}
\underline{Table X. Nuclear Size and Structure Corrections}
\nopagebreak

\begin{tabular}{|l|rl|c|c|}    \hline
     & $ $ &   &$\Delta E(1S)$ kHz &$\Delta E(2S)$ kHz
\\ \hline \hline
&$$&&$$&$$\\
Leading nuclear size
&$$&&$$&$$\\
contribution
&$\frac{2}{3n^3}(Z\alpha)^4m_r^3<r^2>\delta_{l0}$&&$1~162~(32)$&$145~(4)$\\
\hline
Proton form factor contribution  &&&&\\
of order $(Z\alpha)^5$&&&&\\
&$$&&$$&$$\\
Borisoglebsky, &$$&&$$&\\
Trofimenko (1979) \cite{bortr}&&&&\\
Friar (1979) \cite{friar79}&$-\frac{m(Z\alpha)^5}{3n^3}m_r^3<r^3>_{(2)}
\delta_{l0}$&&$
-0.036$&$-0.004$
\\ \hline
Polarizability contribution&&&&\\
$\Delta E(nS)$&&&&\\
&$$&&$$&$$\\
Startsev,Petrun'kin,&&&&\\
Khomkin(1976)\cite{spk}&&&&\\
Khriplovich, &
$-\frac{\alpha(Z\alpha)^3}{\pi n^3}{m_r}^3m$&&&$$
\\
Sen'kov (1997) \cite{khrsen,khrsen2,khrsen3}&$[5\bar\alpha(0,0)
-\bar\beta(0,0)]\ln\frac{\Lambda}{m}$&&$-0.070(11)(7)$&$-0.009(1)(1)$\\
\hline
Polarizability contribution&&&&\\
$\Delta E(nP)$&&&&\\
Ericson, Hufner (1972) \cite{erichuf}&
$-\frac{\alpha(Z\alpha)^4\bar\alpha
m_r^4}{2}\frac{3n^2-l(l+1)}{2n^5(l+\frac{3}{2})
(l+1)(l+\frac{1}{2})l(l-\frac{1}{2})}$&&&$$
\\
\hline
Nuclear size correction &&&&\\
of order $(Z\alpha)^6$ $\Delta E(nS)$&&&&\\
$$&&&&\\
Borisoglebsky, &$$&&$$&\\
Trofimenko (1979) \cite{bortr}&&&&\\
Friar (1979) \cite{friar79}&$-\frac{2(Z\alpha)^6}{3n^3}m_r^3<r^2>
[<\ln\frac{2mrZ\alpha}{n}>$&&&\\
Friar, Payne (1997) \cite{friarmay} &
$+\psi(n)+2\gamma+\frac{9}{4n^2}-\frac{1}{n}-\frac{11}{4}]+\delta
E$&&$0.709~(20)$&$0.095~(3)$ \\
\hline
Nuclear size correction &&&&\\
of order $(Z\alpha)^6$ $\Delta E(nP_j)$&&&&\\
$$&&&&\\
Friar (1979) \cite{friar79}&$\frac{(Z\alpha)^6(n^2-1)}{6n^5}
m_r^3<r^2>\delta_{j\frac{1}{2}}$&&&\\
\hline
Electron-line &&&&\\
radiative correction&&&&\\
&&&$$&$$ \\
Pachucki (1993) \cite{pach93}$$&&&&\\
Eides, Grotch (1997) \cite{eg97rad}&
$-1.985~(1)m_r^3<r^2>\frac{\alpha(Z\alpha)^5}{n^3}\delta_{l0}
$&&$-0.184~(5)$&$-0.023~(1)$\\
\hline
Polarization operator &&&&\\
radiative correction&&&&\\
&&&$$&$$ \\
Friar (1979) \cite{friarpol}&&&$$&$$ \\
Hylton (1985) \cite{hylton} &$$&&$$&$$\\
Pachucki (1993) \cite{pach93}$$&&&&\\
Eides, Grotch (1997) \cite{eg97rad}&
$\frac{1}{2}m_r^3<r^2>\frac{\alpha(Z\alpha)^5}{n^3}\delta_{l0}
$&&$0.046~(1)$&$$0.006\\
\hline
\end{tabular}
\end{center}

\part{Weak Interaction Contribution}

\begin{figure}[ht]
\centerline{\epsfig{file=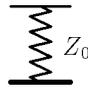}}
\vspace{0.5cm}
\caption{$Z$-boson exchange diagram}
\label{zbosonlambfig}
\end{figure}

The weak interaction contribution to the Lamb shift is generated by the
$Z$-boson exchange in Fig.\ \ref{zbosonlambfig}, which may be described by
the effective local low-energy Hamiltonian

\beq
H^{Z}(L)=
-\frac{16\pi\alpha}{\sin^2\theta_W\cos^2\theta_W}
\left(\frac{1}{4}-\sin^2\theta_W\right)^2\frac{mM}
{M_Z^2}\int
d^3x\left(\psi^+(x)\psi(x)\right)\left(\Psi^+(x)\Psi(x)\right),
\eeq

\noindent
where $M_Z$ is the $Z$-boson mass, $\theta_W$ is the Weinberg angle, and
$\psi$ and $\Psi$ are the two-component wave functions of the light and
heavy particles, respectively.

Then we easily obtain the weak interaction contribution to the Lamb shift in
hydrogen  \cite{eid96}

\beq
\Delta E^{Z}(L)
=-\frac{\alpha(Z\alpha)^3m_r}{\pi n^3}
\frac{8Gm_r^2}{{\sqrt2}\alpha}\left(\frac{1}{4}-\sin^2\theta_W
\right)^2\delta_{l0}
\approx -7.7\cdot 10^{-13}\frac{\alpha(Z\alpha)^3m_r}{\pi n^3}\delta_{l0}.
\eeq

This contribution is too small to be of any phenomenological significance.

\part{Lamb Shift in Light Muonic Atoms}\label{mouniclambgen}

Theoretically, light muonic atoms  have two main special features as
compared with the ordinary electronic hydrogenlike atoms, both of which
are connected with the fact that the muon is about 200 times heavier
than the electron\footnote{Discussing light muonic atoms we will often
speak about muonic hydrogen but almost all results below are valid also
for another phenomenologically interesting case, namely muonic helium.
In the Sections on light muonic atoms, $m$ is the muon mass, $M$ is the
proton mass, and $m_e$ is the electron mass.}. First, the role of the
radiative corrections generated by the closed electron loops is greatly
enhanced, and second, the leading proton size contribution becomes the
second largest individual contribution to the energy shifts after the
polarization correction.

\begin{figure}[ht]
\centerline{\epsfig{file=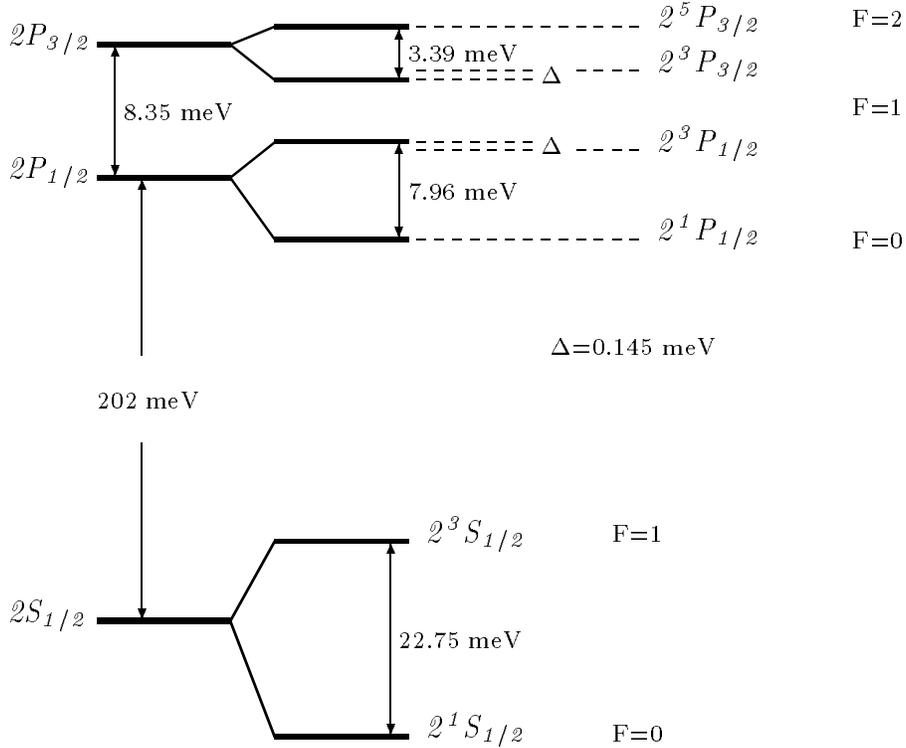,height=10cm}}
\vspace{0.5cm}
\caption{Energy levels in muonic hydrogen}
\label{muonhydlevels}
\end{figure}

The reason for an enhanced contribution of the radiatively corrected
Coulomb potential in Fig.\ \ref{polradfig} may be easily explained.
The characteristic distance at which the Coulomb potential is distorted
by the polarization insertion is determined by the electron Compton
length $1/m_e$ and in the case of electronic hydrogen it is about $137$
times less than the average distance between the atomic electron and
the Coulomb source $1/(m_eZ\alpha)$. This is the reason why even the
leading polarization contribution to the Lamb shift in
\eq{leadcoulpollambnaiv} is so small for ordinary hydrogen. The
situation with muonic hydrogen is completely different. This time
the average radius of the muon orbit is about $r_{at}\approx1/(m
Z\alpha)$ and is of order of the electron Compton length
$r_C\approx1/m_e$, the respective ratio is about $r_{at}/r_C\approx
m_e/(m Z\alpha)\approx0.7$, and the muon spends a significant part of
its life inside the region of the strongly distorted Coulomb potential.
Qualitatively one can say that the muon penetrates deep in the
screening polarization cloud of the Coulomb center, and sees a larger
unscreened charge. As a result the binding becomes stronger, and for
example the $2S$-level in muonic hydrogen in Fig.\ \ref{muonhydlevels}
turns out to be lower than the $2P$-level \cite{wheeler},  unlike the
case of ordinary hydrogen where the order of levels is just the
opposite. In this situation the polarization correction becomes by far
the largest contribution to the Lamb shift in muonic hydrogen.

The relative contribution of the leading proton size contribution to
the Lamb shift interval in electronic hydrogen is about $10^{-4}$. It
is determined mainly by the ratio of the proton size contribution to
the leading logarithmically enhanced Dirac form factor slope
contribution in \eq{diracslope} (which is much larger than the
polarization contribution for electronic hydrogen). The relatively
larger role of the leading proton size contribution in muonic hydrogen
may also be easily understood qualitatively. Technically the
leading proton radius contribution in \eq{chargerad} is of order
$(Z\alpha)^4m^3<~r^2~>$, where $m$ is the mass of the light particle,
electron or muon in the case of ordinary and muonic hydrogen,
respectively. We thus see that the relative weight of the leading
proton charge contribution to the Lamb shift, in comparison with the
standard nonrecoil contributions, is enhanced in muonic hydrogen by the
factor $(m/m_e)^2$ in comparison with the relative weight of the
leading proton charge contribution in ordinary hydrogen, and it becomes
larger than all other standard nonrecoil and recoil contributions.
Overall the weight of the leading proton radius contribution in the
total Lamb shift in muonic hydrogen is determined by the ratio of the
proton size contribution to the leading electron polarization
contribution. In electronic hydrogen the ratio of the proton radius
contribution  and the leading polarization contribution is about
$5\cdot10^{-3}$, and is much larger than the weight of the proton
charge radius contribution in the total Lamb shift. In muonic hydrogen
this ratio is $10^{-2}$, four times larger than the ratio of the
leading proton size contribution and the leading polarization
correction in electronic hydrogen. Both the leading proton size
correction and the leading vacuum polarization contribution are
parametrically enhanced in muonic hydrogen, and an extra factor four
in their ratio is due to an additional accidental numerical
enhancement.

Below we will discuss corrections to the Lamb shift in muonic hydrogen,
with an emphasis on the classic $2P-2S$ Lamb shift, having in mind the
experiment on measurement of this interval which is now under way
\cite{kottman} (see also Section \ref{munichydexp}). Being interested
in theory, we will consider even those corrections to the Lamb shift
which are an order of magnitude smaller than the expected experimental
precision $0.008$ meV. Such corrections could become phenomenologically
relevant for muonic hydrogen in the future. Another reason to consider
these small corrections is that many of them scale as powers of the
parameter $Z$, and produce larger contributions for atoms with higher
$Z$.  Hence, even being too small for hydrogen they become
phenomenologically relevant for muonic helium where $Z=2$.

\section{Closed Electron-Loop Contributions of Order
$\alpha^{\lowercase{n}}(Z\alpha)^2m$}

\subsection{Diagrams with One External Coulomb Line}

\subsubsection{Leading Polarization Contribution of Order
$\alpha(Z\alpha)^2m$}\label{leadpolmuononeelloop}

The effects connected with the electron vacuum polarization
contributions in muonic atoms were first quantitatively discussed in
\cite{galpom}. In electronic hydrogen polarization loops of other
leptons and hadrons considered in Section \ref{heavypartlambpol} played
a relatively minor role, because they were additionally suppressed by
the typical factors $(m_e/m)^2$. In the case of muonic hydrogen we have
to deal with the polarization loops of the light electron, which are
not suppressed at all. Moreover, characteristic exchange momenta
$mZ\alpha$ in muonic atoms are not small in comparison with the
electron mass $m_e$, which determines the momentum scale of the
polarization insertions ($m(Z\alpha)/m_e\approx1.5$). We see that even
in the simplest case the polarization loops cannot be expanded in the
exchange momenta, and the radiative corrections in muonic atoms induced
by the electron loops should be calculated exactly in the parameter
$m(Z\alpha)/m_e$.

Electron polarization insertion in the photon propagator in Fig.\
\ref{polradfig} induces a correction to the Coulomb potential, which
may be easily written in the form\cite{blp}

\beq \label{onelooppoppot}
\delta V_{VP}^C(r)=-\frac{Z\alpha}{r}\frac{2\alpha}{3\pi}\int_1^\infty
d\zeta e^{-2m_er\zeta}\left(1+\frac{1}{2\zeta^2}\right)
\frac{\sqrt{\zeta^2-1}}{\zeta^2}.
\eeq

\noindent
The respective correction to the energy levels is given by the
expectation value of this perturbation potential

\beq  \label{perttheory}
\Delta E_{nl}=<nlm|\delta V|nlm>=\int_0^\infty dr
R_{nl}^2(r)\delta V(r)r^2
\eeq
\[
=-\frac{2\alpha^2Z}{3\pi}\int_0^\infty rdr\int_1^\infty d\zeta
R_{nl}^2(r)e^{-2m_er\zeta}\left(1+\frac{1}{2\zeta^2}\right)
\frac{\sqrt{\zeta^2-1}}{\zeta^2},
\]

\noindent
where

\beq
R_{nl}(r)=2\left(\frac{m_r
Z\alpha}{n}\right)^\frac{3}{2}\sqrt{\frac{(n-l-1)!}
{n[(n+l)!]^3}}\left(\frac{2m_r Z\alpha}{n}r\right)^le^{-\frac{m_r
Z\alpha}{n}r}L^{2l+1}_{n-l-1}(\frac{2m_r Z\alpha}{n}r)
\eeq

\noindent
is the radial part of the Schr\"odinger-Coulomb wave function in
\eq{schreqcoulinf} (but now it depends on the reduced mass), and
$L^{2l+1}_{n-l-1}$  is the associated Laguerre polynomial, defined as
in \cite{landlif,schiff}

\beq
L^{2l+1}_{n-l-1}(x)=\sum_{i=0}^{n-l-1}\frac{(-1)^i[(n+l)!]^2}
{i!(n-l-i-1)! (2l+i+1)!}x^i.
\eeq

\noindent
The radial wave functions depend on radius only via the combination
$\rho=r m_r Z\alpha$ and it is convenient to write it explicitly as a
function of this dimensionless variable

\beq
R_{nl}(r)=2\left(\frac{m_r
Z\alpha}{n}\right)^\frac{3}{2}f_{nl}(\frac{\rho}{n}),
\eeq

\noindent
where

\beq
f_{nl}(\frac{\rho}{n})\equiv \sqrt{\frac{(n-l-1)!}
{n[(n+l)!]^3}}\left(\frac{2\rho}{n}\right)^le^{-\frac{\rho}{n}}
L^{2l+1}_{n-l-1}(\frac{2\rho}{n}).
\eeq

Explicit dependence of the leading polarization correction on the
parameters becomes more transparent after transition to the
dimensionless integration variable $\rho$ \cite{galpom}

\beq \label{galpomshift}
\Delta E_{nl}^{(1)}=-\frac{8\alpha(Z\alpha)^2}{3\pi n^3}
Q_{nl}^{(1)}(\beta)m_r,
\eeq

\noindent
where

\beq  \label{qnlanal}
Q_{nl}^{(1)}(\beta)\equiv
\int_0^\infty \rho d\rho\int_1^\infty d\zeta
f_{nl}^2(\frac{\rho}{n})e^{-2\rho\zeta\beta}
\left(1+\frac{1}{2\zeta^2}\right)
\frac{\sqrt{\zeta^2-1}}{\zeta^2},
\eeq

\noindent
and $\beta=m_e/(m_r Z\alpha)$. The integral $Q_{nl}(\beta)$ may easily
be calculated numerically for arbitrary $n$. It was calculated
analytically for the lower levels $n=1,2,3$ in
\cite{mickel,pustovalov}, and later these results were confirmed
numerically in \cite{giacomo}. Analytic results for all states with
$n=l+1$ were obtained in \cite{glauber}.

The leading electron vacuum polarization contribution to
the Lamb shift in muonic hydrogen in \eq{qnlanal} is of order
$\alpha(Z\alpha)^2m$. Recall that the leading  vacuum polarization
contribution to the Lamb shift in electronic hydrogen in \eq{qpol} is
of order $\alpha(Z\alpha)^4m$. Thus, the relative magnitude of the
leading polarization correction in muonic hydrogen is enhanced by the
factor $1/(Z\alpha)^2\sim (m/m_e)^2$. This means that the electronic
vacuum polarization gives by far the largest contribution to the Lamb
shift in muonic hydrogen.  The magnitude of the energy shift in
\eq{galpomshift} is determined also by the dimensionless integral
$Q_{nl}(\beta)$. At the physical value of $\beta={m_e}/({m_r
Z\alpha})\approx0.7$ this integral is small
($Q_{10}^{(1)}(\beta)\approx0.061$, $Q_{20}^{(1)}(\beta)\approx0.056$,
$Q_{21}^{(1)}(\beta)\approx0.0037$) and suppresses somewhat the leading
electron polarization contribution.

The expression for $Q_{nl}^{(1)}(\beta)$ in \eq{qnlanal} is valid for
any $\beta$, in particular we can consider the case when $m=m_e$. Then
$\beta=m/(m_rZ\alpha)\gg 1$, and it is easy to show that the leading
term in the expansion of the result in \eq{galpomshift} over $1/\beta$
coincides with the leading polarization contribution in electronic
hydrogen in \eq{qpol}.

Numerically, contribution to the $2P-2S$ Lamb shift in muonic hydrogen
is equal to

\beq
\Delta E(2P-2S)=205.0074~ \mbox{meV}.
\eeq

\subsubsection{Two-Loop Electron Polarization Contribution of Order
$\alpha^2(Z\alpha)^2m$}

\begin{figure}[ht]
\centerline{\epsfig{file=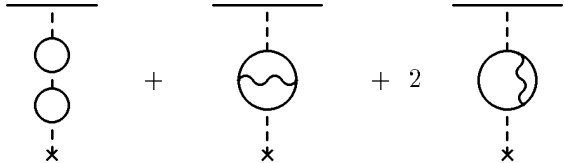}}
\vspace{0.5cm}
\caption{Two-loop polarization insertions in the Coulomb photon}
\label{twolooppolred}
\end{figure}

In electronic hydrogen the leading contribution  generated by the
two-loop irreducible polarization operator in Fig.\ \ref{secondpolfig}
is of order $\alpha^2(Z\alpha)^4m$ (see \eq{pol2}), and is determined
by the leading low-frequency term in the polarization operator. The
reducible diagram in Fig.\ \ref{twolooppolred} with two one-loop
insertions in the Coulomb photon does not generate a correction of the
same order in electronic hydrogen because it vanishes at the
characteristic atomic momenta, which are small in comparison with the
electron mass. In the case of muonic hydrogen atomic momenta are of
order of the electron mass and the two-loop irreducible and reducible
polarization insertions in Fig.\ \ref{twolooppolred} both generate
contributions of order $\alpha^2(Z\alpha)^2m$ and should be considered
simultaneously.

Two-loop electron polarization contribution to the Lamb shift may be
calculated exactly like the one-loop contribution, the only difference
is that one has to use as a perturbation potential the two-loop
correction to the Coulomb potential from \cite{kalsab}. We use it in
the form of the integral representation derived in \cite{blomkwist}
(see also \cite{huang})

\beq
\delta
V^{(2)}(r)=\frac{Z\alpha}{r}\left(\frac{\alpha}{\pi}\right)^2
\int_1^\infty d\zeta
e^{-2m_er\zeta}\left\{\left(\frac{13}{54\zeta^2}+\frac{7}{108\zeta^4}
+\frac{2}{9\zeta^6}\right)\sqrt{\zeta^2-1}
\right.
\eeq
\[
+\left(-\frac{44}{9\zeta}+\frac{2}{3\zeta^3}+\frac{5}{4\zeta^5}
+\frac{2}{9\zeta^7}\right)\ln[\zeta+\sqrt{\zeta^2-1}]
\]
\[
\left.
+\left(\frac{4}{3\zeta^2}+\frac{2}{3\zeta^4}\right)\sqrt{\zeta^2-1}
\ln[8\zeta(\zeta^2-1)]
+\left(-\frac{8}{3\zeta}+\frac{2}{3\zeta^5}\right)F(\zeta)\right\},
\]

\noindent
where

\beq
F(\zeta)=\int_\zeta^\infty
dx[\frac{3x^2-1}{x(x^2-1)}\ln[x+\sqrt{x^2-1}]-
\frac{1}{\sqrt{x^2-1}}\ln[8x(x^2-1)].
\eeq

\noindent
Then we easily obtain

\beq \label{kallensabr}
\Delta E_{nl}^{(2)}=\frac{4\alpha^2(Z\alpha)^2}{\pi^2 n^3}
Q_{nl}^{(2)}(\beta)m_r,
\eeq

\noindent
where

\beq
Q_{nl}^{(2)}(\beta)\equiv
\int_0^\infty \rho d\rho\int_1^\infty d\zeta
f_{nl}^2(\frac{\rho}{n})e^{-2\rho\zeta\beta}
\left\{\left(\frac{13}{54\zeta^2}+\frac{7}{108\zeta^4}
+\frac{2}{9\zeta^6}\right)\sqrt{\zeta^2-1}
\right.
\eeq
\[
+\left(-\frac{44}{9\zeta}+\frac{2}{3\zeta^3}+\frac{5}{4\zeta^5}
+\frac{2}{9\zeta^7}\right)\ln[\zeta+\sqrt{\zeta^2-1}]
\]
\[
\left.
+\left(\frac{4}{3\zeta^2}+\frac{2}{3\zeta^4}\right)\sqrt{\zeta^2-1}
\ln[8\zeta(\zeta^2-1)]
+\left(-\frac{8}{3\zeta}+\frac{2}{3\zeta^5}\right)F(\zeta)\right\}.
\]

Numerically, this correction for the $2P-2S$ Lamb shift was first
calculated in \cite{giacomo}

\beq
\Delta E(2P-2S)=1.5079~\mbox{meV}.
\eeq

\subsubsection{Three-Loop Electron Polarization of
Order $\alpha^3(Z\alpha)^2m$}

\begin{figure}[ht]
\centerline{\epsfig{file=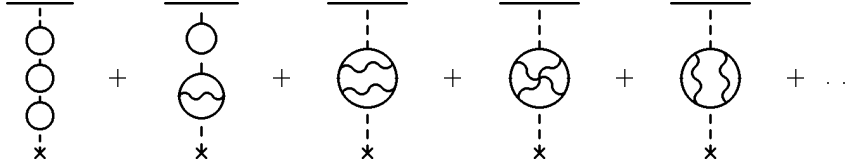}}
\vspace{0.5cm}
\caption{Three-loop polarization insertions in the Coulomb photon}
\label{threelooppolred}
\end{figure}

As in the case of the two-loop electron polarization insertions in the
external Coulomb line, reducible and irreducible three-loop
polarization insertions enter on par in muonic hydrogen, and we have
to consider all respective corrections to the Coulomb potential in
Fig.\ \ref{threelooppolred}. One-, two-, and three-loop polarization
operators were in one form or another calculated in the literature
\cite{blp,kalsab,kinlind1,kinlind2,baibroad}. Numerical calculation of
the respective contribution the $2P-2S$ splitting  in muonic hydrogen
was performed in \cite{kinnio99}

\beq         \label{kinion1}
\Delta E(2P-2S)=0.083~53~(1)\frac{\alpha^3(Z\alpha)^2}{\pi^3}m_r
\approx 0.0053~\mbox{meV}
\eeq

\subsection{Diagrams with Two External Coulomb Lines}

\subsubsection{Reducible Diagrams. Contributions of Order
$\alpha^2(Z\alpha)^2m$}

\begin{figure}[ht]
\centerline{\epsfig{file=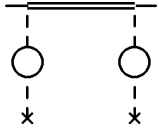}}
\vspace{0.5cm}
\caption{Perturbation theory contribution with two
one-loop polarization insertions}
\label{redtwoonelp}
\end{figure}

In electronic hydrogen characteristic exchanged momenta in the diagram
in Fig.\ \ref{redtwoonelp} were determined by the electron mass, and
since this mass in electronic hydrogen is large in comparison with the
characteristic atomic momenta we could ignore binding and calculate
this diagram in the scattering approximation. As a result the
respective contribution was suppressed in comparison with the leading
polarization contribution, not only by an additional factor $\alpha$
but also by an additional factor $Z\alpha$. The situation is completely
different in the case of muonic hydrogen. This time atomic momenta are
just of order of the electron mass, one cannot neglect binding, and the
additional suppression factor $Z\alpha$ is missing. As a result the
respective correction in muonic hydrogen is of the same order
$\alpha^2(Z\alpha)^2$ as the contributions of the diagrams with
reducible and irreducible two-loop polarization insertions in one and
the same Coulomb line considered above.

Formally the contribution of diagram in Fig.\ \ref{redtwoonelp} is
given by the standard quantum mechanical second order perturbation
theory term. Summation over the intermediate states, which accounts for
binding, is realized with the help of the reduced Green function.
Convenient closed expressions for the reduced Green function in
the lower states were obtained in \cite{laurenzi} and independently
reproduced in \cite{pachucki96muon}. Numerical calculation of the
contribution to the $2P-2S$ splitting leads to the result
\cite{pachucki96muon,pachuckimu99}

\beq
\Delta E(2P-2S)=0.01244~\frac{4\alpha^2(Z\alpha)^2}{9\pi^2}m_r
=0.1509~\mbox{meV}.
\eeq


\subsubsection{Reducible Diagrams. Contributions of order
$\alpha^3(Z\alpha)^2m$}


\begin{figure}[ht]
\centerline{\epsfig{file=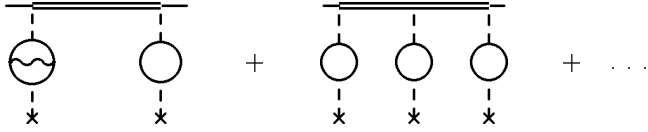}}
\vspace{0.5cm}
\caption{Perturbation theory contribution of  order
$\alpha^3(Z\alpha)^2$ with polarization insertions}
\label{redthreeonelp}
\end{figure}

As in the case of corrections of order $\alpha^2(Z\alpha)^2m$, not only
the diagrams in Fig.\ \ref{threelooppolred} with insertions of
polarization operators in one and the same external Coulomb line but
also the reducible diagrams Fig.\ \ref{redthreeonelp} with polarization
insertions in different external Coulomb lines generate corrections of
order $\alpha^3(Z\alpha)^2m$. Respective contributions were calculated
in \cite{kinnio99} with the help of the subtracted Coulomb Green
function from \cite{pachucki96muon}

\beq                \label{kinion2}
\Delta E(2P-2S)=0.036~506~(4)\frac{\alpha^3(Z\alpha)^2}{\pi^3}m_r
\approx0.0023~\mbox{meV}
\eeq

Total contribution of order $\alpha^3(Z\alpha)^2m$ is a sum of the
contributions in \eq{kinion1} and \eq{kinion2} \cite{kinnio99}

\beq
\Delta E(2P-2S)=0.120~045~(12)\frac{\alpha^3(Z\alpha)^2}{\pi^3}m_r
=0.0076~\mbox{meV}
\eeq

\section{Relativistic Corrections to the Leading Polarization
Contribution with Exact Mass Dependence}

The leading electron polarization contribution in \eq{galpomshift} was
calculated in the nonrelativistic approximation between the
Schr\"odinger-Coulomb wave functions. Relativistic corrections of
relative order $(Z\alpha)^2$ to this contribution may easily be
obtained in the nonrecoil limit. To this end one has to calculate the
expectation value of the radiatively corrected potential in
\eq{onelooppoppot} between the relativistic Coulomb-Dirac wave
functions instead of averaging it with the nonrelativistic
Coulomb-Schr\"odinger wave functions.

Numerical calculations of the Uehling potential contribution to the
energy shift without expansion over $Z\alpha$ (and therefore with
account of the leading nonrecoil relativistic corrections of order
$\alpha(Z\alpha)^4m$) are abundant in the literature, see. e.g.,
\cite{sundar72}, and references in the review \cite{borierink}.
Analytic results without expansion over $Z\alpha$ were obtained for the
states with $n=l+1$, $j=l+1/2$ \cite{karpol99}. All these results might
be very useful for heavy muonic atoms. However, in the case of muonic
hydrogen with a relatively large muon-proton mass ratio recoil
corrections to the nonrecoil relativistic corrections of order
$\alpha(Z\alpha)^4m$ may be rather large, while corrections of higher
orders in $Z\alpha$ are expected to be very small. In such conditions
it is reasonable to adopt another approach to the relativistic
corrections, and try to calculate them from the start in the nonrecoil
approximation with exact dependence on the mass ratio.

In the leading nonrelativistic approximation the one-loop electron
polarization insertion in the Coulomb photon generates a nontrivial
correction \eq{onelooppoppot} to the unperturbed Coulomb binding
potential in muonic hydrogen, which may be written as a weighted
integral of a potential corresponding to an exchange by a massive
photon with continuous mass $\sqrt{t'}\equiv 2m_e\zeta$

\beq
\delta V_{VP}^C(r)=\frac{2\alpha}{3\pi}\int_1^\infty d\zeta
(1+\frac{1}{2\zeta^2})\frac{\sqrt{\zeta^2-1}}{\zeta^2}
(-\frac{Z\alpha}{r}e^{-2m_e\zeta r}).
\eeq

This situation is radically different from the case of electronic
hydrogen where inclusion of the electron loop in the photon propagator
generates effectively a $\delta$-function correction to the Coulomb
potential (compare discussion in Section \ref{originlamb}).

\begin{figure}[ht]
\centerline{\epsfig{file=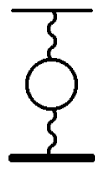}}
\vspace{0.5cm}
\caption{One-photon exchange with one-loop polarization insertion}
\label{breitvaccorponeph}
\end{figure}

Calculation of the leading relativistic corrections to the
nonrelativistic electronic vacuum polarization contribution may be done
in the framework of the Breit approach used in Section
\ref{leadreclambdef} to derive leading relativistic corrections to the
ordinary one-photon exchange. All we need to do now is to
derive an analogue of the Breit potential, which is generated by the
exchange of one photon with electron-loop insertion in Fig.\
\ref{breitvaccorponeph}. This derivation is facilitated by the well
known observation that the dispersion relation for the polarization
operator allows one to represent the one-loop radiatively corrected
photon propagator as an integral over continuous photon mass (see,
e.g., \cite{blp}). Then everything one has to do to derive the analogue
of the Breit potential is to obtain an expression for the Breit
potential corresponding to the exchange by a massive photon and then to
integrate over the effective photon mass. Derivation of the massive
Breit potential proceeds exactly as for the massless case, and one
obtains \cite{pachucki96muon} (as in \eq{breitpot} we omit below all
terms in the massive Breit potential which depend on the spin of the
heavy particle since we do not consider hyperfine structure now)

\beq  \label{intgrandbr}
{\cal V}_{VP}^{Br}(\sqrt{t'}\equiv 2m_e\zeta)
=\frac{Z\alpha}{2}\left(\frac{1}{m^2}+\frac{1}{M^2}\right)
\left(\pi\delta^3({\bf r})-\frac{m_e^2\zeta^2}{r}e^{-2m_e\zeta
r}\right)
\eeq
\[
-\frac{Z\alpha m_e^2\zeta^2}{m M}\frac{e^{-2m_e\zeta
r}}{r}\left(1-m_e\zeta r\right)
-\frac{Z\alpha }{2mM}p^i\frac{e^{-2m_e\zeta
r}}{r}\left(\delta_{ij}+\frac{r^ir^j} {r^2}(1+2m_e\zeta r)\right)p^j
\]
\[
+\frac{Z\alpha
}{r^3}\left(\frac{1}{4m^2}+\frac{1}{2mM}\right)e^{-2m_e\zeta r}
(1+2m_e\zeta r)[{\bf r\times p}]\cdot\bbox\sigma.
\]

\noindent
Then the analogue of the Breit potential induced by the electron vacuum
polarization insertion is given by the integral

\beq         \label{radcorrbreit}
V_{VP}^{Br}
=\frac{2\alpha}{3\pi}\int_1^\infty d\zeta
\left(1+\frac{1}{2\zeta^2}\right)\frac{\sqrt{\zeta^2-1}}{\zeta^2}
{\cal V}_{VP}(2m_e\zeta).
\eeq

\noindent
Calculation of the leading recoil corrections of order
$\alpha(Z\alpha)^4$ becomes now almost trivial. One has to take into
account that in our approximation the analogue of the Breit Hamiltonian
in \eq{totbreitnaiv} has the form \cite{pachucki96muon}

\beq
H=\frac{\bf p^2}{2m}+\frac{\bf p^2}{2M}-\frac{Z\alpha}{r}+V_{Br}
+V_{VP}^C+V_{VP}^{Br},
\eeq

\noindent
where $V_{Br}$ was defined in \eq{breitpot}.

\begin{figure}[ht]
\centerline{\epsfig{file=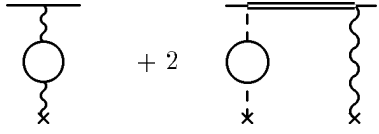}}
\vspace{0.5cm}
\caption{Relativistic corrections to the leading electron polarization
contribution}
\label{relcorroneloopbreit}
\end{figure}

Then the leading relativistic corrections of order
$\alpha(Z\alpha)^4$ may be easily obtained as a sum of the first and
second order perturbation theory contributions corresponding to the
diagrams in Fig.\ \ref{relcorroneloopbreit} \cite{pachucki96muon}

\beq          \label{matrixelmcalc}
\Delta E=<V_{VP}^{Br}>+2<V_{Br}G'(E_n)V_{VP}^C>.
\eeq

\noindent
Numerical calculation of the respective contribution to the $2P-2S$
Lamb shift leads to the result \cite{pachuckimu99}

\beq
\Delta E(2P-2S)=0.0594~\mbox{meV}.
\eeq

\section{~~Higher Order Electron-Loop Polarization Contributions}

\subsection{Wichmann-Kroll Electron-Loop Contribution of
Order $\alpha(Z\alpha)^4m$}

Contribution of the Wichmann-Kroll diagram in Fig.\ \ref{wichkrolfig}
with three external fields attached to the electron loop \cite{wichkr}
may be considered in the same  way as the polarization insertions in
the Coulomb potential, and as we will see below it generates a
correction to the Lamb shift of order $\alpha(Z\alpha)^4m$.

A convenient representation for the Wicmann-Kroll polarization potential
was obtained in \cite{blomkwist}

\beq  \label{blomkwrepr}
\delta
V^{WK}(r)=\frac{Z\alpha}{r}\frac{\alpha(Z\alpha)^2}{\pi}\int_0^\infty
d\zeta
e^{-2m_e\zeta r}\frac{1}{\zeta^4}\left[-\frac{\pi^2}{12}\sqrt{\zeta^2-1}
\theta(\zeta-1)
+\int_0^\zeta dx\sqrt{\zeta^2-x^2}f(x)\right],
\eeq

\noindent
where

\beq
f(x)=-2x \mbox{Li}_2(x^2)-x\ln^2(1-x^2)
+\frac{1-x^2}{x^2}\ln(1-x^2)\ln\frac{1+x}{1-x}
\eeq
\[
+\frac{1-x^2}{4x}\ln^2
\frac{1+x}{1-x}+\frac{2-x^2}{x(1-x^2)}\ln(1-x^2)+\frac{3-2x^2}{1-x^2}
\ln\frac{1+x}{1-x}-3x
\]

\noindent
for $x<1$, and

\beq
f(x)=\frac{1}{x^2} \mbox{Li}_2\left(\frac{1}{x^2}\right)-
\frac{3x^2+1}{2x}\left[\mbox{Li}_2\left(\frac{1}{x}\right)
-\mbox{Li}_2\left(-\frac{1}{x}\right)
\right]-\frac{2x^2-1}{2x^2}\left[\ln^2\left(1-\frac{1}{x^2}\right)
+\ln^2\frac{x+1}{x-1}
\right]
\eeq
\[
-(2x-1)\ln\left(1-\frac{1}{x^2}\right)\ln\frac{x+1}{x-1}
+\frac{3x^2+1}{4x}\ln^2\frac{x+1}{x-1}
-2\ln x\ln\left(1-\frac{1}{x^2}\right)
-\frac{3x^2+1}{2x}\ln x\ln\frac{x+1}{x-1}
\]
\[
+\left[5-\frac{x(3x^2-2)}{x^2-1}\right]\ln\left(1-\frac{1}{x^2}\right)
+\left[\frac{3x^2+2}{x}-\frac{3x^2-2}{x^2-1}\right]\ln\frac{x+1}{x-1}
+3\ln x-3
\]

\noindent
for $x>1$.

This representation allows us to calculate the correction to the Lamb
shift in the same way as we have done above for the Uehling and
K\"allen-Sabry potentials in \eq{galpomshift} and \eq{kallensabr},
respectively. Let us write the potential in the form

\beq
\delta V^{WK}(r)=\frac{Z\alpha}{r}\frac{\alpha(Z\alpha)^2}{\pi}g(m_er).
\eeq

\noindent
Then the contribution to the energy shift is given by the expression

\beq
\Delta E_{nl}^{(WK)}=\frac{4\alpha(Z\alpha)^4}{\pi n^3}
Q_{nl}^{(WK)}(\beta)m_r,
\eeq

\noindent
where

\beq
Q_{nl}^{(WK)}(\beta)\equiv
\int_0^\infty \rho d\rho
f_{nl}^2(\frac{\rho}{n})g(\rho\beta).
\eeq

The Uehling and K\"allen-Sabry potentials are attractive, and shift the
energy levels down. Physically this corresponds to the usual charge
screening in QED, and one can say that at finite distances the muon
sees a larger unscreened charge of the nucleus. From this point of view
the Uehling and K\"allen-Sabry potentials are just the attractive
potentials corresponding to the excess of the bare charge over
physical the charge.

The case of the Wichmann-Kroll potential is qualitatively different.
Due to current conservation the total charge which induces the
Wichmann-Kroll potential is zero \cite{wichkr}. Spatially the
induced charge distribution consists of two components: a
delta-function induced charge at the origin with the  sign opposite to
the sign of the nuclear charge, and a spatially distributed compensating
charge of the same sign as nuclear. The radius of this spatial
distribution is roughly equal to the electron Compton length. As a
result the muon which sees the nucleus from a finite distance
experiences net repulsion, the Wichmann-Kroll potential shifts the
levels up, and gives a positive contribution to the level shift (the
original calculation in \cite{fricke} produced a result with a wrong
sign and magnitude).

Practical calculations of the Wichmann-Kroll contribution are greatly
facilitated by convenient approximate interpolation formulae for the
potential in \eq{blomkwrepr}. One such formula was obtained in
\cite{huang} fitting the results of the numerical calculation of the
potential from \cite{vogel}

\beq
g_(x)=0.361~662~331~\exp\left[0.372~807~9~x - \sqrt{4.416~798~x^2
+ 11.399~11~x +2.906~096 }\right].
\eeq

\noindent
This expression fits the exact potential in the interval $0.01<x<1.0$
with an accuracy of about $1\%$, and due to an exponential decrease of
the wave functions and smallness of the potential at large distances,
it may be safely used for calculations at all $x$. After numerical
calculation with this interpolation formula we obtain for the $2P-2S$
Lamb shift

\beq   \label{wichmkrelemuon}
\Delta E(2P-2S)=-0.0010~\mbox{meV}.
\eeq

\noindent
Another convenient interpolation formula for the potential in
\eq{blomkwrepr} was obtained in \cite{borierink}.  One more way to
calculate the Wichmann-Kroll contribution numerically, is to use for
the small values of the argument an asymptotic expansion of the
potential in \eq{blomkwrepr}, which was obtained in \cite{blomkwist},
and for the large values of the argument the interpolation formula from
\cite{rinker76}. Calculations in both these approaches reproduce the
numerical value for the $2P-2S$ Lamb shift  in
\eq{wichmkrelemuon}\footnote{A result two times smaller than this
contribution was obtained in \cite{pachucki96muon}. We are convinced of
the correctness of the result in \eq{wichmkrelemuon}. Besides
calculations with all three forms of the interpolation formulae, we
also calculated the energy shift for muonic helium with $Z=2$, and
reproduced the well known old helium results
\cite{rinker76,borierink78,borierink}.}.

\subsection{Light by Light Electron-Loop Contribution of
Order $\alpha^2(Z\alpha)^3m$}

Light by light electron-loop contribution to the Lamb shift in Fig.\
\ref{6setscoulfig} $(e)$ in muonic atoms was considered in
\cite{chen,wilets,fujimoto,borie76,calmet}. This is a correction of
order $\alpha^2(Z\alpha)^3$ in muonic hydrogen.  Characteristic momenta
in the electron polarization loop are of the order of the atomic
momenta in muonic hydrogen, and hence, one cannot neglect the atomic
momenta calculating the matrix element of this kernel as it was done in
the case of electronic hydrogen. An initial numerical estimate in
\cite{chen} turned out to be far too large, and consistent much
smaller numerical estimates were obtained in
\cite{wilets,fujimoto,borie76}.

Momentum-space potential generated by the light by light diagrams
and the respective contribution to the energy shifts in heavy atoms
were calculated numerically in \cite{borie76}. Certain approximate
expressions for the effective momentum space potential were obtained in
\cite{fujimoto,borie76,borierink}. After extensive numerical work
electron-loop light by light scattering contribution was calculated
for muonic helium \cite{borierink78,borierink}, and turned out to be
equal to $0.02$ meV for the $2P-2S$ interval.  Scaling this result with
$Z$ we expect that the respective contribution in muonic hydrogen is at
the level of $0.01-0.04$ meV. This is one of the largest still
unknown purely electrodynamic corrections to the $2P-2S$ interval
in muonic hydrogen.

\subsection{Diagrams with Radiative Photon and Electron-Loop
Polarization Insertion in the Colulomb Photon. Contribution of Order
$\alpha^2(Z\alpha)^4m$}

\begin{figure}[ht]
\centerline{\epsfig{file=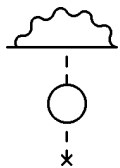}}
\vspace{0.5cm}
\caption{Diagram with radiative photon and electron-loop polarization
insertion in the Coulomb photon}
\label{radphotelpolcoul}
\end{figure}

In electronic hydrogen the leading contributions of diagrams such as
Fig.\ \ref{radphotelpolcoul} were generated at the scale of
the mass of the light constituent. The diagrams effectively looked like
Fig.\ \ref{6setscoulfig} $(c)$, could be calculated in the
scattering approximation, and produced the corrections of order
$\alpha^2(Z\alpha)^5m$. In muonic hydrogen electron polarization
insertion in the Coulomb photon is not suppressed at characteristic
atomic momenta, and respective contribution to the energy shift is only
$\alpha$ times smaller  than the contribution of the diagrams with
insertions of one radiative photon in the muon line (leading diagrams
for the Lamb shift in case of electronic hydrogen). One should expect
that, in the same way as the leading Lamb shift contribution in
electronic hydrogen, this contribution is also logarithmically enhanced
and is of order $\alpha^2(Z\alpha)^4m$. This contribution was never
calculated completely, the leading logarithmic contribution was
obtained in \cite{pachucki96muon}.

The leading logarithmic contribution generated by the diagrams with the
radiative photon spanning any number of the Coulomb photons and one
Coulomb photon with electron-loop polarization insertion, the simplest
of which is resented in Fig.\ \ref{radphotelpolcoul}, may be calculated
by closely following the classical Bethe calculation \cite{bethe} of the
leading logarithmic comtribution to the Lamb shift in electronic
hydrogen. As is well known, in the dipole approximation the
standard subtracted logarithmically divergent (at high frequencies)
expression for the Lamb shift may be written in the form
\cite{pachucki96muon} (compare, e.g., \cite{blp,bd})

\beq   \label{logintvacpol}
\Delta E=\frac{2\alpha}{3\pi m^2}\int d\omega <n|{\bf
p}\frac{H-E_n}{H-E_n+\omega}{\bf p}|n>,
\eeq

\noindent
where $H$ is the nonrelativistic Hamiltonian for the muon in the
external field equal to the sum of the Coulomb field and radiatively
corrected Coulomb field from \eq{onelooppoppot}

\beq
V_{tot}=-\frac{Z\alpha}{r}+V_C^{VP}.
\eeq

\noindent
To obtain the leading contribution generated by the integral in
\eq{logintvacpol}, it is sufficient to integrate over the wide
logarithmic region $m_r (Z\alpha)^2\gg\omega\gg m$, where one can
neglect the terms $H-E_n$ in the denominator. Then one easily obtains
\cite{pachucki96muon}

\beq
\Delta E=\frac{2\alpha}{3\pi m^2}\ln\frac{m}{m_r(Z\alpha)^2}
<n|{\bf p}(H-E_n){\bf p}|n>.
\eeq

\noindent
Using the trivial identity

\beq
<n|{\bf p}(H-E_n){\bf p}|n>=\frac{<n|\Delta(V_C+V_C^{VP})|n>}{2},
\eeq

\noindent
throwing away the standard leading polarization independent
logarithmic correction to the Lamb shift, which is also contained in
this expression, and expanding the state vectors up to first order in
the potential $V_{VP}$ one easily obtains

\beq
\Delta E=\frac{\alpha}{3\pi m^2}\ln\frac{m}{m_r(Z\alpha)^2}
\{<n|\Delta V_C^{VP}|n>+2<n|V_C^{VP}G'(E_n)\Delta V_C|n>\}.
\eeq

\noindent
This contribution to $2P-2S$ splitting was calculated numerically in
\cite{pachucki96muon,pachuckimu99}

\beq
\Delta E(2P-2S)=-0.005~(1)~\mbox{meV}.
\eeq

\noindent
The uncertainty here is due to the unknown nonlogarithmic terms.
Calculation of these nonlogarithmic terms is one of the future
tasks in the theory of muonic hydrogen.

\subsection{Electron-Loop Polarization Insertion in the Radiative
Photon. Contribution of Order
$\alpha^2(Z\alpha)^4m$}\label{electrloppradph}

\begin{figure}[ht]
\centerline{\epsfig{file=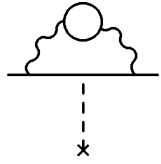}}
\vspace{0.5cm}
\caption{Electron polarization insertion in the radiative photon}
\label{elpolpolradradins}
\end{figure}

Contributions of order $\alpha^2(Z\alpha)^4m$ in muonic hydrogen
generated by the two-loop muon form factors have almost exactly the
same form as the respective contributions in the case of electronic
hydrogen. The only new feature is connected with the contribution to
the muon form factors generated by insertion of one-loop electron
polarization in the radiative photon in Fig.\ \ref{elpolpolradradins}.
Respective insertion of the muon polarization in the electron form
factors in electronic hydrogen is suppressed as $(m_e/m)^2$, but
insertion of a light loop in the muon case is logarithmically enhanced.

\noindent
The graph in Fig.\ \ref{elpolpolradradins} is gauge invariant and
generates a correction to the slope of the Dirac form factor, which was
calculated in \cite{barb}

\beq
\frac{dF^{(2)}_1(-{\bf k}^2)}{d{\bf k}^2}_{|{\bf k}^2=0}=-
\left[\frac{1}{9}\ln^2\frac{m}{m_e}-
\frac{29}{108}\ln\frac{m}{m_e}
+\frac{\pi^2}{54}+\frac{395}{1296}+O(\frac{m_e}{m})\right]
\frac{1}{m^2}
\left(\frac{\alpha}{\pi}\right)^2
\eeq
\[
\approx -\left(2.216~56+O(\frac{m_e}{m})\right)
\frac{1}{m^2}\left(\frac{\alpha}{\pi}\right)^2.
\]

\noindent
Then the contribution to the Lamb shift has the form \cite{barb}

\beq
\Delta E_{F_1}=-4\pi
(Z\alpha)|\Psi_n(0)|^2\frac{dF^{(2)}_1(-{\bf k}^2)}{d{\bf k}^2}_{|{\bf
k}^2=0}
=\left(2.216~56+O(\frac{m_e}{m})\right)\frac{4\alpha^2(Z\alpha)^4}{\pi^2n^3}
\left(\frac{m_r}{m}\right)^3m
\:\delta_{l0}.
\eeq

\noindent
We also have to consider the electron-loop contribution to the muon
anomalous magnetic moment

\beq
F^{(2)}_2(0)=\left[\frac{1}{3}\ln\frac{m}{m_e}
-\frac{25}{36}+\frac{\pi^2}{4}\frac{m}{m_e}
-4\left(\frac{m_\mu}{m_e}\right)^2\ln\frac{m}{m_e}+3\left(\frac{m}{m_e}
\right)^2+O(\left(\frac{m}{m_e}\right)^3)
\right]\left(\frac{\alpha}{\pi}\right)^2
\eeq
\[
\approx\left(1.082~75+O(\frac{m}{m_e})\right)
\left(\frac{\alpha}{\pi}\right)^2.
\]

\noindent
The first two terms in this expression were obtained in
\cite{suura,peterman}, and an exact analytic result without expansion
over $m_e/m$ was calculated in \cite{elend,ericliu}.

\noindent
Then one readily obtains for the Lamb shift contribution
\cite{borierink}

\beq
\Delta E_{|l=0}
=1.082~75\frac{\alpha^2(Z\alpha)^4m}{\pi^2n^3}
\left(\frac{m_r}{m}\right)^3,
\eeq
\[
\Delta E_{|l\neq
0}=1.082~75\frac{\alpha^2(Z\alpha)^4m}{\pi^2n^3}
\frac{j(j+1)-l(l+1)-3/4}{l(l+1)(2l+1)}\left(\frac{m_r}{m}\right)^2
\]

Numerically for the $2P_\frac{1}{2}-2S_\frac{1}{2}$ interval we obtain

\beq
\Delta E(2P_\frac{1}{2}-2S_\frac{1}{2})=-0.0016~\mbox{meV}.
\eeq

\subsection{Insertion of One Electron and One
Muon Loops in the same Coulomb Photon. Contribution of Order
$\alpha^2(Z\alpha)^2(m_e/m)^2m$}

\begin{figure}[ht]
\centerline{\epsfig{file=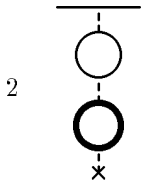}}
\vspace{0.5cm}
\caption{Electron- and muon-loop polarization insertions in the Coulomb
photon}
\label{mixedloopcoul}
\end{figure}

Contribution of the mixed polarization graph with one electron- and one
muon-loop insertions in the Coulomb photon in Fig.\ \ref{mixedloopcoul}
may be easily calculated by the same methods as the contributions of
purely electron loops, and it was first considered in \cite{borie}. The
momentum space perturbation potential corresponding to the mixed loop
diagram is given by the expression (factor 2 is due to two diagrams)

\beq
2\frac{\Pi_\mu(k^2)}{k^4}\frac{\Pi_e(k^2)}{k^2},
\eeq

\noindent
where $\Pi_{\mu}(k^2)$ and $\Pi_{e}(k^2)$ are the muon- and
electron-loop polarization operators, respectively (compare
\eq{leadcoulpollambnaiv}).

The characteristic integration momenta in the matrix element of this
perturbation potential between the Coulomb-Schr\"odinger wave functions
are of the atomic scale $mZ\alpha$, and are small in comparison with
the muon mass $m$. Hence, in the leading approximation the muon
polarization may be approximated by the first term in its low-frequency
expansion

\beq        \label{effctmompotmixed}
\frac{2\alpha}{15\pi m^2}\frac{\Pi_e(k^2)}{k^2}.
\eeq

\noindent
This momentum space potential is similar to the momentum space
potential corresponding to insertion of the electron-loop polarization
in the Coulomb photon, considered in Section
\ref{leadpolmuononeelloop}. The only difference is in the overall
multiplicative constant, and that the respective expression in the case
of the one electron polarization insertion contains $k^4$ in the
denominator instead of $k^2$ in \eq{effctmompotmixed}. This means that
the mixed loop contribution is suppressed in comparison with the purely
electron loops by an additional recoil factor $(m_e/m)^2$.

Similarly to \eq{onelooppoppot} it is easy to write a coordinate space
representation for the perturbation potential corresponding to the
diagram in Fig.\ \ref{mixedloopcoul}

\beq
\delta V(r)
=\frac{Z\alpha}{r}\frac{16}{45}\left(\frac{\alpha}{\pi}\right)^2
\left(\frac{m_e}{m}\right)^2\int_1^\infty d\zeta
e^{-2m_e\zeta r}\left(1+\frac{1}{2\zeta^2}\right)\sqrt{\zeta^2-1}.
\eeq

\noindent
Then we easily obtain

\beq       \label{mixedlopen}
\Delta E_{nl}^{(3)}=\frac{64\alpha^2(Z\alpha)^2}{45\pi^2 n^3}
\left(\frac{m_e}{m_\mu}\right)^2
Q_{nl}^{(3)}(\beta)m_r,
\eeq

\noindent
where

\beq \label{qnl3beta}
Q_{nl}^{(3)}(\beta)\equiv
\int_0^\infty \rho d\rho\int_1^\infty d\zeta
f_{nl}^2(\frac{\rho}{n})e^{-2\rho\zeta\beta}
\left(1+\frac{1}{2\zeta^2}\right)\sqrt{\zeta^2-1}.
\eeq

\noindent
Due to the additional recoil factor $(m_e/m)^2$ this contribution is
suppressed by four orders of magnitude in comparison with the nonrecoil
corrections generated by insertion of two electron loops in the Coulomb
photon (compare \eq{kallensabr}). Numerically, for the $2P-2S$
interval we obtain

\beq   \label{mixedloop2p2s}
\Delta E(2P-2S)=0.00007~{meV}.
\eeq

\section{~~Hadron Loop Contributions}

\subsection{Hadronic Vacuum Polarization Contribution of
Order $\alpha(Z\alpha)^4m$}

\begin{figure}[ht]
\centerline{\epsfig{file=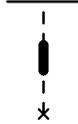}}
\vspace{0.5cm}
\caption{Hadron polarization insertion in the Coulomb
photon}
\label{hadronpol}
\end{figure}

Masses of pions are only slightly larger than the muon mass, and we
should expect that the contribution of the diagram with insertion of
the hadronic vacuum polarization in the Coulomb photon in Fig.\
\ref{hadronpol} is of the same order of magnitude as contribution of
the respective diagrams with muon vacuum polarization. Hadronic
polarization correction is of order $\alpha(Z\alpha)^4m$, it depends
only on the leading low-momentum asymptotic term in the hadronic
polarization operator, and has the same form as in the case of
electronic hydrogen in Section \ref{heavypartlambpol}. It was
considered in the literature many times and consistent results were
obtained in \cite{folomeshkin,sundar,gerdt,borie81,kinnio99}.
According to \cite{fms99}

\beq
\Delta E(nS)=-0.671~(15)~\Delta E_\mu,
\eeq

\noindent
where $\Delta E_\mu$ is the muon-loop polarization contribution to the
Lamb shift in \eq{muonpollmab}\footnote{In the case of muonic hydrogen
$m_r$ in \eq{muonpollmab} is the muon-proton reduced mass.}.

\noindent
The latest treatment of this diagram in \cite{faustov2000} produced

\beq
\Delta E(nS)=-0.638~(22)~\Delta E_\mu.
\eeq

\noindent
Respective results for the $2P-2S$ splitting are

\beq
\Delta E(2P-2S)=0.0113~(3)~\mbox{meV},
\eeq

\noindent
and

\beq
\Delta E(2P-2S)=0.0108~(4)~\mbox{meV}.
\eeq

As in the case of electronic hydrogen this correction may be hidden in
the main proton radius contribution to the Lamb shift and we ignored it
in the phenomenological discussion of the Lamb shift in electronic
hydrogen (see discussion in Section \ref{empnuklrad}). However, we
include the hadronic polarization in the theoretical expression for the
Lamb shift in muonic hydrogen having in mind that in the future all
radiative corrections should be properly taken into account while
extracting the value of the proton charge radius from the scattering
and optical experimental data.

\subsection{Hadronic Vacuum Polarization Contribution of
Order $\alpha(Z\alpha)^5m$}

\begin{figure}[ht]
\centerline{\epsfig{file=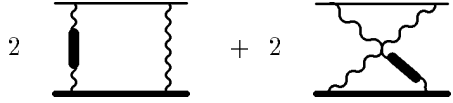}}
\vspace{0.5cm}
\caption{Hadron polarization contribution of order $\alpha(Z\alpha)^5$}
\label{hadronpoltwophotexc}
\end{figure}

Due to the analogy between contributions of the diagrams with muon
and hadron vacuum polarizations, it is easy to see that insertion of
hadron vacuum polarization in one of the exchanged photons in the
skeleton diagrams with two-photon exchanges generates correction of
order $\alpha(Z\alpha)^5$ (see Fig.\ \ref{hadronpoltwophotexc}).
Calculation of this correction is straightforward. One may even take
into account the composite nature of the proton and include the proton
form factors in photon-proton vertices. Such a calculation was
performed in \cite{faustov2000} and produced a very small contribution

\beq
\Delta E(2P-2S)=0.000047~\mbox{meV}.
\eeq

\subsection{Contribution of Order $\alpha^2(Z\alpha)^4m$ induced by
Insertion of the  Hadron Polarization in the Radiative
Photon}

\begin{figure}[ht]
\centerline{\epsfig{file=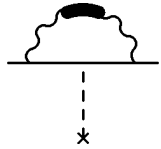}}
\vspace{0.5cm}
\caption{Hadron polarization insertion in the radiative
photon}
\label{hadronpolradph}
\end{figure}

The muon mass is only slightly lower than the pion mass, and
we should expect that insertion of hadronic vacuum polarization in the
radiative photon in Fig.\ \ref{hadronpolradph}  will give a contribution
to the anomalous magnetic moment comparable with the contribution
induced by insertion of the muon vacuum polarization.

Respective corrections are written via the slope of the Dirac form
factor and the anomalous magnetic moment exactly as in Section
\ref{electrloppradph}. The only difference is that the contributions
to the form factors are produced by the hadronic vacuum polarization.

Numerically this contribution to the $2P-2S$ interval was calculated in
\cite{faustov2000}

\beq
\Delta E(2P-2S)=-0.000015~\mbox{meV},
\eeq

\noindent
and is too small to be of any practical significance.

In the case of electronic hydrogen this hadronic insertion in the
radiative photon is additionally suppressed in comparison with the
contribution of the electron vacuum polarization roughly speaking as
$(m_e/m_\pi)^2$.

\subsection{Insertion of One Electron and One
Hadron Loops in the same Coulomb Photon}

\begin{figure}[ht]
\centerline{\epsfig{file=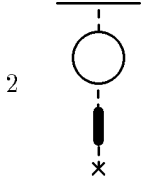}}
\vspace{0.5cm}
\caption{Hadron and electron polarization insertions in the Coulomb
photon}
\label{mixedhadrmuon}
\end{figure}

Due to similarity between the muon and hadron polarizations, such a
correction generated by the diagram in Fig.\ \ref{mixedhadrmuon}
should be of the same order as the respective correction with the muon
loop in \eq{mixedloop2p2s} and thus is too small for any practical
needs. It may be easily calculated.

\section{Standard Radiative, Recoil and Radiative-Recoil Corrections}

All corrections to the energy levels obtained above in the case of
ordinary hydrogen and collected in the Tables II,III,V,VII-IX
are still valid for muonic hydrogen after an obvious substitution of
the muon mass instead of the electron mass in all formulae.
These contributions are included in Table XI.

\section{~~Nuclear Size and Structure Corrections}

Nuclear size and structure corrections for the electronic hydrogen were
considered in Section \ref{nonelectromagnetic} and are collected in
Table X. Below we will consider what happens with these corrections in
muonic hydrogen. The form of the main proton size contribution of order
$(Z\alpha)^4m_r^3<r^2>$ from \eq{chargerad} does not change.

\subsection{~Nuclear Size and Structure Corrections of Order $(Z\alpha)^5
\lowercase{m}$}

\subsubsection{Nuclear Size Corrections of Order $(Z\alpha)^5
\lowercase{m}$}

It is easy to see that neglecting recoil the nuclear size correction of
order $(Z\alpha)^5 \lowercase{m}$ in muonic hydrogen is still given by
\eq{bortrfr}. Calculating this contribution with the same values of
parameters as in \cite{friarmay} for $r_p=0.862~(12)$~fm one obtains

\beq             \label{nonrecmuhydsize}
\Delta E(2P-2S)=0.0247~\mbox{meV}.
\eeq

The muon-proton mass ratio is much larger than the electron-proton mass
ratio, and one could expect a relatively large recoil correction to this
result. The total nuclear size correction of order $(Z\alpha)^5$ with
account for recoil is given by the sum of two-photon diagrams in Figs.\
\ref{ffeltwophotonefig} and \ref{ffeltwophottwofig}. As in the case of
electronic hydrogen, due to large effective integration momenta, it is
sufficient to calculate these diagrams in the scattering approximation.
Respective calculations with realistic form factors and for
$r_p=0.862~(12)$~fm were performed in
\cite{pachucki96muon,faustmart99,pachuckimu99}

\beq
\Delta E(2P-2S)=0.0232~(15)~\mbox{meV}.
\eeq

\noindent
We see that the additional recoil contribution turns out to be not too
important.

Let us notice that using the self-consistent proton radius
$0.891~(18)$~fm from \eq{selfconsrad} we would obtain in  the nonrecoil
limit

\beq
\Delta E(2P-2S)=0.0264~\mbox{meV}
\eeq

\noindent
instead of the result in \eq{nonrecmuhydsize}. Comparing these numbers
we see that the numbers discussed in this Section may be used
for estimates of the proton size contribution of order $(Z\alpha)^5$,
but when the results of the Lamb shift measurements become
available, this correction will require some kind of a self-consistent
consideration. Happily, respective integrals, despite being cumbersome
are relatively simple.

\subsubsection{Nuclear Polarizability Contribution of Order $(Z\alpha)^5
\lowercase{m}$ to $S$-Levels}

Calculation of the nuclear structure corrections of order $(Z\alpha)^5m$
generated by the diagrams in Fig.\ \ref{inelasticfig} follows the same
route as in the case of electronic hydrogen in Section
\ref{nuclstructza5} starting with the forward Compton scattering
amplitude. The only difference is that due to relatively large mass of
the muon the logarithmic approximation is not valid any more, and one
has to calculate the integrals more accurately.  According to
\cite{spk,pachuckimu99}

\beq  \label{polarizabmuza5}
\Delta E=-\frac{0.095~(18)}{n^3}\delta_{l0}~\mbox{meV},
\eeq

\noindent
while the result in \cite{rosenfeld99} is

\beq
\Delta E=-\frac{0.136~(30)}{n^3}\delta_{l0}~\mbox{meV}.
\eeq

We think that the reasons for a minor discrepancy between these results
are the same as for a similar discreapncy in the case of electronic
hydrogen, see discussion in Section \ref{nuclstructza5}. The improved
result in cite \cite{faustmart99} (see footnote \ref{prvkhripl}) is

\beq
\Delta E=-\frac{0.129}{n^3}\delta_{l0}~\mbox{meV}.
\eeq

We will adopt the result in \eq{polarizabmuza5} for further discussion.
Respective contribution to the $2P-2S$ splitting is \cite{pachuckimu99}

\beq
\Delta E(2P-2S)=0.012~(2)~\mbox{meV}.
\eeq

\subsection{~Nuclear Size and Structure Corrections of Order
$(Z\alpha)^6 \lowercase{m}$}

The nuclear polarizability contribution of order $(Z\alpha)^6m$ was
considered above in Section \ref{nuclpolarizbza6}, and we may directly
use the expression for this energy shift in \eq{lpolar} for muonic
hydrogen. In electronic hydrogen the nuclear size correction of order
$(Z\alpha)^6 \lowercase{m}$ is larger than the nuclear size and
structure corrections of order $(Z\alpha)^5\lowercase{m}$. This
enhancement is due to the smallness of the electron mass (see
discussion in Section \ref{nuclsizezalpha6}). The muon mass is much
larger than the electron mass. As a result this hierarchy of the
corrections does not survive in muonic hydrogen, and corrections of
order $(Z\alpha)^6\lowercase{m}$ are smaller than the corrections of
the previous order in $Z\alpha$. Numerically, the nuclear
polarizability contribution of order $(Z\alpha)^6 \lowercase{m}$ to the
$2P-2S$ Lamb shift in muonic hydrogen is about $5\cdot10^{-6}$ meV, and
is negligible.

Nuclear size corrections of order $(Z\alpha)^6 \lowercase{m}$ to
the $S$ levels were calculated in \cite{bortr,friar79} and were
discussed above in Section \ref{nuclsizezalpha6}) for electronic
hydrogen. Respective formulae may be directly used in the case of
muonic hydrogen. Due to the smallness of this correction it is
sufficient to consider only the leading logarithmically enhanced
contribution to the energy shift from \eq{za6leading}
\cite{pachuckimu99}

\beq
\Delta E=-\frac{2(Z\alpha)^6}{3n^3}m_r^3<r^2>\left[<\ln\frac{Z\alpha}{n}>
-\frac{2}{3}m_r^2<r^2>\right].
\eeq

We have restored in this equation a small second term from
\cite{friar79} which, due to the smallness of the electron mass was
omitted in the case of electronic hydrogen in \eq{za6leading}.

Numerically the respective contribution to the $2P-2S$ energy shift is
\cite{pachuckimu99}

\beq
\Delta E(2P-2S)=-0.0009~(3)\mbox{meV}.
\eeq

\noindent
The error of this contribution may easily be reduced if we would use
the total expressions in \eq{za6leading} and \eq{frairaddterms} for its
calculation.

The nuclear size correction of order $(Z\alpha)^6m$ to $P$ levels from
\eq{nuclsizeza6p} gives an additional contribution $4\cdot10^{-5}$
meV to the $2P_\frac{1}{2}-2S_\frac{1}{2}$ energy splitting and may
safely be neglected.

\subsection{Radiative Corrections to the Nuclear Finite Size Effect}

Radiative corrections to the leading nuclear finite size contribution
were considered in Section \ref{radcorrfintitesize}. Respective results
may be directly used for muonic hydrogen, and numerically we obtain

\beq   \label{muonloopradsize}
\Delta E(2P-2S)=0.0006<r^2>=0.0005~\mbox{meV}.
\eeq

\noindent
This contribution is dominated by the diagrams with radiative photon
insertions in the muon line. As usual in muonic hydrogen a much larger
contribution is generated by the electron loop insertions in the
external Coulomb photons. In muonic hydrogen, even after insertion of
the electron loop in the external photon, the effective integration
momenta are still of the atomic scale $k\sim m Z\alpha\sim m_e$, and
the respective contribution to the energy shift is of order
$\alpha(Z\alpha)^4m_r^3<r^2>$, unlike the case with the muonic loop
insertions, when the respective contribution is of higher order
$\alpha(Z\alpha)^5m_r^3<r^2>$ (compare discussion in Section
\ref{radcorrfintitesize}).

\begin{figure}[ht]
\centerline{\epsfig{file=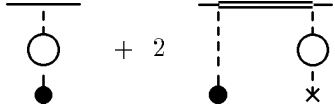}}
\vspace{0.5cm}
\caption{Electron polarization corrections to the leading nuclear size
effect}
\label{elpolfinsize}
\end{figure}

Electron-loop radiative corrections to the leading nuclear finite size
contribution in light muonic atoms were considered in
\cite{friarpol,pachucki96muon}. Two diagrams in Fig.\
\ref{elpolfinsize} give contributions of order
$\alpha(Z\alpha)^4m_r^3<r^2>$. Analytic expression for the first
diagram up to a numerical factor coincides with the expression for the
mixed electron and muon loops in \eq{mixedlopen}, and we obtain

\beq
\Delta E_{nl}=\frac{16\alpha(Z\alpha)^2}{9\pi n^3}m_e^2<r^2>
Q_{nl}^{(3)}(\beta)m_r,
\eeq

\noindent
where $Q_{nl}^{(3)}(\beta)$ is defined in \eq{qnl3beta}.

Numerically, the respective contribution to the $2P-2S$ splitting is
equal to

\beq     \label{radcorsizemu}
\Delta E(2P-2S)=-0.0083~\mbox{meV}.
\eeq

The contribution of the second diagram in Fig.\ \ref{elpolfinsize} may
be written as \cite{pachucki96muon}

\beq
\Delta E=\frac{4\pi Z\alpha<r^2>}{3}\int
d^3\phi(r)V_{VP}G'(r,0)\phi(0).
\eeq

\noindent
The respective contribution to the $2P-2S$ splitting was calculated in
\cite{pachucki96muon,pachuckimu99}

\beq            \label{redradcorsizemu}
\Delta E(2P-2S)=-0.0126~\mbox{meV}.
\eeq

Collecting all terms in \eq{muonloopradsize}, \eq{radcorsizemu} and
\eq{redradcorsizemu} we obtain

\beq
\Delta E(2P-2S)=-0.0275<r^2>\approx-0.0204~(6)~\mbox{meV}.
\eeq

\begin{center}
\underline{Table XI. Lamb Shift in Muonic Hydrogen}
\nopagebreak

\begin{tabular}{|l|r|r|}
\hline
& $\Delta E(nl)$& $\Delta E(2P-2S)$ meV
\\
\hline\hline
One-loop electron polarization&   $$  &\\
   $$ & &\\
Galanin,Pomeranchuk(1952)\cite{galpom}& $-\frac{8\alpha(Z\alpha)^2}{3\pi n^3}
Q^{(1)}_{nl}(\beta)m_r$     &$205.0074$
\\ \hline
Two-loop electron polarization&   $$  &\\
  $$ & &\\
Di Giacomo(1969)\cite{giacomo}&
$\frac{4\alpha^2(Z\alpha)^2}{\pi^2 n^3}Q^{(2)}_{nl}(\beta)m_r$
&$1.5079$
\\ \hline
Three-loop electron polarization&   $$ &\\
contribution, order $\alpha^3(Z\alpha)^2$ & &\\
$$& &\\
Kinoshita,Nio(1999)\cite{kinnio99}
& &$0.0053$\\
\hline
Polarization insertions in two
 &  &\\
Coulomb lines, order $\alpha^2(Z\alpha)^2$  $$ & &\\
&   $$&
\\
Pachucki(1996)\cite{pachucki96muon,pachuckimu99}&
$$& $0.1509$\\
\hline
Polarization insertions in two and three
 &  &\\
Coulomb lines, order $\alpha^3(Z\alpha)^2$& &\\
&   $$&
\\
Kinoshita,Nio(1999)\cite{kinnio99}&
$$& $0.0023$
\\
\hline
Relativistic corrections of order
$\alpha(Z\alpha)^4$ &  &\\
&   $$ &
\\
Pachucki(1996)\cite{pachucki96muon,pachuckimu99}&
$<V_{VP}^{Br}>+2<V_{Br}G'_EV_{VP}^C>$ &$0.0594$
\\
\hline
Wichmann-Kroll,  order $\alpha(Z\alpha)^4$ &  &
\\
&   $$ &\\
Rinker(1976)\cite{rinker76}
&   $$&\\
Borie,Rinker(1978)\cite{borierink78}&$\frac{4\alpha(Z\alpha)^4}{\pi
n^3}m_r Q^{WK}_{nl}(\beta)$ &$-0.0010$\\
\hline
Radiative photon and electron &   $$& \\
polarization in the Coulomb &   $$&
\\
line, order
$\alpha^2(Z\alpha)^4$ &  &\\
&   $$ & \\
Pachucki(1996)\cite{pachucki96muon,pachuckimu99}&
$<\Delta V_C^{VP}>+2<V_C^{VP}G'_E\Delta
V_C>$ &$-0.005~(1)$
\\
\hline
Electron loop in the radiative& &
\\
photon, order $\alpha^2(Z\alpha)^4$&  &
\\
&  &  \\
Barbieri et al(1973)\cite{barb}&&
\\
Suura,Wichmann(1957)\cite{suura}
&   $[1.082~75(-\frac{2}{3}\frac{m}{mr}-1)-4\cdot2.216~56]$&
\\
Peterman(1957)\cite{peterman}&   $
\frac{\alpha^2(Z\alpha)^4}{\pi^2
n^3}(\frac{m_r}{m})^3m$& $-0.0016$ \\
\hline
Mixed electron and  muon loops&   $$& \\
order
$\alpha^2(Z\alpha)^2(\frac{m_e}{m})^2m$&& \\
&   $$ & \\
Borie(1975)\cite{borie}&$\frac{64\alpha^2(Z\alpha)^2}
{45\pi^2 n^3}(\frac{m_e}{m})^2
Q_{nl}^{(3)}(\beta)m_r$&$0.00007$ \\
\hline
\end{tabular}
\end{center}

\newpage
\begin{center}
\underline{Table XI. Lamb Shift in Muonic Hydrogen (continuation)}
\nopagebreak

\begin{tabular}{|l|r|r|}
\hline
Hadronic polarization, order $\alpha(Z\alpha)^4m$ &  &
\\
&   $$&\\
Folomeshkin(1974)\cite{folomeshkin}
& &\\
Friar,Martorell,Sprung(1999)\cite{fms99}&   $$&
\\
Faustov,Martynenko(1999)\cite{faustov2000}&   $-0.638~(22)
\frac{4\alpha(Z\alpha)^4}{15\pi
n^3}(\frac{m_r}{m})^3m\delta_{l0}$&$0.0108~(4)$
\\
\hline
Hadronic polarization, order $\alpha(Z\alpha)^5m$ &  &
\\
&   $$&\\
Faustov,Martynenko(1999)\cite{faustov2000}&   $$&$0.000047$
\\
\hline
Hadronic polarization in the radiative  & &
\\
photon, order $\alpha^2(Z\alpha)^4m$&   $$&
\\
Faustov,Martynenko(1999)\cite{faustov2000}& $$&$-0.000015$ \\
\hline
Recoil contribution of order $(Z\alpha)^4(m/M)^2m$&$$& \\
&   $$ &\\
Barker-Glover(1955)\cite{bg}&$\frac{(Z\alpha)^4m_r^3}{2n^3M^2}
\left(\frac{1}{j+\frac{1}{2}}-\frac{1}{l+\frac{1}{2}}\right)(1-\delta_{l0})$
& $0.0575$\\
\hline
&$$ &\\
Radiative corrections of order
$\alpha^n(Z\alpha)^km$ &Tables II,III,V,VII& $-0.6677$\\
&   $$ &\\
\hline
&   $$ &\\
Recoil corrections of order $(Z\alpha)^n\frac{m}{M}m$&Table VIII&
$-0.0440$\\
&$$ &\\
\hline
&   $$ &\\
Radiative-recoil corrections
&$$ &\\
of order $\alpha(Z\alpha)^n\frac{m}{M}m$&Table IX& $-0.0095$\\
\hline
Leading nuclear size
&&$$\\
contribution
&$\frac{2}{3n^3}(Z\alpha)^4m_r^3<r^2>\delta_{l0}$&$-3.862~(108)$\\
\hline
Nuclear size correction
of order $(Z\alpha)^5$&& $$\\
&$$ &\\
Pachucki(1996)\cite{pachucki96muon,pachuckimu99}&$$& $0.0232~(15)$
\\
Faustov,Martynenko(1999)\cite{faustmart99}&$$& \\
\hline
Nuclear structure correction
of order $(Z\alpha)^5$&& $$\\
&$$ &\\
Startsev,Petrun'kin,&&\\
Khomkin(1976)\cite{spk}&&\\
Rosenfelder(1999)\cite{rosenfeld99}&   $$ &\\
Faustov,Martynenko(1999)\cite{faustmart99}&   $$ &\\
Pachucki(1999)\cite{pachuckimu99}&$-\frac{0.095~(18)}{n^3}\delta_{l0}$
&$0.012~(2)$
\\
\hline
Nuclear size correction
of order $(Z\alpha)^6$&& $$\\
&$$ &\\
Borisoglebsky,Trofimenko(1979)\cite{bortr}&$$ &\\
Friar(1979)\cite{friar79}&
$-\left[<\ln\frac{Z\alpha}{n}>-\frac{2}{3}m_r^2<r^2>\right]$ &\\
Pachucki(1999)\cite{pachuckimu99}&$
\frac{2(Z\alpha)^6}{3n^3}m_r^3<r^2>$& $-0.0009~(3)$
\\
\hline
Radiative corrections to the nuclear &   $$ &\\
finite size effect, order $\alpha(Z\alpha)^4m_r^3<r^2>$&   $$ &\\
&   $$ &\\
Friar(1979)\cite{friarpol}
&   $$ &\\
Pachucki(1996)\cite{pachucki96muon}&   $$ &$-0.0204~(6)$\\
\hline
\end{tabular}
\end{center}


\part{Physical Origin of the Hyperfine Splitting and the Main
Nonrelativistic Contribution}\label{hfsphysor}

The theory of the atomic energy levels developed in the previous sections is
incomplete, since we systematically ignored the nuclear spin which leads to
an additional splitting of the energy levels. This effect will be the
subject of our discussion below.

\begin{figure}[ht]
\centerline{\epsfig{file=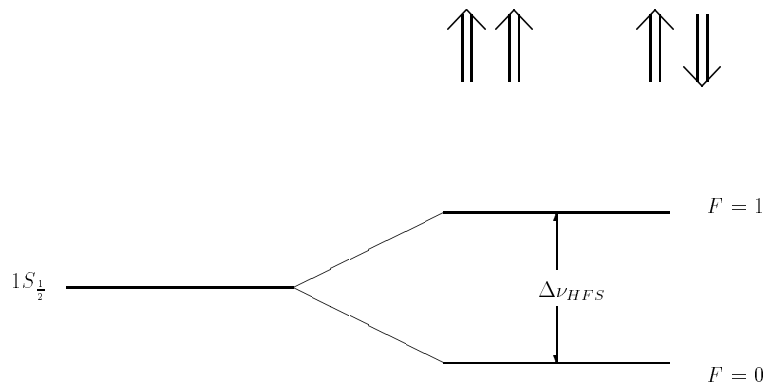,height=5cm}}
\vspace{0.5cm}
\caption{Scheme of hyperfine energy levels in the ground state}
\label{hyperf}
\end{figure}

Unlike the Lamb shift, the hyperfine splitting (HFS) (see Fig.\
\ref{hyperf}) can be readily understood in the framework of
nonrelativistic quantum mechanics. It originates from the interaction
of the magnetic moments of the electron and the nucleus.  The classical
interaction energy between two magnetic dipoles is given by the
expression (see, e.g., \cite{bd,schwinger})

\beq
H=-\frac{2}{3}{\bbox{ \mu_1\mu_2}}\delta({\bf r}).
\eeq

\begin{figure}[ht]
\centerline{\epsfig{file=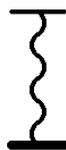,height=2cm}}
\vspace{0.5cm}
\caption{Leading order contribution to hyperfine splitting}
\label{fermifig}
\end{figure}

This effective Hamiltonian for the interaction  of two magnetic
moments may also easily be derived from the one photon exchange diagram
in Fig.\ \ref{fermifig}. In the leading nonrelativistic approximation
the denominator of the photon propagator cancels the exchanged momentum
squared in the numerator, and we immediately obtain the Hamiltonian for
the interaction of two magnetic moments, reproducing the above result
of classical electrodynamics.

The simple calculation of the matrix element of this Hamiltonian
between the nonrelativistic Schr\"odinger-Coulomb wave functions gives
the Fermi result \cite{ef} for the splitting between the
$1^3S_\frac{1}{2}$ and $1^1S_\frac{1}{2}$ states\footnote{In comparison
with the Fermi result, we have restored here the proper dependence of
the hyperfine splitting on the reduced mass.}

\beq           \label{ef}
E_F=\frac{8}{3}(Z\alpha)^4(1+a_\mu)\frac{m}{M}
\left(\frac{m_r}{m}\right)^{3}mc^2
\eeq
\[
=\frac{16}{3}Z^4\alpha^2 (1+a_\mu)\frac{m}{M}
\left(\frac{m_r}{m}\right)^{3}ch\:R_{\infty},
\]

\noindent
where $m$ and $M$ are the electron and muon masses
respectively\footnote{Here we call the heavy particle the muon,
having in mind that the precise theory  of hyperfine splitting finds
its main application in comparison with the highly precise experimental
data on muonium hyperfine splitting. However, the theory of nonrecoil
corrections is valid for any hydrogenlike atom.}, $Z$ is the charge of
the muon in terms of the proton charge\footnote{Of course, $Z=1$ for
muon, but the Fermi formula is valid for any heavy nucleus with arbitrary
$Z$. As in the case of the Lamb shift, it is useful to preserve $Z$ as a
parameter in all formulae for the different contributions to HFS, since it
helps to clarify the origin of different corrections.}, $c$ is the velocity
of light, $a_\mu$ is the muon anomalous magnetic moment, $R_\infty$ is the
Rydberg constant and $h$ is the Planck constant.

The sign of this contribution may easily be understood from purely
classical considerations, if one thinks about the magnetic dipoles in
the context of the Ampere hypothesis about small loops of current.
According to classical electrodynamics parallel currents attract each other
and antiparallel ones repel. Hence, it is clear that the state with
antiparallel magnetic moments (parallel spins) should have a higher
energy than the state with antiparallel spins and parallel magnetic
moments.

As in the case of the Lamb shift, QED provides the framework for
systematic calculation of numerous corrections to the Fermi formula for
hyperfine splitting. We again have the three small parameters, namely,
the fine structure constant $\alpha$, $Z\alpha$ and the small
electron-muon mass ratio $m/M$. Expansion in these parameters
generates relativistic (binding), radiative, recoil, and
radiative-recoil corrections. At a certain level of accuracy the weak
interactions and, for the case of hadronic atoms, the nuclear size
and structure effects also become important. Below we will first
discuss corrections to hyperfine splitting in the case of a
structureless nucleus, having in mind the special case of muonium where
the most precise comparison between theory and experiment is possible.
In a separate Chapter we will also consider the nuclear size and
structure effects which should be taken into account in the case of
hyperfine splitting in hydrogen. We postpone more detailed discussion
of the phenomenological situation to Chapter \ref{hyperineexp}. Even in
the case of muonium, strong interaction contributions generated by the
hadron polarization insertions in the exchanged photons and the weak
interaction contribution induced by the $Z$-boson exchange should be
taken into account at the current level of accuracy. The experimental
value of the hyperfine splitting in muonium is measured with
uncertainty $\pm 53$ Hz \cite{lbdd} (relative accuracy
$1.2\cdot10^{-8}$), and the next task of the theory is to obtain all
corrections which could be as large as $10$ Hz. This task is made even
more challenging by the fact that only a few years ago reduction of the
theoretical error below $1$ kHz was considered as a great success (see,
e.g., discussion in \cite{eid}).

\part{External Field Approximation}

\section{~~Relativistic (Binding) Corrections to HFS}

Relativistic and radiative corrections depend on the electron and muon
masses only via the explicit mass factors in the electron and muon magnetic
moments, and via the reduced mass factor in the Schr\"odinger wave function.
All such corrections may be calculated in the framework of the external
field approximation.

In the external field approximation the heavy particle magnetic moment
factorizes and the relativistic and radiative corrections have the form

\beq                \label{bef}
\Delta E_{HFS}=E_{F}(1+corrections).
\eeq

This factorization of the total muon momentum occurs because the
virtual momenta involved in calculation of the relativistic and radiative
corrections are small in comparison with the muon mass, which sets the
natural momentum scale for corrections to the muon magnetic moment.

Purely relativistic corrections are by far the simplest corrections to
hyperfine splitting. As in the case of the Lamb shift, they essentially
correspond to the nonrelativistic expansion of the relativistic
square root expression for the energy of the light particle in
\eq{einsqroot}, and have the form of a series over $(Z\alpha)^2\approx{\bf
p}^2/m^2$. Calculation of these corrections should be carried out in the
framework of the spinor Dirac equation, since clearly there would not be any
hyperfine splitting for a scalar particle.

The binding corrections to hyperfine splitting as well as the main Fermi
contribution are contained in the matrix element of the interaction
Hamiltonian of the electron with the external vector potential created by
the muon magnetic moment ({\boldmath$A=\nabla\times\mu$}$/(4\pi r)$). This
matrix element should be calculated between the Dirac-Coulomb wave functions
with the proper reduced mass dependence (these wave functions are
discussed at the end of Section \ref{effdir}). Thus we see that the
proper approach to calculation of these corrections is to start with
the EDE (see discussion in Section \ref{effdir}), solve it with the
convenient zero-order potential and obtain the respective Dirac-Coulomb
wave functions.  Then all binding corrections are given by the matrix
element

\beq
\Delta E_{Br}=<n|\gamma_0\mbox{\boldmath$\gamma A$}|n>.
\eeq

As discovered by Breit \cite{br} an exact calculation of this matrix
element is really no more difficult than calculation of the leading
binding correction of relative order $(Z\alpha)^2$. After
straightforward calculation one obtains a closed expression for the
hyperfine splitting of an energy level with an arbitrary principal
quantum number $n$ \cite{br,eryen2}\footnote{The closed expression for
an arbitrary $n$ is calculated in formula (6.14a) in \cite{eryen2},
p.471. While this closed expression is correct, its expansion over
$Z\alpha$ printed in \cite{eryen2} after an equality sign, contains two
misprints. Namely the sign before $(Z\alpha)^2$ in the square brackets
should be changed to the opposite, and the numerical factor inside these
brackets should be $-2$ instead of $-1$. After these corrections the
expansion in formula (6.14a) in \cite{eryen2} does not contradict the exact
expression in the same formula, and also coincides with the result in
\cite{br}.} (see also \cite{schwinger})

\beq \label{brhfs}
\Delta E_{Br}(nS)=\frac{1+2\sqrt{1-\frac{(Z\alpha)^2}{N^2}}}
{N^3\gamma(4\gamma^2-1)}E_F,
\eeq

\noindent
where $N=\sqrt{n^2-2(Z\alpha)^2(n-1)/(1+\gamma)}$,
$\gamma=\sqrt{1-(Z\alpha)^2}$.

Let us emphasize once more that the expression in \eq{brhfs} contains all
binding corrections. Expansion of this expression in $Z\alpha$ gives
explicitly

\beq
\Delta
E_{Br}(nS)=\left[1+\frac{11n^2+9n-11}{6n^2}(Z\alpha)^2
\right.
\eeq
\[
\left.
+\frac{203n^4+225n^3-134n^2-330n+189}{72n^4}(Z\alpha)^4
+\ldots\right]\frac{E_F}{n^3},
\]

\noindent
or

\beq
\Delta E_{Br}(1S)=\frac{E_F}{\sqrt{1-(Z\alpha)^2}(2\sqrt{1-(Z\alpha)^2}-1)}
\eeq
\[
=\left[1+\frac{3}{2}(Z\alpha)^2+\frac{17}{8}(Z\alpha)^4+\ldots\right]E_F,
\]

\noindent
and

\beq
\Delta
E_{Br}(2S)=\left[1+\frac{17}{8}(Z\alpha)^2+\frac{449}{128}(Z\alpha)^4+\ldots
\right]\frac{E_F}{8}.
\eeq

Only the first two terms in the series give contributions larger than
$1$ Hz to the ground state splitting in muonium. As usual, in the
Coulomb problem, expansion in the series for the binding corrections
goes over the parameter $(Z\alpha)^2$ without any factors of $\pi$ in
the denominator. This is characteristic for the Coulomb problem and
emphasizes the nonradiative nature of the relativistic corrections.

The sum of the main Fermi contribution and the Breit correction is given in
the last line of Table XII. The uncertainty of the main Fermi
contribution determines the uncertainty  of the theoretical prediction
of HFS in the ground state in muonium, and is in its turn determined by
the experimental uncertainty of the electron-muon mass ratio.

\begin{center}
\underline{Table XII. Relativistic (Binding) Corrections }
\nopagebreak

\begin{tabular}{|l|rl|c|}
\hline
& $E_F$ &   	& kHz
\\ \hline \hline
&   $$       & &\\
Fermi (1930)\cite{ef}&   $1$    & &$4~459~031.922~(519)$\\
&   $$       & &
\\ \hline
&   $$       & &\\
Breit (1930)\cite{br}&$
\frac{1}{\sqrt{1-(Z\alpha)^2}(2\sqrt{1-(Z\alpha)^2}-1)}-1$& &
$356.201$\\ &   $$       & &
\\ \hline \hline
Total Fermi and Breit
&$$& &\\
contributions&   $\frac{1}{\sqrt{1-(Z\alpha)^2}(2\sqrt{1-(Z\alpha)^2}-1)}$       & &
$4~459~388.123~(519)$
\\ \hline
\end{tabular}
\end{center}

\section{~~~Electron Anomalous Magnetic Moment Contributions
(Corrections of Order $\alpha^{\lowercase{n}}E_F$)}

\begin{figure}[h]
\centerline{\epsfig{file=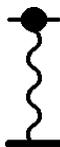,height=2cm}}
\vspace{0.5cm}
\caption{Electron anomalous magnetic moment contribution to HFS. Bold dot
corresponds to the Pauli form factor}
\label{anomomfig}
\end{figure}

Leading radiative corrections to HFS are generated either by the electron or
muon anomalous magnetic moments. The muon anomalous
magnetic moment contribution is already taken into account in  the
expression for the Fermi energy in \eq{ef} and we will not discuss it here.
All corrections of order $\alpha^nE_F$ are generated by the electron
anomalous magnetic moment insertion in the electron photon vertex in Fig.\
\ref{anomomfig}. These are the simplest of the purely radiative corrections
since they are independent of the binding parameter $Z\alpha$.  The value of
the electron anomalous magnetic moment entering in the expression for HFS
coincides with the one for the free electron.  In this situation the
contribution to HFS is given by the matrix element of the electron Pauli
form factor between the wave functions which are the products of the
Schr\"odinger-Coulomb wave functions and the free electron spinors.
Relativistic Breit corrections may also be trivially included in this
calculation by calculating the matrix element between the Dirac-Coulomb
wave functions. However, we will omit here the Breit correction of
order $\alpha(Z\alpha)^2E_F$ to the anomalous moment contribution to
HFS, since we will take it into account below, together with other
corrections of order $\alpha(Z\alpha)^2E_F$. Then the anomalous moment
contribution to HFS has the form

\beq
\Delta E_F=a_eE_F,
\eeq

\noindent
where \cite{sch,karpkroll,pet,somm,knan,laprem} (see
analytic expressions above in \eq{pauli}, \eq{pauli2}, and \eq{pauli3})

\beq
a_e=F_2(0)=\frac{\alpha}{2\pi}-0.328~478~965\ldots~
\left(\frac{\alpha}{\pi}\right)^2
+1.181~241~456\ldots~\left(\frac{\alpha}{\pi}\right)^3.
\eeq

We have omitted here higher order electron-loop contributions as well as the
heavy particle loop contributions to the electron anomalous magnetic moment
(see, e.g., \cite{kin96}) because respective corrections to HFS are smaller
than $0.001$ kHz. Let us note that the electron anomalous magnetic moment
contributions to HFS do not introduce any additional uncertainty in the
theoretical expression for HFS (see also Table XIII).

In analogy with the case of the Lamb shift discussed in Section
\ref{originlamb} one could expect that the polarization insertion in the
one-photon exchange would also generate corrections of order $\alpha E_F$.
However, due to the short-distance nature of the main contribution to HFS,
the leading small momentum (large distance) term in the polarization
operator expansion does not produce any contribution to HFS. Only the
higher momentum (smaller distance) part of the polarization operator
generates a contribution to HFS and such a contribution inevitably
contains, besides the factor $\alpha$, an extra binding factor of
$Z\alpha$. This contribution will be discussed in the next section.

\begin{center}
\underline{Table XIII. Electron AMM Contributions}
\nopagebreak

\begin{tabular}{|l|rl|c|}
\hline
& $E_F$ &   	& kHz
\\ \hline  \hline
&   $$       & &\\
Schwinger (1948)\cite{sch}&$\frac{\alpha}{2\pi}$ & & $5~178.763$\\
&   $$       & &
\\ \hline
&   $$       & &\\
Sommerfield (1957)\cite{somm}&$\left[\frac{3}{4}\zeta(3)-\frac{\pi^2}{2}\ln2
+\frac{\pi^2}{12}+\frac{197}{144}\right](\frac{\alpha}{\pi})^2$ &&\\
&   $$       & &\\
Peterman (1957)\cite{pet}
&$\approx-0.328~478~965\ldots(\frac{\alpha}{\pi})^2$& &$-7.903$ \\ &   $$ &
& \\ \hline &   $$       & &\\
Kinoshita(1990)\cite{knan}&$\{\frac{83}{72}\pi^2\zeta(3)
-\frac{215}{24}\zeta(5)+\frac{100}{3}[(a_4+\frac{1}{24}\ln^4 2)$ &&\\
&   $$       & &\\
Laporta,Remiddi(1996)\cite{laprem}&
$-\frac{1}{24}\pi^2\ln^22]-\frac{239}{2~160}\pi^4
+\frac{139}{18}\zeta(3)$
&&\\
&   $$       & &\\
&$-\frac{298}{9}\pi^2\ln2+\frac{17~101}{810}\pi^2
+\frac{28~259}{5~184}\}(\frac{\alpha}{\pi})^3
$&&\\
&   $$       & &\\
&$
\approx 1.181~241~456\ldots~(\frac{\alpha}{\pi})^3$&&$0.066$\\
&   $$       & &
\\ \hline
\hline
Total electron AMM				 &$$& &\\
contribution&   $$       & &$5~170.926$ \\
\hline
\end{tabular}
\end{center}

\section{~Radiative Corrections of Order
$\alpha^{\lowercase{n}}(Z\alpha)E_F$}

\subsection{Corrections of Order $\alpha(Z\alpha)E_F$}

Nontrivial interplay between radiative corrections and binding effects
first arises in calculation of the combined expansion over $\alpha$ and
$Z\alpha$. The simplest contribution of this type is of order
$\alpha(Z\alpha) E_F$  and was calculated a long time
ago in classical papers \cite{kp,kks1,kk}.

\begin{figure}[ht]
\centerline{\epsfig{file=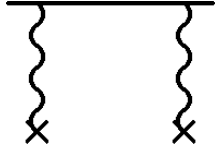,height=2cm}}
\vspace{0.5cm}
\caption{Skeleton two-photon diagram for HFS in the external field
approximation}
\label{skelhfs}
\end{figure}

The crucial observation, which greatly facilitates the calculations, is
that the scattering approximation (skeleton integral approach) is
adequate for calculation of these corrections (see, e.g., a detailed
proof in \cite{eksann1}). As in the case of the radiative corrections
to the Lamb shift discussed in Section \ref{skeletonbas}, radiative
corrections to HFS of order $\alpha(Z\alpha)E_F$ are given by the
matrix elements of the diagrams with all external electron lines on the
mass shell calculated between free electron spinors. The external
spinors should be projected on the respective spin states and
multiplied by the square of the Schr\"{o}dinger-Coulomb
wave function at the origin. One may easily understand the physical
reasons beyond this recipe. Radiative insertions in the skeleton
two-photon diagrams in Fig.\ \ref{skelhfs} suppress low integration
momenta (of atomic order $mZ\alpha$) in the exchange loops and the
effective loop integration momenta are of order $m$.  Account of off
mass shell external lines would produce an additional factor $Z\alpha$
and thus generate a higher order correction. Let us note that
suppression of low intermediate momenta in the loops takes place only
for gauge invariant sets of radiative insertions, and does not happen
for all individual diagrams in an arbitrary gauge. Only in the Yennie
gauge is the infrared behavior of individual diagrams not worse than
the infrared behavior of their gauge invariant sums.  Hence, use of the
Yennie gauge greatly facilitates the proof of the validity of the
skeleton diagram approach \cite{eksann1}. In other gauges the
individual diagrams with on mass shell external lines often contain
apparent infrared divergences, and an intermediate infrared
regularization (e.g., with the help of the infrared photon mass of the
radiative photons) is necessary. Due to the above mentioned theorem about
the infrared behavior of the complete gauge invariant set of diagrams, the
auxiliary infrared regularization may safely be lifted after calculation of
the sum of all contributions. Cancellation of the infrared divergent terms
may be used as an additional test of the correctness of calculations.

The contribution to HFS induced by the skeleton diagram with two external
photons in Fig.\ \ref{skelhfs} is given by the infrared divergent integral

\beq                                \label{skel}
\frac{8Z\alpha}{\pi n^3}E_F\int_0^\infty \frac{d{ k}}{{k}^2}.
\eeq

Insertion in the integrand of the factor  $F({ k})$ which describes
radiative corrections, turns the infrared divergent skeleton integral into a
convergent one. Hence, the problem of calculating  contributions of order
$\alpha(Z\alpha)E_F$ to HFS turns into the problem of calculating the
electron factor describing radiative insertions in the electron line.
Calculation of the radiative corrections induced by the polarization
insertions in the external photon is straightforward since the explicit
expression for the polarization operator is well known.

\subsubsection{Correction Induced by the Radiative Insertions in the
Electron Line}

\begin{figure}[h]
\centerline{\epsfig{file=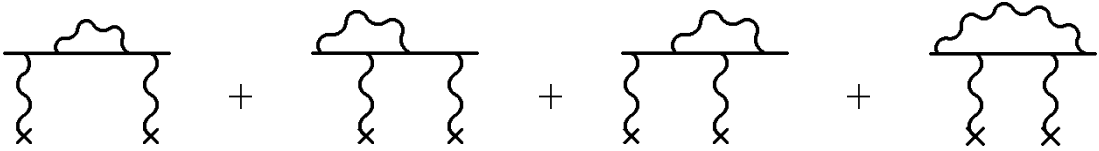,height=2cm}}
\vspace{0.5cm}
\caption{Diagrams with radiative insertions in the electron line}
\label{ellineradhfsfig}
\end{figure}

\begin{figure}[h]
\centerline{\epsfig{file=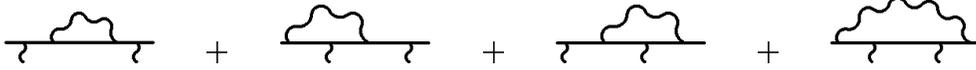,height=1cm}}
\vspace{0.5cm}
\caption{One-loop electron factor}
\label{electfacthfsfig}
\end{figure}

For calculation of the contribution to HFS of order $\alpha(Z\alpha)E_F$
induced by the one-loop radiative insertions in the electron line in Fig.\
\ref{ellineradhfsfig} we have to substitute in the integrand in
\eq{skel} the gauge invariant electron factor $F(k)$. This electron
factor is equal to the one loop correction to the amplitude of the
forward Compton scattering in Fig.\ \ref{electfacthfsfig}. Due to
absence of bremsstrahlung in the forward scattering the electron factor
is infrared finite.

Convergence of the integral for the contribution to HFS is determined  by
the asymptotic behavior of the electron factor at small and large momenta.
The ultraviolet (with respect of the large momenta of the external
photons) asymptotics of the electron factor is proportional to the
ultraviolet asymptotics of the skeleton graph for the Compton amplitude.
This may easily be understood in the Landau gauge when all individual
radiative insertions in the electron line do not contain logarithmic
enhancements \cite{blp}. One may also prove the absence of logarithmic
enhancement with the help of the Ward identities.  This means that insertion
of the electron factor in \eq{skel} does not spoil the ultraviolet
convergence of the integral.  More interesting is the low momentum behavior
of the electron factor. Due to the generalized low energy theorem for the
Compton scattering (see, e.g., \cite{eksann1}), the electron factor has a
pole at small momenta and the residue at this pole is completely determined
by the one loop anomalous magnetic moment. Hence, naive substitution of the
electron factor in \eq{skel} (see \eq{electrfhfs} below) would produce a
linearly infrared divergent contribution to HFS. This would be infrared
divergence of the anomalous magnetic moment contribution should be expected.
As discussed in the previous section, the contribution connected with the
anomalous magnetic moment does not contain an extra factor $Z\alpha$ present
in our skeleton diagram. The contribution is generated by the region
of small (atomic scale) intermediate momenta, and the linear divergence
would be cutoff by the wave function at the scale $k\sim mZ\alpha$ and will
produce the correction of the previous order in $Z\alpha$ induced by the
electron anomalous magnetic moment and already considered above.  Hence, to
obtain corrections of order $\alpha(Z\alpha)E_F$ we simply have to subtract
from the electron factor its part generated by the anomalous magnetic
moment. This subtraction reduces to subtraction of the leading pole term in
the infrared asymptotics of the electron factor. A closed analytic
expression for this subtracted electron factor as a function of momentum $k$
was obtained in \cite{eks1}.  This electron factor was normalized according
to the relationship

\beq          \label{electrfhfs}
\frac{1}{k^2}\rightarrow \frac{\alpha}{2\pi}F(k).
\eeq

The subtracted electron factor generates a finite radiative correction after
substitution in the integral in \eq{skel}.  The contribution to HFS
is equal to

\beq
\Delta E_{elf}=\frac{4Z\alpha}{\pi^2 n^3}E_F\int_0^\infty d{ k}F({ k})
=(\ln2-\frac{13}{4})\alpha(Z\alpha)E_F,
\eeq

\noindent
and was first obtained in a different way in \cite{kp,kks1,kk}.

This expression should be compared with the correction of order
$\alpha(Z\alpha)^5$ to the Lamb shift in \eq{aza5}. Both expressions have
the same physical origin, they correspond to the radiative insertions
in the diagrams with two external photons, may be calculated in the
skeleton diagram approach, and do not contain a factor $\pi$ in the
denominators. The reasons for its absence were discussed in the end of
Section \ref{skeletonbas}.

\subsubsection{Correction Induced by the
Polarization Insertions in the External Photons}\label{hfsradrecpol}

\begin{figure}[h]
\centerline{\epsfig{file=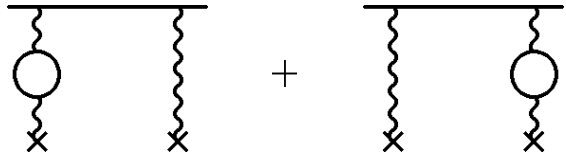,height=2cm}}
\vspace{0.5cm}
\caption{Diagrams with electron-loop polarization insertions in the external
photon lines}
\label{photlineradhfsfig}
\end{figure}

Explicit expression for the electron loop polarization contribution to HFS
in Fig.\ \ref{photlineradhfsfig} is obtained from the skeleton integral in
\eq{skel} by the standard substitution in \eq{sub}.  One also has to take
into account an additional factor 2 which corresponds to two possible
insertions of the polarization operator in either of the external photon
lines. The final integral may easily be calculated and the polarization
operator insertion leads to the correction \cite{kp,kks1,kk}

\beq           \label{polarnonrechfs}
\Delta E_{pol}=\frac{16\alpha(Z\alpha)}{\pi^2 n^3}E_F
\int_0^\infty dk I_1(k^2)
=\frac{3}{4}\alpha(Z\alpha)E_F.
\eeq

\begin{figure}[ht]
\centerline{\epsfig{file=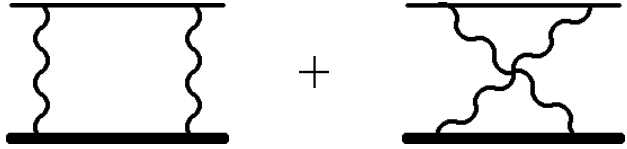,height=2cm}}
\vspace{0.5cm}
\caption{Skeleton two-photon diagrams for HFS }
\label{skeltothfsrad}
\end{figure}

There is one subtlety in this result, which should be addressed here. The
skeleton integral in \eq{skel} may be understood as the heavy muon pole
contribution in the diagrams with two exchanged photons in Fig.\
\ref{skeltothfsrad}. In such a case an exact calculation will produce an
extra factor $1/(1+m/M)$ before the skeleton integral in \eq{skel}. We
have considered above only the nonrecoil contributions, and so we have
ignored an extra factor of order $m/M$, keeping in mind that it would
be considered together with other recoil corrections of order
$\alpha(Z\alpha)E_F$. This strategy is well suited for consideration of
the recoil and nonrecoil corrections generated by the electron factor,
but it is less convenient in the case of the polarization insertion.

\begin{figure}[h]
\centerline{\epsfig{file=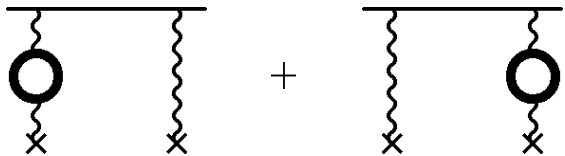,height=2cm}}
\vspace{0.5cm}
\caption{Diagrams with muon-loop polarization insertions in the external
photon lines}
\label{photlineradmuonhfsfig}
\end{figure}

In the case of the polarization insertions the calculations may  be
simplified by simultaneous consideration of the insertions of both the
electron and muon polarization loops \cite{ty,sty}. In such an approach
one explicitly takes into account internal symmetry of the problem at
hand with respect to both particles. So, let us preserve the factor
$1/(1+m/M)$ in \eq{skel}, even in calculation of the nonrecoil
polarization operator contribution.  Then we will obtain an extra
factor $m_r/m$ on the right hand side in \eq{polarnonrechfs}. To
facilitate further recoil calculations we could simply declare that the
polarization operator contribution with this extra factor $m_r/m$ is
the result of the nonrecoil calculation but there exists a better
choice. Insertion in the external photon lines of the polarization loop
of a heavy particle with mass $M$ generates correction to HFS
suppressed by an extra recoil factor $m/M$ in comparison with the
electron loop contribution. Corrections induced by such heavy particles
polarization loop insertions clearly should be discussed together with
other radiative-recoil corrections.  However, as was first observed in
\cite{ty,sty}, the muon loop plays a special role. Its contribution to HFS
differs from the result in \eq{polarnonrechfs} by an extra recoil factor
$m_r/M$, and, hence, the sum of the electron loop contribution in Fig.\
\ref{photlineradhfsfig} (with the extra factor $m_r/m$ taken into account)
and of the muon loop contribution in Fig.\ \ref{photlineradmuonhfsfig} is
exactly equal to the result in \eq{polarnonrechfs}, which we will call below
the nonrecoil polarization operator contribution.  We have considered here
this cancellation of part of the  radiative-recoil correction in order to
facilitate consideration of the total radiative-recoil correction generated
by the polarization operator insertions below. Let us emphasize that there
was no need to restore the factor $1/(1+m/M)$ in the consideration of the
electron line radiative corrections, since in the analytic calculation of
the respective radiative-recoil corrections to be discussed below we do not
use any subtractions and recalculate the nonrecoil part of these corrections
explicitly.

Recoil corrections induced by the polarization loops containing other heavy
particles will be considered below in Section \ref{hfsradrec} together with
other radiative-recoil corrections.

\subsection{Corrections of Order $\alpha^2(Z\alpha)E_F$}

\begin{figure}[ht]
\centerline{\epsfig{file=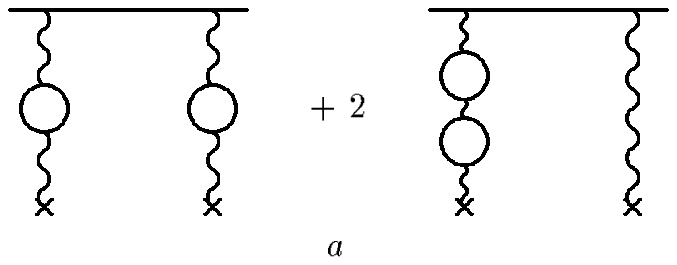,height=2.5cm,width=6.5cm}
\hskip1cm\epsfig{file=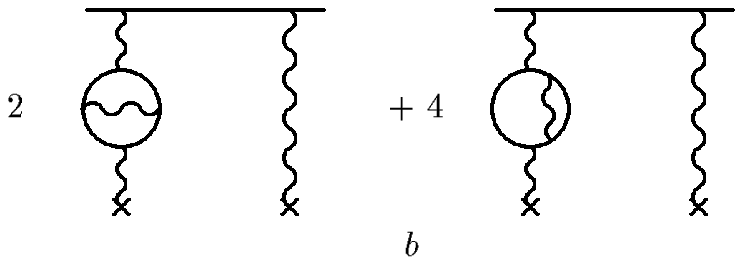,height=2.5cm,width=7cm}}
\vspace{0.8cm}
\centerline{\epsfig{file=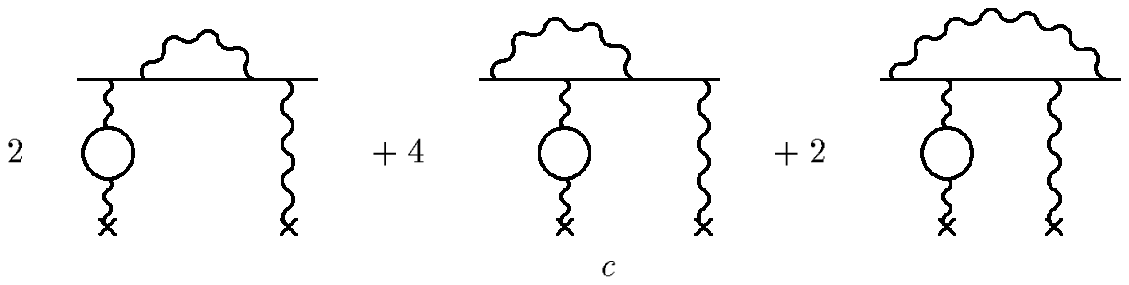,height=3cm}}
\vspace{0.8cm}
\centerline{\epsfig{file=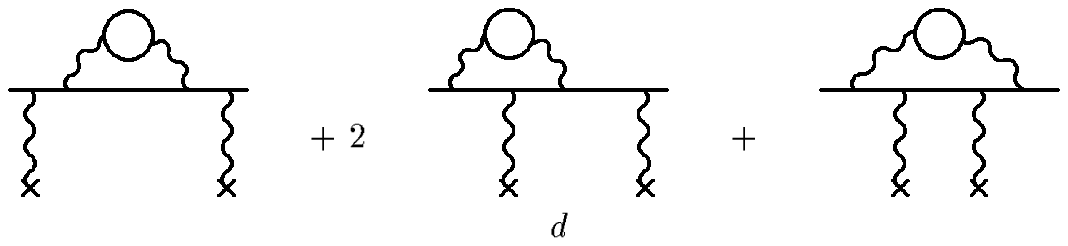,height=2.8cm,width=14cm}}
\vspace{0.8cm}
\centerline{\epsfig{file=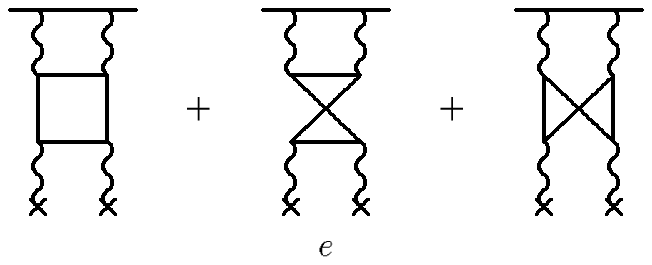,height=3cm}\hskip3cm
\epsfig{file=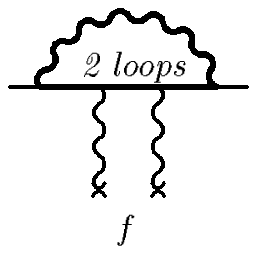,height=3cm}}
\vspace{0.5cm}
\caption{Six gauge invariant sets of diagrams for corrections of order
$\alpha^2(Z\alpha)E_F$}
\label{6setshfsfig}
\end{figure}

Calculation of the corrections of order $\alpha^2(Z\alpha)E_F$ goes in
principle along the same lines as the calculation of the corrections of the
previous order in $\alpha$ in the preceding section. Once again
the scattering approximation is adequate for calculation of these
corrections. There exist six gauge invariant sets of graphs in Fig.\
\ref{6setshfsfig} which produce corrections of order $\alpha^2(Z\alpha)/\pi
E_F$ to HFS \cite{eks1}.  Respective contributions once again may be
calculated with the help of the skeleton integral in \eq{skel}
\cite{eks1,eks2,eks3}.

Some of the diagrams in Fig.\ \ref{6setshfsfig} also generate corrections of
the previous order in $Z\alpha$, which would naively induce infrared
divergent contributions after substitution in the skeleton integral in
\eq{skel}.

The physical nature of these contributions is quite transparent. They
correspond to the anomalous magnetic moment which is hidden in the
two-loop electron factor. The true order in $Z\alpha$ of these anomalous
magnetic moment contributions is lower than their apparent order and they
should be subtracted from the electron factor prior to calculation of the
contributions to HFS. We have already encountered a similar situation
above in the case of the correction of order $\alpha(Z\alpha)E_F$
induced by the electron factor, and the remedy is the same. Let us
mention that the analogous problem was also discussed in connection
with the Lamb shift calculations in Section \ref{skeletonbas}.

Technically the lower order contributions to HFS are produced by the
constant terms in the low-frequency asymptotic expansion of the electron
factor. These lower order contributions are connected with integration
over external photon momenta of the characteristic atomic scale $mZ\alpha$
and the approximation based on the skeleton integrals in \eq{skel} is
inadequate for their calculation. In the skeleton integral approach
these previous order contributions arise as the infrared divergences
induced by the low-frequency terms in the electron factors. We subtract
leading low-frequency terms in the low-frequency asymptotic expansions
of the electron factors, when necessary, and thus get rid of the
previous order contributions.

Let us discuss in more detail calculation of different contributions of
order $\alpha^2(Z\alpha)E_F$. The reader could notice that the discussion
below is quite similar to the discussion of calculation of the corrections
of order $\alpha^2(Z\alpha)^5m$ to the Lamb shift in Section \ref{a2za5chap}.

\subsubsection{One-Loop Polarization Insertions in the External Photons}

The simplest correction is induced by the diagrams in Fig.\
\ref{6setshfsfig} $(a)$ with two insertions of the one-loop vacuum
polarization in the external photon lines. The respective contribution
to HFS is obtained from the skeleton integral in \eq{skel} by the
substitution of the polarization operator squared

\beq
\frac{1}{k^2}\rightarrow \left(\frac{\alpha}{\pi}\right)^2k^2I_1(k).
\eeq

\noindent
Taking into account the multiplicity factor 3 one easily obtains \cite{eks1}

\beq     \label{onelooppolhfs}
\Delta E=\frac{24Z\alpha}{\pi n^3}\left(\frac{\alpha}{\pi}\right)^2E_F
\int_0^\infty dk k^2I^2_1(k)
=\frac{36}{35}\frac{\alpha^2(Z\alpha)}{\pi n^3}E_F.
\eeq

\subsubsection{Insertions of the Irreducible Two-Loop Polarization in the
External Photons}

Expression for the two-loop vacuum polarization contribution to HFS in Fig.\
\ref{6setshfsfig} $(b)$ is obtained from the skeleton integral in \eq{skel}
by the substitution

\beq
\frac{1}{k^2}\rightarrow \left(\frac{\alpha}{\pi}\right)^2I_2(k).
\eeq

\noindent
With account of the multiplicity factor 2 one  obtains \cite{eks1}

\beq     \label{twolooppolhfs}
\Delta E=\frac{16Z\alpha}{\pi n^3}\left(\frac{\alpha}{\pi}\right)^2E_F
\int_0^\infty dk I_2(k)
=\left(\frac{224}{15}\ln2-\frac{38}{15}\pi-\frac{118}{225}\right)
\frac{\alpha^2(Z\alpha)}{\pi n^3}E_F
\eeq
\[
\approx 1.87\ldots
\frac{\alpha^2(Z\alpha)}{\pi n^3}E_F.
\]

\subsubsection{Insertion of One-Loop Electron Factor in the Electron Line
and of the One-Loop Polarization in the External Photons}

The next correction of order $\alpha^2(Z\alpha)E_F$ is generated by the
gauge invariant set of diagrams in Fig.\ \ref{6setshfsfig} $(c)$. The
respective analytic expression is obtained from the skeleton integral
by simultaneous insertion in the integrand of the one-loop polarization
function $I_1(k)$ and of the electron factor $F(k)$.

Then taking  into account the multiplicity factor 2 corresponding to two
possible insertions of the one-loop polarization, one obtains

\beq     \label{onelooppolelfhfs}
\Delta E=\frac{8Z\alpha}{\pi n^3}\left(\frac{\alpha}{\pi}\right)^2E_F
\int_0^\infty dk k^2F(k)I_1(k)
\eeq
\[
=\left(-\frac{4}{3}\ln^2\frac{1+\sqrt{5}}{2}-\frac{20}{9}\sqrt{5}
\ln\frac{1+\sqrt{5}}{2}
-\frac{64}{45}\ln2+\frac{\pi^2}{9}+\frac{1~043}{675}\right)
\frac{\alpha^2(Z\alpha)}{\pi n^3}E_F
\]

\noindent
We used in \eq{onelooppolelfhfs} subtracted electron factor. However, it is
easy to see that the one-loop anomalous magnetic moment  term in the
electron factor generates a correction of  order $\alpha^2(Z\alpha)E_F$ in
the diagrams in Fig, and also should be taken into account. An easy direct
calculation of the anomalous magnetic moment contribution leads to the
correction

\beq
\Delta E=\frac{3}{8}\frac{\alpha^2(Z\alpha)}{\pi n^3}E_F,
\eeq

\noindent
which may be also obtained multiplying the result in \eq{polarnonrechfs} by
the one-loop anomalous magnetic moment $\alpha/(2\pi)$.

Hence, the total correction of order $\alpha^2(Z\alpha)E_F$ generated
by the diagrams in Fig.\ \ref{6setshfsfig} $(c)$ is equal to

\beq     \label{onelooppolelfthfs}
\Delta E=\frac{8Z\alpha}{\pi n^3}\left(\frac{\alpha}{\pi}\right)^2E_F
\int_0^\infty dk k^2F(k)I_1(k)
\eeq
\[
=\left(-\frac{4}{3}\ln^2\frac{1+\sqrt{5}}{2}-\frac{20}{9}\sqrt{5}
\ln\frac{1+\sqrt{5}}{2}
-\frac{64}{45}\ln2+\frac{\pi^2}{9}+\frac{3}{8}+\frac{1~043}{675}\right)
\frac{\alpha^2(Z\alpha)}{\pi n^3}E_F
\]
\[
\approx 2.23\ldots
\frac{\alpha^2(Z\alpha)}{\pi n^3}E_F.
\]

\subsubsection{One-Loop Polarization Insertions in the Radiative Electron
Factor}

This correction is induced by the gauge invariant set of diagrams in Fig.\
\ref{6setshfsfig} $(d)$ with the polarization operator insertions in the
radiative photon. The two-loop anomalous magnetic moment generates
correction of order $\alpha^2E_F$ to HFS and the respective leading
pole term in the infrared asymptotics of the electron factor should be
subtracted to avoid infrared divergence and double counting.

The subtracted radiatively corrected electron factor may be obtained
from the subtracted one-loop electron factor in \eq{electrfhfs}. To
this end, one should restore the radiative photon mass in the one-loop
electron factor, and then the polarization operator insertion in the
photon line is taken into account with the help of the dispersion
integral like one in \eq{polloopinsrt} for the spin-independent
electron factor. In terms of the electron factor $F(k,\lambda)$ with a
massive radiative photon with mass $\lambda=2/\sqrt{1-v^2}$ the
contribution to HFS has the form \cite{eks2}

\beq     \label{polradhfs}
\Delta E=\frac{4\alpha^2(Z\alpha)}{\pi^2 n^3}E_F
\int_0^\infty dk\int_0^1dv\frac{v^2(1-\frac{v^2}{3})}{1-v^2}L(k,\lambda)
\eeq

This integral was analytically simplified to a one-dimensional integral of
a complete elliptic integral, which admits numerical evaluation with an
arbitrary precision \cite{eks2}

\beq     \label{polradnumhfs}
\Delta E=-0.310~742\ldots\frac{\alpha^2(Z\alpha)}{\pi n^3}E_F
\eeq

\subsubsection{Light by Light Scattering Insertions in the External
Photons}

The diagrams in Fig.\ \ref{6setshfsfig} $(e)$ with the light by light
scattering insertions in the external photons do not generate
corrections of the previous order in $Z\alpha$. As is well known, the
light by light scattering diagrams are apparently logarithmically
ultraviolet divergent, but due to gauge invariance the diagrams are
really  ultraviolet convergent. Also, as a result of gauge invariance
the light by light scattering tensor is strongly suppressed at small
momenta of the external photons. The contribution to HFS can easily be
expressed in terms of a weighted integral of the light by light
scattering tensor \cite{eks3}, and further calculations are in
principle quite straightforward though technically involved. This
integral was analytically simplified to a three-dimensional integral
which may be calculated with high accuracy \cite{eks3}

\beq  \label{lblhfs}
\Delta E=-0.472~514(1)\frac{\alpha^2(Z\alpha)}{\pi n^3}E_F.
\eeq

The original result in \cite{eks3} differed from the one in \eq{lblhfs}
by two percent. A later purely numerical calculation of the light by
light contribution in \cite{kn1,kn10} produced a less precise result
which, however, differed from the original result in \cite{eks3} by two
percent.  After a throughout check of the calculations in \cite{eks3} a
minor arithmetic mistake in one of the intermediate expressions in the
original version of \cite{eks3} was discovered. After correction of
this mistake, the semianalytic calculations in \cite{eks3} lead to the
result in \eq{lblhfs} in excellent agreement with the somewhat less
precise purely numerical result in \cite{kn1,kn10}.

\subsubsection{Diagrams with Insertions of Two Radiative Photons in the
Electron Line}

\begin{figure}
\centerline{\epsfig{file=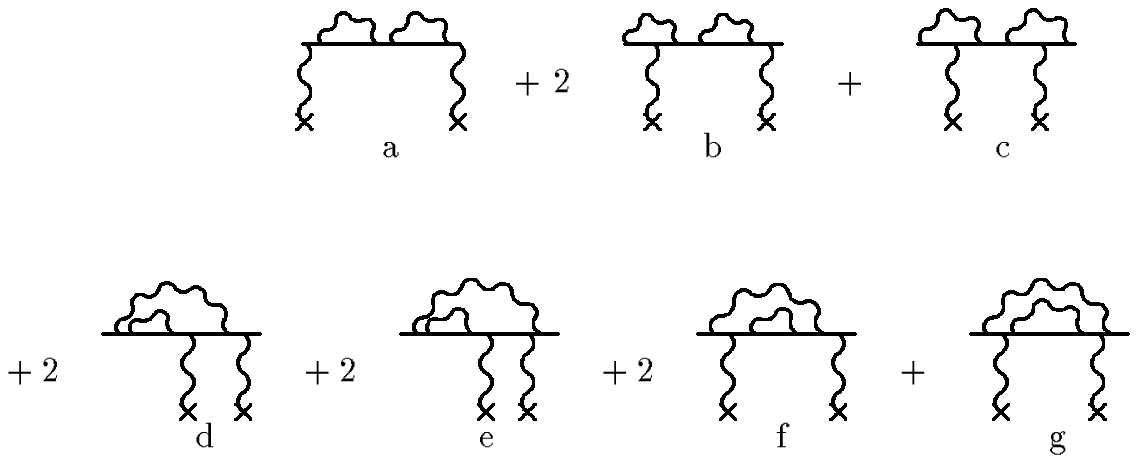,height=6.5cm}}
\vspace{1cm}
\centerline{\epsfig{file=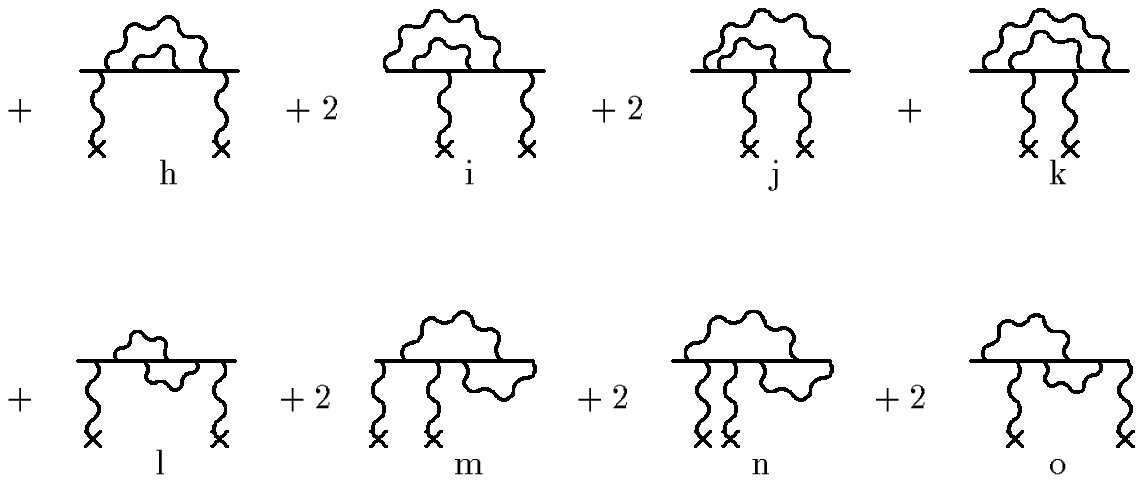,height=6.5cm}}
\vspace{1cm}
\centerline{\epsfig{file=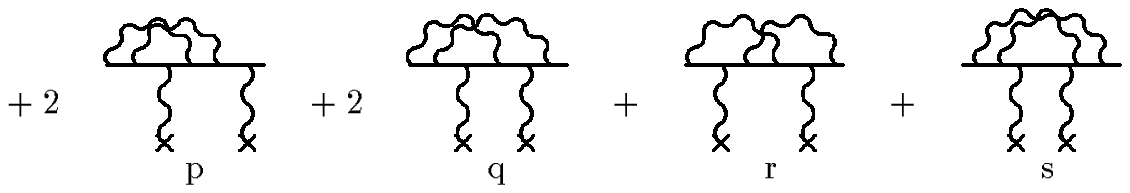,height=2.7cm}}
\vspace{0.5cm}
\caption{Nineteen topologically different diagrams with two
radiative photons insertions in the electron line}
\label{twoloopradhfsfig}
\end{figure}

By far the most difficult task in calculations of corrections of order
$\alpha^2(Z\alpha)E_F$ to HFS is connected with the last gauge
invariant set of diagrams in Fig.\ \ref{6setshfsfig} $(f)$, which
consists of nineteen topologically different diagrams \cite{eks1}
presented in Fig.\ \ref{twoloopradhfsfig} (compare a similar set of
diagrams in Fig.\ \ref{twoloopradcoulfig} in the case of the Lamb
shift). These nineteen graphs may be obtained from the three graphs for
the two-loop electron self-energy by insertion of two external photons
in all possible ways. The graphs in Fig.\ \ref{twoloopradhfsfig}
$(a-c)$ are obtained from the two-loop reducible electron self-energy
diagram, graphs in Fig.\ \ref{twoloopradhfsfig} $(d-k)$ are the result
of all possible insertions of two external photons in the rainbow
self-energy diagram, and diagrams in Fig.\ \ref{twoloopradhfsfig}
$(l-s)$ are connected with the overlapping two-loop self-energy graph.

Calculation of the respective contribution  to HFS in the skeleton
integral approach was initiated in \cite{eksl1,eksl2}, where contributions
induced by the diagrams in Fig.\ \ref{twoloopradhfsfig} $(a-h)$ and Fig.\
\ref{twoloopradhfsfig} $(l)$ were obtained. In order to avoid spurious
infrared divergences in the individual diagrams the semianalytic
calculations in \cite{eksl1,eksl2} were performed in the Yennie gauge. The
diagrams under consideration contain anomalous magnetic moment contributions
which were subtracted before taking the scattering approximation integrals.

The total contribution of all nineteen diagrams to HFS was first calculated
purely numerically in the Feynman gauge in the NRQED framework  in
\cite{kn1,kn10}. The semianalytic skeleton integral calculation in the Yennie
gauge was completed a bit later in \cite{esjetp,es}

\beq         \label{tworadhfs}
\Delta E=-0.672~6(4)\ldots\frac{\alpha^2\left(Z\alpha\right)}{\pi
n^3}E_F.
\eeq

This semianalytic result is consistent with the  purely numerical result in
\cite{kn1,kn10} but more than an order of magnitude more precise. It is
remarkable that the results of two complicated calculations performed in
completely different approaches turned out to be in excellent agreement.

\subsubsection{Total Correction of Order $\alpha^2(Z\alpha)E_F$}

The total contribution of order $\alpha^2(Z\alpha)E_F$ is given by the sum
of contributions in  \eq{onelooppolhfs}, \eq{twolooppolhfs},
\eq{onelooppolelfthfs}, \eq{polradnumhfs}, \eq{lblhfs}, and \eq{tworadhfs}

\beq
\Delta E=\bigg[-\frac{4}{3}\ln^2\frac{1+\sqrt5}{2}-\frac{20}{9}\sqrt{5}
\ln\frac{1+\sqrt5}{2}+\frac{608}{45}\ln2+\frac{\pi^2}{9} -\frac{38}{15}\pi
+\frac{91~639}{37~800}
\eeq
\[
-0.310~742-0.472~514(1)-0.672~6(4) \bigg]
\frac{\alpha^2(Z\alpha)}{\pi}E_F
\approx 0.771~7(4)\frac{\alpha^2(Z\alpha)}{\pi}E_F,
\]

\noindent
or numerically

\beq
\Delta E=0.425~6(2)~\mbox{kHz}.
\eeq

As we have already mentioned consistent results for this correction were
obtained independently in different approaches by two groups in
\cite{eks1,eks2,eks3,eksl1,eksl2,esjetp,es} and \cite{kn1,kn10}.

\subsection{Corrections of Order $\alpha^3(Z\alpha)E_F$}

The corrections of order $\alpha^3(Z\alpha)E_F$ were never considered
in the literature. Their natural scale is determined by the factor
$\alpha^3(Z\alpha)/\pi^2E_F$, which is equal about $10^{-3}$ kHz. Hence,
this correction is too small to be of any interest now. Note that the
uncertainty of the total contribution in the last line in Table XIV is
determined not,by the uncertainties of any of the entries in the upper
lines of this Table, which are too small, but just by the uncalculated
contribution of order $\alpha^3(Z\alpha)E_F$.

\begin{center}
\underline{Table XIV. Radiative Corrections of Order
$\alpha^n(Z\alpha)E_F$}
\nopagebreak

\begin{tabular}{|l|rl|c|}
\hline &
$\alpha(Z\alpha)E_F$ &   & kHz \\ \hline \hline
Electron-Line Insertions&&&\\&&&\\
Kroll, Pollock (1951)\cite{kp}&       & & \\
Karplus, Klein,&$(\ln 2-\frac{13}{4})$&&$-607.123$\\
Schwinger (1951)\cite{kks1,kk}&         & &
\\ \hline
Polarization Insertion&&&\\
&&&\\
Kroll, Pollock (1951)\cite{kp}&       & & \\
Karplus, Klein,&$\frac{3}{4}$&&$178.087$\\
Schwinger (1951)\cite{kks1,kk}&         & &
\\ \hline
One-Loop Polarization&&&\\
&$$&& \\
Eides, Karshenboim,&$$&&\\
Shelyuto (1989)\cite{eks1}&$\frac{36}{35}\frac{\alpha}{\pi}$& & $0.567$
\\ \hline
Two-Loop Polarization&&&\\
&$$&& \\
Eides, Karshenboim,&$$& &\\
Shelyuto (1989)\cite{eks1}&$(\frac{224}{15}\ln 2-\frac{38\pi}{15}
-\frac{118}{225})\frac{\alpha}{\pi}$& &$1.030$
\\
&$$&& \\
\hline
One-Loop Polarization&&&\\
and Electron Factor&&&\\
&$$&& \\
Eides,Karshenboim, &$\{-\frac{4}{3}\ln^2\frac{1+\sqrt{5}}{2}
-\frac{20}{9}\sqrt 5\ln\frac{1+\sqrt{5}}{2}$&&\\
Shelyuto (1989)\cite{eks1}&$-\frac{64}{45}\ln 2+\frac{\pi^2}{9}+\frac{3}{8}
+\frac{1~043}{675}\}\frac{\alpha}{\pi}$
&&$-0.369$\\
&$$&&
\\ \hline
Polarization insertion  &&&\\
in the Electron Factor&&&\\
&$$&& \\
Eides, Karshenboim,&   $$       & & \\
Shelyuto (1990)\cite{eks2}&$-0.310~742\ldots\frac{\alpha}{\pi}$& &$-0.171$
\\ &$$&&
\\
\hline
Light by Light Scattering&&&\\
&$$&& \\
Eides, Karshenboim,&   $$       & & \\
Shelyuto (1991,1993)\cite{eks3}& $-0.472~514(1)\frac{\alpha}{\pi}$&
&$-0.261$\\
Kinoshita,Nio (1994,1996)\cite{kn1,kn10}  &   $$       & &
\\ \hline
Insertions of Two Radiative &&&\\
Photons in the Electron Line&&&\\
&&&\\
Kinoshita,Nio (1994,1996)\cite{kn1,kn10}  &$$       & & \\
Eides,Shelyuto(1995)\cite{esjetp,es}
&$-0.672~6(4)\frac{\alpha}{\pi}$& &$-0.371$
\\ &   $$       & &
\\
\hline
\hline
Total correction of	 &$$& &\\
order $\alpha^n(Z\alpha)E_F$&   $$       & &$-428.611(1)$ \\
\hline
\end{tabular}
\end{center}

\section{~Radiative Corrections of Order
$\alpha^{\lowercase{n}}(Z\alpha)^2E_F$}

\subsection{Corrections of Order $\alpha(Z\alpha)^2E_F$}


\subsubsection{Electron-Line Logarithmic Contributions }\label{hfselalzal2}

\begin{figure}[ht]
\centerline{\epsfig{file=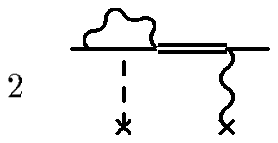,height=2cm}}
\vspace{0.5cm}
\caption{Leading logarithm squared contribution of order
$\alpha(Z\alpha)^2E_F$ to HFS}
\label{leadalpha2zal2hfsfig}
\end{figure}

Binding effects are crucial in calculation of the corrections of order
$\alpha(Z\alpha)^2E_F$. Unlike the corrections of the first
order in the binding parameter $Z\alpha$, in this case the exchanged photon
loops with low (of order $\sim mZ\alpha$) exchanged momenta give
a significant contribution and the external wave functions at the origin do
not factorize as in the scattering approximation. The anticipated low
momentum logarithmic divergence in the loop integration is cut off by the
wave functions at the atomic scale and, hence, the contribution of order
$\alpha(Z\alpha)^2E_F$ is enhanced by the low-frequency logarithmic terms
$\ln^2Z\alpha$ and $\ln Z\alpha$.  The situation for this
calculation resembles the case of the main contribution to the
Lamb shift of order $\alpha(Z\alpha)^4$ and also corrections to the Lamb
shift of order $\alpha(Z\alpha)^6$. Once again, the factors before
logarithmic terms originate from the electron form factor and from the
logarithmic integration over the loop momenta in the diagrams with three
exchanged photons. The leading logarithm squared term is state independent
and can easily be calculated by the same methods as the leading logarithm
cube contribution of order $\alpha^2(Z\alpha)^6$ (see Section
\ref{lambcube}). The only difference is that this time it is necessary to
take as one of the perturbation potentials the potential responsible for the
main Fermi contribution to HFS

\beq         \label{fermipothfseff}
V_F=\frac{8}{3}\frac{\pi Z\alpha}{mM}(1+a_\mu).
\eeq

This calculation of the leading logarithm squared term \cite{kar93}
(see Fig.\ \ref{leadalpha2zal2hfsfig}) also produces a recoil
correction to the nonrecoil logarithm squared contribution. We will
discuss this radiative-recoil correction below in the Section
\ref{hfsleadlogradrecza2} dealing with other radiative-recoil corrections,
and we will consider in this section only the nonrecoil part of the
logarithm squared term.

All logarithmic terms for the $1S$ state were originally calculated in
\cite{l,z}. The author of \cite{l} also calculated the logarithmic
contribution to the $2S$ hyperfine splitting

\beq             \label{leadloghfsaza2}
\Delta E(1S)=\left[-\frac{2}{3}\ln^2(Z\alpha)^{-2}
-\left(\frac{8}{3}\ln 2-\frac{37}{72}\right) \ln(Z\alpha)^{-2}\right]
\frac{\alpha(Z\alpha)^2}{\pi}E_F,
\eeq

\beq
\Delta E(2S)=\left[-\frac{2}{3}\ln^2(Z\alpha)^{-2}-\left(\frac{16}{3}
\ln 2-4-\frac{1}{72}\right)\ln(Z\alpha)^{-2}\right]
\frac{\alpha(Z\alpha)^2}{8\pi}E_F.
\eeq

\subsubsection{Nonlogarithmic Electron-Line Corrections}

Calculation of the nonlogarithmic part of the contribution of order
$\alpha(Z\alpha)^2E_F$ is a more complicated task than obtaining the leading
logarithmic terms. The short distance leading logarithm squared
contributions cancel in the difference $\Delta E(1S)-8\Delta E(2S)$ and it
is this difference, containing both logarithmic  and nonlogarithmic
corrections, which was calculated first in \cite{zwanz1}. An estimate of the
nonlogarithmic terms for $n=1$ and $n=2$ in accordance with \cite{zwanz1}
was obtained in \cite{brer}. This work also confirmed the results of
\cite{l,z} for the logarithmic terms. The first complete calculation of the
nonlogarithmic terms was done in a purely numerical approach in \cite{s}.
The idea of this calculation is similar to the one used for calculation of
the $\alpha(Z\alpha)^6$ contribution to the Lamb shift in \cite{sapir}. In
this calculation the electron-line radiative corrections were written in the
form of the diagrams with the electron propagator in the external field, and
then an approximation scheme for the relativistic electron propagator in
the Coulomb field was set with the help of the well known representation
\cite{hostler,schprop} for the nonrelativistic propagator in the Coulomb
field. In view of the significant theoretical progress achieved recently in
calculation of contributions to HFS of order $\alpha^2(Z\alpha)E_F$,
insufficient accuracy in the calculation in \cite{s} (about $0.2$ kHz)
became the main source of uncertainty in the theoretical expression for
muonium HFS. Two new independent calculations of this correction were
performed recently \cite{pach96,kn2}. The author of \cite{pach96} used his
approach developed for calculation of corrections of order
$\alpha(Z\alpha)^6$ to the Lamb shift. The main ideas of this approach were
discussed above in Section \ref{pachappr}. Calculation in \cite{kn2}  was
performed in the completely different framework of nonrelativistic QED (see,
e.g., \cite{caswelllep,kn1,kn10}). Results of both calculations are in
remarkable agreement, the result of \cite{pach96} being equal to

\beq
\Delta E=17.122\frac{\alpha(Z\alpha)^2}{\pi}E_F,
\eeq

\noindent
and the result of \cite{kn2} is

\beq
\Delta E=17.122~7(11)\frac{\alpha(Z\alpha)^2}{\pi}E_F.
\eeq

\subsubsection{Logarithmic Contribution induced by the Polarization
Operator}

\begin{figure}[ht]
\centerline{\epsfig{file=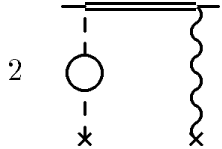,height=2cm}}
\vspace{0.5cm}
\caption{Leading logarithm polarization contribution of order
$\alpha(Z\alpha)^2E_F$ to HFS}
\label{logpolhfsfig}
\end{figure}

Calculation of the leading logarithmic contribution induced by the
polarization operator insertion (see Fig.\ \ref{logpolhfsfig}) proceeds
along the same lines as the calculation of the logarithmic polarization
operator contribution of order $\alpha(Z\alpha)^6$ to the Lamb shift
(compare Section \ref{lambalzal6pol}).  Again, only the leading term in
the small momentum expansion of the polarization operator (see
\eq{qpol}, \eq{2ndorderpol}) produces a contribution of the order under
consideration and only the state-independent logarithmic corrections to
the Schr\"odinger-Coulomb function are relevant.  The only difference
is that the correction to the wave function is produced, not by the
Darwin term as in the case of the Lamb shift (compare \eq{darwincor})
but, by the external magnetic moment perturbation in
\eq{fermipothfseff}.  This correction to the wave function was
calculated in \cite{zwanz1,broder}, and with its help one immediately
obtains the state-independent logarithmic contribution
$\alpha(Z\alpha)^2E_F$ to HFS induced by the polarization operator
insertion \cite{l,z}

\beq
\Delta E_{pol}=\frac{4}{15}\ln(Z\alpha)^{-2}\frac{\alpha(Z\alpha)^2}{\pi
n^3}E_F.
\eeq


\subsubsection{Nonlogarithmic Corrections Induced by the Polarization
Operator}

\begin{figure}[ht]
\centerline{\epsfig{file=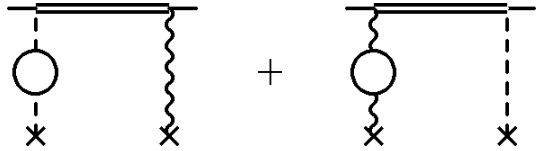,height=2cm}}
\vspace{0.5cm}
\caption{Corrections of order $\alpha(Z\alpha)^2E_F$ induced by the
polarization operator}
\label{nonlogpolaza2hfsfig}
\end{figure}

Calculation of the nonlogarithmic polarization operator contribution is
quite straightforward. One simply has to calculate two terms given by
ordinary perturbation theory, one is the matrix element of the
radiatively corrected external magnetic field, and another is the matrix
element of the radiatively corrected external Coulomb field between
wave functions corrected by the external magnetic field (see Fig.\
\ref{nonlogpolaza2hfsfig}). The first calculation of the respective matrix
elements was performed in \cite{brer}.  Quite recently a number of
inaccuracies in \cite{brer} were uncovered
\cite{kn1,kn10,kn2,sapirstein,brodskyerr,sgs} and the correct result for the
nonlogarithmic contribution of order $\alpha(Z\alpha)^2E_F$ to HFS is given
by

\beq
\Delta E=\left(-\frac{8}{15}\ln2+\frac{34}{225}\right)
\frac{\alpha(Z\alpha)^2}{\pi}E_F.
\eeq

\subsection{Corrections of Order $\alpha^2(Z\alpha)^2E_F$}

\subsubsection{Leading Double Logarithm Corrections }

\begin{figure}[ht]
\centerline{\epsfig{file=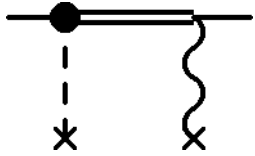,height=2cm}}
\vspace{0.5cm}
\caption{Leading double logarithm contribution of order
$\alpha^2(Z\alpha)^2E_F$}
\label{laedllogal2zal2hfsfig}
\end{figure}

Corrections of order $\alpha^2(Z\alpha)^2E_F$ are again enhanced by a
logarithm squared term and one should expect that they are smaller by the
factor $\alpha/\pi$ than the corrections of order $\alpha(Z\alpha)^2E_F$
considered above. Calculation of the leading logarithm squared contribution
to HFS may be performed in exactly the same way as the calculation of the
leading logarithm cube contribution of order $\alpha^2(Z\alpha)^6$ to the
Lamb shift considered above in Section \ref{lambcube}. Both results were
originally obtained in one and the same work \cite{kar93}. The logarithm
cube term is missing in the case of HFS, since now at least one of the
perturbation operators in Fig.\ \ref{laedllogal2zal2hfsfig} should contain a
magnetic exchange photon and the respective anomalous magnetic moment is
infrared finite. It is easy to realize that as the result of this
calculation one obtains simply the leading logarithmic contribution of order
$\alpha(Z\alpha)^2E_F$ in \eq{leadloghfsaza2}, multiplied by the electron
anomalous magnetic moment $\alpha/(2\pi)$ \cite{kar93}

\beq
\Delta E=-\frac{1}{3}\ln^2(Z\alpha)^{-2}\frac{\alpha^2(Z\alpha)^2}{\pi^2}E_F.
\eeq

Numerically this correction is about $-0.04$ kHz, and this contribution is
large enough to justify calculation of the single-logarithmic contributions.

\subsubsection{Single-Logarithmic and Nonlogarithmic Contributions}

Terms linear in the large logarithm were recently calculated in
the NRQED framework \cite{kin98}

\beq
\Delta
E=\left[\left(\frac{9}{4}\zeta(3)-\frac{3}{2}\pi^2\ln2+\frac{10}{27}\pi^2
+\frac{4~358}{1~296}\right)
-\frac{3}{4}\left(\frac{3}{4}\zeta(3)-\frac{\pi^2}{2}\ln2+\frac{\pi^2}{12}
+\frac{197}{144}\right)
\right.
\eeq
\[
\left.
-\frac{4}{3}\left(\ln2-\frac{3}{4}\right)
+\frac{1}{4}\left(-\frac{11}{9}-1+\frac{8}{15}\right)\right]
\ln(Z\alpha)^{-2}\frac{\alpha^2(Z\alpha)^2}{\pi^2}E_F
\]
\[
\approx-0.639~000~544\ldots\frac{\alpha^2(Z\alpha)^2}{\pi^2}E_F.
\]

The nonlogarithmic contributions of order $\alpha^2(Z\alpha)^2E_F$ are also
estimated in \cite{kin98}

\beq
\Delta E=(10\pm2.5)\frac{\alpha^2(Z\alpha)^2}{\pi^2}E_F.
\eeq

The numerical uncertainty of the last contribution is about $0.003$ kHz.

Corrections of order $\alpha^3(Z\alpha)^2E_F$ are suppressed by an extra
factor $\alpha/\pi$ in comparison with the leading contributions of order
$\alpha^2(Z\alpha)^2E_F$ and are too small to be of any phenomenological
interest now. All corrections of order $\alpha^n(Z\alpha)^2E_F$ are
collected in Table XV, and their total uncertainty is determined by the
error of the nonlogarithmic contribution of order
$\alpha^2(Z\alpha)^2E_F$.


\begin{center}
\underline{Table XV. Radiative Corrections of
Order $\alpha^n(Z\alpha)^2E_F$}
\nopagebreak

\begin{tabular}{|l|rl|c|}
\hline
& $\frac{\alpha(Z\alpha)^2}{\pi}E_F$ &   & kHz
\\ \hline    \hline
Logarithmic Electron-Line &&$$& \\
Contribution &&$$&
\\ &&&
\\
Zwanziger (1964)\cite{z}&$$& &
\\
Layzer (1964)\cite{l}&$-\frac{2}{3}\ln^2(Z\alpha)^{-2}
-(\frac{8}{3}\ln 2-\frac{37}{72})
\ln (Z\alpha)^{-2}
$&&$-42.850$
\\ &&&
\\ \hline
Nonlogarithmic Electron-Line &&$$& \\
Contribution &&$$&
\\ &&&
\\
Pachucki(1996)\cite{pach96}&&&\\
Kinoshita,Nio(1996)\cite{kn2}&$17.122~7(11)$&&$9.444$
\\ &&&
\\ \hline
Logarithmic Polarization&&&\\
Operator Contribution&&&\\
&&&\\
Zwanziger (1964)\cite{z}&$$& &
\\
Layzer (1964)\cite{l}&$\frac{4}{15}\ln(Z\alpha)^{-2}
$&&$1.447$
\\ &&&
\\ \hline
Nonlogarithmic Polarization&&&\\
Operator Contribution&&&\\
&&&\\
Kinoshita,Nio(1996)\cite{kn1,kn10,kn2}&&&\\
Sapirstein(1996)\cite{sapirstein},Brodsky(1996)\cite{brodskyerr}
&$(-\frac{8}{15}\ln2+\frac{34}{225})$
&&$-0.121$\\
Schneider,Greiner,Soff(1994)\cite{sgs}&&&
\\
&&&  \\
\hline
Leading Logarithmic &&$$& \\
Contribution of order $\alpha^2(Z\alpha)^2E_F$&&$$&
\\ &&&\\
Karshenboim(1993)\cite{kar93}&$
-\frac{1}{3}\ln^2(Z\alpha)^{-2}\frac{\alpha}{\pi}$&&$-0.041$
\\
&&&  \\
\hline
Single-logarithmic &&$$& \\
Contributions of order $\alpha^2(Z\alpha)^2E_F$&&$$&
\\ &&&\\
Kinoshita,Nio(1998)\cite{kin98}&$
[(\frac{9}{4}\zeta(3)-\frac{3}{2}\pi^2\ln2+\frac{10}{27}\pi^2
+\frac{4358}{1296})$&&
\\
&$-\frac{3}{4}(\frac{3}{4}\zeta(3)-\frac{\pi^2}{2}\ln2+\frac{\pi^2}{12}
+\frac{197}{144})$ &&  \\
&$-\frac{4}{3}(\ln2-\frac{3}{4})
+\frac{1}{4}(-\frac{11}{9}-1+\frac{8}{15})]
\ln(Z\alpha)^{-2}\frac{\alpha}{\pi}$&&$-0.008$
\\ &&&  \\
\hline
Nonlogarithmic Contributions &&$$& \\
of order $\alpha^2(Z\alpha)^2E_F$&&$$&
\\
Kinoshita,Nio(1998)
&$(10\pm2.5)\frac{\alpha}{\pi}$&&$0.013(3)$
\\
\hline
\hline
Total correction of	 &$$& &\\
order $\alpha^n(Z\alpha)^2E_F$&   $$       & &$-32.115~(3)$ \\
\hline
\end{tabular}
\end{center}

\section{~Radiative Corrections of Order $\alpha(Z\alpha)^3E_F$ and of
Higher Orders}

As we have repeatedly observed corrections to the energy
levels suppressed by an additional power of the binding parameter $Z\alpha$
are usually numerically larger than the corrections suppressed by
an additional power of $\alpha/\pi$, induced by radiative insertions.
In this perspective one could expect that the corrections of order
$\alpha(Z\alpha)^3E_F$ would be numerically larger than considered above
corrections of order  $\alpha^2(Z\alpha)^2E_F$.

\subsection{Corrections of Order $\alpha(Z\alpha)^3E_F$}

\subsubsection{Leading Logarithmic Contributions Induced by the Radiative
Insertions in the Electron Line}

Calculation of the leading logarithmic corrections of order
$\alpha(Z\alpha)^3E_F$ to HFS parallels the calculation of the leading
logarithmic corrections of order $\alpha(Z\alpha)^7$ to the Lamb shift,
described above in Section \ref{lambalphazalpha7}. Again all leading
logarithmic contributions may be calculated with the help of second order
perturbation theory (see \eq{secoredr}).

\begin{figure}[ht]
\centerline{\epsfig{file=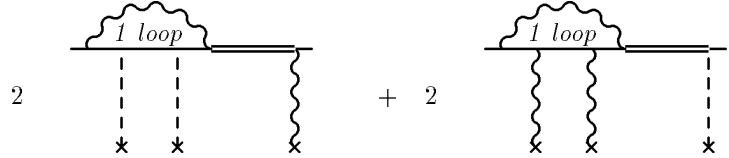}}
\vspace{0.5cm}
\caption{Leading logarithmic electron-line contributions of order
$\alpha(Z\alpha)^3E_F$}
\label{logaza3hfsfig}
\end{figure}

\begin{figure}[ht]
\centerline{\epsfig{file=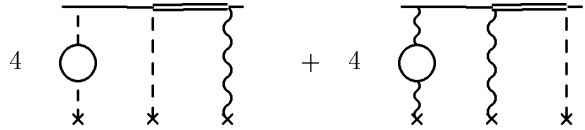}}
\vspace{0.5cm}
\caption{Leading logarithmic photon-line contributions of order
$\alpha(Z\alpha)^3E_F$}
\label{logaza3phothfsfig}
\end{figure}

It is easy to check that the leading contribution is linear in the
large logarithm. Due to the presence of the local potential in
\eq{fermipothfseff} which corresponds to the main contribution to HFS,
and of the potential

\beq    \label{lpeffpotel}
V_{KP}=\frac{8}{3}\left(\ln2-\frac{13}{4}\right)\alpha(Z\alpha)\frac{\pi
\alpha(Z\alpha)^2}{mM}(1+a_\mu),
\eeq

\noindent
which corresponds to the contribution of order $\alpha(Z\alpha)E_F$, we
now have two combinations of local potentials $V_1$ (\eq{aza5lambpot}),
$V_F$ (\eq{fermipothfseff}), and $V_2$ (\eq{darwin}), $V_{KP}$
(\eq{lpeffpotel}) in Fig.\ \ref{logaza3hfsfig}, which generate
logarithmic contributions.  This differs from the case of the Lamb
shift when only one combination of local operators was relevant for
calculation of the leading logarithmic contribution. An easy
calculation produces \cite{pl,kar96}

\beq
\Delta E=4\left(\frac{1}{2}\ln
2-1-\frac{11}{128}\right)\ln(Z\alpha)^{-2}
\frac{\alpha(Z\alpha)^3}{n^3}\left(\frac{m_r}{m}\right)^2E_F,
\eeq

\noindent
and \cite{kar96}

\beq
\Delta E=\frac{1}{2}\left(\ln2-\frac{13}{4}\right)
\ln(Z\alpha)^{-2}\frac{\alpha(Z\alpha)^3}{n^3}E_F,
\eeq

\noindent
for the first and second combinations of the perturbation potentials,
respectively.

\subsubsection{Leading Logarithmic Contributions Induced by the
Polarization Insertions in the External Photon Lines}

The leading logarithmic correction induced by the radiative insertions
in the external photon in Fig.\ \ref{logaza3phothfsfig} is calculated
in exactly the same way as was done above in the case of radiative
insertions in the electron line. The only difference is that instead of
the potential $V_1$ in \eq{aza5lambpot} we have to use the respective
potential in \eq{uelpoteff}, and instead of the potential $V_{KP}$ in
\eq{lpeffpotel} we have to use the potential

\beq
V_{KP}=2\alpha(Z\alpha)\frac{\pi\alpha(Z\alpha)^2}{mM}(1+a_\mu).
\eeq

\noindent
connected with the external photons. Then we immediately obtain
\cite{pl,kar96}

\beq
\Delta E=-\frac{5}{48}\ln(Z\alpha)^{-2}
\frac{\alpha(Z\alpha)^3}{n^3}(\frac{m_r}{m})^2E_F,
\eeq

\noindent
and \cite{kar96}

\beq
\Delta E=\frac{3}{8}
\ln(Z\alpha)^{-2}\frac{\alpha(Z\alpha)^3}{n^3}E_F,
\eeq

\noindent
for the first and second combinations of the perturbation potentials.

\subsubsection{Nonlogarithmic Contributions of Order
$\alpha(Z\alpha)^3E_F$ and of Higher Orders in $Z\alpha$}

The large magnitude of the leading logarithmic contributions of order
$\alpha(Z\alpha)^3E_F$ generated by the radiative photon insertions in the
electron line warrants consideration of the respective nonlogarithmic
corrections. Numerical calculations of radiative corrections to HFS
generated by insertion of one radiative photon in the electron line
without expansion over $Z\alpha$, which are routinely used for the high
$Z$ atoms, were recently extended for $Z=1$. Initial disagreements
between the results of \cite{psgsl} and \cite{bcs} were resolved in
\cite{spssls}, which confirmed (but with an order of magnitude  worse
accuracy) the result of \cite{bcs}

\beq   \label{blchsap}
\Delta E=-3.82(63)\alpha(Z\alpha)^3E_F.
\eeq

Let us emphasize that this result describes not only nonlogarithmic
corrections of order  $\alpha(Z\alpha)^3E_F$ but also includes
contributions of all terms of the form $\alpha(Z\alpha)^nE_F$ with
$n\geq4$. The magnitude of the nonlogarithmic coefficient in
\eq{blchsap} seems to be quite reasonable qualitatively. Numerically
the contribution to HFS in \eq{blchsap} is about $0.09$ of the leading
logarithmic contribution.

Let us also mention nonlogarithmic contributions induced by the
insertions in the external photon. In analogy to the case of the
insertions in the electron line it is reasonable to expect that
magnitude of these nonlogarithmic terms is less than one tenth of the
respective leading logarithmic contribution, and is thus smaller than
the uncertainty of the electron line contribution.  Numerical
calculations without expansion in $Z\alpha$ in \cite{spssls} are
consistent with these expectations.

Thus all corrections of order $\alpha(Z\alpha)^nE_F$ collected in Table
XVI are now known with an uncertainty of about $0.008$ kHz. Of all these
corrections only the nonlogarithmic contribution of order
$\alpha(Z\alpha)^3E_F$ was not calculated independently by at least two
groups. In view of the complexity of the numerical calculations an
independent consideration of this correction would be helpful.

\begin{center}
\underline{Table XVI. Radiative Corrections of Order
$\alpha(Z\alpha)^3E_F$}
\nopagebreak

\begin{tabular}{|l|rl|c|}
\hline
& $\alpha(Z\alpha)^3E_F$ &   	& kHz
\\ \hline  \hline
Logarithmic Electron-Line &&$$& \\
Contribution &&$$&
\\ &&&
\\
Lepage(1994)\cite{pl},Karshenboim (1996)\cite{kar96}
&$(\frac{5}{2}\ln2-\frac{191}{32})\ln(Z\alpha)^{-2}$&&$-0.527$\\
&&& \\
\hline
Logarithmic Polarization&&&\\
Operator Contribution&&&\\
&&&\\
Lepage(1994)\cite{pl},Karshenboim(1996)\cite{kar96}&
$\frac{13}{48}\ln(Z\alpha)^{-2}$&&$0.034$
\\
&&&  \\
\hline
&&&  \\
Nonlogarithmic Electron-Line &&$$& \\
Contribution &&$$&
\\ &&&\\
Blundell,Cheng,Sapirstein(1997)\cite{bcs}&$-3.82(63)$&&$-0.048(8)$\\
&&&  \\
\hline
\hline
Total correction of	 &$$& &\\
order $\alpha(Z\alpha)^3E_F$&   $$       & &$-0.542~(8)$ \\
\hline
\end{tabular}
\end{center}

\subsection{Corrections  of Order $\alpha^2(Z\alpha)^3E_F$ and of Higher
Orders in $\alpha$}

One should expect that corrections of order $\alpha^2(Z\alpha)^3E_F$ are
suppressed relative to the contributions of order
$\alpha(Z\alpha)^3E_F$ by the factor $\alpha/\pi$. This means that at
the present level of experimental accuracy one may safely neglect these
corrections, as well as corrections of even higher orders in $\alpha$.

\part{Essentially Two-Body Corrections to HFS}

\section{~~Recoil Corrections to HFS}\label{recoilhfsmuon}

The very presence of the recoil factor $m/M$ emphasizes that the
external field approach is inadequate  for calculation of recoil
corrections and, in principle, one needs the complete machinery of the
two-particle equation in this case. However, many results may be
understood without a cumbersome formalism.

Technically, the recoil factor $m/M$ arises because the integration over
the exchanged momenta in the diagrams which generate the recoil corrections
goes over a large interval up to the muon mass, and not just up to the
electron mass, as was the case of the nonrecoil radiative corrections.
Due to large intermediate momenta in the general expression for the recoil
corrections only the Dirac magnetic moment of the muon factorizes naturally
in the general expression for the recoil corrections

\beq
\Delta E=\widetilde{E}_{F}(1+corrections),
\eeq

\noindent
where

\beq      \label{baremuonfermi}
\widetilde{E}_{F}=\frac{16}{3}Z^4\alpha^2
\frac{m}{M} \left(\frac{m_r}{m}\right)^{3}ch\:R_{\infty}.
\eeq

Here $\widetilde{E}_{F}$ does not include, unlike \eq{ef}, the muon
anomalous magnetic moment $a_\mu$ which should now be considered on the
same grounds as other corrections to hyperfine splitting.
Nonfactorization of the muon anomalous magnetic moment is a natural
consequence of the presence of the large integration region mentioned
above. It is worth mentioning that the expression for the Fermi energy
$\widetilde{E}_{F}$ is symmetric with respect to the light and heavy
particles, and does not change under exchange of the particles
$m\leftrightarrow M$.

\subsection{Leading Recoil Correction}\label{leadrecmuonhfs}

\begin{figure}[ht]
\centerline{\epsfig{file=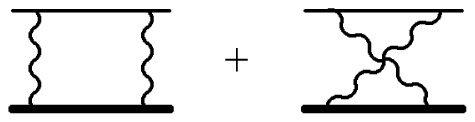,height=2cm}}
\vspace{0.5cm}
\caption{Diagrams with two-photon exchanges}
\label{twophothfsfig}
\end{figure}

The leading recoil correction of order $Z\alpha(m/M)\widetilde
E_F$ is generated by the graphs with two exchanged photons in Fig.\
\ref{twophothfsfig}, similar to the case of the recoil correction to the
Lamb shift of order $(Z\alpha)^5(m/M)m$ considered in Section
\ref{lambreczalpha5}.  However, calculations in the case of hyperfine
splitting are much simpler in comparison with the Lamb shift, since the
region of extreme nonrelativistic exchange momenta
$m(Z\alpha)^2<k<m(Z\alpha)$ does not generate any correction of order
$Z\alpha(m/M)\widetilde E_F$. This is almost obvious in the
nonrelativistic perturbation theory framework, which is quite
sufficient for calculation of all corrections generated at such small
momenta. Unlike the case of the Lamb shift the leading contribution
which is due to the one-transverse quanta exchange in the
nonrelativistic dipole approximation is given by the Breit potential.
This contribution is simply the Fermi energy, and all nonrelativistic
corrections to the Fermi energy are suppressed at least by the
additional factor $(Z\alpha)^2$. Then the leading recoil correction to
hyperfine splitting may reliably be calculated in the scattering
approximation, ignoring even the wave function momenta of order
$mZ\alpha$. The formal proof has been given, e.g., in \cite{eksann1},
but this may easily be understood at the qualitative level.  The
skeleton integral is linearly infrared divergent and this divergence
has a clear origin since it corresponds to the classical Fermi
contribution to HFS.  This divergence is produced by the heavy particle
pole contribution and after subtraction (note that we effectively
subtract the skeleton integral in \eq{skel} with restored factor
$1/(1+m/M)$, see discussion in Section \ref{hfsradrecpol}) we obtain
the convergent skeleton integral

\beq   \label{rechfssk}
\Delta E= {\widetilde E}_F
\frac{mM}{M^2-m^2}\frac{Z\alpha}{2\pi}\int_0^\infty\frac{dk}{k}
[f(k)-f(\mu k)],
\eeq

\noindent
where

\beq
f(k)=\frac{k^4-4k^2-32}{k\sqrt{4+k^2}}-k^2+\frac{16}{k}
\eeq

\noindent
and $\mu=m/M$. One may easily perform the momentum integration in this
infrared finite integral and obtain \cite{arn,fm,newsalp}

\beq              \label{arnnewcsalp}
\Delta E_{rec}=-\frac{3mM}{M^2-m^2}\frac{Z\alpha}{\pi}\ln\frac{M}{m}
{\widetilde E}_F.
\eeq

The subtracted heavy pole (Fermi) contribution is generated by the exchange
of a photon with a small (atomic scale $\sim mZ\alpha$) momentum
and after subtraction of this contribution only high loop momenta
$k$ ($m<k<M$) contribute to the integral for the recoil correction. Then
the exchange loop momenta are comparable to the virtual momenta determining
the anomalous magnetic moment of the muon and there are no reasons to expect
that the anomalous magnetic moment will enter as a factor in the formula for
the recoil corrections. It is clear that the contribution of the muon
anomalous magnetic moment in this case cannot be separated from
contributions of other radiative-recoil corrections.

Let us emphasize that, unlike the other cases in this review where we
encountered the logarithmic contributions, the result in
\eq{arnnewcsalp} is exact in the sense that this is a complete contribution
of order $Z\alpha(m/M)\widetilde E_F$. There are no nonlogarithmic
contributions of this order.

Despite its nonsymmetric appearance the recoil correction in
\eq{arnnewcsalp} is symmetric with respect to the electron and muon masses.
As in the case of the leading recoil corrections to the Lamb shift coming
from one-photon exchanges, this formula generated by the two-photon exchange
is exact in the electron-muon mass ratio. This is crucial from the
phenomenological point of view, taking into account the large value of the
correction under consideration and the high precision of the current
experimental results.

\subsection{Recoil Correction of Relative Order
$(Z\alpha)^2(\lowercase{m}/M)$} \label{rechfsza2muon}

\begin{figure}[ht]
\centerline{\epsfig{file=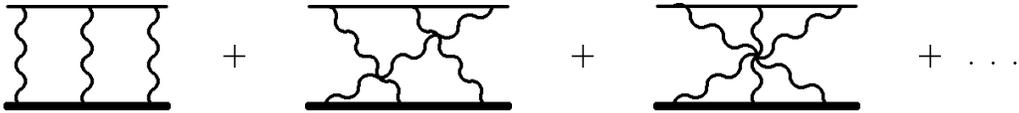,height=1.5cm}}
\vspace{0.5cm}
\caption{Recoil corrections of order $(Z\alpha)^2(m/M)E_F$}
\label{recza2hfsfig}
\end{figure}

Recoil corrections of relative order $(Z\alpha)^2(m/M)$  are generated
by the kernels with three exchanged photons (see Fig.\ \ref{recza2hfsfig}).
One might expect, similar to the case of the leading recoil correction,
emergence of a recoil contribution of order $(Z\alpha)^2(m/M)\widetilde E_F$
logarithmic in the mass ratio. The logarithm of the mass ratio could
originate only from the integration region $m\ll~k\ll~M$, where one can
safely omit electron masses in the integrand. The integrand simplifies, and
it turns out that despite the fact that the individual diagrams produce
logarithmic contributions, these contributions sum to zero
\cite{byg78,cp}. It is not difficult to understand the technical reason
for this effect which is called the Caswell-Lepage cancellation. For
each exchanged diagram there exists a pair diagram where the exchanged
photons are attached to the electron line in an opposite order.
Respective electron line contributions to the logarithmic integrands
generated by these diagrams differ only by sign \cite{cp} and, hence,
the total contribution logarithmic in the mass ratio vanishes.

This means that all pure recoil corrections of relative order
$(Z\alpha)^2(m/M)$ originate from the exchanged momenta of order of
the electron mass and smaller. One might think that as a result the
muon anomalous magnetic moment would enter the expression  for these
corrections in a factorized form, and the respective corrections should
be written in terms of $E_F$ and not of $\widetilde E_F$, as was in the
case of the leading recoil correction. However, if we include the muon
anomalous magnetic moment in the kernels with three exchanged photons,
they would generate the contributions proportional not only to the muon
anomalous magnetic moment but to the anomalous magnetic moment squared.
This makes any attempt to write the recoil correction of relative order
$(Z\alpha)^2(\lowercase{m}/M)$ in terms of $E_F$ unnatural, and it is
usually written in terms of $\widetilde E_F$ (see however discussion of
this correction in the case of hydrogen in Section
\ref{recoilhfshydza2}). The corrections to this result due to the muon
anomalous magnetic moment should be considered separately as the
radiative-recoil corrections of order
$(Z^2\alpha)\alpha(Z\alpha)^2(m/M)\widetilde E_F$. A naive attempt to
write the recoil correction of relative order
$(Z\alpha)^2(\lowercase{m}/M)$ in terms of $E_F$ would shift the
magnitude of this correction by about $10$ Hz. This shift could be
understood as an indication that calculation of the radiative-recoil
corrections of order $(Z^2\alpha)\alpha(Z\alpha)^2(m/M)\widetilde E_F$
could be of some phenomenological interest at the present level of
experimental accuracy.

Leading terms logarithmic in $Z\alpha$ were first considered in
\cite{for}, and the complete logarithmic contribution was obtained in
\cite{lep,by}

\beq                       \label{logrechfs}
\Delta E=2\ln(Z\alpha)^{-1}(Z\alpha)^2\frac{m_r^2}{mM} \widetilde E_F.
\eeq

As usual this logarithmic contribution is state-independent.
Calculation of the nonlogarithmic contribution turned out to be a much
more complicated task and the whole machinery of the relativistic
two-particle equations was used in this work. First, the difference
$\Delta E(1S)-8\Delta E(2S)$ was calculated \cite{stern}, then some
nonlogarithmic contributions of this order for the $1S$ level were
obtained \cite{cp78}, and only later the total nonlogarithmic
contribution \cite{byg,bygrev} was obtained

\beq                      \label{nonlogrechfs}
\Delta E=\left(-8\ln2+\frac{65}{18}\right)(Z\alpha)^2\frac{m_r^2}{mM}
\widetilde E_F.
\eeq

This result was later confirmed in a purely numerical calculation
\cite{caswelllep} in the framework of NRQED.

Recoil contributions in \eq{logrechfs}, and \eq{nonlogrechfs} are symmetric
with respect to masses of the light and heavy particles. As in the case of
the leading recoil correction, they were obtained without expansion in
the mass ratio, and hence an exact dependence on the mass ratio is
known (not just the first term in the expansion over $m/M$). Let us
mention that while for the nonrecoil nonlogarithmic contributions of
order $(Z\alpha)^6$, both to HFS and the Lamb shift, only numerical
results were obtained, the respective recoil contributions are known
analytically in both cases (compare discussion of the Lamb shift
contributions in Section \ref{elkfootn}).

\subsection{Recoil Corrections of Order
$(Z\alpha)^3(\lowercase{m}/M)\widetilde E_F$}

\begin{figure}[ht]
\centerline{\epsfig{file=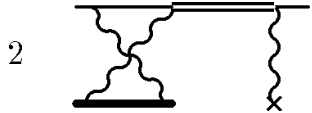,height=1.5cm}}
\vspace{0.5cm}
\caption{Leading logarithm squared
contribution of order $(Z\alpha)^3(\lowercase{m}/M)\widetilde E_F$ }
\label{lag2za3hfsfig}
\end{figure}

There are two different double logarithm contributions of order
$(Z\alpha)^3(m/M)\widetilde E_F$, one contains a regular low-frequency
logarithm squared, and the second depends on the product of the
low-frequency logarithm  and the logarithm of the mass ratio. Calculation
of these contributions is quite straightforward and goes along the same
lines as calculation of the leading logarithmic contribution of order
$\alpha^2(Z\alpha)^6$ to the Lamb shift (see Section \ref{lambcube}).
Taking as one of the perturbation potentials the potential corresponding to
the logarithmic recoil contribution of order $(Z\alpha)^5$ to the Lamb shift
in \eq{totrecza5lamb} and as the other the potential responsible for
the main Fermi contribution to HFS in \eq{fermipothfseff} (see Fig.\
\ref{lag2za3hfsfig}), one obtains \cite{kar93} a small logarithm squared
contribution

\beq
\Delta
E=-\frac{2}{3}\frac{(Z\alpha)^3}{\pi}\frac{m_r^2}{mM}\ln^2(Z\alpha)^{-1}
E_F.
\eeq

\begin{figure}[ht]
\centerline{\epsfig{file=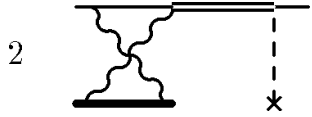,height=1.5cm}}
\vspace{0.5cm}
\caption{Leading mixed double logarithm
contribution of order $(Z\alpha)^3(\lowercase{m}/M)\widetilde E_F$ }
\label{lag2za3mixhfsfig}
\end{figure}

A significantly larger double mixed logarithm correction is generated
by the potential corresponding to the leading recoil correction to hyperfine
splitting in \eq{arnnewcsalp}  and the leading logarithmic Dirac
correction to the Coulomb-Schr\"odinger wave function
\cite{kn1,kn10,kar96} (see Fig.\ \ref{lag2za3mixhfsfig})

\beq       \label{mixedloghfs}
\Delta
E=-{3}\frac{(Z\alpha)^3}{\pi}\frac{m}{M}\ln(Z\alpha)^{-1}
\ln\frac{M}{m} {\widetilde E}_F.
\eeq

Single-logarithmic recoil corrections of relative order $(Z\alpha)^3$
as well as an estimate of nonlogarithmic contributions were also
obtained recently \cite{kin98}

\beq
\Delta
E_{log}=\left[-2C_S+\frac{32}{3}\left(-\ln2+\frac{3}{4}\right)\right]
\frac{(Z\alpha)^3}{\pi}\frac{m}{M}\ln(Z\alpha)^{-1}E_F,
\eeq

\noindent
where

\beq
C_S=-\frac{1}{9}+\frac{7}{3}(2\ln2+3)-\frac{2}{1-\left(\frac{m}{M}\right)^2}
\left[\ln\left(1+\frac{m}{M}\right)-\left(\frac{m}{M}\right)^2
\ln\left(1+\frac{M}{m}\right)\right]
\eeq

\noindent
and

\beq
\Delta
E_{nonlog}=(40\pm 22)
\frac{(Z\alpha)^3}{\pi}\frac{m}{M}\ln(Z\alpha)^{-1}E_F.
\eeq

The error bars of the last result are still rather large, and more
accurate calculation would be necessary in pursuit of all corrections
of the scale of $10$ Hz. The relatively large magnitude of the mixed
logarithm contribution in \eq{mixedloghfs} warrants calculation of the
contribution which is linear in the logarithm of the mass ratio and
nonlogarithmic in $Z\alpha$.  All recoil corrections are collected in
Table XVII. The  uncertainty of the total recoil correction in the last
line of this Table includes an estimate of the magnitude of the yet
unknown recoil contributions.

\begin{center}
\underline{Table XVII. Recoil Corrections}
\nopagebreak

\begin{tabular}{|l|rl|c|}    \hline
     & $\widetilde E_F$ &   & kHz
\\ \hline  \hline
Leading recoil correction &&$$& \\
&&$$& \\
Arnowitt
(1953)\cite{arn}&&&\\
Fulton,Martin(1954)\cite{fm}&$-\frac{3(Z\alpha)}{\pi}\frac{mM}{M^2-m^2}
\ln\frac{M}{m}$&&$-800.304$  \\
Newcomb,Salpeter(1955)\cite{newsalp}&&&
\\&&&
\\ \hline
Leading logarithmic recoil correction, &&$$& \\
relative order $(Z\alpha)^2$&&$$& \\
Lepage (1977)\cite{lep}&$$&&$$\\
Bodwin, Yennie (1978)\cite{by}&$2(Z\alpha)^2\frac{m_r^2}{mM}
\ln(Z\alpha)^{-1}$&&$11.179$
\\
&&& \\
\hline
Nonlogarithmic recoil correction, &&$$& \\
relative order $(Z\alpha)^2$&&$$& \\
&&& \\
Bodwin, Yennie, &$$& & $$ \\
Gregorio (1982)\cite{byg,bygrev}&$(Z\alpha)^2
\frac{m_r^2}{mM}(-8\ln
2+\frac{65}{18})$&&$-2.197$
\\ \hline
Logarithm squared correction, &&$$& \\
relative order $(Z\alpha)^3$&&$$& \\
&&& \\
Karshenboim
(1993)\cite{kar93}&$-\frac{2}{3}\frac{(Z\alpha)^3}{\pi}
\frac{m_r^2}{mM}
\ln^2 (Z\alpha)^{-1}(1+a_\mu)$& & $-0.043$
\\
&&& \\
\hline
Mixed logarithm correction, &&$$& \\
relative order $(Z\alpha)^3$&&$$& \\
&&& \\
Kinoshita, Nio (1994)\cite{kn1,kn10}&& & $$
\\Karshenboim(1996)\cite{kar96}
&$-{3}\frac{(Z\alpha)^3}{\pi}\frac{m}{M}\ln(Z\alpha)^{-1}
\ln\frac{M}{m}$& & $-0.210$ \\
&&& \\
\hline
Single-logarithmic correction, &&$$& \\
relative order $(Z\alpha)^3$&&$$&$-0.257$ \\
&&& \\
Kinoshita(1998)\cite{kin98}&
$-19.621~9\ldots\frac{(Z\alpha)^3}{\pi}\frac{m}{M}\ln(Z\alpha)^{-1}$ & &\\
&&& \\
\hline
Nonlogarithmic correction, &&$$& \\
relative order $(Z\alpha)^3$&&$$& \\
&&& \\
Kinoshita(1998)\cite{kin98}
&$(40\pm22)\frac{(Z\alpha)^3}{\pi}\frac{m}{M}$& & $0.107~(59)$ \\
&&& \\
\hline
\hline
&   $$       & & \\
Total recoil correction 	 &$$& &$-791.714~(80)$\\
\hline
\end{tabular}
\end{center}

\section{~~~~Radiative-Recoil Corrections to HFS} \label{hfsradrec}

\subsection{Corrections of Order $\alpha(Z\alpha)(\lowercase{m}/M)\widetilde
E_F$ and $(Z^2\alpha)(Z\alpha)(\lowercase{m}/M)\widetilde
E_F$}

\begin{figure}[ht]
\centerline{\epsfig{file=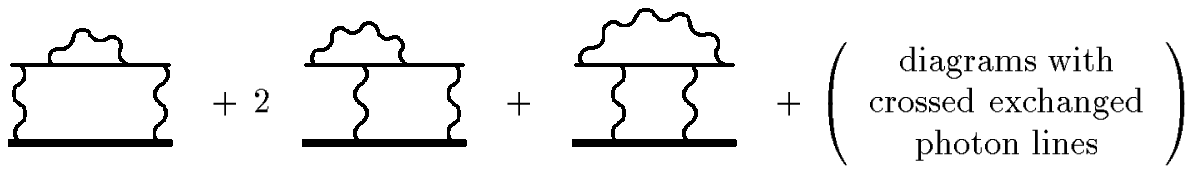,height=2cm}}
\vspace{0.5cm}
\caption{Electron-line radiative-recoil corrections}
\label{ellineradrechfsfir}
\end{figure}

As in the case of the purely radiative corrections of order
$\alpha(Z\alpha)E_F$, all diagrams relevant for calculation of
radiative-recoil corrections of order $\alpha(Z\alpha)(m/M)\widetilde
E_F$ may be obtained by radiative insertions in the skeleton diagrams.
The only difference is that now the heavy particle line is also
dynamical. The skeleton diagrams for this case coincide with the
diagrams for the leading recoil corrections  in Fig.\
\ref{twophothfsfig}. Note that even the leading recoil correction to
HFS may be calculated in the scattering approximation. Insertion of
radiative corrections in the skeleton diagrams emphasizes the high
momenta region even more and, hence, the radiative-recoil correction to
HFS splitting may be calculated in the scattering approximation. The
diagrams for contributions of order $\alpha(Z\alpha)(m/M)\widetilde
E_F$ are presented in Fig.\ \ref{ellineradrechfsfir}, in Fig.\
\ref{muonlineradrechfsfir} and in Fig.\ \ref{photlineradrechfsfir}, and
they coincide topologically with the set of diagrams used for
calculation of the radiative-recoil corrections to the Lamb shift
(formal selection of relevant diagrams and proof of validity of the
scattering approximation based on the relativistic two-particle
equation, see, e.g., in \cite{sty,kes,eksann1}).  We will consider below
separately the corrections generated by the three types of diagrams:
polarization insertions in the exchanged photons, radiative insertions in
the electron line, and radiative insertions in the muon line.

\subsubsection{Electron-Line Logarithmic Contributions of Order
$\alpha(Z\alpha)(\lowercase{m}/M)\widetilde E_F$}

The leading recoil correction to hyperfine splitting is generated in
the broad integration region over exchanged momenta $m\ll k\ll M$, and
one might expect that insertion of radiative corrections in the
skeleton diagrams in Fig.\ \ref{twophothfsfig} would produce double
logarithmic contributions since the radiative insertions are themselves
logarithmic when the characteristic momentum is larger than the
electron mass.  However, this is only partially true, since the sum of
the radiative insertions in the electron line does not have a
logarithmic asymptotic behavior. The simplest way to see this is to
work in the Landau gauge where respective radiative insertions are
nonlogarithmic \cite{blp}. In other gauges, individual radiative
insertions have leading logarithmic terms, but it is easy to see that
due to the Ward identities these logarithmic terms cancel. The first
time this cancellation was observed as a result of direct calculation
in \cite{cp}. In the absence of the leading logarithmic contribution
to the electron factor the logarithmic contribution to HFS is equal to
the product of the leading constant term $-5\alpha/4\pi$ in the
electron factor \cite{eks89} and the leading recoil correction
\eq{arnnewcsalp}, as calculated in \cite{ty}

\beq         \label{electronaslog}
\Delta
E=\frac{15}{4}\frac{\alpha(Z\alpha)}{\pi^2}\frac{m}{M}\ln\frac{M}{m}
{\widetilde E}_F.
\eeq

\subsubsection{Electron-Line Nonlogarithmic Contributions of Order
$\alpha(Z\alpha)(\lowercase{m}/M)\widetilde E_F$}\label{hfsradreceleline}

The validity of the scattering approximation for calculation of  all
radiative and radiative-recoil corrections of order
$\alpha(Z\alpha)\widetilde E_F$ greatly facilitates the calculations.
One may obtain a compact general expression for all  such corrections
(both logarithmic and nonlogarithmic) induced by the radiative
insertions in the electron line in Fig.\ \ref{ellineradrechfsfir} (see,
e.g., \cite{beks})

\begin{equation}      \label{electrgen}
\Delta E^{\text{e-line}} ~=~ \frac{Z\alpha}{\pi} E_F~ \Big(- \frac{3}{16\mu}
\Big) ~ \int \frac{d^4 k}{i \pi^2}~ \frac{1}{(k^2 + i0)^2}~ \biggl(
\frac{1}{k^2 + \mu^{-1}k_0 + i0}
\end{equation}
\[ ~+~ \frac{1}{k^2 - \mu^{-1}k_0 +
i0}   \biggr) \langle \gamma^{\mu} \hat {k} \gamma^{\nu} \rangle _{(\mu)} ~
L_{\mu\nu},
\]

\noindent
where the electron factor $L_{\mu\nu}$ describes all radiative
corrections in Fig.\ \ref{electfacthfsfig}, $\langle \gamma^{\mu} \hat
{k} \gamma^{\nu} \rangle _{(\mu)}$ is the projection of the muon-line
numerator on the spinor structure relevant to HFS, and $\mu=m/(2M)$.

The integral in \eq{electrgen} contains nonrecoil radiative
corrections of order $\alpha(Z\alpha)E_F$, as well as radiative-recoil
corrections of all orders in the electron-muon mass ratio generated by
the radiative insertions in the electron line. This integral admits in
principle a direct brute force numerical calculation. The complicated
structure of the integrand makes analytic extraction of the corrections
of definite order in the mass ratio more involved. Direct application
of the standard Feynman parameter methods leads to integrals for the
radiative-recoil corrections which do not admit expansion of the
integrand over the small mass ratio prior to integration, thus making
the analytic calculation virtually impossible. Analytic results may be
obtained with the help of the approach developed in \cite{eks5} (see
also review in \cite{eksann1}). The idea is to perform integration over
the exchanged momentum directly in spherical coordinates.  Following
this route, we come to the expression of the type

\begin{equation}  \label{polphi}
\Delta E ~
=\alpha (Z\alpha) E_F~
\int_0^1 {dx} \int_0^x {dy} \int_0^{\infty} {d k^2} ~
\frac{\Phi (k,x,y)}{~(k^2 + a^2)^2 ~-~ 4 \mu^2 b^2 k^4~},
\end{equation}

where $a(x,y)$, $b(x,y)$, and  $\Phi(k)$ are explicitly known functions.

The crucial property of the integrand in Eq.\ (\ref{polphi}), which
facilitates calculation, is that the denominator admits  expansion in
the small parameter $\mu$ prior to momentum integration. This is true
due to the inequality  $4\mu^2b^2k^4/(k^2+a^2)^2\leq 4\mu^2$, which is
valid according to the definitions of the functions $a$ and $b$. In
this way, we may easily reproduce the nonrecoil skeleton integral in
\eq{skel}, and obtain once again the nonrecoil corrections induced by
the radiative insertions in the electron line \cite{kp,kks1,kk}. This
approach admits also an analytic calculation of the radiative-recoil
corrections of the first order in the mass ratio.

Nonlogarithmic radiative-recoil corrections to HFS were first
calculated numerically in the Yennie gauge \cite{stylett,sty} and then
analytically in the Feynman gauge \cite{eks5}

\beq \label{radrechfs}
\Delta
E=\left(6\zeta(3)+3\pi^2\ln2+\frac{\pi^2}{2}+\frac{17}{8}\right)
\frac{\alpha(Z\alpha)}{\pi^2}\frac{m}{M}\ln\frac{M}{m} {\widetilde E}_F,
\eeq

\noindent
where $\zeta(3)$ is the Riemann $\zeta$-function.

This expression contains all characteristic structures ($\zeta$-function,
$\pi^2\ln2$, $\pi^2$ and a rational number) which one usually
encounters in the results of the loop calculations. Let us emphasize
that the relative scale of these subleading terms is rather large, of
order $\pi^2$, which is  just what one should expect for the constants
accompanying the large logarithm.

Numerically there is a certain discrepancy between the analytic result
in the Feynman gauge \cite{eks5}
$(6\zeta(3)+3\pi^2\ln2+{\pi^2}/{2}+{17}/{8})/\pi^2=3.526$ and the
numerical result in the Yennie gauge \cite{sty} $3.335\pm0.058$. When
both works were completed this discrepancy which is as large as
three standard deviations of the accuracy of the numerical integration
in \cite{sty} was purely academic. But nowadays, when the accuracy of
the experimental data has achieved $0.053$ kHz, the discrepancy of
about $0.22$ kHz has a phenomenological significance. In order to
resolve this discrepancy, an independent analytic calculation of the
electron-line contribution in the Yennie gauge was undertaken
\cite{beks}. Let us emphasize that despite being partially performed by
the same authors as \cite{eks5}, this new work was logically
independent of \cite{eks5}. It was performed in the Yennie gauge, and
the expression for the electron factor from \cite{stylett,sty} was used
as the initial point of the calculation.  The result of \cite{beks}
confirmed the earlier analytic result in \cite{eks5}, and thus we are
convinced that the discrepancy mentioned above is resolved in favor of
the result in \eq{radrechfs}\footnote{See one more comment on this
discrepancy below in Section \ref{secordermassradrec} where the
radiative-recoil correction of order $\alpha(Z\alpha)(m/M)^2E_F$ is
discussed.}.

\subsubsection{Muon-Line Contribution of Order
$(Z^2\alpha)(Z\alpha)(\lowercase{m}/M)\widetilde E_F$}

\begin{figure}[ht]
\centerline{\epsfig{file=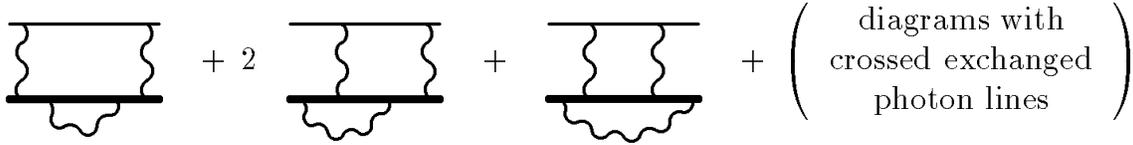,height=1.9cm}}
\vspace{0.5cm}
\caption{Muon-line radiative-recoil corrections}
\label{muonlineradrechfsfir}
\end{figure}

Radiative-recoil correction to HFS generated by the diagrams in Fig.\
\ref{muonlineradrechfsfir} with insertions of the radiative photons in
the muon line is given by an expression similar to the one in
\eq{electrgen} \cite{eks6,eksann1}, the only difference is that now we
have to insert the muon factor instead of the heavy line numerator and
to preserve the skeleton electron-factor numerator. Unlike the case of
the electron line, radiative insertions in the heavy muon line do not
generate nonrecoil corrections.  This is easy to realize if one recalls
that the nonrecoil electron factor contribution is generated by the
muon pole, which is absent in the diagrams with two exchanged photons
and the muon-line fermion factor (muon anomalous magnetic moment is
subtracted from the muon factor, since in the same way as in the case
of the electron factor it generates corrections of lower order in
$\alpha$). Radiative insertions in the heavy fermion line do not
generate logarithmic terms \cite{ty} either.  This can be understood
with the help of the low energy theorem for the Compton scattering.
The effective momenta in the integral for the radiative-recoil
corrections are smaller than the muon mass and, hence, the muon-line
factor enters the integral in the low momenta limit. The classical low
energy theorem for Compton scattering cannot be used directly in this
case since the exchanged photons are virtual, but nevertheless it is
not difficult to prove the validity of a generalized low energy theorem
in this case \cite{ty,eksann1}.  Then we see that the logarithmic
skeleton integrand gets an extra factor $k^2$ after insertion of the
muon factor in the integrand, and this extra factor changes the
logarithmic nature of the integral.

Analytic calculations of the muon-line radiative-recoil correction are
carried out in the same way as in the electron-line case and the purely
numerical \cite{stylett,sty} and analytic \cite{eks6,eksann1} results  for
this contribution are in excellent agreement

\beq
\Delta
E=\left(\frac{9}{2}\zeta(3)-3\pi^2\ln2+\frac{39}{8}\right)
\frac{(Z^2\alpha)(Z\alpha)}{\pi^2}\frac{m}{M}\ln\frac{M}{m} {\widetilde
E}_F.
\eeq

\subsubsection{Leading Photon-Line Double Logarithmic Contribution of Order
$\alpha(Z\alpha)(\lowercase{m}/M)\widetilde E_F$}

The only double-logarithmic radiative-recoil contribution of order
$\alpha(Z\alpha)(m/M)\widetilde E_F$ is generated by the leading logarithmic
term in the polarization operator.  Substitution of this leading
logarithmic term in the logarithmic skeleton integral in \eq{rechfssk}
immediately leads to the double logarithmic contribution \cite{cp}

\beq    \label{logsqarpol}
\Delta E=-2\frac{\alpha(Z\alpha)}{\pi^2}\frac{m}{M}\ln^2\frac{M}{m}
{\widetilde E}_F.
\eeq

As was noted in \cite{eks89} this contribution may be obtained without
any calculations at all. It is sufficient to realize that with
logarithmic accuracy the characteristic momenta in the leading recoil
correction in \eq{rechfssk} are of order $M$ and, in order to account
for the leading logarithmic contribution generated by the polarization
insertions, it is sufficient to substitute in \eq{arnnewcsalp} the
running value of $\alpha$ at the muon mass instead of the fine
structure $\alpha$. This algebraic operation immediately reproduces the
result above.

\subsubsection{Photon-Line Single-Logarithmic and Nonlogarithmic
Contributions of Order $\alpha(Z\alpha)(\lowercase{m}/M)\widetilde E_F$}

\begin{figure}[ht]
\centerline{\epsfig{file=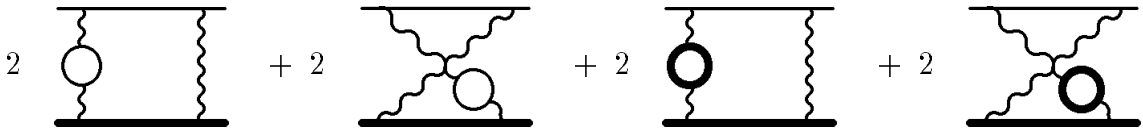,height=1.7cm}}
\vspace{0.5cm}
\caption{Photon-line radiative-recoil corrections}
\label{photlineradrechfsfir}
\end{figure}

Calculation of the nonrecoil radiative correction of order
$\alpha(Z\alpha)\widetilde E_F$ was facilitated by simultaneous
consideration of the electron and muon loop polarization insertions in the
exchanged photons. Similarly calculation of the radiative-recoil
corrections generated by the diagrams in Fig.\ \ref{photlineradrechfsfir}
with insertions of the vacuum polarizations, is technically simpler if one
considers simultaneously both electron and muon vacuum polarizations. All
corrections may be obtained by substituting the explicit expression for the
sum of vacuum polarizations in the skeleton integral \eq{rechfssk}. In this
skeleton integral, part of the recoil correction corresponding to the factor
$1/(1+m/M)$ is subtracted and this explains why we have restored this factor
in consideration of the nonrecoil part of the vacuum polarization.
Technically, consideration of the sum of the electron and muon vacuum
polarizations leads to simplification of the integrand for the
radiative-recoil corrections, and after an easy calculation one
obtains the single-logarithmic and nonlogarithmic contributions to the
total radiative-recoil correction induced by the sum of the electron
and muon vacuum polarizations \cite{ty,sty}

\beq
\Delta
E=\left(-\frac{8}{3}\ln\frac{M}{m}-\frac{28}{9}-\frac{\pi^2}{3}\right)
\frac{\alpha(Z\alpha)}{\pi^2}\frac{M}{m}
{\widetilde E}_F.
\eeq

Note that in the parenthesis we have parted with our usual practice of
considering the muon as a particle with charge $Ze$, and assumed $Z=1$.
Technically this is inspired by the cancellation of certain contributions
between the electron and muon polarization loops mentioned above, and from
the physical point of view it is not necessary to preserve a nontrivial
factor $Z$ here, since we need it only as a reference to an interaction with
the "constituent" muon and not with the one emerging in the polarization
loops.

As was explained above, the coefficient before the leading logarithm
squared term in \eq{logsqarpol} may easily be obtained almost without
any calculations. Simultaneous account for the electron and muon loops
does not effect this contribution, since all logarithmic contributions
are generated only by the insertion of the electron loop. It may be
shown also that the coefficient before the subleading logarithm
originates from the first two terms in the asymptotic expansion of the
polarization operator $2\alpha/(3\pi)\ln(k/m)-5/9$ \cite{eks89}.
Substituting the polarization operator asymptotics in the skeleton
integral in \eq{rechfssk} and, multiplying the result by the factor 2
in order to take into account all possible insertion of the
polarization loop in the exchanged photons, one obtains $-8/3$ for the
coefficient before the single-logarithmic term, in accordance with the
result above. The factor $-6$ comes from the leading logarithmic term
in the polarization operator expansion, and the factor $10/3$ is
generated by the subleading constant, their sum being equal to $-8/3$.

\subsubsection{Heavy Particle Polarization Contributions of Order
$\alpha(Z\alpha)\widetilde E_F$}\label{heavypolmuonhfs}

\begin{figure}[ht]
\centerline{\epsfig{file=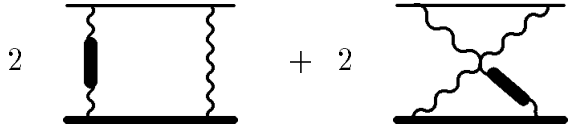,height=1.7cm}}
\vspace{0.5cm}
\caption{Heavy-particle polarization contribution to HFS}
\label{heavypartpolhfsfir}
\end{figure}

The contribution of the muon polarization operator was already
considered above. One might expect that contributions of the diagrams
in Fig.\ \ref{heavypartpolhfsfir} with the heavy particle polarization
loops are of the same order of magnitude as the contribution of the
muon loop, so it is natural to consider this contribution here.
Respective corrections could easily be calculated by substituting the
expressions for the heavy particle polarizations in the unsubtracted
skeleton integral in \eq{rechfssk}.  However, only the polarization
operator of the heavy lepton $\tau$ may be calculated analytically.
Polarization contributions due to the loops of pions and other hadrons
cannot be calculated with the help of the QCD perturbation theory, and
the best approach for their evaluation is to use some low energy
effective theory and experimental data. Respective calculations were
performed in \cite{sty,kf1,kf1} and currently the most accurate result
for the hadron polarization operator contribution to HFS is \cite{kf2}

\beq
\Delta
E=3.598~8~(104~5)\frac{\alpha(Z\alpha)}{\pi^2}\frac{mM}{m_{\pi}^2}
{\widetilde E}_F.
\eeq

\subsection{Leading Logarithmic Contributions of Order
$\alpha^2(Z\alpha)(\lowercase{m}/M)\widetilde E_F$}

\begin{figure}[ht]
\centerline{\epsfig{file=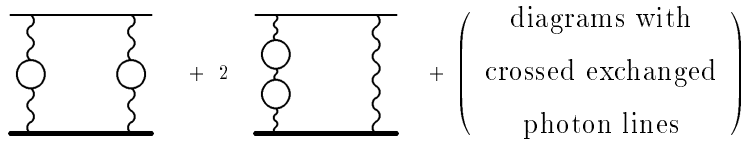,height=1.8cm}}
\vspace{0.5cm}
\caption{Graphs with two one-loop polarization insertions}
\label{onelooppolrecfhsfig}
\end{figure}

\begin{figure}[ht]
\centerline{\epsfig{file=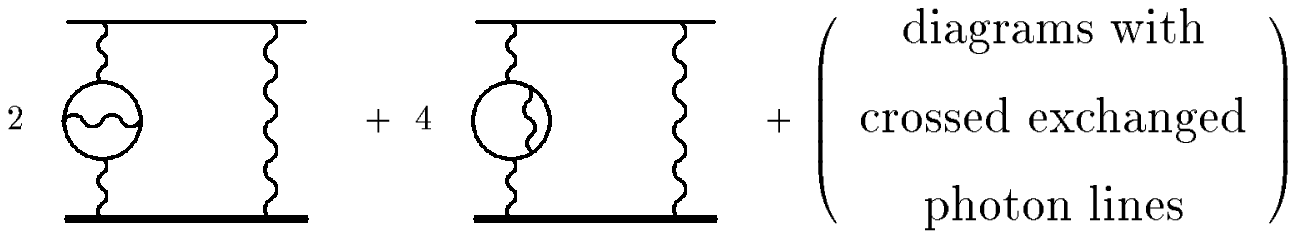,height=1.8cm}}
\vspace{0.5cm}
\caption{Graphs with two--loop polarization insertions}
\label{twolooppolrecfhsfig}
\end{figure}

\begin{figure}[ht]
\centerline{\epsfig{file=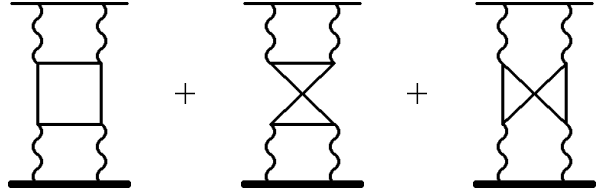,height=2.5cm}}
\vspace{0.5cm}
\caption{Graphs with light by light scattering insertions}
\label{lihlightrechfsfig}
\end{figure}

\begin{figure}[ht]
\centerline{\epsfig{file=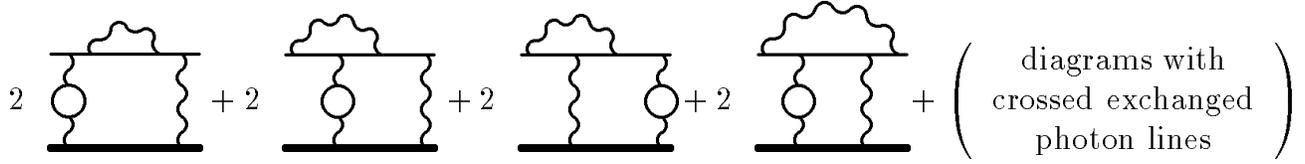,height=2.2cm}}
\vspace{0.5cm}
\caption{Graphs with radiative photon insertions}
\label{ellineinsrechfsfig}
\end{figure}

\begin{figure}[ht]
\centerline{\epsfig{file=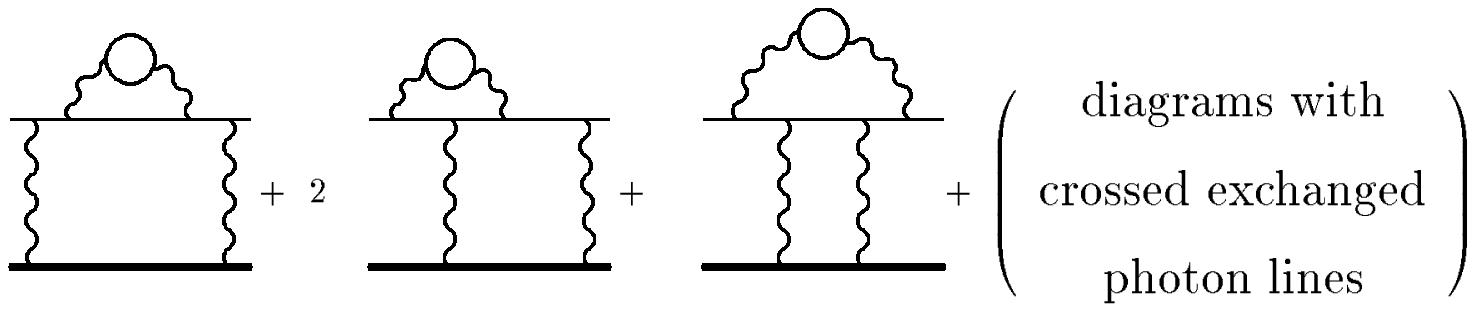,height=2.1cm,width=13cm}}
\vspace{0.5cm}
\caption{Graphs with polarization insertions in the radiative photon}
\label{polellineinsrechfsfig}
\end{figure}

The leading contribution of order $\alpha^2(Z\alpha)(m/M)\widetilde
E_F$ is enhanced by the cube of the large logarithm of the
electron-muon mass ratio. One could expect that logarithm cube terms
would be generated by a few types of radiative insertions in the
skeleton graphs with two exchanged photons:  insertions of the first
and second order polarization operators in the exchanged photons in
Fig.\ \ref{onelooppolrecfhsfig} and in Fig.\ \ref{twolooppolrecfhsfig},
insertions of the light-by-light scattering contributions in Fig.\
\ref{lihlightrechfsfig}, insertions of two radiative photons in the
electron line in Fig.\ \ref{ellineinsrechfsfig}, and insertions of
polarization operator in the radiative photon in Fig.\
\ref{polellineinsrechfsfig}. In \cite{es0}, where the leading logarithm
cube contribution was calculated explicitly, it was shown that only the
graphs in Fig.\ \ref{onelooppolrecfhsfig} with insertions of the
one-loop polarization operators generate logarithm cube terms. This
leading contribution may be obtained without any calculations by simply
substituting the effective charge $\alpha(M)$ defined at the
characteristic scale $M$ in the leading recoil correction of order
$(Z\alpha)(m/M)E_F$ instead of the fine structure constant $\alpha$ and
expanding the resulting expression in the power series over $\alpha$
\cite{eks89} (compare with a similar remark above concerning the
leading logarithm squared term of order $\alpha(Z\alpha)E_F$).

\begin{figure}[ht]
\centerline{\epsfig{file=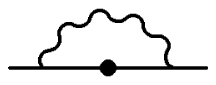,height=1.5cm}}
\vspace{0.5cm}
\caption{Renormalization of the fifth current}
\label{fifthcurhfsfig}
\end{figure}

Calculation of the logarithm squared term of order
$\alpha^2(Z\alpha)(m/M)\widetilde E_F$ is more challenging
\cite{eks89}. All graphs in Figs.\ \ref{onelooppolrecfhsfig},
\ref{twolooppolrecfhsfig}, \ref{lihlightrechfsfig},
\ref{ellineinsrechfsfig}, and \ref{polellineinsrechfsfig} generate
corrections of this order. The contribution induced by the irreducible
two-loop vacuum polarization in Fig.\ \ref{twolooppolrecfhsfig} is
again given by the effective charge expression.  Subleading logarithm
squared terms generated by the one-loop polarization insertions in
Fig.\ \ref{onelooppolrecfhsfig} may easily be calculated with the help
of the two leading asymptotic terms in the polarization operator
expansion and the skeleton integral. An interesting effect takes place
in calculation of the logarithm squared term generated by the
polarization insertions in the radiative photon in Fig.\
\ref{polellineinsrechfsfig}. One might expect that the high energy
asymptote of the electron factor with the polarization insertion is
given by the product of the leading constant term of the electron
factor $-5\alpha/(4\pi)$ and the leading polarization operator term.
However, this expectation turns out to be wrong.  One may check
explicitly that instead of the naive factor above one has to multiply the
polarization operator by the factor $-3\alpha/(4\pi)$. The reason for
this effect may easily be understood.  The factor $-3\alpha/(4\pi)$ is
the asymptote of the electron factor in massless QED and it gives a
contribution to the logarithmic asymptotics only after the polarization
operator insertion. This means that in massive QED the part
$-2\alpha/(4\pi)$ of the constant electron factor originates from the
integration region where the integration momentum is of order of the
electron mass. Naturally this integration region does not give any
contribution to the logarithmic asymptotics of the radiatively
corrected electron factor. The least trivial logarithm squared
contribution is generated by the three-loop diagrams in Fig.\
\ref{lihlightrechfsfig} with the insertions of light by light
scattering block. Their contribution was calculated explicitly in
\cite{eks89}. Later it was realized that these contributions are
intimately connected with the well known anomalous renormalization of
the axial current in QED \cite{kes90}. Due to the projection on the HFS
spin structure in the logarithmic integration region the heavy particle
propagator effectively shrinks to an axial current vertex, and in this
situation calculation of the respective contribution to HFS reduces to
substitution of the well known two-loop axial renormalization factor in
Fig.\ \ref{fifthcurhfsfig} \cite{adler} in the recoil skeleton diagram.
Of course, this calculation reproduces the same contribution as
obtained by direct calculation of the diagrams with light by light
scattering expressions. From the theoretical point of view it is
interesting that having sufficiently accurate experimental data one can
in principle measure anomalous two-loop renormalization of the axial
current in the atomic physics experiment.

The sum of all logarithm cubed and logarithm squared  contributions of
order $\alpha^2(Z\alpha)(m/M)\widetilde E_F$ is given by the expression
\cite{es0,eks89}

\beq
\Delta
E=\left(-\frac{4}{3}\ln^3\frac{M}{m}+\frac{4}{3}\ln^2\frac{M}{m}\right)
\frac{\alpha^2(Z\alpha)}{\pi^3}\frac{m}{M} {\widetilde E}_F.
\eeq

It was also shown in \cite{eks89} that there are no other
contributions with the large logarithm of the mass ratio squared
accompanied by the factor $\alpha^3$, even if the factor $Z$ enters in
another manner than in the equation above.

\begin{figure}[ht]
\centerline{\epsfig{file=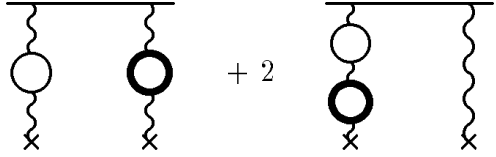,height=2cm}}
\vspace{0.5cm}
\caption{Graphs with simulataneous insertions of the electron
and muon loops}
\label{singllogrechsffig}
\end{figure}

Unlike the logarithm cube and logarithm squared terms which are
generated only by a small number of diagrams discussed above, there are
numerous sources of the single-logarithmic terms. All these terms were
never calculated and only a partial result for one of the
gauge-invariant  sets of such graphs in Fig.\ \ref{singllogrechsffig}
is known now \cite{lse}

\beq
\Delta E=0.645~5~\ln\frac{M}{m}\frac{\alpha^2(Z\alpha)}{\pi}\frac{m}{M}
{\widetilde E}_F.
\eeq

This contribution may be used only as an indication of the scale of
other uncalculated single-logarithmic terms.

\subsection{Corrections of Order
$\alpha(Z\alpha)(\lowercase{m}/M)^2E_F$}\label{secordermassradrec}

Radiative-recoil corrections of order $\alpha(Z\alpha)(m/M)^2E_F$ are
generated by the same set of diagrams in Fig.\
\ref{ellineradrechfsfir}, in Fig.\ \ref{muonlineradrechfsfir} and in
Fig.\ \ref{photlineradrechfsfir} with the radiative insertions in the
electron and muon lines, and with the polarization insertions in the
photon lines, as the respective corrections of the previous order in
the mass ratio.  Analytic calculation proceeds as in that case, the
only difference is that now one has to preserve all contributions which
are of second order in the small mass ratio. It turns out that all such
corrections are generated at the scale of the electron mass, and one
obtains for the sum of all corrections \cite{egs98}

\begin{equation}
\Delta E_{\text{rad-rec}}=
\left[\left(-6  \ln2 ~-~ \frac{3}{4}~\right)~
\alpha (Z\alpha)
- \frac{17}{12}~
(Z^2 \alpha) (Z\alpha)\right] \biggl(\frac{m}{M} \biggr)^2  E_F.
\end{equation}

The electron-line contribution

\begin{equation}   \label{hfssecordmass}
\Delta E_{\text{ el}}=\left(-6  \ln2 ~-~ \frac{3}{2}~\right)
\alpha (Z\alpha) \biggl(\frac{m}{M}\biggr)^2E_F\approx-0.03~\mbox{kHz},
\end{equation}

is of special interest in view of the discrepancy between the results
for the electron line contributions of order $\alpha(Z\alpha)(m/M)E_F$
in \cite{eks5,beks} and in \cite{sty} (see discussion in Section
\ref{hfsradreceleline}). The result for the electron line
contribution in \cite{sty} was obtained without expansion in $m/M$, so
one could try to ascribe the discrepancy to the contribution of these
higher order terms.  However, we see that the contribution of the
second order in the mass ratio in \eq{hfssecordmass} is by far too
 small to explain the discrepancy. We would like to emphasize once
again that the coinciding results in \cite{eks5} and \cite{beks} were
obtained completely independently in different gauges, so there is
little, if any, doubt that this result is correct.

\subsection{Corrections of Order
$\alpha(Z\alpha)^2(\lowercase{m}/M)E_F$}\label{hfsleadlogradrecza2}

\begin{figure}[h]
\centerline{\epsfig{file=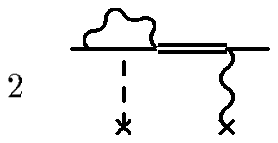,height=2cm}}
\vspace{0.5cm}
\caption{Leading logarithm squared
contribution of order $\alpha(Z\alpha)^2 E_F$ }
\label{leadlogaza2rechfsfir}
\end{figure}

Radiative-recoil corrections of order $\alpha(Z\alpha)^2(m/M)E_F$ were
never calculated completely. As we have mentioned in Section
\ref{hfselalzal2} the leading logarithm squared contribution of order
$\alpha(Z\alpha)^2E_F$  may easily be calculated if one takes as one of
the perturbation potentials the potential corresponding to the electron
electric form factor and as the other the potential responsible for the
main Fermi contribution to HFS (see Fig.\ \ref{leadlogaza2rechfsfir}).
Then one obtains the leading logarithm squared contribution in the form
\cite{kar93}

\beq           \label{redmassfact}
\Delta
E=-\frac{2}{3}\ln^2(Z\alpha)^{-2}\frac{\alpha(Z\alpha)^2}{\pi}
\left(\frac{m_r}{m}\right)^2
E_F,
\eeq

\noindent
which differs from the leading logarithm squared term in \eq{leadloghfsaza2}
by the recoil factor $(m_r/m)^2$. Preserving only the linear term in the
expansion of this result over the mass ratio one obtains

\beq           \label{redmassfactexp}
\Delta
E=\frac{4}{3}\ln^2(Z\alpha)^{-2}\frac{\alpha(Z\alpha)^2}{\pi}\frac{m}{M}E_F.
\eeq

Numerically this contribution is about $0.3$ kHz, and clearly has to be
taken into account in comparison of the theory with the experimental data.
In this situation it is better simply to use in the theoretical formulae the
leading logarithm squared contribution of order $\alpha(Z\alpha)^2E_F$ in
the form in \eq{redmassfact} instead of the expression for this logarithmic
term in \eq{leadloghfsaza2}.

Similar single-logarithmic

\beq
\Delta
E=\frac{16}{3}\left(\ln2-\frac{3}{4}\right)
\ln(Z\alpha)^{-2}\frac{\alpha(Z\alpha)^2}{\pi}\frac{m}{M}E_F
\eeq

\noindent
and nonlogarithmic

\beq
\Delta
E=-4(10\pm2.5)\frac{\alpha(Z\alpha)^2}{\pi}\frac{m}{M}E_F
\eeq

\noindent
radiative-recoil corrections of order  $\alpha(Z\alpha)^2(m/M)E_F$
associated with the reduced mass factors in the nonrecoil corrections of
order $\alpha(Z\alpha)^2E_F$ were obtained recently \cite{kin98}.

The relatively large magnitude of the correction in \eq{redmassfactexp}
demonstrates that a calculation of all radiative-recoil corrections of
order $\alpha(Z\alpha)^2(m/M)E_F$ is warranted. The error of the total
radiative-recoil correction in the last line in Table XVIII includes,
besides the errors of individual contributions  in the upper lines of
this Table, also an educated guess on the magnitude of yet uncalculated
contributions.

\begin{center}
\underline{Table XVIII. Radiative-Recoil Corrections}
\nopagebreak

\begin{tabular}{|l|rl|c|}    \hline
     & $\widetilde E_F$ &   & kHz
\\ \hline  \hline
Leading logarithmic  &&$$& \\
electron-line correction&&$$& \\
&&$$& \\
Terray,Yennie (1982)\cite{ty}&$
\frac{15}{4}\frac{\alpha(Z\alpha)}{\pi^2}\frac{m}{M}\ln\frac{M}{m}$&&$2.324$
\\ \hline
Nonlogarithmic  &&$$& \\
electron-line correction&&$$& \\
&&$$& \\
Eides,Karshenboim,Shelyuto (1986)\cite{eks5} &$(6\zeta(3)+3\pi^2\ln
2+\frac{\pi^2}{2}+\frac{17}{8})\frac{\alpha(Z\alpha)}{\pi^2}
\frac{m}{M}$& & $4.044$ \\
\hline
Muon-line contribution&&$$& \\
&&$$& \\
Eides,Karshenboim,Shelyuto (1988)\cite{eks6} &$(\frac{9}{2}\zeta(3)
-3\pi^2\ln 2+\frac{39}{8})\frac{(Z^2\alpha)(Z\alpha)}{\pi^2}
\frac{m}{M}$& & $$ $-1.190$
\\ \hline
Logarithm squared  &&$$& \\
polarization contribution&&$$& \\
&&$$& \\
Caswell,Lepage (1978)\cite{cp}
&$-2\ln^2\frac{M}{m}\frac{\alpha(Z\alpha)}{\pi^2}
\frac{m}{M}$&&$-6.607$\\
\hline
Single-logarithmic and nonlogarithmic  &&$$& \\
polarization contributions&&$$& \\
&&$$& \\
Terray,Yennie (1982)\cite{ty}
&$(-\frac{8}{3}\ln\frac{M}{m}-\frac{\pi^2}{3}-\frac{28}{9})
\frac{\alpha(Z\alpha)}{\pi^2}\frac{m}{M}$&&$-2.396$\\
\hline
Hadron  polarization contribution&&$$& \\
&&$$& \\
Faustov&&$$& \\
Karimkhodzhaev,Martynenko (1999)\cite{kf2}
&$3.599(105)\frac{\alpha(Z\alpha)}{\pi^2}\frac{mM}{m_{\pi}^2}$&&$0.240(7)$\\
\hline
Leading log cube correction &&$$& \\
&&$$& \\
Eides, Shelyuto (1984)\cite{es0}&$-\frac{4}{3}\frac{\alpha^2
(Z\alpha)}{\pi^3}\frac{m}{M}\ln^3\frac{M}{m}$& & $-0.055$ \\
\hline
Log square correction &&$$& \\
&&$$& \\
Eides,Karshenboim,Shelyuto (1989)\cite{eks89}
&$\frac{4}{3}\frac{\alpha^2(Z\alpha)}{\pi^3}\frac{m}{M}
\ln^2\frac{M}{m}$& & $0.010$ \\
\hline
One of linear in log corrections &&$$& \\
&&$$& \\
Li,Samuel,Eides (1993)\cite{lse} &$0.645~5~\frac{\alpha^2(Z\alpha)}{\pi}
\frac{m}{M}\ln\frac{M}{m}$& & $0.009$
\\ \hline
Second order in mass ratio &$$&&$$\\
contribution&$$&&$$\\
&$$&&$$\\
Eides,Grotch,Shelyuto (1998)\cite{egs98} &$[(-6\ln2-\frac{3}{4})
\alpha(Z\alpha)-\frac{17}{12}(Z^2\alpha)(Z\alpha))](\frac{m}{M})^2$& &
$-0.035$
\\
\hline
Leading logarithmic $\alpha(Z\alpha)^2$ &&$$& \\
radiative-recoil correction&$$&&$$\\
&$$&&$$\\
Karshenboim (1993)\cite{kar93}&$\frac{4}{3}\frac{\alpha(Z\alpha)^2}{\pi}
\frac{m}{M}\ln^2(Z\alpha)^{-2}$& & $0.344$
\\
\hline
\end{tabular}
\end{center}

\begin{center}
\underline{Table XVIII. Radiative-Recoil Corrections (Continuation)}
\nopagebreak

\begin{tabular}{|l|rl|c|}    \hline
     & $\widetilde E_F$ &   & kHz
\\ \hline
Single-logarithmic $\alpha(Z\alpha)^2$ &&$$& \\
radiative-recoil correction&$$&&$$\\
&$$&&$$\\
Kinoshita (1998)\cite{kin98}&$\frac{16}{3}(\ln2-\frac{3}{4})
\frac{\alpha(Z\alpha)^2}{\pi}
\frac{m}{M}\ln(Z\alpha)^{-2}$& & $-0.008$\\
\hline
Nonlogarithmic $\alpha(Z\alpha)^2$ &&$$& \\
radiative-recoil correction&$$&&$$\\
&$$&&$$\\
Kinoshita (1998)\cite{kin98}&$-4(10\pm2.5)\frac{\alpha(Z\alpha)^2}{\pi}
\frac{m}{M}$& & $-0.107(27)$\\
\hline
\hline
&   $$       & & \\
Total radiative-recoil correction  &$$& &$-3.427~(70)$\\
\hline
\end{tabular}
\end{center}

\part{Weak Interaction Contribution}

The weak interaction contribution to hyperfine splitting is due to
$Z$-boson exchange between the electron and muon in Fig.\
\ref{zbosonlambfig}. Due to the large mass of the $Z$-boson this exchange is
effectively described by the local four-fermion interaction Hamiltonian

\beq
H_{int}=-\frac{1}{2M^2_Z}\int d^3x~({\bf j\cdot j}),
\eeq

\noindent
where ${\bf j}$ is the spatial part of the weak current $(j^0,{\bf
j})$, weak charge is included in the definition of the weak current,
and $M_Z$ is the $Z$-boson mass.

The weak interaction contribution to HFS was calculated many years ago
\cite{ba,bf}, and even radiative corrections to the leading term were
discussed in the literature \cite{repko,grotch,alcotra}. However, quite
recently it turned out that the weak contribution to HFS is cited in
the literature with different signs \cite{by,eid,kn1,kn10}.  This
happened probably because the weak correction was of purely academic
interest for early researchers. This discrepancy in sign was subjected
to scrutiny in a number of recent works
\cite{kn1,kn10,sapirstein,eides96} which all produced the result in
agreement with \cite{bf}

\beq  \label{wekahfscontr}
\Delta E=-\frac{G_F}{\sqrt 2} \frac{3mM}{4\pi Z\alpha} E_F
\approx -0.065~{\rm kHz}.
\eeq

\part{Hyperfine Splitting in Hydrogen}


Hyperfine splitting in the ground state of hydrogen is one of the most
precisely measured quantities in modern physics \cite{hvl,ed} (see
for more details Section \ref{hydhfsexp} below), and to describe it
theoretically we need to consider  additional contributions to HFS
connected with the bound state nature of the proton.

Dominant nonrecoil contributions to the hydrogen hyperfine splitting
are essentially the same as in the case of muonium. The only
differences are that now in all formulae the proton mass
replaces the muon mass, and we have to substitute the proton anomalous
magnetic moment $\kappa=1.792~847~386~(63)$ measured in nuclear
magnetons instead of the muon anomalous magnetic moment $a_\mu$
measured in the Bohr magnetons in the expression for the hydrogen Fermi
energy in \eq{ef}.  After this substitution one can use the nonrecoil
corrections collected in Tables XII-XVI for the case of hydrogen.

As in the case of the Lamb shift the composite nature of the nucleus
reveals itself first of all via a relatively large finite size
correction. It is also necessary to reconsider all recoil and
radiative-recoil corrections. Due to existence of the proton
anomalous magnetic moment and nontrivial proton form factors,
simple minded insertions of the hydrogen Fermi energy instead of the
muonium Fermi energy in the muonium expressions for these corrections
leads to the wrong results. As we have seen in Sections
\ref{recoilhfsmuon}-\ref{hfsradrec} leading recoil and radiative-recoil
corrections originate from distances small on the atomic scale,
between the characteristic Compton lengths of the heavy nucleus $1/M$
and the light electron $1/m$. Proton contributions to hyperfine
splitting coming from these distances cannot be satisfactory described
only in terms of such global characteristics as its electric and
magnetic form factors. Notice that the leading recoil and
radiative-recoil contributions to the Lamb shift are softer, they come
from larger distances (see Sections
\ref{lambreczalpha5}-\ref{recgenaza6}), and respective formulae are
valid both for elementary and composite nuclei. Theoretical
distinctions between the case of elementary and composite nucleus are
much more important for HFS than for the Lamb shift.

Despite the difference between the two cases, discussion of the
proton size and structure corrections to HFS in hydrogen below is in
many respects parallel to the discussion of the respective corrections
to the Lamb shift in Chapter \ref{nonelectromagnetic}.

\section{~~~~Nuclear Size, Recoil and Structure Corrections of Orders
$(Z\alpha)E_F$ and $(Z\alpha)^2E_F$} \label{hfssizestr}

Nontrivial nuclear structure beyond the nuclear anomalous magnetic
moment first becomes important for corrections to hyperfine splitting
of order  $(Z\alpha)^5$ (compare respective corrections to the Lamb
shift in Section \ref{lambsizestr}). Corrections of order
$(Z\alpha)E_F$ connected with the nonelementarity of the nucleus are
generated by the diagrams with two-photon exchanges.  Insertion of the
perturbation corresponding to the magnetic or electric form factors in
one of the legs of the skeleton diagram in Fig.\ \ref{skelhfs}
described by the infrared divergent integral in \eq{skel} makes the
integral infrared convergent and pushes characteristic integration
momenta to the high scale determined by the characteristic scale of the
hadron form factor.  Due to the composite nature of the nucleus,
besides intermediate elastic nuclear states, we also have to consider
the contribution of the diagrams with inelastic intermediate states.

We will first consider the contributions generated only by the elastic
intermediate nuclear states. This means that calculating this
correction we will treat the nucleus as a particle which interacts with
the photons via its nontrivial Sachs electric and magnetic form
factors in \eq{sachsform}.

\subsection{Corrections of Order $(Z\alpha)E_F$}

\subsubsection{Correction of Order
$(Z\alpha)(\lowercase{m}/\Lambda)E_F$ (Zemach
Correction)}\label{zemchhfsprot}

As usual we start consideration of the contributions of order
$(Z\alpha)E_F$ with the infrared divergent integral \eq{skel}
corresponding to the two-photon skeleton diagram in Fig.\
\ref{skelhfs}. Insertion of factors $G_E(-k^2)-1$ or
$G_M(-k^2)/(1+\kappa)-1$ in one of the external proton legs corresponds
to the presence of a nontrivial proton form factor\footnote{Subtraction
is necessary in order to avoid double counting since the diagrams with
the subtracted term correspond to the pointlike proton contribution,
already taken into account in the expression for the Fermi energy in
\eq{ef}.}.

\begin{figure}[h]
\centerline{\epsfig{file=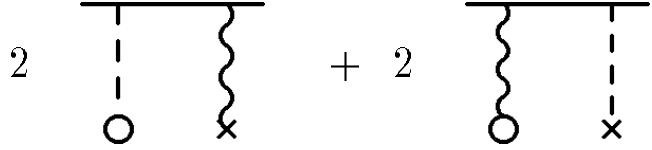,height=2cm}}
\vspace{0.5cm}
\caption{Elastic nuclear size corrections of order
$(Z\alpha)E_F$ with one form factor insertion. Empty dot
corresponds either to $G_E(-k^2)-1$ or $G_M(-k^2)/(1+\kappa)-1$}
\label{hfsffeltwophotonefig}
\end{figure}

\begin{figure}[ht]
\centerline{\epsfig{file=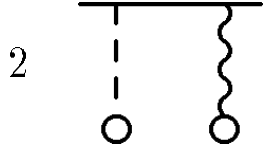,height=2cm}}
\vspace{0.5cm}
\caption{Elastic nuclear size correction of order
$(Z\alpha)E_F$ with two form factor insertions. Empty dot
corresponds either to $G_E(-k^2)-1$ or $G_M(-k^2)/(1+\kappa)-1$}
\label{hfsffeltwophottwofig}
\end{figure}

We need to consider diagrams in Fig.\ \ref{hfsffeltwophotonefig} with
insertion of one factor $G_{E(M)}(-k^2)-1$ in the proton
vertex\footnote{Dimensionless integration momentum in \eq{skel} is
measured in electron mass. We return here to dimensionful integration
momenta, which results in an extra factor $m$ in the numerators in
\eq{hfsoneff} and \eq{hfstwoff}. Notice also the minus sign before the
momentum in the arguments of form factors, it arises because in the
equations below $k=|{\bf k}|$.}

\beq            \label{hfsoneff}
\Delta E=\frac{8(Z\alpha)m}{\pi n^3}E_F
\int_0^\infty\frac{dk}{k^2}\left\{\left[G_E(-k^2)-1\right]
+\left[\frac{G_M(-k^2)}{1+\kappa}-1\right]\right\},
\eeq

\noindent
and the diagram in Fig.\ \ref{hfsffeltwophottwofig} with
simultaneous insertion of both factors $G_E(-k^2)-1$ and $G_M(-k^2)-1$
in two proton vertices

\beq                 \label{hfstwoff}
\Delta E=\frac{8(Z\alpha)m}{\pi n^3}E_F
\int_0^\infty\frac{dk}{k^2}\left[G_E(-k^2)-1\right]
\left[\frac{G_M(-k^2)}{1+\kappa}-1\right].
\eeq

Effectively the integration in \eq{hfstwoff}, and \eq{hfsoneff} goes up
to the form factor scale. This scale is much higher than the electron
mass and high momenta, in this section, mean momenta much higher than
the electron mass. In earlier sections high momenta often meant momenta
of the scale of the electron mass.

The total proton size dependent contribution of order $(Z\alpha)E_F$,
which is often called the Zemach correction, has the form

\beq   \label{hfszema}
\Delta E=\frac{8(Z\alpha)m}{\pi n^3}E_F
\int_0^\infty\frac{dk}{k^2}\left[\frac{G_E(-k^2)G_M(-k^2)}
{1+\kappa}-1\right],
\eeq

\noindent
or in the coordinate space \cite{zemach}

\beq  \label{hfsbortrfr}
\Delta E=-2(Z\alpha)m<r>_{(2)}E_F,
\eeq

\noindent
where $<r>_{(2)}$ is the first Zemach moment (or the Zemach radius)
\cite{zemach}, defined via the weighted convolution of the electric and
magnetic densities $\rho_{E(M)}(r)$ corresponding to the respective
form factors (compare with the third Zemach moment in \eq{thirdzem})

\beq
<r>_{(2)}\equiv\int d^3rd^3{\tilde r}\rho_E(r)\rho_M({\tilde r})
|{\bf r}+{\tilde{\bf r}}|.
\eeq

Parametrically the result in \eq{hfsbortrfr} is of order
$(Z\alpha)(m/\Lambda)E_F$, where $\Lambda$ is the form factor scale.
This means that this correction should be considered together with
other recoil corrections, even though it was obtained from a nonrecoil
skeleton integral.

The simple coordinate form of the result in \eq{hfsbortrfr} suggests
that it might have an intuitively clear interpretation. This is the
case and the expression for the Zemach correction was originally
derived from simple nonrelativistic quantum mechanics without any field
theory \cite{zemach}. Let us describe the main steps of this quite
transparent derivation. Recall first that the main Fermi contribution
to hyperfine splitting in \eq{ef} is simply a matrix element of the
one-photon exchange which, due to the local nature of the magnetic
interaction, is simply proportional to the value of the Schr\"odinger
wave function squared at the origin $|\psi(0)|^2$. However, the nuclear
magnetic moment is not pointlike, but distributed over a finite
region described by the magnetic moment density $\rho_M(r)$.  This
effect can be taken into account in the matrix element for the leading
contribution to HFS  with the help of the obvious substitution

\beq  \label{psi0spr}
|\psi(0)|^2\rightarrow \int d^3r\rho_M(r)|\psi(r)|^2.
\eeq

Hence, the correction to the leading contribution to HFS depends
on the behavior of the bound-state wave function near the origin. The
ordinary Schr\"odinger-Coulomb wave function for the ground state
behaves near the origin as $\exp(-m_rZ\alpha r)\approx
1-m_rZ\alpha r$. For a very short-range nonlocal source of the
electric field, the wave function behaves as

\beq        \label{elndistr}
\psi(r)\approx 1-{m_rZ\alpha}\int d^3{\tilde r}|{\bf
r}-\widetilde{\bf r}|\rho_E({\tilde {\bf r}}),
\eeq

\noindent
as may easily be checked using the Green function of the Laplacian
operator \cite{zemach}. Substituting \eq{elndistr} in \eq{psi0spr} we
again come to the Zemach correction but with the reduced mass factor
instead of the electron mass in \eq{hfsbortrfr}. The difference between
these two results is of order $m/M$, and might become important in a
systematic treatment of the corrections of second order in the
electron-proton mass ratio.

The quantum mechanical derivation also explains the sign of the
Zemach correction. The spreading of the magnetic and electric
charge densities weakens the interaction and consequently diminishes
hyperfine splitting in accordance with the analytic result in
\eq{hfsbortrfr}.

The Zemach correction is essentially a nontrivial weighted integral of
the product of electric and magnetic densities, normalized to unity.
It cannot be measured directly, like the rms proton charge radius
which determines the main proton size correction to the Lamb shift
(compare the case of the proton size correction to the Lamb shift of
order $(Z\alpha)^5$ in \eq{bortrfr} which depends on the third Zemach
moment). This means that the correction in \eq{hfsbortrfr} may only
conditionally be called the proton size contribution.

The Zemach correction was calculated numerically in a number of
recent papers \cite{gy2,bodyen88,karshhfs97}. The most straightforward
approach is to use the phenomenological dipole fit for the Sachs form
factors of the proton

\beq
G_E(k^2)=\frac{G_M(k^2)}{1+\kappa}=\frac{1}
{\left(1-\frac{k^2}{\Lambda^2}\right)^2},
\eeq

\noindent
with the parameter $\Lambda=0.898~(13)M$, where $M$ is the
proton mass\footnote{We used the same value of $\Lambda$ in Section
\ref{radprot} for calculation of the correction to the Lamb shift in
hydrogen generated by the radiative insertions in the proton line. Due
to the logarithmic dependence of this correction on $\Lambda$ small
changes of its value do not affect the result for the proton line
contribution to the Lamb shift.}.  Substituting this parametrization in
the integral in \eq{hfszema} one obtains $\Delta
E=-38.72~(56)\cdot10^{-6}E_F$ \cite{bodyen88} for the Zemach
correction. The uncertainty in the brackets accounts only for the
uncertainty in the value of the parameter $\Lambda$ and the uncertainty
introduced by the approximate nature of the dipole fit is
completely ignored. This last uncertainty could be significantly
underestimated in such approach. As was emphasized in \cite{karshhfs97}
the integration momenta which are small in comparison with the proton
mass play an important role in the integral in \eq{hfszema}. About
fifty percent of the integral comes from the integration region where
$k<\Lambda/2$. But the dipole form factor parameter $\Lambda$ is
simply related to the rms proton radius $r^2_p=12/\Lambda^2$, and
one can try to use the empirical value of the proton radius as an input
for calculation of the low momentum contribution to the Zemach
correction. Such an approach, which assumes validity of the dipole
parametrization for both form factors at small momenta transfers, but
with the parameter $\Lambda$ determined by the proton radius leads to
the Zemach correction $\Delta E=-41.07~(75)\cdot10^{-6}E_F$
\cite{karshhfs97} for $r_p=0.862$ fm ($\Lambda=0.845~(12)M$). As we
will see below in Section \ref{phen1s} (see \eq{selfconsrad}), modern
spectroscopic data indicates an even higher value of the proton
charge radius $r_p=0.891~(18)~\mbox{fm}$. The respective Zemach
correction is $\Delta E=-42.4~(1.1)\cdot10^{-6}E_F$. We will use this
last value of the Zemach correction for further numerical estimates.
New experimental data on the proton charge radius, and more numerical
work with the existing experimental data on the proton form factors
could result in a more accurate value of the Zemach correction.

\subsubsection{Recoil Correction of Order $(Z\alpha)(m/M)E_F$}
\label{recoiltozemach}

\begin{figure}[ht]
\centerline{\epsfig{file=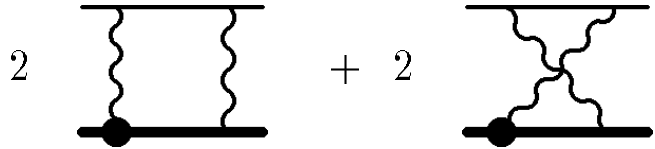,height=2cm}}
\vspace{0.5cm}
\caption{Diagrams with one form factor insertion for total elastic
nuclear size corrections of order $(Z\alpha)E_F$ }
\label{hfsonfffig}
\end{figure}

\begin{figure}[ht]
\centerline{\epsfig{file=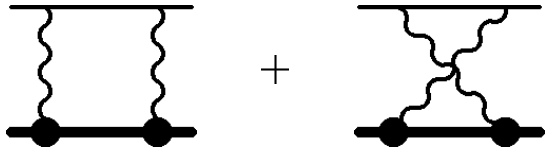,height=2cm}}
\vspace{0.5cm}
\caption{Diagrams with two form factor insertions for total elastic
nuclear size corrections of order $(Z\alpha)E_F$ }
\label{hfstwofffig}
\end{figure}

For muonium the skeleton two-photon exchange diagrams in
Fig.\ \ref{twophothfsfig} generated, after subtraction of the heavy
pole contribution, a recoil correction of order $(Z\alpha)(m/M)E_F$
to hyperfine splitting. Calculation of the respective recoil
corrections for hydrogen does not reduce to substitution of the proton
mass instead of the muon mass in \eq{arnnewcsalp}, but requires an
account of the proton anomalous magnetic moment and the proton form
factors. As we have seen, insertions of the nontrivial proton form
factors in the external field skeleton diagram pushes the
characteristic integration momenta into the region determined by the
proton size and, as a result respective contribution to HFS in
\eq{hfsbortrfr} contains the small proton size factor $m/\Lambda$. The
scale  of the proton form factor is of order of the proton mass and
thus the Zemach correction in Fig.\ \ref{hfsffeltwophotonefig} and
Fig.\ \ref{hfsffeltwophottwofig} is of the same order as the recoil
contributions in Fig.\ \ref{hfsonfffig} and Fig.\ \ref{hfstwofffig}
generated by the form factor insertions in the skeleton diagrams in
Fig.\ \ref{twophothfsfig}. It is natural to consider all contributions
of order $(Z\alpha)F_F$ together and to call the sum of these
corrections the total proton size correction. However, we
have two different parameters $m/\Lambda$ and $m/M$, and the Zemach and
the recoil corrections admit separate consideration.

The recoil part of the  proton size correction of order $(Z\alpha)E_F$
was first considered in \cite{arn,newsalp}. In these works existence of
the nontrivial nuclear form factors was ignored and the proton was
considered as a heavy particle without nontrivial momentum dependent
form factors but with an anomalous magnetic moment. The result of such
a calculation is most conveniently written in terms of the "elementary"
proton Fermi energy ${\widetilde E}_{F}$ which does not include the
contribution of the proton anomalous magnetic moment (compare
\eq{baremuonfermi} in the muonium case). Calculation of this correction
coincides almost exactly with the one in the case of the leading
muonium recoil correction in \eq{arnnewcsalp} \footnote{Really the
original works \cite{arn,newsalp} contain just the elementary proton
ultraviolet divergent result in \eq{hydarnnewcsalp} which turns into
the ultraviolet finite muonium result in \eq{arnnewcsalp} if the
anomalous magnetic moment $\kappa$ is equal zero.} and generates an
ultraviolet divergent result \cite{arn,newsalp,drellsul}

\beq              \label{hydarnnewcsalp}
\Delta E=-\frac{3mM}{M^2-m^2}\frac{Z\alpha}{\pi}
\left\{\left(1-\frac{\kappa^2}{4}\right)\ln\frac{M}{m}
+\frac{\kappa^2}{4}\left(-\frac{1}{6}+3\ln\frac{\cal
M}{M}\right)\right\} {\widetilde E}_F,
\eeq

\noindent
where ${\cal M}$ is an arbitrary ultraviolet cutoff.

The ultraviolet divergence is generated by the diagrams with insertions
of two anomalous magnetic moments in the heavy particle line. This
should be expected since quantum electrodynamics of elementary
particles with nonvanishing anomalous magnetic moments is
nonrenormalizable.

For the real proton we have to include in the vertices the proton form
factor, which decays fast enough at large momenta transfer and
neutralizes the divergence. Insertion of the form factors will
effectively cut off momentum integration at the form factor scale
$\Lambda$, which is slightly smaller than the proton mass. Precise
calculation needs an accurate rederivation of the recoil correction
with account for the form factors, but one important feature of the
expected result is obvious immediately. The factor in the braces in
\eq{hydarnnewcsalp} is numerically small just for the physical value of
the proton anomalous magnetic moment $\kappa=1.792~847~386~(63)$
measured in nuclear magnetons and the ultraviolet cutoff of the order
of the proton mass \cite{bodyen88}. We should expect that this
accidental suppression of the recoil correction would survive account
for the form factors, and this correction will at the end of the day be
numerically much smaller than the Zemach correction, even though these
two corrections are parametrically of the same order.

Since the Zemach and recoil corrections are parametrically of the
same order of magnitude only their sum was often considered in the
literature. The first calculation of the total proton size correction of
order $(Z\alpha)E_F$ with form factors was done in \cite{iddpl},
followed by the calculations in \cite{iddings,drellsul}. Separately the
Zemach and recoil corrections were calculated in \cite{gy2,bodyen88}.
Results of all these works essentially coincide, but some minor
differences are due to the differences in the parameters of the dipole
nucleon form factors used for numerical calculations.

We will present here results in the form obtained in
\cite{iddings,drellsul}. They are independent of the specific
parametrization of the form factors and especially convenient for
further consideration of the inelastic polarizability contributions.
The total proton size correction of order $(Z\alpha)E_F$ generated by
the diagrams with the insertions of the total proton form factors in
Fig.\ \ref{hfsonfffig} and in Fig.\ \ref{hfstwofffig} may easily be
calculated. The resulting integral contains two contributions of the
pointlike proton with the anomalous magnetic moment which were already
taken into account. One is the infrared divergent nonrecoil
contribution corresponding to the external field skeleton integral in
Fig.\ \ref{skelhfs} with insertion of the anomalous magnetic moment,
the other is the pointlike proton recoil correction in
\eq{hydarnnewcsalp}.  After subtraction of these contributions we
obtain an expression for the remaining proton size correction in the
form of the Euclidean four-dimensional integral \cite{iddings,drellsul}

\beq     \label{idddrsul}
\Delta E=
\int
\frac{d^4k}{k^6}\left\{\frac{2k^2(2k^2+k_0^2)\left[F_1(F_1+F_2)
-(1+\kappa)\right]
+6k^2k_0^2\left[F_2(F_1+F_2)-\kappa(1+\kappa)\right]}{4k_0^2+\frac{k^4}{M^2}}
\right.
\eeq
\[
\left.
-\frac{(2k^2+k_0^2)(F_2^2-\kappa^2)}{2}\right\}
\frac{Z\alpha}{\pi}\frac{m}{M}{\widetilde E}_F.
\]

The last term in the braces is ultraviolet divergent, but it exactly
cancels in the sum with the point proton contribution in
\eq{hydarnnewcsalp}. The sum of contributions in \eq{hydarnnewcsalp}
and \eq{idddrsul} is the total proton size correction, including
the Zemach correction. According to the numerical calculation in
\cite{bodyen88} this is equal to $\Delta E=-33.50~(55)\cdot10^{-6}E_F$.
As was discussed above, the Zemach correction included in this result
strongly depends on the precise value of the proton radius, while
numerically the much smaller recoil correction is less sensitive
to the small momenta behavior of the proton formfactor and has smaller
uncertainty. For further numerical estimates we will use the estimate
$\Delta E=5.22~(01)\cdot10^{-6}E_F$ of the recoil correction obtained
in \cite{bodyen88}.

\subsubsection{Nuclear Polarizability Contribution of Order
$(Z\alpha)E_F$} \label{protpolhfs}

\begin{figure}[ht]
\centerline{\epsfig{file=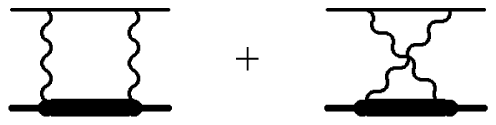,height=2cm}}
\vspace{0.5cm}
\caption{Diagrams for nuclear polarizability correction
of order $(Z\alpha)E_F$}
\label{inelasticfighfs}
\end{figure}

Up to now we considered only the contributions of order
$(Z\alpha)E_F$ to hyperfine splitting in hydrogen generated by the
elastic intermediate nuclear states. As was first realized by Iddings
\cite{iddings} inelastic contributions in Fig.\ \ref{inelasticfighfs}
admit a nice representation in terms  spin-dependent proton structure
functions $G_1$ and $G_2$ \cite{iddings,drellsul}

\beq    \label{hfspolariz}
\Delta E=(\Delta_1+\Delta_2)
\frac{Z\alpha}{2\pi}\frac{m}{M}{\widetilde E}_F\equiv \delta_{pol}E_F,
\eeq

\noindent
where

\beq
\Delta_1=\int_0^\infty\frac{dQ^2}{Q^2}\left\{\frac{9}{4}F_2^2(Q^2)
+4M^2\int_{\nu_0(Q^2)}^\infty\frac{d\nu}{\nu^2}\beta_1(\frac{\nu^2}{Q^2})
G_1(Q^2,\nu)\right\}
\eeq
\beq
\Delta_2=12M^2\int_0^\infty\frac{dQ^2}{Q^2}\int_{\nu_0(Q^2)}^\infty
\frac{d\nu}{\nu^2}\beta_2(\frac{\nu^2}{Q^2})
G_2(Q^2,\nu).
\eeq

The inelastic pion-nucleon threshold $\nu_0$ may be written as

\beq
\nu_0(Q^2)=m_\pi+\frac{m_\pi^2+Q^2}{2M},
\eeq

\noindent
and the auxiliary functions $\beta_i$ have the form

\beq
\beta_1(x)=-3x+2x^2+2(2-x)\sqrt{x(1+x)},
\eeq
\beq
\beta_2(x)=1+2x-2\sqrt{x(1+x)}.
\eeq

The structure functions $G_1$ and $G_2$ may be measured in
inelastic scattering of polarized electrons on polarized protons. The
difference between the spin antiparallel and spin parallel cross
sections has the form

\beq
\frac{d^2\sigma^{\uparrow\downarrow}}{dq^2dE'}
-\frac{d^2\sigma^{\uparrow\uparrow}}{dq^2dE'}
=\frac{4\pi\alpha^2}{E^2Q^2}\left[\frac{E+E'\cos\theta}{\nu}G_1(Q^2,\nu)+
G_2(Q^2,\nu)\right],
\eeq

\noindent
where $E,E'$ are the initial and final electron energies,
$\nu=E-E'$, and $\theta$ is the electron scattering angle in the
laboratory frame.

Only partial experimental data is known today for the proton
spin-dependent structure functions, and there is not enough information
to calculate the integrals in \eq{hfspolariz} directly. First estimates
of the polarizability correction were obtained a long time ago
\cite{iddplat59,iddings,vergazwa,drellsul,guerin,zinovev}. One popular
approach is to consider the polarizability correction directly as a
sum of contributions of all intermediate nuclear states. Then the
leading contributions to this correction are generated by the low lying
states, and the contribution of the $\Delta$ isobar was estimated many
times \cite{iddplat59,iddings,vergazwa,drellsul,guerin}. The latest
result \cite{faustmartsal} consistent with the earlier estimates is

\beq
\delta_{pol}(\Delta)= -0.12\cdot10^{-6}.
\eeq

The total polarizability contribution in this approach  may be obtained
after summation over contributions of all relevant intermediate states.

About thirty years ago the general properties of the structure functions
and the known experimental data were used to set rigorous bounds on the
polarizability contribution in \eq{hfspolariz} \cite{rafael,gnkuti}
(see also reviews in \cite{hugheskuti,bodyen88})

\beq \label{rigboundpol}
|\delta_{pol}|\leq 4\cdot10^{-6}.
\eeq

The problem of the polarizability contribution clearly requires new
consideration which takes into account more recent experimental data.


\subsection{Recoil Corrections of Order
$(Z\alpha)^2(m/M)E_F$}\label{recoilhfshydza2}

\begin{figure}[h]
\centerline{\epsfig{file=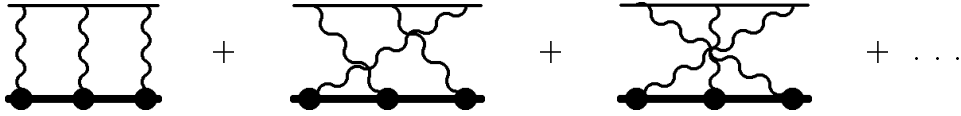,height=1.5cm}}
\vspace{0.5cm}
\caption{Diagrams for recoil corrections of order
$(Z\alpha)^2E_F$}
\label{hfsrechydza2fig}
\end{figure}

Recoil corrections of relative order $(Z\alpha)^2(m/M)$  are connected
with the diagrams with three exchanged photons (see Fig.\
\ref{hfsrechydza2fig}). Due to the Caswell-Lepage cancellation
\cite{byg78,cp} recoil corrections of order $(Z\alpha)^2(m/M)\widetilde
E_F$ in muonium (see discussion in Section \ref{rechfsza2muon})
originate from the exchanged momenta of order of the electron mass and
smaller. The same small exchanged momenta are also relevant in the case
of hydrogen.  This means that unlike the case of the recoil correction
of order $(Z\alpha)(m/M)\widetilde E_F$ considered above, the proton
structure is irrelevant in calculation of corrections of order
$(Z\alpha)^2(m/M)\widetilde E_F$. However, we cannot simply use the
muonium formulae for hydrogen because the muonium calculations ignored
the anomalous magnetic moment of the heavy particle. A new
consideration \cite{bodyen88} of the recoil corrections of order
$(Z\alpha)^2(m/M)\widetilde E_F$ in the case of a heavy particle with
an anomalous magnetic moment resulted in the correction

\beq        \label{rechfshydza2tot}
\Delta
E=\left\{\left[2(1+\kappa)+\frac{7\kappa^2}{4}\right]\ln(Z\alpha)^{-1}
-\left[8(1+\kappa)-\frac{\kappa(12-11\kappa)}{4}\right]\ln2\right.
\eeq
\[
\left.
+\frac{65}{18}
+\frac{\kappa(11+31\kappa)}{36}
\right\}(Z\alpha)^2\frac{m_r^2}{mM}
\widetilde E_F.
\]

For a vanishing anomalous magnetic moment of the heavy particle
($\kappa=0$) this correction turns into the muonium result in
\eq{logrechfs} and \eq{nonlogrechfs}.

Calculation of the respective radiative-recoil correction of order
$\alpha(Z\alpha)^2(m/M)\widetilde E_F$ in the skeleton integral
approach is quite straightforward and may readily be done. However,
numerically the correction in \eq{rechfshydza2tot} is smaller than the
uncertainty of the Zemach correction, and calculation of corrections to
this result does not seem to be an urgent task.

\subsection{Correction of Order $(Z\alpha)^2\lowercase{m^2<r^2>}E_F$}

The leading nuclear size  correction of order $(Z\alpha)^2m^2<r^2>E_F$
may easily be calculated in the framework of nonrelativistic
perturbation theory if one takes as one of the perturbation potentials
the potential corresponding to the main proton size contribution to the
Lamb shift in \eq{chargerad}. The other perturbation potential is the
potential in \eq{fermipothfseff} responsible for the main Fermi
contribution to HFS (compare calculation of the leading logarithmic
contribution of order $\alpha(Z\alpha)^2(\lowercase{m}/M)E_F$ in
Section \ref{hfsleadlogradrecza2}). The result is \cite{karshhfs97}

\beq          \label{za2r2hydhfs}
\Delta E=-\frac{2}{3}(Z\alpha)^2\ln(Z\alpha)^{-2}m^2<r^2>E_F
=-1.7\cdot10^{-9}E_F.
\eeq

This tiny correction is too small to be of any phenomenological interest
for hydrogen.

\subsection{Correction of Order $(Z\alpha)^3(\lowercase{m}/\Lambda)E_F$}

The logarithmic nuclear size correction of order $(Z\alpha)^3E_F$ may
simply be obtained from the Zemach correction if one takes into account
the Dirac correction to the Schr\"odinger-Coulomb wave function in
\eq{darwincor} \cite{karshhfs97}

\beq  \label{za3hfshydr}
\Delta E=-(Z\alpha)^3\ln(Z\alpha)^{-2}m<r>_{(2)}E_F.
\eeq

The corrections in \eq{za2r2hydhfs} and \eq{za3hfshydr} are
negligible for ground state hyperfine splitting in
hydrogen. However, it is easy to see that these corrections are state
dependent and give contributions to the difference of hyperfine
splittings in the $2S$ and $1S$ states $8\Delta E(2S)-\Delta E(1S)$.
Respective formulae were obtained in \cite{trof79,karshhfs97} and are
of phenomenological interest in the case of HFS splitting in
the $2S$ state in hydrogen \cite{heberle1,rothery}, in deuterium
\cite{heberle2}, and in the $^3He^+$ ion \cite{priorwang}, and also for
HFS in the $2P$ state \cite{ljp} (see also review in \cite{ramsey}).

\section{~~Radiative Corrections to Nuclear Size and Recoil Effects}


\subsection{Radiative-Recoil Corrections of
Order $\alpha(Z\alpha)(\lowercase{m}/\Lambda)E_F$}\label{radcorrzemach}

\begin{figure}[ht]
\centerline{\epsfig{file=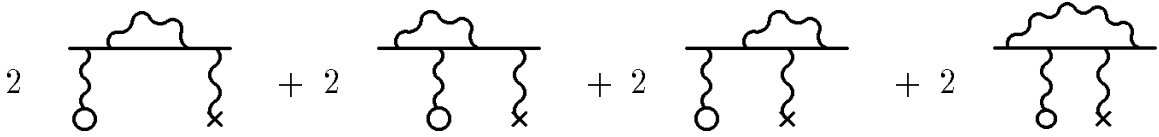,height=1.8cm}}
\vspace{0.5cm}
\caption{Electron-line radiative correction to the Zemach contribution}
\label{radzemelectr}
\end{figure}

\begin{figure}[ht]
\centerline{\epsfig{file=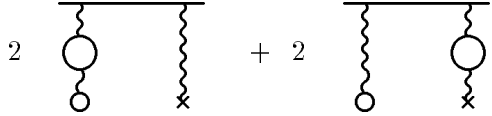,height=1.5cm}}
\vspace{0.5cm}
\caption{Photon-line radiative correction to the Zemach contribution}
\label{radzemphoton}
\end{figure}

Diagrams for the radiative corrections to the Zemach contribution in
Fig.\ \ref{radzemelectr} and in Fig.\ \ref{radzemphoton} are obtained
from the diagrams in Fig.\ \ref{hfsffeltwophotonefig} and in Fig.\
\ref{hfsffeltwophottwofig} by insertions of the radiative photons in
the electron line or of the polarization operator in the external
photon legs. Analytic expressions for the nuclear size corrections of
order $\alpha(Z\alpha)E_F$ are obtained from the integral for the
Zemach correction in \eq{hfszema} by insertions of the electron factor
or the one-loop polarization operator in the integrand in \eq{hfszema}.
Effective integration momenta in \eq{hfszema} are determined by the
scale of the proton form factor, and so we need only the leading terms
in the high-momentum expansion of the polarization operator and the
electron factor for calculation of the radiative corrections to the
Zemach correction. The leading term in the high-momentum asymptotic
expansion of the electron factor is simply a constant (see the text
above \eq{electronaslog}) and the correction to hyperfine splitting is
the product of this constant and the Zemach correction
\cite{karshhfs97}

\beq
\Delta E=\frac{5}{2}\frac{\alpha(Z\alpha)}{\pi}m<r>_{(2)}E_F.
\eeq

The contribution of the polarization operator is logarithmically
enhanced due to the logarithmic asymptotics of the polarization
operator. This logarithmically enhanced contribution of the
polarization operator is equal to the doubled product of the Zemach
correction and the leading term in the polarization operator expansion
(an extra factor two is necessary to take into account two ways to
insert the polarization operator in the external photon legs in Fig.\
\ref{hfsffeltwophotonefig} and in Fig.\ \ref{hfsffeltwophottwofig})

\beq
\Delta
E=-\frac{4}{3}\left(\ln\frac{\Lambda^2}{m^2}\right)
\frac{\alpha(Z\alpha)}{\pi}m<r>_{(2)}E_F .
\eeq

Calculation of the nonlogarithmic part of the polarization operator
insertion requires more detailed information on the proton form
factors, and using the dipole parametrization one obtains
\cite{karshhfs97}

\beq
\Delta
E=-\frac{4}{3}\left(\ln\frac{\Lambda^2}{m^2}-\frac{317}{105}\right)
\frac{\alpha(Z\alpha)}{\pi}
m<r>_{(2)}E_F.
\eeq

\subsection{Radiative-Recoil Corrections of Order
$\alpha(Z\alpha)(m/M)E_F$}

Radiative-recoil corrections of order $\alpha(Z\alpha)(m/M)E_F$ are
similar to the radiative corrections to the Zemach contribution, and in
principle admit a straightforward calculation in the framework of the
skeleton integral approach. Leading logarithmic contributions of this
order were considered in \cite{bodyen88,karshhfs97}. The logarithmic
estimate in \cite{karshhfs97} gives

\beq
\Delta E= 0.11~(2)\cdot10^{-6} E_F,
\eeq

\noindent
for the contribution of the electron-line radiative insertions, and

\beq
\Delta E= -0.02\cdot10^{-6} E_F,
\eeq

\noindent
for the contribution of the vacuum polarization insertions in the
exchanged photons.

Numerically these contributions are much smaller than the uncertainty
of the Zemach correction.

\subsection{Heavy Particle Polarization Contributions}

Muon and heavy particle polarization contributions to hyperfine
splitting in muonium were considered in Sections \ref{hfsradrecpol} and
\ref{heavypolmuonhfs}.

In the external field approximation the skeleton integral with the muon
polarization insertion coincides with the respective integral for
muonium (compare   \eq{polarnonrechfs} and the discussion after this
equation) and one easily obtains \cite{karmupol}

\beq
\Delta E=\frac{3}{4}\alpha(Z\alpha)\frac{m}{m_\mu}E_F.
\eeq

This result gives a good idea of the magnitude of the muon
polarization contribution since muon is relatively light in comparison
to the scale of the proton form factor which was ignored in this
calculation.

The total muon polarization contribution may be calculated without
great efforts but due to its small magnitude such a calculation is of
minor phenomenological significance and was never done. Only an
estimate of the total muon polarization contribution exists in the
literature \cite{karshhfs97}

\beq
\Delta E=0.07~(2)\cdot10^{-6}E_F.
\eeq

Hadronic vacuum polarization in the external field approximation
for the pointlike proton also was calculated in \cite{karmupol}.
Such a calculation may serve only as an order of magnitude estimate
since both the external field approximation and the neglect of the
proton form factor are not justified in this case, because the scale of
the hadron polarization contribution is determined by the same
$\rho$-meson mass which determines the scale of the proton form factor.
Again a more accurate calculation is feasible but does not seem to be
warranted, and only an estimate of the hadronic polarization
contribution appears in the literature \cite{karshhfs97}

\beq
\Delta E=0.03~(1)\cdot10^{-6}E_F.
\eeq

\section{~~~Weak Interaction Contribution}

The weak interaction contribution to hyperfine splitting in hydrogen is
easily obtained by generalization of the muonium result in
\eq{wekahfscontr}

\beq
\Delta E=\frac{g_A}{1+\kappa}\frac{G_F}{\sqrt 2} \frac{3mM}{4\pi
Z\alpha} E_F \approx 5.8\cdot10^{-8}~{\rm kHz}.
\eeq

Two features of this result deserve some comment. First, the axial
coupling constant for the composite proton is renormalized by the
strong interactions and its experimental value is $g_A=1.267$, unlike
the case of the elementary muon when it was equal unity. Second the
signs of the weak interaction correction are different in the case of
muonium and hydrogen \cite{eid96}.

\begin{center}
\underline{Table XIX. Hyperfine Splitting in Hydrogen}
\nopagebreak

\begin{tabular}{|l|rl|c|}    \hline
     & $E_F=1~418~840.11~(3)~(1)~\mbox{kHz}$ &   & kHz
\\ \hline  \hline
Total nonrecoil contribution &&$$& \\
Tables XII-XV&$1.001~136~089~6~(19)$&&$1~420~452.04~(3)~(1)$
\\ \hline
Proton size correction, &&$$& \\
relative order $(Z\alpha)(m/\Lambda)$&&$$& \\
Zemach (1956)\cite{zemach}&$-2(Z\alpha)m<r>_{(2)}
=-42.4~(1.1)\cdot10^{-6}$&&$-60.2~(1.6)$
\\
\hline
Recoil correction, &&$$& \\
relative order $(Z\alpha)(m/M)$&&$$& \\
Arnowitt (1953) \cite{arn}&&& \\
Newcomb,Salpeter (1955) \cite{newsalp}&$$& & $$ \\
Iddings,Platzman (1959) \cite{iddpl}&$5.22~(1)\cdot10^{-6}$&&$7.41~(2)$
\\ \hline
Recoil correction, &&$$& \\
relative order $(Z\alpha)^2(m/M)$
&$\left\{\left[2(1+\kappa)+\frac{7\kappa^2}{4}\right]
\ln(Z\alpha)^{-1}\right.$&& \\
&$\left.
-\left[8(1+\kappa)-\frac{\kappa(12-11\kappa)}{4}\right]\ln2\right.$
&&$$\\
Bodwin,Yennie (1988)
\cite{bodyen88}&$\left.
+\frac{65}{18}
+\frac{\kappa(11+31\kappa)}{36}
\right\}\frac{(Z\alpha)^2}{1+\kappa}\frac{m_r^2}{mM}$
&&$$\\
&$=0.4585\cdot10^{-6}$& & $0.65$
\\
\hline
Leading logarithmic correction, &&$$& \\
relative order $(Z\alpha)^2m^2r_p^2$&&$$& \\
Karshenboim (1997) \cite{karshhfs97}&
$-\frac{2}{3}(Z\alpha)^2\ln(Z\alpha)^{-2}m^2<r^2>
=-0.002\cdot10^{-6}$& & $$-0.002$$\\
\hline
Leading logarithmic correction, &&$$& \\
relative order $(Z\alpha)^3(\lowercase{m}/\Lambda)$
&&&$$ \\
Karshenboim (1997) \cite{karshhfs97}&
$-(Z\alpha)^3\ln(Z\alpha)^{-2}m<r>_{(2)}
=-0.01\cdot10^{-6}$ & &$-0.016$\\
\hline
Electron-line correction, &&$$& \\
relative order $\alpha(Z\alpha)(\lowercase{m}/\Lambda)$&&$$& \\
Karshenboim (1997) \cite{karshhfs97}
&$\frac{5}{2}\frac{\alpha(Z\alpha)}{\pi}m<r>_{(2)}
=0.12\cdot10^{-6}$& & $0.17$ \\
\hline
Photon-line correction, &&$$& \\
relative order $\alpha(Z\alpha)(\lowercase{m}/\Lambda)$&&$$& \\
Karshenboim (1997) \cite{karshhfs97}
&$-\frac{4}{3}\left(\ln\frac{\Lambda^2}{m^2}-\frac{317}{105}\right)
\frac{\alpha(Z\alpha)}{\pi}
m<r>_{(2)}
=-0.77\cdot10^{-6}$& & $-1.10$ \\
\hline
Leading electron-line correction, &&$$& \\
relative order $\alpha(Z\alpha)(\lowercase{m}/M)$&&$$& \\
Karshenboim (1997) \cite{karshhfs97}
&$0.11~(2)\cdot10^{-6}$& & $0.16$ \\
\hline
Leading photon-line correction, &&$$& \\
relative order $\alpha(Z\alpha)(\lowercase{m}/M)$&&$$& \\
Karshenboim (1997) \cite{karshhfs97}
&$-0.02\cdot10^{-6}$ &&$-0.03$\\
\hline
Muon vacuum polarization, &&$$& \\
relative order $\alpha(Z\alpha)({m}/m_\mu)$&&$$& \\
Karshenboim (1997) \cite{karshhfs97,karmupol}
&$0.07~(2)\cdot10^{-6}$ &&$0.10~(3)$\\
\hline
Hadron vacuum polarization, &&$$& \\
Karshenboim (1997) \cite{karshhfs97,karmupol}
&$0.03~(1)\cdot10^{-6}$ &&$0.04~(1)$\\
\hline
Weak interaction contribution, &&$$& \\
Beg,Feinberg (1975) \cite{bf}
&$\frac{g_A}{1+\kappa}\frac{G_F}{\sqrt 2} \frac{3mM}{4\pi
Z\alpha}=0.06\cdot10^{-6}$ &&$0.08$\\
\hline
\hline
&   $$       & & \\
Total theoretical HFS &$$& &$1~420~399.3~(1.6)$\\
\hline
\end{tabular}
\end{center}

\part{Hypefrine Splitting in Muonic Hydrogen}

We have considered level shifts in muonic hydrogen in
Section \ref{mouniclambgen} neglecting hyperfine structure.
However, future measurements (see discussion in Section
\ref{munichydexp}) will be done on the components of hyperfine
structure, and knowledge of this hyperfine structure is crucial for
comparison of the theoretical predictions for the Lamb shift in muonic
hydrogen with the experimental data.  We will consider below hyperfine
structure  in the states $2S$ and $2P$.

\section{~~~~~Hyperfine Structure of the $2S$ State}

\begin{figure}[ht]
\centerline{\epsfig{file=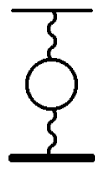}}
\vspace{0.5cm}
\caption{Electron vacuum polarization insertion in the exchanged photon}
\label{onephtpolhfs}
\end{figure}

Due to enhancement of the light electron loops in muonic hydrogen they
produce the largest contribution to the Lamb shift in muonic hydrogen
(see Section \ref{mouniclambgen}). Unlike the Lamb shift, where the
leading contribution is a radiative (loop) correction, the leading
contribution to hyperfine splitting already exists at the tree level
(see discussion in Section \ref{hfsphysor}). Hence, the Fermi
contribution in \eq{ef} (with the natural substitution of the heavy
particle mass and anomalous magnetic moment instead of the respective
muon characteristics) remains by far the largest contribution to HFS in
muonic atoms. The leading electron vacuum polarization contribution to
HFS generated by the exchange of one-photon with polarization insertion
in Fig.\ \ref{onephtpolhfs} is enhanced in muonic atoms, and becomes
the next largest individual contribution to HFS after the Fermi
contribution. The reason for this enhancement is the same as for the
respective enhancement in the case of the Lamb shift: electron vacuum
polarization distorts the field of an external source at distances of
about $1/m_e$ and for muonic atoms the wave function is concentrated in
a region of comparable size determined by the Bohr radius
$1/(mZ\alpha)$. Hence, the effect of the electronic vacuum polarization
on the HFS is much stronger in muonic atoms than in the electronic
atoms where the region where the external potential is distorted by the
vacuum  potential is negligible in comparison with the effective radius
of the wave function. As a result the electron vacuum polarization
contribution to HFS in light muonic atoms is of order $\alpha E_F$
\cite{sternheim65}, to be compared with the leading polarization
contribution in electronic hydrogen of order $\alpha(Z\alpha)E_F$ in
\eq{polarnonrechfs}. Thus, in order to translate the hyperfine results
for ordinary hydrogen into the results for muonic hydrogen we have to
consider additional contributions which are due to the polarization
insertions.

Notice first of all that the nonrecoil results in Tables XII-XVII may be
directly used in the case of muonic hydrogen. Since we are interested
in the contribution to HFS in $2S$ state we should properly restore the
dependence on the principal quantum number $n$, which was sometimes
omitted above. This dependence is known for the results in Table XII and
all corrections in Tables XIII-XIV are state-independent.  For the
corrections in Tables XV-XVII only the leading logarithmic
contributions are state-independent, and the exact $n$ dependence of
the other terms is often unknown. However, all corrections in these
Tables are of order $1\cdot 10^{-4}$ meV and are too small for
phenomenological goals (see discussion in Section \ref{munichydexp}).
The sum of all corrections in  Tables XII-XVII for the $2S_\frac{1}{2}$
state is equal to

\beq
\Delta E=22.8322~\mbox{meV}.
\eeq

Vacuum polarization corrections of order $\alpha E_F$ to HFS may be
easily calculated with the help of the spin-spin term in the
Breit-like potential corresponding to an exchange of a radiatively
corrected photon in Fig.\ \ref{onephtpolhfs}. This spin-spin potential
$V_{VP}^{spin}$ can be obtained by substituting the spin-spin term
corresponding to a massive exchange \cite{pachucki96muon}

\beq
{\cal
V}_{VP}^{spin}(2m_e\zeta)=\frac{8}{3}\frac{Z\alpha}{mM}(1+a_\mu)(1+\kappa)
({\bf s_\mu\cdot s_p})\left(\pi\delta({\bf
r})-\frac{e^{-2m_e\zeta r}}{4r^3}(2m_e\zeta r)^2\right)
\eeq

in the integral in \eq{radcorrbreit} instead of ${\cal
V}_{VP}(2m_e\zeta)$.

\begin{figure}[ht]
\centerline{\epsfig{file=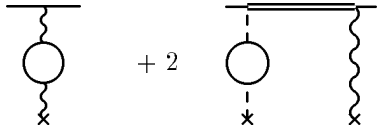}}
\vspace{0.5cm}
\caption{Leading electron vacuum polarization correction to HFS in
muonic hydrogen}
\label{leadpolelhfs}
\end{figure}

\noindent
Correction to HFS is then given by a sum of the first and second order
perturbation theory contributions similar to the respective
contribution to the Lamb shift in \eq{matrixelmcalc} (see Fig.\
\ref{leadpolelhfs})

\beq
\Delta E=<V_{VP}^{spin}>+2<V_{Br}G'(E_n)V_{VP}^C>,
\eeq

\noindent
and one obtains \cite{pachucki96muon}

\beq
\Delta E=0.058~\mbox{meV}.
\eeq

We should also consider other new corrections with insertions of
vacuum polarizations, but they all contain at least one extra factor
$\alpha$, should be less or about $0.001$ meV and at the present stage
may safely be ignored.

Next we turn to the nuclear size, recoil and structure corrections,
where one cannot ignore the composite nature of the proton. As in the
case of ordinary hydrogen the main contribution is connected with the
proton size corrections of order $(Z\alpha)E_F$ considered in Sections
\ref{zemchhfsprot} and \ref{recoiltozemach}. Respective considerations
may literally be repeated for muonic hydrogen, the only difference is
that due to a larger ratio of the muon to proton mass, separate
consideration of the nonrecoil (Zemach) and recoil corrections of order
$(Z\alpha)E_F$ makes even less sense than in the case of electronic
hydrogen. Hence, one should consider the total contribution of order
$(Z\alpha)E_F$ given by the sum of the contributions
in \eq{hydarnnewcsalp} and \eq{idddrsul}. Numerical calculations
for the $2S_\frac{1}{2}$ state lead to the result \cite{pachucki96muon}

\beq
\Delta E=-0.145~\mbox{meV}.
\eeq

\noindent
As was discussed in Sections \ref{zemchhfsprot} and
\ref{recoiltozemach} this contribution depends on the dipole
parametrization of the proton form factors, the value of the proton
radius, and can probably be improved as a result of dedicated
analysis.

Proton polarizability contributions of order $(Z\alpha)E_F$
discussed for electronic hydrogen in Section \ref{protpolhfs} are
notoriously difficult to evaluate. Comparing the results for the upper
boundary for the inelastic contribution in \eq{rigboundpol} and the
elastic contribution from \cite{bodyen88} discussed in Section
\ref{recoiltozemach} we see that the polarizability contribution is at
the level of $10\%$ of the elastic contribution in electronic hydrogen.
As a conservative estimate it was suggested in
\cite{pachucki96muon} to assume that the same estimate is valid for
muonic hydrogen. This assumption means that the polarizability
contribution in muonic hydrogen does not exceed $0.15$ meV.
Collecting all contributions above we obtain the total HFS splitting in
the $2S_\frac{1}{2}$ state in muonic hydrogen \cite{pachucki96muon}

\beq
\Delta E=22.745~(15)~\mbox{meV}.
\eeq

As was discussed at the end of Section \ref{recoiltozemach} we
expect that the uncertainty of this result determined by the unknown
polarizability contribution can be reduced as a result of a new analysis.

\section{~~~~~~Fine and Hyperfine Structure of the $2P$ States}

The main contribution to the fine and hyperfine structure of the $2P$
states is described by the spin-orbit and spin-spin terms in the Breit
Hamiltonian in \eq{breitpot} (spin-spin terms were omitted in
\eq{breitpot}). For proper description of the fine and hyperfine
structure we have to include in the Breit potential anomalous magnetic
moments of both constituents and restore all terms which depend on the
heavy particle spin. These were omitted in \eq{breitpot}. Then relevant
terms in the Breit potential have the form

\beq  \label{breitspin}
V_{Br}=\frac{Z\alpha}{r^3}\left(\frac{1+2a_\mu}{2m^2}+\frac{1+a_\mu}{mM}
\right)({\bf L\cdot s_\mu})
+\frac{Z\alpha}{r^3}\left(\frac{1+2\kappa}{2M^2}+\frac{1+\kappa}{mM}
\right)({\bf L\cdot s_p})
\eeq
\[
-\frac{Z\alpha}{r^3}\frac{(1+\kappa)(1+a_\mu)}{mM}\left(({\bf
s_\mu\cdot s_p})- 3\frac{({\bf s_\mu\cdot r})({\bf s_p\cdot
r})}{r^2}\right)
\]

\noindent
The first term in this potential describes spin-orbit interaction of
light particles and its matrix element determines the fine structure
splitting between $2P_\frac{3}{2}$ and $2P_\frac{1}{2}$ states. Two
other terms depend on the heavy particle spin. It is easy to see that
these terms mix the states with the same total angular momentum ${\bf
F}={\bf J+s_p}$ and different  $\bf J$ \cite{romanov}, in our case
these are the states $2^3P_\frac{3}{2}$ and $2^3P_\frac{1}{2}$ (see
Fig.\ \ref{muonhydlevels}). Thus to find the fine and hyperfine
structure of the $2P$ states we have to solve an elementary
quantum-mechanical problem of diagonalizing a simple four by four
Hamiltonian, where only a two by two submatrix is nondiagonal. Before
solving this problem we have to consider if there are any other
contributions to the Hamiltonian besides the terms in the Breit
potential in \eq{breitspin}. It is easy to realize that the only other
contribution to the effective potential is given by the radiatively
corrected one-photon exchange. The respective Breit-like potential  may
be obtained exactly as in \eq{radcorrbreit} by integrating the
spin-orbit potential corresponding the exchange of a particle with mass
$\sqrt{t'}=2m_e\zeta$

\beq  \label{spinorbdens}
{\cal
V}_{VP}^{spin-orbit}(2m_e\zeta)=
\frac{Z\alpha}{r^3}\left(\frac{1+2a_\mu}{2m^2}+\frac{1+a_\mu}{mM}
\right)e^{-2m_e\zeta r}(1+2m_e\zeta r)({\bf L\cdot s_\mu})
\eeq
\[
+\frac{Z\alpha}{r^3}\left(\frac{1+2\kappa}{2M^2}+\frac{1+\kappa}{mM}
\right)e^{-2m_e\zeta r}(1+2m_e\zeta r)({\bf L\cdot s_p})
\]
\[
-\frac{Z\alpha}{r^3}\frac{(1+\kappa)(1+a_\mu)}{mM}\left[\left(({\bf
s_\mu\cdot s_p})- 3\frac{({\bf s_\mu\cdot r})({\bf s_p\cdot
r})}{r^2}\right)(1+2m_e\zeta r)
\right.
\]
\[
\left.
+\left(({\bf s_\mu\cdot s_p})- \frac{({\bf s_\mu\cdot
r})({\bf s_p\cdot
r})}{r^2})(2m_e\zeta r)^2\right)\right]e^{-2m_e\zeta r}.
\]

\noindent
All that is left to obtain the fine and hyperfine structure of the $2P$
states is to diagonalize the four by four Hamiltonian with
interaction which is the sum of the Breit potential in \eq{breitspin}
and the respective integral of the potential density in
\eq{spinorbdens}. This problem was solved in \cite{pachucki96muon},
where it was obtained

\beq
\Delta E(2P_\frac{1}{2})=7.963~\mbox{meV},
\eeq
\[
\Delta E(2P_\frac{3}{2})=3.393~\mbox{meV}.
\]

\noindent
Due to mixing of the states $2^3P_\frac{3}{2}$ and $2^3P_\frac{1}{2}$
(see Fig.\ \ref{muonhydlevels})  they are additonally shifted by
$\Delta=0.145$ meV.

\part{Comparison of Theory and Experiment} \label{compthexp}

In numerical calculations below we will use the most precise modern
values of the fundamental physical constants. The value of the
Rydberg constant is \cite{schwob}

\beq
R_\infty=10~973~731.568~516~(84)~\mbox{m}^{-1}\qquad\delta=7.7\cdot
10^{-12},
\eeq

\noindent
the fine structure constant is equal to \cite{hk}

\beq    \label{alpha}
\alpha^{-1}=137.035~999~58~(52)\qquad \delta=3.8\cdot 10^{-9},
\eeq

\noindent
the proton-electron mass ratio is equal to \cite{fds}

\beq  \label{eprmassrexp}
\frac{M}{m}=1~836.152~666~5~(40)\qquad \delta=2.2\cdot 10^{-9},
\eeq

\noindent
the muon-electron mass ratio is equal to \cite{lbdd}

\beq  \label{emmassrexp}
\frac{M}{m}=206.768~277~(24)\qquad \delta=1.2\cdot 10^{-7},
\eeq

\noindent
and the deuteron-proton mass ratio is equal to \cite{audi}

\beq  \label{dpmassrexp}
\frac{M}{m}=1.999~007~5013~(14)\qquad \delta=7.0\cdot 10^{-10}.
\eeq

\section{~~~~Lamb Shifts of the Energy Levels}

From the theoretical point of view the accuracy of calculations is limited
by the magnitude of the yet uncalculated contributions to the Lamb shift.
Corrections to the $P$ levels are known now with a higher accuracy
than the corrections to the $S$ levels, and do not limit the results of
the comparison between theory and experiment.

\subsection{Theoretical Accuracy of $S$-state Lamb Shifts}

Corrections of order $\alpha^2(Z\alpha)^6$ are the largest uncalculated
contributions to the energy levels for $S$-states. The correction of
this order is a polynomial in $\ln(Z\alpha)^{-2}$, starting with the
logarithm cubed term. Both the logarithm cubed term and the
contribution of the logarithm squared terms to the difference $\Delta
E_L(1S)-8\Delta E_L(2S)$ are known (for more details on these
corrections see discussion in Section \ref{alpha2zalpha6}). However,
the calculation of the respective contributions to the individual
energy levels is still missing. With this circumstance it is reasonable
to take one half of the logarithm cubed term (which has roughly the
same magnitude as the logarithm squared contribution to the interval
$\Delta E_L(1S)-8\Delta E_L(2S)$) as an estimate of the scale of all
yet uncalculated logarithm squared contributions. We thus assume that
uncertainties induced by the uncalculated contributions of order
$\alpha^2(Z\alpha)^6$ constitute $14$ kHz and $2$ kHz for the $1S$- and
$2S$-states, respectively.

All other unknown theoretical contributions to the Lamb shift are much
smaller, and $14$ kHz for the $1S$-state and $2$ kHz for the
$2S$-state are reasonable estimates of the total theoretical
uncertainty of the expression for the Lamb shift. Theoretical
uncertainties for the higher $S$ levels may be obtained from the
$1S$-state uncertainty ignoring its state-dependence and scaling it
with the principal quantum number $n$.

\subsection{Theoretical Accuracy of $P$-state Lamb
Shifts}\label{pstaac}

The Lamb shift theory of $P$-states is in a better shape than the
theory of $S$-states. The largest unknown corrections to the $P$-state
energies are the single logarithmic contributions of the form
$\alpha^2(Z\alpha)^6\ln(Z\alpha)^{-2}m$, like the one in
\eq{singllogaza6pol}, induced by radiative insertions in the electron and
external photon lines, and the uncertainty of the nonlogarithmic
contributions $G_{SE,7}$ of order $\alpha(Z\alpha)^7m$ (see Table VI).
One half of the double logarithmic contribution in \eq{twolooplod2}
can be taken as a fair an estimate of the magnitude of the uncalculated
single logarithmic contributions of the form
$\alpha^2(Z\alpha)^6\ln(Z\alpha)^{-2}m$. An estimate of the theoretical
accuracy of the $2P$ Lamb shift is then about $0.08$ kHz. Theoretical
uncertainties for the higher non-$S$ levels may be obtained from the
$2P$-state uncertainty ignoring its state-dependence and scaling it
with the principal quantum number $n$.

\subsection{Theoretical Accuracy of the Interval
$L(1S)-8L(2S)$}\label{l1s82s}

State-independent contributions to the Lamb shift scale as $1/n^3$ and
vanish in the difference $E_L(1S)-8E_L(2S)$, which may be
calculated more accurately than the positions of the individual
energy levels (see discussion in Section \ref{cancellation}). All main
sources of theoretical uncertainty of the individual energy levels,
namely, proton charge radius contributions and yet uncalculated
state-independent corrections to the Lamb shift vanish in this
difference. This observation plays an important role in
extracting the precise value of the $1S$ Lamb shift from modern
highly accurate experimental data (see discussion below in
Section \ref{phen1s}). Earlier the practical usefulness of the
theoretical value of the interval $E_L(1S)-8E_L(2S)$ for extraction of
the experimental value of the $1S$ Lamb shift was impeded by the
insufficient theoretical accuracy of this interval and by the
insufficient accuracy of the frequency measurement. Significant
progress was achieved recently in both respects, especially on the
experimental side. On the theoretical side the last relatively large
contribution to $E_L(1S)-8E_L(2S)$ of order
$\alpha^2(Z\alpha)^6\ln^2(Z\alpha)^{-2}$ was calculated in
\cite{kar96zetp,karjp96,karyf95,karjetp94} (see \eq{1s82scanc} above),
and the theoretical uncertainty of this interval was reduced to $5$ kHz

\begin{equation} \label{selfcons}
\Delta\equiv E_L(1S)-8E_L(2S)=-187~231~(5)~\mbox{kHz}.
\end{equation}

\subsection{Classic Lamb Shift
$2S_{\frac{1}{2}}-2P_{\frac{1}{2}}$}\label{classlamb2s2p}

Discovery of the classic Lamb shift, i.e. splitting of the
$2S_{\frac{1}{2}}$ and the $2P_{\frac{1}{2}}$ energy levels  in
hydrogen triggered a new stage in the development of modern physics. In
the terminology accepted in this paper the classic Lamb shift is equal
to the difference of Lamb shifts in the respective states $\Delta
E(2S_{\frac{1}{2}}-2P_{\frac{1}{2}})
=L(2S_{\frac{1}{2}})-L(2P_{\frac{1}{2}})$). Unlike the much larger Lamb
shift in the $1S$ state, the classic Lamb shift is directly observable
as a small splitting of energy levels which should be degenerate
according to Dirac theory. This greatly simplifies comparison
between the theory and experiment for the classic Lamb shift, since the
theoretical predictions are practically independent of the exact value
of the Rydberg constant, which can be measured independently.

Many experiments on precise measurement of the classic Lamb shift
were performed since its experimental discovery in 1947. We have
collected modern post 1979 experimental results in Table XX. Two
entries in this Table are changed compared to the original published
experimental results \cite{sok,hp}. These alterations reflect recent
improvements of the theory used for extraction of the Lamb shift value
from the raw experimental data.

The magnitude of the Lamb shift in \cite{sok} was derived from the
ratio of the $2P_\frac{1}{2}$ decay width and the $\Delta
E(2S_{\frac{1}{2}}-2P_{\frac{1}{2}})$ energy splitting which was directly
measured by the atomic-interferometer method. The theoretical
expression for the $2P_\frac{1}{2}$-state lifetime was used for
extraction of the magnitude of the Lamb shift. An additional leading
logarithmic correction to the width of the $2P_\frac{1}{2}$ state of
relative order $\alpha(Z\alpha)^2\ln(Z\alpha)^{-2}$, not taken into
account in the original analysis of the experiment, was obtained
recently in \cite{karjetp94}. This correction slightly changes the
original experimental result \cite{sok} $\Delta E=1~057~851.4(1.9)$
kHz, and we cite this corrected value in Table XX. The magnitude of
the new correction \cite{karjetp94} triggered a certain discussion in
the literature \cite{psy97,kphysscr98}. From the phenomenological point
of view the new correction \cite{karjetp94} is so small that neither of
our conclusions below about the result in \cite{sok} is affected by
this correction.

The Lamb shift value \cite{hp} was obtained from the measurement of
interval $2P_{\frac{3}{2}}-2S_{\frac{1}{2}}$, and the value of the
classical Lamb shift was extracted by subtraction of this energy
splitting from the theoretical value of the fine structure interval
$2P_{\frac{3}{2}}-2P_{\frac{1}{2}}$. As was first noted in
\cite{jenpach}, recent progress in the Lamb shift theory for $P$-states
requires reconsideration of the original value $\Delta E=1~057~839(12)$
kHz  of the classical Lamb shift obtained in \cite{hp}. We assume that
the total theoretical uncertainty of the fine structure interval is
about $0.08$ kHz (see discussion of the accuracy of $P$-state Lamb
shift above in Section \ref{pstaac}). Comparable contribution of $0.08$
kHz to the uncertainty of the fine structure interval originates from
the uncertainty of the most precise modern value of the fine structure
constant in \cite{hk} (see \eq{alpha}). Calculating the theoretical
value of the fine structure interval we obtain $\Delta
E(2P_{\frac{3}{2}}-2P_{\frac{1}{2}})=10~969~041.52~(11)$ kHz, which is
different from the value $\Delta
E(2P_{\frac{3}{2}}-2P_{\frac{1}{2}})=10~969~039.4~(2)$ kHz, used in
\cite{hp}. As a consequence the original experimental value \cite{hp}
of the classic Lamb shift changes to the one cited in Table XX. Due
to a relatively large uncertainty of the result this change does not
alter the conclusions below on comparison of the theory and
experiment.

Accuracy of the radiofrequency measurements of the classic $2S-2P$ Lamb
shift \cite{nau,lp,sok,hp,wijn98} is limited by the large (about $100$
MHz) natural width of the $2P$ state, and cannot be significantly
improved. New perspectives in reducing the experimental error bars of
the classic $2S-2P$ Lamb shift were opened with the development of the
Doppler-free two-photon laser spectroscopy for measurements of the
transitions between the energy levels with different principal quantum
numbers. Narrow linewidth of such transitions allows very precise
measurement of the respective transition frequencies, and indirect
accurate determination of $2S-2P$ splitting from this data\footnote{See
more discussion of this method below in Section \ref{phen1s}.}. The
latest experimental value \cite{schwob} in the  fifth line of  Table
XX was obtained by such methods.

Both the theoretical and experimental data for the classic
$2S_{1/2}-2P_{1/2}$ Lamb shift are collected in Table XX.
Theoretical results for the energy shifts in this Table contain errors
in the parenthesis where the first error is determined by the yet
uncalculated contributions to the Lamb shift, discussed
above\footnote{We have used in the calculations the result in
\eq{bgradrec} for the radiative-recoil correction of order
$\alpha(Z\alpha)^5$.  Competing result in \eq{pachradrec} would shift
the value of the classic Lamb shift by $0.78$ kHz, and would not effect
our conclusions on the comparison between theory and experiment below.}
and the second reflects the experimental uncertainty in the measurement
of the proton rms charge radius.  An immediate conclusion from the data
in Table XX is that the value of the  proton rms radius as measured
in \cite{dhr} is by far too small to accommodate the experimental data
on the Lamb shift. Even the larger value of the proton charge
radius obtained in \cite{ssbw} is inconsistent with the result of the
apparently most precise measurement of the $2S_{1/2}-2P_{1/2}$
splitting in \cite{sok}. The respective discrepancy is more than five
standard deviations. Results of four other direct measurements of the
classic Lamb shift collected in Table XX are compatible with the
theory if one uses the proton radius from \cite{ssbw}. Unfortunately,
these results are rather widely scattered and have rather large
experimental errors. Their internal consistency as well as their
consistency with theory leaves much to be desired. Taken at face value
the experimental results on the $2S_{1/2}-2P_{1/2}$ splitting indicate
of an even larger value of the proton charge radius than measured in
\cite{ssbw}. The situation with the experimental values of the proton
charge radius is unsatisfactory and a new measurement is clearly
warranted.

We will return to the numbers in the five last lines in Table XX
below.

\begin{center}
\underline{Table XX. Classic $2S_{1/2}-2P_{1/2}$ Lamb Shift}
\nopagebreak

\begin{tabular}{|l|l|l|}
\hline
&$\Delta E\mbox{ (kHz)}$&
\\ \hline  \hline
Newton,Andrews,& &\\
Unsworth (1979)\cite{nau}&$1~057~862~(20)$&Experiment\\
\hline
Lundeen,Pipkin (1981)\cite{lp}&$1~057~845~(9)$&Experiment\\
\hline
Palchikov,Sokolov,& &\\
Yakovlev (1983)\cite{sok}&$1~057~857.~6~(2.1)
$&Experiment \\
\hline
Hagley,Pipkin (1994)\cite{hp} & $1~057~842~(12)~$&Experiment \\
\hline
Wijngaarden,Holuj,& &\\
Drake (1998)\cite{wijn98}&
$1~057~852~(15)~$&Experiment \\
\hline
Schwob,Jozefovski,& &\\
de Beauvoir et al (1999)\cite{schwob}&
$1~057~845~(3)$&Exp.,
\cite{beauv,weitz,bosh,bourz,udem,lp,hp,wijn98} \\
\hline \hline
&$1~057~814~(2)~(4)$&Theory, $r_{p}=0.805~(11)$~fm \cite{dhr}\\
\hline
&$1~057~833~(2)~(4)$&Theory, $r_{p}=0.862~(12)$~fm \cite{ssbw}\\
\hline\hline
Weitz,Huber,Schmidt-& &\\
Kahler et al (1995)\cite{weitz}&
$1~057~857~(12)$&Self-consistent value
\\ \hline
Berkeland,Hinds,& &\\
Boshier (1995)\cite{bosh}&$1~057~842~(11)$&Self-consistent value
\\ \hline
Bourzeix,de Beauvoir,& &\\
Nez et al (1996)\cite{bourz}&$1~057~836~(8)$&
Self-consistent value
\\ \hline
Udem,Huber,& &\\
Gross et al (1997)\cite{udem}&$1~057~848~(5)$&Self-consistent value
\\
\hline
Schwob,Jozefovski,& &\\
de Beauvoir et al (1999)\cite{schwob}&
$1~057~841~(5)$& Self-consistent value\\
\hline \hline
&$1~057~843~(2)~(6)$&Theory, $r_{p}=0.891~(18)$~fm \\ \hline
\end{tabular}
\end{center}

\subsection{$1S$ Lamb Shift}\label{phen1s}

Unlike the case of the classic Lamb shift above, the Lamb shift in the
$1S$ is not amenable to a direct measurement as a splitting between
certain energy levels and in principle could be extracted from the
experimental data on the transition frequencies between the energy
levels with different principal quantum numbers. Such an approach
requires very precise measurement of the gross structure intervals, and
became practical only with the recent development of Doppler-free
two-photon laser spectroscopy. These methods allow very precise
measurements of the gross structure intervals in hydrogen with an
accuracy which is limited in principle only by the small natural
linewidths of respective transitions. For example, the $2S-1S$
transition in hydrogen is banned as a single photon process in the
electric dipole and quadrupole approximations, and also in the
nonrelativistic magnetic dipole approximation. As a result the natural
linewidth of this transition is determined by the process with
simultaneous emission of two electric dipole photons \cite{bs,blp},
which leads to the natural linewidth of the $2S-1S$ transition in
hydrogen about $1.3$ Hz. Many recent spectacular experimental successes
where achieved in an attempt to achieve an experimental accuracy
comparable with this extremely small natural linewidth.

The intervals of gross structure are mainly determined by the Rydberg
constant, and the same transition frequencies should be used both
for measurement of the Rydberg constant and for measurement of the
$1S$ Lamb shift. The first experimental task is to obtain an
experimental value of the $1S$ Lamb shift which is independent of the
precise value of the Rydberg constant. This goal may be achieved by
measuring two intervals with different principal quantum numbers. Then
one constructs a linear combination of these intervals which is
proportional $\alpha^2R$ (as opposed to $\sim R$ leading contributions
to the intervals themselves). Due to the factor $\alpha^2$ the precise
magnitude of the Lamb shift extracted from the above mentioned linear
combination of measured frequencies practically does not depend on the
exact value of the Rydberg constant.

For example, in one of the  most recent experiments \cite{udem}
measurement of the $1S$ Lamb shift is disentangled from the measurement
of the Rydberg constant by using the experimental data on two different
intervals of the hydrogen gross structure \cite{udem}

\beq   \label{1s2sudem}
f_{1S-2S}=2~466~061~413~187.34~(84)~\mbox{kHz}\qquad
\delta=3.4\cdot10^{-13},
\eeq

\noindent
and \cite{beauv,schwob}\footnote{The original experimental value
$f_{2S_\frac{1}{2}-8D_\frac{5}{2}}=770~649~561~585.0~(4.9)$~kHz
\cite{beauv} used in \cite{udem} was revised in \cite{schwob}, and we
give in \eq{2s8dbeau} this later value. The values of the Lamb shifts
obtained in \cite{udem} change respectively and Tables XX
and XXI contain these revised values.\label{revision}}

\beq    \label{2s8dbeau}
f_{2S_\frac{1}{2}-8D_\frac{5}{2}}=770~649~561~581.1~(5.9)~\mbox{kHz}\qquad
\delta=7.7\cdot10^{-12}.
\eeq

Theoretically these intervals are given by the expression in
\eq{lambdef}

\beq              \label{linearcomb}
E_{1S-2S}=[E_{2S_\frac{1}{2}}^{DR}-E_{1S_\frac{1}{2}}^{DR}]
+L_{2S_\frac{1}{2}}-L_{1S_\frac{1}{2}},
\eeq
\[
E_{2S-8D}=[E_{8D_\frac{5}{2}}^{DR}-E_{2S_\frac{1}{2}}^{DR}]
+L_{8D_\frac{5}{2}}-L_{2S_\frac{1}{2}},
\]

\noindent
where $E_{nl_j}^{DR}$  is the leading Dirac and recoil contribution to
the position of the respective energy level (first two terms in
\eq{lambdef}).

The first differences on the right hand side in equations
(\ref{linearcomb}) are proportional to the Rydberg constant, which thus
can be simply excluded from this system of two equations. Then we
obtain an equality between a linear combination of the $1S$, $2S$ and
$8D_\frac{5}{2}$ Lamb shifts and a linear combination of the
experimentally measured frequencies. This relationship admits direct
comparison with the Lamb shift theory without any further
complication. However, to make a comparison between the results of
different experiments feasible (different intervals of the
hydrogen gross structure are measured in different experiments) the
final experimental results are usually expressed in terms of the $1S$
Lamb shift measurement\footnote{The value of the $1S$ Lamb shift is
also often needed for extraction of the precise value of the Rydberg
constant from the experimental data, see Section \ref{rydberg} below.}.
The bulk contribution to the Lamb shift scales as $1/n^3$ which allows
one to use the theoretical value $L_{8D{_\frac{5}{2}}}=71.5$ kHz for the
$D$-state Lamb shift without loss of accuracy. Then a linear
combination of the Lamb shifts in $1S$ and $2S$ states may be directly
expressed in terms of the experimental data.  All other recent
measurements of the $1S$ Lamb shift \cite{schwob,weitz,bosh,bourz} also
end up with an experimental number for a linear combination of the
$1S$, $2S$  and higher level Lamb shifts. An unbiased extraction of the
$1S$ Lamb shift from the experimental data remains a problem even after
an experimental decoupling of the Lamb shift measurement from the
measurement of the Rydberg constant.

Historically the most popular approach to extraction of the value of
the $1S$ Lamb shift was to use the experimental value of
the classic $2S-2P$ Lamb shift (see three first lines in Table XXI).
Due to the large natural width of the $2P$ state the experimental
values of the classical Lamb shift  have relatively large experimental
errors (see Table XX), and unfortunately different results are not too
consistent. Such a situation clearly warrants another approach
to extraction of the $1S$ Lamb shift, one which should be independent
of the magnitude of the classic Lamb shift. A natural way to obtain a
self-consistent value of the $1S$ Lamb shift independent of the
experimental data on the $2S-2P$ splitting, is provided by the
theoretical relation between the $1S$ and $2S$ Lamb shifts discussed
above in Section \ref{l1s82s}. An important advantage of the
self-consistent method is that it produces an unbiased value of the
$L(1S)$ Lamb shift independent of the widely scattered experimental
data on the $2S-2P$ interval. Spectacular experimental progress in
the frequency measurement now allows one to obtain self-consistent
values of $1S$ Lamb shift from the experimental data\cite{udem}, with
comparable or even better accuracy (see five lines in Table XXI
below the theoretical values in the middle of the Table) than in the
method based on experimental results of the classic Lamb shift in
\cite{lp,hp}. The original experimental numbers from
\cite{udem,schwob} in the fourth and fifth lines in Table XXI are
averages of the self-consistent values and the values based on the
classic Lamb shift. The result in \cite{udem} is based on the
$f_{1S-2S}$ frequency measurement,  $f_{2S-8D/S}$ frequency from
\cite{beauv,schwob}, and the classic Lamb shift measurements
\cite{lp,hp}, while the result in \cite{schwob} is based on the
$f_{2S-12D}$ frequency measurement, as well as on the
frequencies measured in \cite{beauv,schwob,weitz,bosh,bourz,udem}, and
the classic Lamb shift measurements \cite{lp,hp,wijn98}. The respective
value of the classic $2S-2P$ Lamb shift is presented in the sixth line
of Table XX. Unlike other experimental numbers in Table XX, this
value of the classic Lamb shift depends on other experimental results
in this Table.

The experimental data on the $1S$ Lamb shift  should be compared with
the theoretical prediction

\begin{equation}    \label{1sth}
\Delta E_L(1S)=8~172~754~(14)~(32)~\mbox{kHz},
\end{equation}

calculated for $r_{p}=0.862~(12)$~fm \cite{ssbw}. The first error in
this result is determined by the yet uncalculated contributions to the
Lamb shift and the second reflects the experimental uncertainty in the
measurement of the proton rms charge radius.

The  experimental results in the first five lines in Table XXI seem to
be systematically higher even than the theoretical value in \eq{1sth}
calculated with the higher experimental value for the proton charge
radius \cite{ssbw}. One is tempted to come to the conclusion that the
experimental data give an indication of an even higher value of the
proton charge radius than the one measured in \cite{ssbw}. However, it
is necessary to remember that the "experimental" results in the first
five lines in the Table are "biased", namely they depend on the
experimental value of the $2S_{1/2}-2P_{1/2}$ Lamb shift
\cite{lp,hp,wijn98}. In view of a rather large scattering of the
results for the classic Lamb shift such dependence is unwelcome. To
obtain unbiased results we have calculated self-consistent values of
the $1S$ Lamb shift which are collected in Table XXI. These values
being formally consistent are rather widely scattered. Respective
self-consistent values of the classic Lamb shift obtained from the
experimental data in \cite{weitz,bosh,bourz,udem} are presented in
Table XX. All experimental results (both original and
self-consistent) in both Tables are systematically larger than the
respective theoretical predictions. The only plausible explanation is
that the true value of the proton charge radius is even larger than the
one measured in \cite{ssbw}. At this point we can invert the problem
and obtain the value of the proton charge radius

\beq    \label{selfconsrad}
r_p=0.891~(18)~\mbox{fm}
\eeq

\noindent
comparing the average $L(1S)=8~172~832~(25)$ kHz of the self-consistent
values of the $1S$ Lamb shift in Table XXI based on the most precise
recent frequency measurements \cite{beauv,schwob,weitz,bosh,bourz,udem}
with theory. Major contribution to the uncertainty of the proton charge
radius in \eq{selfconsrad} is due to the uncertainty of the
self-consistent Lamb shift.

A new analysis of the low momentum transfer electron scattering data
with account of the Coulomb and recoil corrections \cite{rosenfeld99}
resulted in  the proton radius value

\beq    \label{rosenrad}
r_p=0.880~(15)~\mbox{fm},
\eeq

\noindent
in good agreement with the self-consistent value in \eq{selfconsrad}.
Another recent analysis of the elastic $e^\pm p$ scattering data
resulted in an even higher value of the proton charge radius
\cite{ezhela}

\beq        \label{ezelarad}
r_p=0.897~(2)~(1)~(3)~\mbox{fm},
\eeq

\noindent
where the error in the first brackets is due to statistics, the second
error is due to normalization effects, and the third error reflects the
model dependence. Comparing results of these two analysis one has to
remember that the Coulomb corrections which played the most important
role in \cite{rosenfeld99} were ignored in \cite{ezhela}.  The results
of \cite{ezhela} depend also on specific parametrization of the nucleon
form factors. Under these conditions, despite the superficial agreement
between the results in \eq{rosenrad} and \eq{ezelarad}, the extraction
of the precise value of the proton charge radius from the scattering
data cannot be considered satisfactory, and further work in this field
is required.

Theoretical values of the classic $2S-2P$ Lamb shift and of the $1S$
Lamb shift corresponding to the proton radius in \eq{selfconsrad} are
given in the last lines of Tables XX and XXI, respectively. It is
clear that there is much more consistency between these theoretical
predictions and the mass of experimental data on the Lamb
shifts than between the predictions based on the proton charge radius
from \cite{ssbw} (to say nothing about the radius from \cite{dhr}) and
experiment. We expect that future experiments on measurement of the
proton charge radius will confirm the hydrogen Lamb shift prediction
of the value of the proton charge radius in \eq{selfconsrad}. Precise
measurements of the Lamb shift in muonic hydrogen (see discussion in
Section \ref{munichydexp}) provide the best approach to measurement of
the proton charge radius, and would allow reduction of error bars
in \eq{selfconsrad} by at least an order of magnitude.

\begin{center}
\underline{Table XXI. $1S$ Lamb Shift}
\nopagebreak

\begin{tabular}{|l|l|l|}    \hline
& $\Delta E\mbox{ (kHz)}$   &
\\ \hline \hline
Weitz,Huber,Schmidt-& &\\
Kahler et al (1995)\cite{weitz}&$8~172~874~(60)$ &
Exp., $L_{2S2P}$ \cite{lp}  \\  \hline
Berkeland,Hinds,& &\\
Boshier (1995)\cite{bosh}     &   $8~172~827~(51)~$ &
Exp., $L_{2S2P}$ \cite{lp,hp}\\ \hline
Bourzeix,de Beauvoir,& &\\
Nez et al (1996)\cite{bourz}  &   $8~172~798~(46)$
& Exp., $L_{2S2P}$ \cite{lp,hp}\\
\hline
Udem,Huber,& &\\
Gross et al (1997)\cite{udem}&$8~172~851~(30)$ &
Exp., $L_{2S2P}$ \cite{lp,hp}\\
\hline
Schwob,Jozefovski,& &\\
de Beauvoir et al
(1999)\cite{schwob}&
$8~172~837~(22)$&Exp.,
\cite{beauv,weitz,bosh,bourz,udem,lp,hp,wijn98} \\
\hline\hline
&$8~172~605~(14)~(28)$&Theory, $r_{p}=0.805~(11)$~fm \cite{dhr}\\
\hline
&$8~172~754~(14)~(32)$&Theory, $r_{p}=0.862~(12)$~fm \cite{ssbw}\\
\hline \hline
Weitz,Huber,Schmidt-& &\\
Kahler et al (1995)\cite{weitz}&$8~172~937~(99)$&
Self-consistent value
\\  \hline
Berkeland,Hinds,& &\\
Boshier (1995)\cite{bosh}&$8~172~819~(89)$&
Self-consistent value\\
\hline
Bourzeix,de Beauvoir,& &\\
Nez et al (1996)\cite{bourz}&$8~172~772~(66)$&
Self-consistent value\\
\hline
Udem,Huber,& &\\
Gross et al (1997)\cite{udem}&$8~172~864~(40)$&
Self-consistent value\\
\hline
Schwob,Jozefovski,& &\\
de Beauvoir et al (1999)\cite{schwob}&
$8~172~805~(43)$&
Self-consistent value\\
\hline\hline
&$8~172~832~(14)~(51)$&Theory, $r_{p}=0.891~(18)$~fm \\
\hline
\end{tabular}
\end{center}

\subsection{Isotope Shift} \label{isshift}

The methods of Doppler-free two-photon laser spectroscopy allow very
precise comparison of the frequencies of the $1S-2S$ transitions in
hydrogen and deuterium. The frequency difference

\beq
\Delta E=[E(2S)-E(1S)]_D-[E(2S)-E(1S)]_H
\eeq

\noindent
is called the hydrogen-deuterium isotope shift. Experimental accuracy
of the isotope shift measurements was improved by three orders of
magnitude during the period from 1989 to 1998 (see Table XXII) and the
uncertainty of the most recent experimental result \cite{huber} was
reduced to $0.15$ kHz.

The main contribution to the hydrogen-deuterium isotope shift is a pure
mass effect and is determined by the term $E_{nj}^{DR}$ in
\eq{lambdef}. Other contributions coincide with the respective
contributions to the Lamb shifts in Tables I-X.  Deuteron specific
corrections discussed in Section \ref{nonelectromagnetic} and
collected in \eq{deuterza5}, \eq{friaraprissh}, \eq{khrsen2}, and
\eq{deutradza6} also should be included in the theoretical expression
for the isotope shift.

All yet uncalculated nonrecoil corrections to the Lamb shift almost
cancel in the formula for the isotope shift, which is thus
much more accurate than the theoretical expressions for the Lamb
shifts.  Theoretical uncertainty of the isotope shift is mainly
determined by the unknown single logarithmic and
nonlogarithmic contributions of order $(Z\alpha)^7(m/M)$ and
$\alpha(Z\alpha)^6(m/M)$ (see Sections \ref{recoilgenza7} and
\ref{recgenaza6}), and also by the uncertainties of the deuteron size
and structure contributions discussed in Section
\ref{nonelectromagnetic}. Overall theoretical uncertainty of all
contributions to the isotope shift, besides the leading proton
and deuteron size corrections does not exceed $0.8$ kHz.

Theoretical predictions for the isotope shift strongly depend on the
magnitude of the radiative-recoil corrections of order
$\alpha(Z\alpha)^5(m/M)m$. Unfortunately, there is still an unresolved
discrepancy between the theoretical results on these corrections
obtained in \cite{bg85,bg87,bg871} and in \cite{pach95} (for more
detail see discussion in Section \ref{lambradrecel}), and the
difference between the respective values of the isotope shift is about
$2.7$ kHz, to be compared with the uncertainty $0.15$ kHz
of the most recent experimental result\cite{huber}. Discrepancy between
the theoretical predictions for the radiative-recoil corrections of
order $\alpha(Z\alpha)^5(m/M)m$ is one of the outstanding theoretical
problems, and efforts for its resolution are necessary.

Numerically the sum of all theoretical contributions to the isotope
shift, besides the leading nuclear size contributions in
\eq{chargerad}, is equal to

\beq
\Delta E=670~999~566.~1~(1.5)~(0.8)~\mbox{kHz},
\eeq

\noindent
for the $\alpha(Z\alpha)^5(m/M)m$ contribution from
\cite{bg85,bg87,bg871}, and

\beq
\Delta E=670~999~568.~9~(1.5)~(0.8)~\mbox{kHz},
\eeq

\noindent
for the $\alpha(Z\alpha)^5(m/M)m$ contribution from \cite{pach95}. The
uncertainty in the first parenthesis is defined by the experimental
error of the electron-proton and proton-deuteron mass ratios, and the
uncertainty in the second parenthesis is the theoretical uncertainty
discussed above.

Individual uncertainties of the proton and deuteron charge radii
introduce by far the largest contributions in the uncertainty of the
theoretical value of the isotope shift. Uncertainty of the charge radii
are much larger than the experimental error of the isotope shift
measurement or the uncertainties of other theoretical contributions. It
is sufficient to recall that uncertainty of the $1S$ Lamb shift due to
the experimental error of the proton charge radius is as large as $32$
kHz (see \eq{1sth}), even if ignore all problems connected with the
proton radius contribution (see discussion in Sections
\ref{classlamb2s2p},\ref{phen1s}).

In such a situation it is natural to invert the problem and
to use the high accuracy of the optical measurements and isotope
shift theory for determination of the difference of charge
radii squared of the deuteron and proton. We obtain

\beq     \label{chergesquare1}
r_D^2-r_p^2=3.819~3~(01)~(11)~(04)~\mbox{fm}^2,
\eeq

\noindent
using the \cite{bg85,bg87,bg871} value of the $\alpha(Z\alpha)^5(m/M)m$
corrections, and

\beq     \label{chergesquare2}
r_D^2-r_p^2=3.821~3~(01)~(11)~(04)~\mbox{fm}^2,
\eeq

\noindent
for the \cite{pach95} value of the $\alpha(Z\alpha)^5(m/M)m$
corrections. Here the first contribution to the uncertainty is due to
the experimental error of the isotope shift measurement, the second
uncertainty is due to the experimental error of the electron-proton
mass ratio determination, and the third is generated by the
theoretical uncertainty of the isotope shift. An improvement of the
precision of the electron-proton mass ratio measurement is crucial for
a more accurate determination of the difference of the charge radii
squared of the deuteron and proton from the isotope shift measurements.

The difference of the deuteron and proton charge radii squared is
connected to the so called deuteron mean square matter radius
(see, e.g., \cite{klarsfeld,friarmartspr}), which may be extracted
on one hand from the experimental data on the low energy
nucleon-nucleon interaction, and on the other hand from the experiments
on low energy elastic electron-deuteron scattering. These two kinds
of experimental data used to generate  inconsistent results for the
deuteron matter radius as was first discovered in \cite{klarsfeld}.
The discrepancy was resolved in \cite{sitr}, where the Coulomb
distortion in the second order Born approximation was taken into
account in the analysis of the  electron-deuteron elastic scattering.
This analysis was further improved in \cite{hermrosen} where also the
virtual excitations of the deuteron in the electron-deuteron scattering
were considered. Now the values of the deuteron matter radius
extracted from the low energy nucleon-nucleon interaction
\cite{friarmartspr} and from the low energy elastic electron-deuteron
scattering  \cite{sitr,hermrosen} are in agreement, and do not
contradict the optical data in \eq{chergesquare1},  \eq{chergesquare2}.
The isotope shift measurements are today the source of the most precise
experimental data on the charge radii squared difference, and the
deuteron matter radius. In view of the unsatisfactory situation with
the proton charge radius measurements, more experimental work is
clearly warranted.

\begin{center}
\underline{Table XXII. Isotope Shift}
\nopagebreak

\begin{tabular}{|l|l|l|}    \hline
& $\Delta E\mbox{ (kHz)}$   &
\\
\hline \hline
& &\\
Boshier,Baird,Foot et al (1989)\cite{boshier}&$670~994~33~(64)$ &
\\
\hline
& &\\
Schmidt-Kaler,Leibfried,Weitz et al (1993)\cite{schmidt}&$670~994~414~(22)$ &
\\
\hline
& &\\
Huber,Udem, Gross et al (1998)\cite{huber}&$670~994~334.~64~(15)$ &
\\
\hline
\end{tabular}
\end{center}

\subsection{Lamb Shift in Helium Ion $He^+$}

The theory of high order corrections to the Lamb shift described above
for $H$ and $D$ may also be applied to other light hydrogenlike ions.
The simplest such ion for which experimental data on the classic
$2S_\frac{1}{2}-2P_\frac{1}{2}$ Lamb shift exists  is $He^+$.  As
measured in \cite{wkd} by the quenching-anisotropy method,
$L(2S_\frac{1}{2}-2P_\frac{1}{2},He^+)=14~042.52~(16)$ MHz. A new
measurement of the classic Lamb shift in $He^+$ by the anisotropy
method has been completed recently \cite{whd2000}. In the process of
this work the authors have discovered a previously unsuspected source
of systematic error in the earlier experiment \cite{wkd}. The result of
the new experiment \cite{whd2000} is
$L(2S_\frac{1}{2}-2P_\frac{1}{2},He^+)=14~041.13~(17)$ MHz. Besides the
experimental data this result depends also on the theoretical value of
the fine structure interval $\Delta
E(2P_\frac{3}{2}-2P_\frac{1}{2})=175~593.50~(2)$ MHz, which may be
easily obtained from the theory described in this paper.

Theoretical calculation of the $He^+$ Lamb shift is straightforward
with all the formulae given above. It is only necessary to recall
that all contributions scale with the power of $Z$, and the terms with
high power of $Z$ are enhanced in comparison with the hydrogen case.
This is particularly important for the contributions of order
$\alpha(Z\alpha)^n$. One can gain in accuracy using in the theoretical
formulae high-$Z$ results for the functions $G_{SE}(Z\alpha)$ and
$G_{VP}(Z\alpha) \cite{mohr96,jms99}$\footnote{The function
$G_{SE}(Z\alpha)$ is defined in footnote \ref{footng7} in Section
\ref{lambalphazalpha7}. The function $G_{VP}(Z\alpha)$ is defined
similarly to the function $G_{VP,7}(Z\alpha)$ in \eq{gudef}, but like
the function $G_{SE}(Z\alpha)$ also includes nonlogarithmic
contributions of order $\alpha(Z\alpha)^6$.}, extrapolated to $Z=2$,
instead of nonlogarithmic terms of order $\alpha(Z\alpha)^6$ from Table
V and of the terms of order $\alpha(Z\alpha)^7$ from Table VII.
Theoretical uncertainty may be estimated by scaling with $Z$ the
uncertainty of the hydrogen formulae.  After calculation we obtain
$L_{th}(2S-2P,He^+)=14~041.18~(13)$ MHz, in excellent agreement with
the latest experimental result \cite{whd2000}. Thus as a result of the
new experiment \cite{whd2000} the only discrepancy between the Lamb
shift theory and experiment which existed in recent years has been
successfully eliminated.

\subsection{Rydberg Constant}  \label{rydberg}

The leading contribution to the energy levels in hydrogen
in \eq{lambdef} is clearly sensitive to the value of the Rydberg
constant, and, hence, any measurement of the gross structure interval
in hydrogen and deuterium may be used for determination of the value of
the Rydberg constant, if the magnitudes of the Lamb shifts of
respective energy levels are known. In practice only the data on the
$1S$ and $2S$ (or classic $2S-2P$) Lamb shifts limits the accuracy of
the determination of the Rydberg constant. Higher order Lamb shifts are
known theoretically with sufficient accuracy. All recent values of the
Rydberg constant are derived from experimental data on at least two
gross structure intervals in hydrogen and/or deuterium. This allows
simultaneous experimental determination of both the $1S$ Lamb shift and
the Rydberg constant from the experimental data, and makes the obtained
value of the Rydberg virtually independent of the Lamb shift theory
and, what is more important on the controversial experimental data on
the proton charge radius. Either self-consistent values of both the
$1S$ and $2S$ Lamb shifts, or direct experimental value of the classic
$2S-2P$ and respective $2S$ dependent value of $1S$ Lamb shift are
usually used for determination of the precise value of the Rydberg
constant.

Recent experimental results for the Rydberg constant are collected
in Table XXIII. A few comments are due on the latest results. The value
in \cite{beauv} is based on the measurement of the $f_{2S-8D/S}$
frequency, $f_{1S-2S}$ frequency from \cite{and}, and the classic Lamb
shift measurements \cite{lp,hp}. This result should be changed
due to recent revision \cite{schwob} of the $f_{2S-8D}$ frequency
$^{\ref{revision}}$. The result in \cite{udem} is based on the
$f_{1S-2S}$ frequency measurement,  $f_{2S-8D/S}$ frequency from
\cite{beauv}, and the classic Lamb shift measurements
\cite{lp,hp}, and also should be revised $^{\ref{revision}}$.
The result in \cite{schwob} is based on the $f_{2S-12D}$ frequency
measurement, as well as on the frequencies measured in
\cite{beauv,schwob,weitz,bosh,bourz,udem}, and the classic Lamb shift
measurements \cite{lp,hp,wijn98}. The results in
\cite{beauv,udem,schwob}  are averages, obtained from experimental data
on different measured frequencies and their linear combinations
in hydrogen and deuterium. In principle they depend both on the
measured and self-consistent values of the Lamb shifts.

To get a better idea of the effect of the rather widely spread
experimental data on the classic Lamb shift on the value of the Rydberg
constant and on the balance of uncertainties one can
compare the experimental results in the upper part of Table XXIII with
the value of the Rydberg constant, which may be calculated from the
experimental frequencies and self-consistent values of Lamb shifts.
Values of the Rydberg constant calculated from the experimental
transition frequencies in hydrogen in \cite{udem,schwob} and the average
self-consistent values of the $1S$ and $2S-2P$ Lamb shifts (see Tables
XX and XXI) are presented in the middle of Table XXIII. The first
error of these values of the Rydberg constant is determined by the
accuracy of the average self-consistent Lamb shifts, the second is
defined by the experimental error of the frequency measurement, and the
third is determined by the accuracy of the electron-proton mass
ratio. We see that the  results in the lower part of Table XXIII are
compatible with the results of the least square adjustments of all
experimental data in the upper half of the Table which thus do not
depend crucially on the somewhat uncertain experimental data on the
$2S-2P$ Lamb shift. We also see that the uncertainties of the Lamb
shift determination and frequency measurements give the largest
contributions to the Rydberg constant uncertainty in most experiments.

High accuracy of the modern experimental data and theory could allow
Rydberg constant determination from direct comparison between the
theory and experiment, without appeal to the Lamb shift results.
Respective values of the Rydberg constant, calculated with the
self-consistent proton radius from \eq{selfconsrad} are presented in
the lower part of Table XXIII.  The first error of these values of the
Rydberg constant is determined by the accuracy of the theoretical
formula, the second is defined by the experimental error of the
frequency measurement, the third is determined by the accuracy of the
electron-proton mass ratio, and the last one depends on the proton
radius uncertainty. The values of the Rydberg constant in the last three
lines in Table XXIII are rather accurate, and would be able to complete
with the other methods of determination of the Rydberg constant from
the experimental data after the current controversial situation with
the precise value of the proton charge radius will be resolved. It is
appropriate to emphasize once again that the experimental values of
the Rydberg constant in the upper part of Table XXIII are based on
measurements of at least two intervals of the hydrogen and/or deuterium
gross structure and are thus independent of the uncertain value of the
proton charge radius.

\begin{center}
\underline{Table XXIII. Rydberg Constant}
\nopagebreak

\begin{tabular}{|l|l|}
\hline
&$R_\infty~\mbox{(cm$^{-1}$)}$
\\ \hline  \hline
Andreae,K\"onig,Wynands& \\
 et al (1992)\cite{and}
&$109~737.~315~684~1~(42)$
\\
\hline
Nez,Plimmer,Bourzeix& \\
 et al (1992)\cite{nezpb}
&$109~737.~315~683~0~(31)$
\\
\hline
Nez,Plimmer,Bourzeix& \\
 et al (1993)\cite{nez}
&$109~737.~315~683~4~(24)$
\\
\hline
Weitz,Huber,Schmidt-& \\
Kahler et al
(1994)\cite{weitz94}&$109~737.~315~684~4~(31)$
\\
\hline
Weitz,Huber,Schmidt-& \\
Kahler et al
(1995)\cite{weitz}&$109~737.~315~684~9~(30)$
\\
\hline
de Beauvoir,Nez,& \\
Julien et al
(1997)\cite{beauv}&$109~737.~315~685~9~(10)$
\\
\hline
Udem,Huber,& \\
Gross et al (1997)\cite{udem}&$109~737.~315~686~39~(91)$
\\
\hline
Schwob,Jozefovski,de &\\
Beauvoir et al (1999)\cite{schwob}&
$109~737.~315~685~16~(84)$
\\
\hline\hline
Self-consistent Lamb,
\cite{beauv,schwob}&$109~737.~315~685~8~(5)~(8)~(1)$
\\
\hline
Self-consistent Lamb, \cite{udem}&$109~737.~315~685~2~(11)~(0)~(1)$
\\  \hline
Self-consistent Lamb, \cite{schwob}&$109~737.~315~684~6~(4)~(9)~(1)$
\\  \hline\hline
Proton radius \eq{selfconsrad}, \cite{beauv,schwob}&
$109~737.~315~684~3~(3)~(8)~(1)~(9)$
\\ \hline
Proton radius \eq{selfconsrad}, \cite{udem}&
$109~737.~315~682~2~(6)~(0)~(1)~(22)$
\\ \hline
Proton radius \eq{selfconsrad}, \cite{schwob}&
$109~737.~315~683~1~(3)~(9)~(1)~(9)$
\\   \hline
\end{tabular}
\end{center}

\subsection{$1S-2S$ Transition in Muonium}

Starting with the pioneering work \cite{chu8889} Doppler-free two-photon
laser spectroscopy was also applied for measurements of the  gross
structure interval in muonium. Experimental results
\cite{chu8889,jung91,maas,meyer99} are collected in Table XXIV, where
the error in the first brackets is due to statistics and the second
error is due to systematic effects. The highest accuracy was
achieved in the latest experiment \cite{meyer99}

\beq            \label{1s2smuonexp}
\Delta E=2~455~528~941.~0~(9.8)~\mbox{MHz}.
\eeq

Theoretically, muonium differs from hydrogen in two main respects.
First, the nucleus in the muonium atom is an elementary structureless
particle unlike the composite proton  which is a quantum chromodynamic
bound state of quarks. Hence nuclear size and structure
corrections  in Table X do not contribute to the muonium energy
levels. Second, the muon is about ten times lighter than the proton, and
recoil and radiative-recoil corrections are numerically much more
important for muonium than for hydrogen. In almost all other respects,
muonium looks exactly like hydrogen with a somewhat lighter nucleus,
and the theoretical expression for the $1S-2S$ transition frequency may
easily be obtained from the leading external field contribution in
\eq{lambdef} and different contributions to the energy levels collected
in Tables I-IX, after a trivial substitution of the muon mass. Unlike
the case of hydrogen, for muonium we cannot ignore
corrections in the two last lines of Table II, and we have to
substitute the classical elementary particle contributions in
\eq{diracnucl} and \eq{pauliprpton} instead of the composite proton
contribution in the fifth line in Table IX.  After these modifications
we obtain a theoretical prediction for the frequency of the $1S-2S$
transition in muonium\footnote{Even though there is an enhanced role of
the recoil corrections for muonium, discrepancy between the results for
the radiative-recoil corrections of order $\alpha(Z\alpha)^5(m/M)m$
discussed in Section \ref{lambradrecel} is too small, in comparison
with the uncertainty originating from the mass ratio, to affect the
theoretical prediction for the gross structure interval.}

\beq                 \label{1s2smuonth}
\Delta E=2~455~528~934.~9~(0.3)~\mbox{MHz}.
\eeq

The dominant contribution to the uncertainty of this theoretical result
is generated by the uncertainty of the muon-electron mass ratio, and
we have used the most precise value of this mass ratio \cite{lbdd} (see
for more details the next Section \ref{hyperineexp}) in our
calculations. All other contributions to the uncertainty of the
theoretical prediction:  uncertainty of the Rydberg constant,
uncertainty of the theoretical expression, etc., are at least an order
of magnitude smaller.

There is a complete agreement between the experimental and
theoretical results for the $1S-2S$ transition frequency in
\eq{1s2smuonexp} and \eq{1s2smuonth}, but clearly further improvement
of the  experimental data is warranted.

\begin{center}
\underline{Table XXIV. $1S-2S$ Transition in Muonium}
\nopagebreak

\begin{tabular}{|l|l|}
\hline
&$\Delta E~\mbox{(MHz)}$
\\ \hline  \hline
& \\
Danzman,Fee,Chu et al (1989)\cite{chu8889}&$2~455~527~936~(120)~(140)$
\\
\hline
& \\
Jungmann,Baird, Barr et al (1991) \cite{jung91}&$2~455~528~016~(58)~(43)$
\\
\hline
& \\
Maas,Braun,Geerds et al (1994) \cite{maas}&$2~455~529~002~(33)~(46)$
\\
\hline
& \\
Meyer,Bagaev,Baird et al (1999)
\cite{meyer99}&$2~455~528~941.0~(9.8)$
\\
\hline\hline
& \\
Theory &$2~455~528~934.9~(0.3)$
\\
\hline
\end{tabular}
\end{center}

\subsection{Phenomenology of Light Muonic Atoms}\label{munichydexp}

There are very few experimental results on the energy levels in
light hydrogenlike muonic atoms. The classic
$2P_{\frac{1}{2}(\frac{3}{2})}-2S_\frac{1}{2}$ Lamb shift in muonic
helium ion $(\mu~^4He)^+$ was measured at CERN many years ago
\cite{bertin,carboni1,carboni2,carboni3} and the experimental data
was found to be in agreement with the existing theoretical predictions.
A comprehensive theoretical review of these experimental results was
given in \cite{borierink}, and we refer the interested reader to this
review. It is necessary to mention, however, that a recent new
experiment \cite{hauser} failed to confirm the old experimental
results. This leaves the problem of the experimental measurement of the
Lamb shift in muonic helium in an uncertain situation, and further
experimental efforts in this direction are clearly warranted. The
theoretical contributions  to the Lamb shift were discussed above in
Section \ref{mouniclambgen} mainly in connection with muonic hydrogen,
but the respective formulae may be used for muonic helium as well. Let
us mention that some of these contributions were obtained a long time
after publication of the review \cite{borierink}, and should be used in
the comparison of the results of the future helium experiments with
theory.

There also exists a proposal on measurement of the hyperfine
splitting in the ground state of muonic hydrogen with the accuracy
about $10^{-4}$ \cite{bakalov}. Inspired by this proposal the hadronic
vacuum polarization contribution of the ground state hyperfine
splitting in muonic hydrogen was calculated in \cite{faustmart97},
where it was found that it gives relative contribution about
$2\cdot10^{-5}$ to hyperfine splitting. We did not include this
correction in our discussion of hyperfine splitting in muonic hydrogen
mainly because it is smaller than the theoretical errors due to the
polarizability contribution.

The current surge of interest in muonic hydrogen is mainly inspired by
the desire to obtain a new more precise value of the proton charge
radius as a result of measurement of the $2P-2S$ Lamb shift
\cite{kottman}. As we have seen in Section \ref{mouniclambgen} the
leading proton radius contribution is about $2\%$ of the total $2P-2S$
splitting, to be compared with the case of electronic hydrogen where
this contribution is relatively two orders of magnitude smaller, about
$10^{-4}$ of the total $2P-2S$. Any measurement of the $2P-2S$ Lamb
shift in muonic hydrogen with relative error comparable with the
relative error of the Lamb shift measurement in electronic hydrogen is
much more sensitive to the value of the proton charge radius.

The natural linewidth of the $2P$ states in muonic hydrogen and
respectively of the $2P-2S$ transition is determined by the linewidth
of the $2P-1S$ transition, which is equal $\hbar \Gamma=0.077$ meV. It
is planned \cite{kottman} to measure $2P-2S$ Lamb shift with an
accuracy at the level of $10\%$ of the natural linewidth, or with
an error about $0.008$ meV, what means measuring the $2P-2S$ transition
with relative error about $4\cdot10^{-5}$.

The total $2P-2S$ Lamb shift in muonic hydrogen calculated according to
the formulae in Table XI for $r_p=0.862~(12)$ fm, is equal

\beq
\Delta E(2P-2S)=202.225~(108)~\mbox{meV}.
\eeq

\noindent
We can write this result as a difference of a theoretical number and a
term proportional to the proton charge radius squared

\beq     \label{workradform}
\Delta E(2P-2S)=206.108~(4)-5.2250<r^2>~\mbox{meV}.
\eeq

\noindent
We see from this equation that when the experiment achieves the planned
accuracy of about $0.008$ meV \cite{kottman} this would
allow determination of the proton charge radius with relative accuracy
about $0.1\%$ which is about an order of magnitude better than the
accuracy of the available experimental results.

Uncertainty in the sum of all theoretical contributions which are not
proportional to the proton charge raduis squared in \eq{workradform}
may be somewhat reduced. This uncertainty is determined by the
uncertainties  of the purely electrodynamic contributions and by the
uncertainty of the nuclear polarizability contribution of order
$(Z\alpha)^5m$. Purely electrodynamic uncertainties are introduced by
the uncalculated nonlogarithmic contribution of order
$\alpha^2(Z\alpha)^4$ corresponding to the diagrams with radiative
photon insertions in the graph for leading electron polarization in
Fig.\ \ref{radphotelpolcoul}, and by the uncalculated light by light
contributions in Fig.\ \ref{6setscoulfig} $(e)$, and may be as large as
$0.004$ meV. Calculation of these contributions and elimination of the
respective uncertainties is the most immediate theoretical problem in
the theory of muonic hydrogen.

After calculation of these corrections, the uncertainty in the sum of
all theoretical contributions except those which are directly
proportional to the proton radius squared will be determined by the
uncertainty of the proton polarizability contribution of order
$(Z\alpha)^5$. This uncertainty of the proton polarizability
contribution is currently about $0.002$ meV, and it will be difficult
to reduce it in the near future. If the experimental error of
measurement $2P-2S$ Lamb shift in hydrogen will be reduced to a
comparable level, it would be possible to determine the proton radius
with relative error smaller that $3\cdot 10^{-4}$ or with absolute
error about $2\cdot10^{-4}$ fm, to be compared with the current
accuracy of the proton radius measurements producing the results with
error on the scale of $0.01$ fm.

\section{~~~Hyperfine Splitting}\label{hyperineexp}

\subsection{Hyperfine Splitting in Hydrogen}\label{hydhfsexp}

Hyperfine splitting in the ground state of hydrogen was measured
precisely about thirty year ago \cite{hvl,ed}

\beq
\Delta E_{HFS}(H) = 1~420~405.751~766~7~(9)~{\rm kHz}\qquad
\delta=6\cdot 10^{-13}.
\eeq

For many years, this hydrogen maser measurement remained the most
accurate experiment in modern physics. Only recently the accuracy of
the Doppler-free two-photon spectroscopy achieved comparable precision
\cite{udem} (see the result for the $1S-2S$ transition frequency in
\eq{1s2sudem}).

The theoretical situation for the hyperfine splitting in hydrogen
always remained less satisfactory due to the uncertainties connected
with the proton structure.

The scale of hyperfine splitting in hydrogen is determined by the Fermi
energy in \eq{ef}

\beq
E_{F}(H)=1~418~840.11~(3)~(1)~\mbox{kHz},
\eeq

\noindent
where the uncertainty in the first brackets is due to the uncertainty
of the proton anomalous magnetic moment $\kappa$ measured in nuclear
magnetons, and the uncertainty in the second brackets is due to the
uncertainty of the fine structure constant in \eq{alpha}.

The sum of all nonrecoil corrections to hyperfine splitting collected
in Tables XII-XV is equal to

\beq   \label{hfsnonrechydtheory}
\Delta E_{HFS}(H)=1~420~452.04~(3)~(1)~\mbox{kHz},
\eeq

\noindent
where the first error comes from the experimental error of the
proton anomalous magnetic moment $\kappa$, and the second comes from
the error in the value of the fine structure constant $\alpha$. The
experimental error of $\kappa$ determines the uncertainty of the sum of
all nonrecoil contributions to the hydrogen hyperfine splitting.

The theoretical error of the sum of all nonrecoil contributions is about
$3$ Hz, at least an order of magnitude smaller than the uncertainty
introduced by the proton anomalous magnetic monent $\kappa$, and we did
not write it explicitly in \eq{hfsnonrechydtheory}. In relative
units this theoretical error is about $2\cdot10^{-9}$, to be compared
with the estimate of the same error $1.2\cdot10^{-7}$ made in
\cite{bodyen88}.  Reduction of the theoretical error by two orders of
magnitude emphasizes the progress achieved in calculations of nonrecoil
corrections during the last decade.

The real stumbling block on the road to a more precise theory of
hydrogen hyperfine splitting is the situation with the proton
structure, polarizability and recoil corrections, and there was little
progress in this respect during recent years.

Following tradition \cite{bodyen88} let us compare the theoretical
result without the unknown proton polarizability correction with the
experimental data in the form

\beq             \label{exptheohfshyd}
\frac{\Delta E_{th}(H)-\Delta
E_{exp}(H)}{E_F(H)}=-4.5~(1.1)\cdot10^{-6}.
\eeq

The difference between the numbers and estimates of errors on the RHS in
\eq{exptheohfshyd} and the respective numbers in \cite{bodyen88} is due
mainly to different treatment of the form factor parametrizations and
the values of the proton radius. New recoil and structure corrections
collected in the lower part of Table XIX had relatively small effect
on the numbers on the RHS in \eq{exptheohfshyd}.

The uncertainty in \eq{exptheohfshyd} is dominated by the uncertainty
of the Zemach correction in \eq{hfsbortrfr}. As we discussed in Section
\ref{zemchhfsprot}, this uncertainty is connected with the accuracy of
the dipole fit for the proton formfactor and contradictory experimental
data on the proton radius. It is fair to say that the estimate of this
uncertainty is to a certain extent subjective and reflects the
prejudices of the authors. One might hope that new experimental data
on the proton radius and the proton form factor would provide more
solid ground for consideration of the Zemach correction and would allow
a more reliable estimate of the difference on the LHS in
\eq{exptheohfshyd}.

The result in \eq{exptheohfshyd} does not contradict a rigorous upper
bound on the proton polarizability correction in \eq{rigboundpol}. It
could be understood as an indication of the relatively large magnitude
of the polarizability contribution, and as a challenge to theory to
obtain a reliable estimate of the polarizability contribution on the
basis of the new experimental data.

\subsection{Hyperfine Splitting in Deuterium}\label{deuthfsexp}

The hyperfine splitting in the ground state of deuterium was measured
with very high accuracy a long time ago \cite{winerams,ramsey}

\beq       \label{deuthfssexp}
\Delta E_{HFS}(D) = 327~384.352~521~9~(17)~{\rm kHz}\qquad
\delta=5.2\cdot 10^{-12}.
\eeq

The expression for the Fermi energy in \eq{ef}, besides the trivial
substitutions similar to the ones in the case of hydrogen, should be
also multiplied by an additional factor $3/4$ corresponding to the
transition from a spin one half nucleus in the case of hydrogen and
muonium to the spin one nucleus in the case of deuterium. The final
expression for the deuterium Fermi energy has the form

\beq           \label{efdeu}
E_F(D)=\frac{4}{9}\alpha^2 \mu_d\frac{m}{M_p}
\left(1+\frac{m}{M_d}\right)^{-3}ch~R_{\infty},
\eeq

\noindent
where $\mu_d=0.857~438~2284~(94)$ is the deuteron magnetic moment in
nuclear magnetons, $M_d$ is the deuteron mass, and $M_p$ is the proton
mass. Numerically

\beq
E_F(D)=326~967.678~(4)~\mbox{kHz},
\eeq

\noindent
where the main contribution to the uncertainty is introduced by the
uncertainty of the deuteron anomalous magnetic moment measure in
nuclear magnetons.

As in the case of hydrogen, after trivial modifications, we can use all
nonrecoil corrections in Tables XII-XVI for calculations in deuterium.
The sum of all nonrecoil corrections is numerically equal to

\beq          \label{norecdeut}
\Delta E_{nrec}(D)=327~339.143~(4)~\mbox{kHz}.
\eeq

Unlike the proton, the deuteron is a weakly bound system so one
cannot simply use the results for the hydrogen recoil and structure
corrections for deuterium. What is needed in the case of deuterium is a
completely new consideration. Only one minor nuclear structure
correction \cite{bohr,low50,lowsal51,greenbfol} was discussed in the
literature for many years, but it was by far too small to explain the
difference between the experimental result in \eq{deuthfssexp} and the
sum of nonrecoil corrections in \eq{norecdeut}

\beq          \label{difexpnordeut}
\Delta E^{exp}_{HFS}(D)-\Delta E_{nrec}(D)=45.2~\mbox{kHz}.
\eeq

A breakthrough was achieved a few years ago when it was realized
that an analytic calculation of the deuterium recoil, structure and
polarizability corrections is possible in the zero range approximation
\cite{milkhr}. An analytic result for the difference in
\eq{difexpnordeut}, obtained as a result of a nice calculation in
\cite{milkhr}, is numerically equal $44$ kHz, and within the accuracy of
the zero range approximation perfectly explains the difference between
the experimental result and the sum of the nonrecoil corrections. More
accurate calculations of the nuclear effects in the deuterium hyperfine
structure beyond the zero range approximation are feasible, and
comparison of such results with the experimental data on the deuterium
hyperfine splitting may be used as a test of the deuteron models.

\subsection{Hyperfine Splitting in Muonium}\label{muonhfsexp}

Being a purely electrodynamic bound state, muonium is the best system
for comparison between the hyperfine splitting theory and experiment.
Unlike the  case of hydrogen the theory of hyperfine splitting in
muonium is free from uncertainties generated by the hadronic nature of
the proton, and is thus much more precise. The scale of hyperfine
splitting is determined by the Fermi energy in \eq{ef}

\beq
E_{F}(Mu)=4~459~031.922~(518)~(34)~\mbox{kHz},
\eeq

\noindent
where the uncertainty in the first brackets is due to the uncertainty
of the muon-electron mass ratio in \eq{emmassrexp} and the uncertainty
in the second brackets is due to the uncertainty of the fine structure
constant in \eq{alpha}.

Theoretical prediction for the hyperfine splitting interval in the
ground state in muonium may easily be obtained collecting all
contributions to HFS displayed in Tables XII-XVIII

\beq   \label{hfstheory}
\Delta E_{HFS}(Mu)=4~463~302.565~(518)~(34)~(100)~\mbox{kHz},
\eeq

\noindent
where the first error comes from the experimental error of the
electron-muon mass ratio $m/M$, the second comes from the error in the
value of the fine structure constant $\alpha$, and the third is an
estimate of the yet unknown theoretical contributions. We see that the
uncertainty of the muon-electron mass ratio gives by far the largest
contributions both in the uncertainty of the Fermi energy and the
theoretical value of the ground state hyperfine splitting.

On the experimental side, hyperfine splitting in the ground state of
muonium admits very precise determination due to its small natural
linewidth.  The lifetime of the higher energy hyperfine state with the
total angular momentum $F=1$ with respect to the $M1$-transition to the
lower level state with $F=0$ is extremely large $\tau=1\cdot
10^{13}~{\rm s}$ and gives negligible contribution to the linewidth.
The natural linewidth $\Gamma_\mu/h=72.3~{\rm kHz}$ is completely
determined by the muon lifetime $\tau_\mu\approx 2.2\cdot 10^{-6}~{\rm
s}$.

A high precision value of the muonium hyperfine splitting was
obtained many years ago \cite{mbb}

\beq
\Delta E_{HFS}(Mu) = 4~463~302.88~(16)~{\rm kHz}\qquad \delta=3.6\cdot
10^{-8}.
\eeq

In the latest measurement \cite{lbdd} this value was improved by a
factor of three

\beq   \label{hfsexperiment}
\Delta E_{HFS}(Mu)=4~463~302.776~(51)~\mbox{kHz},\qquad \delta=1.1\cdot
10^{-8},
\eeq

The new value has an experimental error which corresponds to
measuring the hyperfine energy splitting at the level of
$\Delta\nu_{exp}/(\Gamma_\mu/h) \approx 7\cdot 10^{-4}$ of the natural
linewidth. This is a remarkable experimental achievement.

The agreement between theory and experiment is excellent. However, the
error bars of the theoretical value are apparently about an order of
magnitude larger than respective error bars of the experimental result.
This is a deceptive impression. The error of the theoretical prediction
in \eq{hfstheory} is dominated by the experimental error of the value
of the electron-muon mass ratio.  As a result of the new experiment
\cite{lbdd} this error was reduced threefold but it is still by far the
largest source of error in the theoretical value for the  muonium
hyperfine splitting.

The estimate of the theoretical uncertainty is only two times larger
than the experimental error. The largest source of theoretical error is
connected with the yet uncalculated theoretical contributions to hyperfine
splitting, mainly with the unknown recoil and radiative-recoil corrections.
As we have already mentioned, reducing the theoretical uncertainty by
an order of magnitude to about $10$ Hz is now a realistic aim for the
theory.

One may extract electron-muon mass ratio from the experimental value of HFS
and the most precise value of $\alpha$

\beq  \label{masssepar}
\frac{M}{m}=206.768~267~2~(23)~(16)~(46),
\eeq

\noindent
where the first error comes from the experimental error of the
hyperfine splitting measurement, the second comes from the error in the
value of the fine structure constant $\alpha$, and the third is an
estimate of the yet unknown theoretical contributions.

Combining all errors we  obtain the mass ratio

\beq
\frac{M}{m}=206.768~267~2~(54)\qquad \delta=2.6\cdot 10^{-8},
\eeq

\noindent
which is almost five times more accurate than the best earlier
experimental value in \eq{emmassrexp}.

We see from \eq{masssepar} that the error of this indirect value of
the mass ratio is dominated by the theoretical uncertainty. This sets a
clear task for the theory to reduce the contribution of the theoretical
uncertainty in the error bars in \eq{masssepar} to the level below two other
contributions to the error bars. It is sufficient to this end to calculate
all contributions to HFS which are larger than $10$ Hz. This would lead to
reduction of the uncertainty of the indirect value of the muon-electron mass
ratio by factor two. There is thus a real incentive for improvement of the
theory of HFS to account for all corrections to HFS of order $10$ Hz,
created by the recent experimental and theoretical achievements.

Another reason to improve the HFS theory is provided by the perspective of
reducing the experimental uncertainty of hyperfine splitting below the weak
interaction contribution in \eq{wekahfscontr}. In such a case, muonium
could become the first atom where a shift of atomic energy levels due
to weak interaction would be observed \cite{jungmann}.

\section{~~~Summary}

High precision experiments with hydrogenlike systems have
achieved a new level of accuracy in recent years and further dramatic
progress is still expected. The experimental errors of measurements of
many energy shifts in hydrogen and muonium were reduced by orders of
magnitude. This rapid experimental progress was matched by
theoretical developments as discussed above. The accuracy of the
quantum electrodynamic theory of such classical effects as Lamb shift
in hydrogen and hyperfine splitting in muonium has increased in many
cases by one or two orders of magnitudes. This was achieved due to
intensive work of many theorists and development of new ingenious
original theoretical approaches which can be applied to the theory of
bound states, not only in QED but also in other field theories, such as
quantum chromodynamics. From the phenomenological point of view recent
developments opened new perspectives for precise determination of many
fundamental constants (the Rydberg constant, electron-muon mass ratio,
proton charge radius, deuteron structure radius, etc.), and for
comparison of the experimental and theoretical results on the Lamb
shifts and hyperfine splitting.

Recent progress also poses new theoretical challenges. Reduction
of the theoretical error in prediction of the value of the $1S$ Lamb
shift in hydrogen to the level of 1 kHz (and, respectively, of the $2S$
Lamb shift to several tenth of kHz) should be considered as a next
stage of the theory. The theoretical error of the hyperfine splitting
in muonium should be reduced the theoretical error to about $10$ Hz.
Achievement of these goals will require hard work and a considerable
resourcefulness, but results which years ago hardly seemed possible are
now within reach.

\acknowledgements

Many friends and colleagues for many years discussed with us the bound
state problem, collaborated on different projects, and shared with us their
vision and insight. We are especially deeply grateful to the late D.
Yennie and M. Samuel, to G. Adkins, M. Braun, M. Doncheski, R. Faustov,
G. Drake, S.  Karshenboim, Y.  Khriplovich, T.  Kinoshita, L.
Labzowsky, P.  Lepage, A. Martynenko, A. Milshtein, P.  Mohr, D.  Owen,
K.  Pachucki, V.  Pal'chikov, J.  Sapirstein, V.  Shabaev, B. Taylor,
and A.  Yelkhovsky.

This work was supported by the NSF grants PHY-9120102, PHY-9421408, and
PHY-9900771.

\end{document}